\documentstyle[12pt]{report}
\begin{document}
\thispagestyle{empty}
\begin{center}
A DISORIENTED CHIRAL CONDENSATE SEARCH AT THE \\
\vskip2ex
FERMILAB TEVATRON \\
\vspace{1in}
by \\
\vskip3ex
MARY ELIZABETH CONVERY \\
\vspace{1in}
Submitted in partial fulfillment of the requirements \\
\vskip2ex
for the degree of Doctor of Philosophy \\
\vspace{.5in}
Thesis Adviser: Prof. Cyrus Taylor \\
\vspace{.5in}
Department of Physics \\
\vskip2ex
CASE WESTERN RESERVE UNIVERSITY \\
\vskip2ex
May, 1997
\end{center}
\vfill\newpage
\pagenumbering{roman}
\setcounter{page}{2}
\begin{center}
A DISORIENTED CHIRAL CONDENSATE SEARCH AT THE FERMILAB TEVATRON \\
\vspace{.5in}
Abstract \\
\vskip2ex
by \\
\vskip2ex
MARY ELIZABETH CONVERY \\
\end{center}
\vspace{.5in}
MiniMax (Fermilab T-864) was a small test/experiment at the Tevatron
designed to search for disoriented chiral condensates (DCC)
in the forward direction.

Relativistic quantum field theory treats the vacuum as a medium, with bulk 
properties characterized by long-range order parameters.  This has led to 
suggestions that regions of ``disoriented vacuum'' might be formed in
high-energy collision processes.  In particular, the approximate chiral 
symmetry 
of QCD could lead to regions of vacuum which have chiral order parameters
disoriented to directions which have non-zero isospin, i.e. disoriented 
chiral condensates.  A signature of DCC is the resulting distribution of the 
fraction of produced pions which are neutral.

The MiniMax detector at the C0 collision region of the Tevatron was a 
telescope of 24 multi-wire proportional chambers (MWPC's) with a lead
converter behind the eighth MWPC, allowing the detection of
charged particles and photon conversions in an acceptance 
approximately a circle of radius 0.6 in pseudorapidity--azimuthal-angle space, 
centered on pseudorapidity $\eta\approx 4$.
An electromagnetic calorimeter was located behind the MWPC telescope,
and hadronic calorimeters and scintillator were located in the upstream
anti-proton direction to tag diffractive events.

The use of standard Monte Carlo simulations for high-energy collisions of 
elementary particles (PYTHIA) and for interactions of particles in the detector 
(GEANT) is described, along with the simulation created by the MiniMax
Collaboration to generate DCC domains.

A description of the data analysis software is given,
including detailed studies of its performance on data from the simulations.

A set of robust observables is derived.  These are insensitive to many 
efficiencies and to the details of the modeling of the parent pion production 
mechanisms, yet have distinguishable values for DCC and generic charged-neutral
distributions.
Simulations show that the robust observables are insensitive to
detector efficiencies and to systematic errors in the data analysis software.

The resulting values for robust observables for approximately $1.5\times 10^6$
events are shown to be consistent with production by only generic
mechanisms.  Results from samples of diffractive-tagged events and of 
high-multiplicity events also show no evidence for DCC.
\vfill\newpage
\begin{center} {\Large{\bf Acknowledgments}} \end{center}
\vskip2ex
I owe many thanks, of course, to all my fellow MiniMax collaboration members:
T.\,C.~Brooks, W.\,L.~Davis, K.\,W.~Del\,\,Signore, T.\,L.~Jenkins,
E.~Kangas, M.\,G.~Knepley, K.\,L.~Kowalski, C.\,C.~Taylor,
S.\,H.~Oh, W.\,D.~Walker, 
P.\,L.~Colestock, B.~Hanna, M.~Martens, J.~Streets,
R.~Ball, H.\,R.~Gustafson, L.\,W.~Jones, M.\,J.~Longo,
J.\,D.~Bjorken, 
A.~Abashian, N.~Morgan, 
and C.\,A.~Pruneau.  
I have learned much from all my collaborators, and have especially benefited
from working with Bj (James Bjorken), Cyrus Taylor, Ken Kowalski, and 
Tom Jenkins.
\vskip1ex
Special thanks go to Jon Streets for all his help with computing,
to Ken Del\,\,Signore for help with the tracker, and to Travis Brooks
for many useful conversations.
\vskip1ex
I am eternally grateful for the opportunity to earn my graduate degree at CWRU 
which I owe to Robert Brown and Cyrus Taylor.
\vskip1ex
This work would not have been possible were it not for the support of the
Timken Foundation and the CWRU Physics Department.

\vfill\newpage
\tableofcontents
\listoffigures
\listoftables
\vfill\newpage
\pagenumbering{arabic}
\makeatletter
 \def\ps@headings{\let\@mkboth\markboth
  \def\@oddfoot{}\def\@evenfoot{}
  \def\@evenhead{\hspace{2.625in}\rm \thepage \hfil}
  \def\@oddhead{\hspace{2.625in}\rm\thepage \hfil}}
\def\footnoterule{\kern-3\p@
  \hrule width \columnwidth
  \kern 2.6\p@}
\makeatother
\pagestyle{headings}
\markright{\hspace{-3in}}
\ifx\undefined\psfig\else\endinput\fi

%
\edef\psfigRestoreAt{\catcode`@=\number\catcode`@\relax}
\catcode`\@=11\relax
\newwrite\@unused
\def\ps@typeout#1{{\let\protect\string\immediate\write\@unused{#1}}}
\ps@typeout{psfig/tex 1.8}


\def\figurepath{./}
\def\psfigurepath#1{\edef\figurepath{#1}}

%
%
\def\@nnil{\@nil}
\def\@empty{}
\def\@psdonoop#1\@@#2#3{}
\def\@psdo#1:=#2\do#3{\edef\@psdotmp{#2}\ifx\@psdotmp\@empty \else
    \expandafter\@psdoloop#2,\@nil,\@nil\@@#1{#3}\fi}
\def\@psdoloop#1,#2,#3\@@#4#5{\def#4{#1}\ifx #4\@nnil \else
       #5\def#4{#2}\ifx #4\@nnil \else#5\@ipsdoloop #3\@@#4{#5}\fi\fi}
\def\@ipsdoloop#1,#2\@@#3#4{\def#3{#1}\ifx #3\@nnil 
       \let\@nextwhile=\@psdonoop \else
      #4\relax\let\@nextwhile=\@ipsdoloop\fi\@nextwhile#2\@@#3{#4}}
\def\@tpsdo#1:=#2\do#3{\xdef\@psdotmp{#2}\ifx\@psdotmp\@empty \else
    \@tpsdoloop#2\@nil\@nil\@@#1{#3}\fi}
\def\@tpsdoloop#1#2\@@#3#4{\def#3{#1}\ifx #3\@nnil 
       \let\@nextwhile=\@psdonoop \else
      #4\relax\let\@nextwhile=\@tpsdoloop\fi\@nextwhile#2\@@#3{#4}}
%
\ifx\undefined\fbox
\newdimen\fboxrule
\newdimen\fboxsep
\newdimen\ps@tempdima
\newbox\ps@tempboxa
\fboxsep = 3pt
\fboxrule = .4pt
\long\def\fbox#1{\leavevmode\setbox\ps@tempboxa\hbox{#1}\ps@tempdima\fboxrule
    \advance\ps@tempdima \fboxsep \advance\ps@tempdima \dp\ps@tempboxa
   \hbox{\lower \ps@tempdima\hbox
  {\vbox{\hrule height \fboxrule
          \hbox{\vrule width \fboxrule \hskip\fboxsep
          \vbox{\vskip\fboxsep \box\ps@tempboxa\vskip\fboxsep}\hskip 
                 \fboxsep\vrule width \fboxrule}
                 \hrule height \fboxrule}}}}
\fi
%
%
\newread\ps@stream
\newif\ifnot@eof       
\newif\if@noisy        
\newif\if@atend        
\newif\if@psfile       
%
%
{\catcode`\%=12\global\gdef\epsf@start{
\def\epsf@PS{PS}
\def\epsf@getbb#1{%
%
%
\openin\ps@stream=#1
\ifeof\ps@stream\ps@typeout{Error, File #1 not found}\else
%
%
   {\not@eoftrue \chardef\other=12
    \def\do##1{\catcode`##1=\other}\dospecials \catcode`\ =10
    \loop
       \if@psfile
	  \read\ps@stream to \epsf@fileline
       \else{
	  \obeyspaces
          \read\ps@stream to \epsf@tmp\global\let\epsf@fileline\epsf@tmp}
       \fi
       \ifeof\ps@stream\not@eoffalse\else
%
%
       \if@psfile\else
       \expandafter\epsf@test\epsf@fileline:. \\%
       \fi
%
%
          \expandafter\epsf@aux\epsf@fileline:. \\%
       \fi
   \ifnot@eof\repeat
   }\closein\ps@stream\fi}%
%
%
\long\def\epsf@test#1#2#3:#4\\{\def\epsf@testit{#1#2}
			\ifx\epsf@testit\epsf@start\else
\ps@typeout{Warning! File does not start with `\epsf@start'.  It may not be a PostScript file.}
			\fi
			\@psfiletrue} 
%
%
{\catcode`\%=12\global\let\epsf@percent=
%
%
%
\long\def\epsf@aux#1#2:#3\\{\ifx#1\epsf@percent
   \def\epsf@testit{#2}\ifx\epsf@testit\epsf@bblit
	\@atendfalse
        \epsf@atend #3 . \\%
	\if@atend	
	   \if@verbose{
		\ps@typeout{psfig: found `(atend)'; continuing search}
	   }\fi
        \else
        \epsf@grab #3 . . . \\%
        \not@eoffalse
        \global\no@bbfalse
        \fi
   \fi\fi}%
%
%
\def\epsf@grab #1 #2 #3 #4 #5\\{%
   \global\def\epsf@llx{#1}\ifx\epsf@llx\empty
      \epsf@grab #2 #3 #4 #5 .\\\else
   \global\def\epsf@lly{#2}%
   \global\def\epsf@urx{#3}\global\def\epsf@ury{#4}\fi}%
%
%
\def\epsf@atendlit{(atend)} 
\def\epsf@atend #1 #2 #3\\{%
   \def\epsf@tmp{#1}\ifx\epsf@tmp\empty
      \epsf@atend #2 #3 .\\\else
   \ifx\epsf@tmp\epsf@atendlit\@atendtrue\fi\fi}


\chardef\letter = 11
\chardef\other = 12

\newif \ifdebug 
\newif\ifc@mpute 
\c@mputetrue 

\let\then = \relax
\def\r@dian{pt }
\let\r@dians = \r@dian
\let\dimensionless@nit = \r@dian
\let\dimensionless@nits = \dimensionless@nit
\def\internal@nit{sp }
\let\internal@nits = \internal@nit
\newif\ifstillc@nverging
\def \Mess@ge #1{\ifdebug \then \message {#1} \fi}

{ 
	\catcode `\@ = \letter
	\gdef \nodimen {\expandafter \n@dimen \the \dimen}
	\gdef \term #1 #2 #3%
	       {\edef \t@ {\the #1}
		\edef \t@@ {\expandafter \n@dimen \the #2\r@dian}%
		\t@rm {\t@} {\t@@} {#3}%
	       }
	\gdef \t@rm #1 #2 #3%
	       {{%
		\count 0 = 0
		\dimen 0 = 1 \dimensionless@nit
		\dimen 2 = #2\relax
		\Mess@ge {Calculating term #1 of \nodimen 2}%
		\loop
		\ifnum	\count 0 < #1
		\then	\advance \count 0 by 1
			\Mess@ge {Iteration \the \count 0 \space}%
			\Multiply \dimen 0 by {\dimen 2}%
			\Mess@ge {After multiplication, term = \nodimen 0}%
			\Divide \dimen 0 by {\count 0}%
			\Mess@ge {After division, term = \nodimen 0}%
		\repeat
		\Mess@ge {Final value for term #1 of 
				\nodimen 2 \space is \nodimen 0}%
		\xdef \Term {#3 = \nodimen 0 \r@dians}%
		\aftergroup \Term
	       }}
	\catcode `\p = \other
	\catcode `\t = \other
	\gdef \n@dimen #1pt{#1} 
}

\def \Divide #1by #2{\divide #1 by #2} 

\def \Multiply #1by #2
       {{
	\count 0 = #1\relax
	\count 2 = #2\relax
	\count 4 = 65536
	\Mess@ge {Before scaling, count 0 = \the \count 0 \space and
			count 2 = \the \count 2}%
	\ifnum	\count 0 > 32767 
	\then	\divide \count 0 by 4
		\divide \count 4 by 4
	\else	\ifnum	\count 0 < -32767
		\then	\divide \count 0 by 4
			\divide \count 4 by 4
		\else
		\fi
	\fi
	\ifnum	\count 2 > 32767 
	\then	\divide \count 2 by 4
		\divide \count 4 by 4
	\else	\ifnum	\count 2 < -32767
		\then	\divide \count 2 by 4
			\divide \count 4 by 4
		\else
		\fi
	\fi
	\multiply \count 0 by \count 2
	\divide \count 0 by \count 4
	\xdef \product {#1 = \the \count 0 \internal@nits}%
	\aftergroup \product
       }}

\def\r@duce{\ifdim\dimen0 > 90\r@dian \then   
		\multiply\dimen0 by -1
		\advance\dimen0 by 180\r@dian
		\r@duce
	    \else \ifdim\dimen0 < -90\r@dian \then  
		\advance\dimen0 by 360\r@dian
		\r@duce
		\fi
	    \fi}

\def\Sine#1%
       {{%
	\dimen 0 = #1 \r@dian
	\r@duce
	\ifdim\dimen0 = -90\r@dian \then
	   \dimen4 = -1\r@dian
	   \c@mputefalse
	\fi
	\ifdim\dimen0 = 90\r@dian \then
	   \dimen4 = 1\r@dian
	   \c@mputefalse
	\fi
	\ifdim\dimen0 = 0\r@dian \then
	   \dimen4 = 0\r@dian
	   \c@mputefalse
	\fi
	\ifc@mpute \then
		\divide\dimen0 by 180
		\dimen0=3.141592654\dimen0
		\dimen 2 = 3.1415926535897963\r@dian 
		\divide\dimen 2 by 2 
		\Mess@ge {Sin: calculating Sin of \nodimen 0}%
		\count 0 = 1 
		\dimen 2 = 1 \r@dian 
		\dimen 4 = 0 \r@dian 
		\loop
			\ifnum	\dimen 2 = 0 
			\then	\stillc@nvergingfalse 
			\else	\stillc@nvergingtrue
			\fi
			\ifstillc@nverging 
			\then	\term {\count 0} {\dimen 0} {\dimen 2}%
				\advance \count 0 by 2
				\count 2 = \count 0
				\divide \count 2 by 2
				\ifodd	\count 2 
				\then	\advance \dimen 4 by \dimen 2
				\else	\advance \dimen 4 by -\dimen 2
				\fi
		\repeat
	\fi		
			\xdef \sine {\nodimen 4}%
       }}

\def\Cosine#1{\ifx\sine\UnDefined\edef\Savesine{\relax}\else
		             \edef\Savesine{\sine}\fi
	{\dimen0=#1\r@dian\advance\dimen0 by 90\r@dian
	 \Sine{\nodimen 0}
	 \xdef\cosine{\sine}
	 \xdef\sine{\Savesine}}}	      

\def\psdraft{
	\def\@psdraft{0}
}
\def\psfull{
	\def\@psdraft{100}
}

\psfull

\newif\if@scalefirst
\def\psscalefirst{\@scalefirsttrue}
\def\psrotatefirst{\@scalefirstfalse}
\psrotatefirst

\newif\if@draftbox
\def\psnodraftbox{
	\@draftboxfalse
}
\def\psdraftbox{
	\@draftboxtrue
}
\@draftboxtrue

\newif\if@prologfile
\newif\if@postlogfile
\def\pssilent{
	\@noisyfalse
}
\def\psnoisy{
	\@noisytrue
}
\psnoisy
\newif\if@bbllx
\newif\if@bblly
\newif\if@bburx
\newif\if@bbury
\newif\if@height
\newif\if@width
\newif\if@rheight
\newif\if@rwidth
\newif\if@angle
\newif\if@clip
\newif\if@verbose
\def\@p@@sclip#1{\@cliptrue}

\newif\if@decmpr


\def\@p@@sfigure#1{\def\@p@sfile{null}\def\@p@sbbfile{null}
	        \openin1=#1.bb
		\ifeof1\closein1
	        	\openin1=\figurepath#1.bb
			\ifeof1\closein1
			        \openin1=#1
				\ifeof1\closein1%
				       \openin1=\figurepath#1
					\ifeof1
					   \ps@typeout{Error, File #1 not found}
						\if@bbllx\if@bblly
				   		\if@bburx\if@bbury
			      				\def\@p@sfile{#1}%
			      				\def\@p@sbbfile{#1}%
							\@decmprfalse
				  	   	\fi\fi\fi\fi
					\else\closein1
				    		\def\@p@sfile{\figurepath#1}%
				    		\def\@p@sbbfile{\figurepath#1}%
						\@decmprfalse
	                       		\fi%
			 	\else\closein1%
					\def\@p@sfile{#1}
					\def\@p@sbbfile{#1}
					\@decmprfalse
			 	\fi
			\else
				\def\@p@sfile{\figurepath#1}
				\def\@p@sbbfile{\figurepath#1.bb}
				\@decmprtrue
			\fi
		\else
			\def\@p@sfile{#1}
			\def\@p@sbbfile{#1.bb}
			\@decmprtrue
		\fi}

\def\@p@@sfile#1{\@p@@sfigure{#1}}

\def\@p@@sbbllx#1{
		\@bbllxtrue
		\dimen100=#1
		\edef\@p@sbbllx{\number\dimen100}
}
\def\@p@@sbblly#1{
		\@bbllytrue
		\dimen100=#1
		\edef\@p@sbblly{\number\dimen100}
}
\def\@p@@sbburx#1{
		\@bburxtrue
		\dimen100=#1
		\edef\@p@sbburx{\number\dimen100}
}
\def\@p@@sbbury#1{
		\@bburytrue
		\dimen100=#1
		\edef\@p@sbbury{\number\dimen100}
}
\def\@p@@sheight#1{
		\@heighttrue
		\dimen100=#1
   		\edef\@p@sheight{\number\dimen100}
}
\def\@p@@swidth#1{
		\@widthtrue
		\dimen100=#1
		\edef\@p@swidth{\number\dimen100}
}
\def\@p@@srheight#1{
		\@rheighttrue
		\dimen100=#1
		\edef\@p@srheight{\number\dimen100}
}
\def\@p@@srwidth#1{
		\@rwidthtrue
		\dimen100=#1
		\edef\@p@srwidth{\number\dimen100}
}
\def\@p@@sangle#1{
		\@angletrue
		\edef\@p@sangle{#1} 
}
\def\@p@@ssilent#1{ 
		\@verbosefalse
}
\def\@p@@sprolog#1{\@prologfiletrue\def\@prologfileval{#1}}
\def\@p@@spostlog#1{\@postlogfiletrue\def\@postlogfileval{#1}}
\def\@cs@name#1{\csname #1\endcsname}
\def\@setparms#1=#2,{\@cs@name{@p@@s#1}{#2}}
%
%
\def\ps@init@parms{
		\@bbllxfalse \@bbllyfalse
		\@bburxfalse \@bburyfalse
		\@heightfalse \@widthfalse
		\@rheightfalse \@rwidthfalse
		\def\@p@sbbllx{}\def\@p@sbblly{}
		\def\@p@sbburx{}\def\@p@sbbury{}
		\def\@p@sheight{}\def\@p@swidth{}
		\def\@p@srheight{}\def\@p@srwidth{}
		\def\@p@sangle{0}
		\def\@p@sfile{} \def\@p@sbbfile{}
		\def\@p@scost{10}
		\def\@sc{}
		\@prologfilefalse
		\@postlogfilefalse
		\@clipfalse
		\if@noisy
			\@verbosetrue
		\else
			\@verbosefalse
		\fi
}
%
%
\def\parse@ps@parms#1{
	 	\@psdo\@psfiga:=#1\do
		   {\expandafter\@setparms\@psfiga,}}
%
%
\newif\ifno@bb
\def\bb@missing{
	\if@verbose{
		\ps@typeout{psfig: searching \@p@sbbfile \space  for bounding box}
	}\fi
	\no@bbtrue
	\epsf@getbb{\@p@sbbfile}
        \ifno@bb \else \bb@cull\epsf@llx\epsf@lly\epsf@urx\epsf@ury\fi
}	
\def\bb@cull#1#2#3#4{
	\dimen100=#1 bp\edef\@p@sbbllx{\number\dimen100}
	\dimen100=#2 bp\edef\@p@sbblly{\number\dimen100}
	\dimen100=#3 bp\edef\@p@sbburx{\number\dimen100}
	\dimen100=#4 bp\edef\@p@sbbury{\number\dimen100}
	\no@bbfalse
}
\newdimen\p@intvaluex
\newdimen\p@intvaluey
\def\rotate@#1#2{{\dimen0=#1 sp\dimen1=#2 sp
		  \global\p@intvaluex=\cosine\dimen0
		  \dimen3=\sine\dimen1
		  \global\advance\p@intvaluex by -\dimen3
		  \global\p@intvaluey=\sine\dimen0
		  \dimen3=\cosine\dimen1
		  \global\advance\p@intvaluey by \dimen3
		  }}
\def\compute@bb{
		\no@bbfalse
		\if@bbllx \else \no@bbtrue \fi
		\if@bblly \else \no@bbtrue \fi
		\if@bburx \else \no@bbtrue \fi
		\if@bbury \else \no@bbtrue \fi
		\ifno@bb \bb@missing \fi
		\ifno@bb \ps@typeout{FATAL ERROR: no bb supplied or found}
			\no-bb-error
		\fi
		%
%
		\count203=\@p@sbburx
		\count204=\@p@sbbury
		\advance\count203 by -\@p@sbbllx
		\advance\count204 by -\@p@sbblly
		\edef\ps@bbw{\number\count203}
		\edef\ps@bbh{\number\count204}
		\if@angle 
			\Sine{\@p@sangle}\Cosine{\@p@sangle}
	        	{\dimen100=\maxdimen\xdef\r@p@sbbllx{\number\dimen100}
					    \xdef\r@p@sbblly{\number\dimen100}
			                    \xdef\r@p@sbburx{-\number\dimen100}
					    \xdef\r@p@sbbury{-\number\dimen100}}
%
                        \def\minmaxtest{
			   \ifnum\number\p@intvaluex<\r@p@sbbllx
			      \xdef\r@p@sbbllx{\number\p@intvaluex}\fi
			   \ifnum\number\p@intvaluex>\r@p@sbburx
			      \xdef\r@p@sbburx{\number\p@intvaluex}\fi
			   \ifnum\number\p@intvaluey<\r@p@sbblly
			      \xdef\r@p@sbblly{\number\p@intvaluey}\fi
			   \ifnum\number\p@intvaluey>\r@p@sbbury
			      \xdef\r@p@sbbury{\number\p@intvaluey}\fi
			   }
			\rotate@{\@p@sbbllx}{\@p@sbblly}
			\minmaxtest
			\rotate@{\@p@sbbllx}{\@p@sbbury}
			\minmaxtest
			\rotate@{\@p@sbburx}{\@p@sbblly}
			\minmaxtest
			\rotate@{\@p@sbburx}{\@p@sbbury}
			\minmaxtest
			\edef\@p@sbbllx{\r@p@sbbllx}\edef\@p@sbblly{\r@p@sbblly}
			\edef\@p@sbburx{\r@p@sbburx}\edef\@p@sbbury{\r@p@sbbury}
		\fi
		\count203=\@p@sbburx
		\count204=\@p@sbbury
		\advance\count203 by -\@p@sbbllx
		\advance\count204 by -\@p@sbblly
		\edef\@bbw{\number\count203}
		\edef\@bbh{\number\count204}
}
%
%
\def\in@hundreds#1#2#3{\count240=#2 \count241=#3
		     \count100=\count240	
		     \divide\count100 by \count241
		     \count101=\count100
		     \multiply\count101 by \count241
		     \advance\count240 by -\count101
		     \multiply\count240 by 10
		     \count101=\count240	
		     \divide\count101 by \count241
		     \count102=\count101
		     \multiply\count102 by \count241
		     \advance\count240 by -\count102
		     \multiply\count240 by 10
		     \count102=\count240	
		     \divide\count102 by \count241
		     \count200=#1\count205=0
		     \count201=\count200
			\multiply\count201 by \count100
		 	\advance\count205 by \count201
		     \count201=\count200
			\divide\count201 by 10
			\multiply\count201 by \count101
			\advance\count205 by \count201
		     \count201=\count200
			\divide\count201 by 100
			\multiply\count201 by \count102
			\advance\count205 by \count201
		     \edef\@result{\number\count205}
}
\def\compute@wfromh{
		\in@hundreds{\@p@sheight}{\@bbw}{\@bbh}
		\edef\@p@swidth{\@result}
}
\def\compute@hfromw{
	        \in@hundreds{\@p@swidth}{\@bbh}{\@bbw}
		\edef\@p@sheight{\@result}
}
\def\compute@handw{
		\if@height 
			\if@width
			\else
				\compute@wfromh
			\fi
		\else 
			\if@width
				\compute@hfromw
			\else
				\edef\@p@sheight{\@bbh}
				\edef\@p@swidth{\@bbw}
			\fi
		\fi
}
\def\compute@resv{
		\if@rheight \else \edef\@p@srheight{\@p@sheight} \fi
		\if@rwidth \else \edef\@p@srwidth{\@p@swidth} \fi
}
%
\def\compute@sizes{
	\compute@bb
	\if@scalefirst\if@angle
	\if@width
	   \in@hundreds{\@p@swidth}{\@bbw}{\ps@bbw}
	   \edef\@p@swidth{\@result}
	\fi
	\if@height
	   \in@hundreds{\@p@sheight}{\@bbh}{\ps@bbh}
	   \edef\@p@sheight{\@result}
	\fi
	\fi\fi
	\compute@handw
	\compute@resv}

%
%
\def\psfig#1{\vbox {
	%
	\ps@init@parms
	\parse@ps@parms{#1}
	\compute@sizes
	\ifnum\@p@scost<\@psdraft{
		\special{ps::[begin] 	\@p@swidth \space \@p@sheight \space
				\@p@sbbllx \space \@p@sbblly \space
				\@p@sbburx \space \@p@sbbury \space
				startTexFig \space }
		\if@angle
			\special {ps:: \@p@sangle \space rotate \space} 
		\fi
		\if@clip{
			\if@verbose{
				\ps@typeout{(clip)}
			}\fi
			\special{ps:: doclip \space }
		}\fi
		\if@prologfile
		    \special{ps: plotfile \@prologfileval \space } \fi
		\if@decmpr{
			\if@verbose{
				\ps@typeout{psfig: including \@p@sfile.Z \space }
			}\fi
			\special{ps: plotfile "`zcat \@p@sfile.Z" \space }
		}\else{
			\if@verbose{
				\ps@typeout{psfig: including \@p@sfile \space }
			}\fi
			\special{ps: plotfile \@p@sfile \space }
		}\fi
		\if@postlogfile
		    \special{ps: plotfile \@postlogfileval \space } \fi
		\special{ps::[end] endTexFig \space }
		\vbox to \@p@srheight true sp{
			\hbox to \@p@srwidth true sp{
				\hss
			}
		\vss
		}
	}\else{
		\if@draftbox{		
			\hbox{\frame{\vbox to \@p@srheight true sp{
			\vss
			\hbox to \@p@srwidth true sp{ \hss \@p@sfile \hss }
			\vss
			}}}
		}\else{
			\vbox to \@p@srheight true sp{
			\vss
			\hbox to \@p@srwidth true sp{\hss}
			\vss
			}
		}\fi

	}\fi
}}
\psfigRestoreAt

\newdimen\rotdimen
\def\vspec#1{\special{ps:#1}}
\def\rotstart#1{\vspec{gsave currentpoint currentpoint translate
   #1 neg exch neg exch translate}}
\def\rotfinish{\vspec{currentpoint grestore moveto}}
\def\rotr#1{\rotdimen=\ht#1\advance\rotdimen by\dp#1%
   \hbox to\rotdimen{\hskip\ht#1\vbox to\wd#1{\rotstart{90 rotate}%
   \box#1\vss}\hss}\rotfinish}
\def\rotl#1{\rotdimen=\ht#1\advance\rotdimen by\dp#1%
   \hbox to\rotdimen{\vbox to\wd#1{\vskip\wd#1\vskip .5in
          \rotstart{270 rotate}%
   \box#1\vss}\hss}\rotfinish}%
\def\rotu#1{\rotdimen=\ht#1\advance\rotdimen by\dp#1%
   \hbox to\wd#1{\hskip\wd#1\vbox to\rotdimen{\vskip\rotdimen
   \rotstart{-1 dup scale}\box#1\vss}\hss}\rotfinish}%
\def\rotf#1{\hbox to\wd#1{\hskip\wd#1\rotstart{-1 1 scale}%
   \box#1\hss}\rotfinish}%

\chapter{Introduction}
\section{Physics motivation}
\subsection{Disoriented Chiral Condensates}

The Lagrangian of QCD (Quantum Chromodynamics \cite{QCD}, the theory of strong 
interactions) for two quarks (u and d) has isospin symmetry SU(2).
In the limit that the quarks are massless, the Lagrangian also has chiral
symmetry SU(2)$_L \times $SU(2)$_R$.  That is, if we write the isospin doublet 
\newline $\Psi=\left( \begin{array}{c} \mbox{u} \\ \mbox{d} \end{array}\right)$
in terms of left- and right-handed fields,
\begin{equation}
\Psi={1\over 2}(1-\gamma_5)\Psi+ {1\over 2}(1+\gamma_5)\Psi=\Psi_L+\Psi_R,
\end{equation}
where $\gamma_5$ is the usual product of gamma matrices of Dirac theory,
then the Lagrangian is symmetric under $\Psi_L\leftrightarrow\Psi_R$.
The spontaneous breaking of this chiral symmetry in the QCD ground state 
is accompanied by a massless Goldstone boson \cite{goldst}, the pion.  
In the real world, the
quarks are light but not massless, and the QCD Lagrangian has an
approximate chiral symmetry which is explicitly broken, giving the pion a 
small mass.  

Spontaneous chiral symmetry breaking is often 
described by the linear sigma model \cite{sigma}, 
in which there is a scalar field
$\sigma$ of isospin 0 and a vector pion field $\vec{\pi}$ of total isospin 1.
The Lagrangian can be written as 
\begin{equation}
{\cal L}={1\over 2}(\partial_\mu\vec{\pi}\partial^\mu\vec{\pi}+
  \partial_\mu\sigma\partial^\mu\sigma)-{\lambda\over 2}
  (\vec{\pi}^2+\sigma^2-{f_\pi}^2)^2,
\end{equation}
and the potential is minimized for $\sigma^2+\vec{\pi}^2={f_\pi}^2$.
The symmetry is broken when a particular minimum is chosen.
In order to break chiral symmetry, but not isospin symmetry, the minimum chosen
is that for which
the $\sigma$ field acquires a vacuum expectation value: 
$\left<\sigma\right>=f_\pi$, $\left<\vec{\pi}\right>=0$.  
A term can be added to the Lagrangian in order to break the chiral symmetry 
explicitly, such as $f_\pi m_\pi^2\sigma$ or $-{1\over 2}m_\pi^2{\vec{\pi}}^2$,
but it is unclear experimentally which of these terms is realized in nature
\cite{kapusta}.

The ordinary vacuum has chiral order
parameter in the $\sigma$ direction (it has no isospin).  A disoriented
chiral condensate (DCC) is a piece of vacuum which is disoriented from the
$\sigma$ direction to a direction with $\vec{\pi}$ components.  
When the DCC domain makes contact with the ordinary vacuum, it coherently
radiates pions with isospin determined by the direction of disorientation
in order to restore the $\sigma$ direction.  For example, if the 
disorientation were in the $\pi^0$ direction, $\pi^0$'s would be emitted.
There has been much theoretical work done recently on DCC and other mechanisms
for coherent semiclassical radiation of pions in high-energy collisions of
hadrons and of heavy ions \cite{dcc1}-\cite{raj}.  

The distinctive signature of DCC is that the pions produced when the DCC
domain makes contact with the outside vacuum have a neutral fraction
$f=N_{\pi^0}/(N_{\pi^+}+N_{\pi^-}+N_{\pi^0})$ distributed according to
\begin{equation}P(f) d\!f = {1\over (2\sqrt{f})} d\!f\end{equation}
in the limit of large numbers of pions.
This neutral fraction distribution is common to some other mechanisms
which produce coherent final states \cite{dcc7}-\cite{dcc8},
\cite{sqrtf1}-\cite{sqrtf3}.
The proof of this has been given in terms of quantum mechanical coherent
state arguments \cite{whitepaper}, but can be seen easily with a geometrical 
argument.  The vacuum condensate has equal probability of having an order 
parameter oriented in any direction in $(\sigma,\vec{\pi})$ space.
If we define $\theta$ as the polar angle relative to the $\pi^0$-axis,
quantum-mechanical arguments give $f=\cos^2{\theta}$, and 
$P(f)\,d\!f=P(\cos^2{\theta})\,d(\cos{\theta})$.
We have $d\!f/d(\cos{\theta})=2\sqrt{f}$, so that
$P(f)\,d\!f=(1/2\sqrt{f})\,d\!f$.  In generic particle production, that is, 
production by (observed) mechanisms other than DCC,
producing a pion of any given charge is equally likely due to isospin symmetry,
so that $f$ is binomially distributed with mean
$\left< f\right>={1\over 3}$.  Note that the $1/(2\sqrt{f})$ distribution also
has a mean of $\left< f\right>={1\over 3}$, but that the distributions are quite
different; in particular, the probability of producing pions with a small or
large neutral fraction is much higher for DCC than for generic production
(Fig. \ref{f:sqrtf}).

It is conjectured that such a condensate may be formed in hadron-hadron 
collisions with high transverse energy and a large multiplicity (number of 
particles produced).  As the collision debris expands outward at almost the 
speed of light, it may form a hot, thin shell, the cool interior of which
would be separated from the
outside vacuum and could conceivably have a chiral order parameter which is
disoriented from the $\sigma$ direction.  When the shell hadronizes, the 
interior condensate makes contact with the outside vacuum and 
radiates pions.  


\subsection{Centauro/anti-Centauro}
\label{sec:cent}
Cosmic ray experiments have found evidence for hadronic events which 
can be interpreted as having 
an anomalously large or small fraction of pions which are neutral and
may therefore be related to DCC.
Centauro events have been observed in emulsion chambers
by the Chacaltaya-Pamir Collaboration 
\cite{centauro,hasegawa} and are characterized by a large number 
($\sim 100$) of charged particles and almost no electromagnetic energy, which 
implies no $\pi^0$'s, since their immediate decay produces two photons.
The center-of-mass (cm) energies are on the order of a TeV or larger.
Further interpretation of the events is controversial. 
What is measured can be related to the transverse momentum, e.g.,
$k_\gamma\left< p_T\right>=0.35\pm 0.15\:$GeV for Centauro I, but the
gamma ray inelasticity $k_\gamma$ is not known.  The value is usually quoted
as $k_\gamma\sim 0.2-0.4$ with the lower and upper values preferred for 
nucleons and pions, respectively.
Most analyses assume that the hadrons produced in Centauros events are
nucleons.  However, if we want to interpret Centauros as being related to DCC,
the hadrons should be pions, and therefore it is possible that the $p_T$
is low ($\left< p_T\right>\approx 0.875\pm 0.375\:$GeV with a large systematic
uncertainty), which is referred to as Chiron behavior.
It has been suggested (Ref. \cite{Goulianos} based on data from 
Ref. \cite{hasegawa}) that these events may be 
diffractive.\footnote{Diffractive processes \cite{diffr} are thought of as 
involving the exchange of a colorless object called a pomeron, but are not well 
understood.  In single diffraction, either the incoming proton or anti-proton 
is dissociated, while the other remains intact and typically has a low 
transverse momentum and a longitudinal momentum almost that of the initial 
beam.  Both hadrons are dissociated in double diffraction. \hfill\newline}
Taking the view that Centauros are diffractive fireballs recoiling against a 
proton or anti-proton, but that pions rather than nucleons are produced,
boosting these events to the lab frame of Fermilab Tevatron collisions,
Centauros would be expected to occur
at $\eta\sim 3.5 - 4.5$.\footnote{We work in what is called ``lego space'' 
where $\phi$ is the azimuthal angle and $\eta$ is the pseudorapidity 
which is defined in terms of $\theta$, the angle from the beam axis, as
{\bf $\eta=-\ln\tan{\theta\over 2}$}.  Therefore, the region explored by
central detectors which look transverse to the beam covers small $\eta$,
while $\eta$ gets infinitely large in the beam direction.
An advantage to using the pseudorapidity is that $\eta$ has a simple 
transformation under boosts in the beam direction.  Also, the density of 
particles produced in a collision is uniform over a large region of lego space.}
(As will be discussed in Sec. \ref{sec:detect}, the
acceptance of the MiniMax detector covers precisely this region.)

Events with a large neutral fraction are referred to as anti-Centauros.
Such cosmic ray events have been reported by the JACEE Collaboration 
\cite{jacee}, which uses balloon-borne detectors.  
An example is the event shown in Fig. \ref{f:jacee}.
The leading cluster contains about
32 photons and only 1 charged particle.  A possibly distinct cluster has about 
54 photons and 17 charged particles.  The collision occurred within the detector
with a cm energy greater than 200 GeV and Chiron behavior was exhibited.

However, the interpretation of these cosmic ray events is controversial,
and observations of this type of event under controlled conditions, such as 
in a collider environment, where they could be better understood, would be
very useful.  Several collider experiments have unsuccessfully looked for large 
Centauro-like domains.  CDF has conducted such a search at the Tevatron 
($\sqrt{s}=1.8\:$TeV), and has reported no evidence for Centauros or
anti-Centauros \cite{CDFcent}.  However, the CDF search only covered the region
$|\,\eta\,|\,\raisebox{-.75ex}{$\stackrel{\textstyle{<}}{\sim}$}\, 3$
and only looked at particles with 
$p_T\,\raisebox{-.75ex}{$\stackrel{\textstyle{>}}{\sim}$}\, 400\:$MeV, 
and therefore might not be expected to observe such phenomena.  
The UA5 Collaboration ruled out large Centauro domains going out to larger 
$\eta$ \cite{UA5cent,UA5rep},
but at cm energies no greater than $900\:$GeV, whereas the necessary cm 
energy is expected to be larger.  UA1 also conducted a search at relatively
low energies and in the central region, and found no evidence \cite{UA1cent}.

\subsection{Multiplicity distributions}
Very little is known about particle production in the forward direction because
this region has not been well studied.  The distribution of charged 
particles in lego space has been reported by experiments such as 
CDF \cite{CDFmult} and UA1 \cite{UA1mult} 
for the central region 
($|\,\eta\,|\,\raisebox{-.75ex}{$\stackrel{\textstyle{<}}{\sim}$}\, 3$).
Measurements of $dN_{ch}/d\eta$ were made for larger $\eta$ by 
UA5 \cite{UA5rep}, and by P238 \cite{p238}
at the CERN Sp$\bar{\mbox{p}}$S collider for proton-anti-proton collisions 
at center-of-mass energy $\sqrt{s}=630\:$GeV.
The distribution of photons $dN_\gamma/d\eta$ in the forward direction
is even less well known; UA5 observed photons at $\sqrt{s}\leq 900\:$GeV
\cite{UA5rep,UA5mult}.
MiniMax, which was able to observe both charged particles and photons in the
region $3.3<\eta <4.5$ at $\sqrt{s}=1.8\:$TeV, therefore had the potential
to make measurements in a previously unexplored region.

Figure \ref{f:dndetach} shows a plot of $dN_{ch}/d\eta$, 
averaged over typical collisions at Tevatron energies, from the event
generator PYTHIA.  The distribution in the central region is taken from 
measured values, and these are extrapolated to the forward region.
Note that the mean number of charged particles varies by only
about 0.5 (about 13\%) in the region $|\,\eta\,| <4$.  Since there is no 
preferred azimuthal direction, the distribution in $\phi$ is also uniform.

\section{Conceptual design of MiniMax}
With the primary goal to search for DCC, the MiniMax detector should be able
to observe both charged particles and photons simultaneously.  
If we take seriously the interpretation of Centauros and anti-Centauros
given in Sec. 
\ref{sec:cent}, the detector should cover the forward region $3.5<\eta < 4.5$,
and be sensitive to low-$p_T$ particles.  The smallest coverage which would
be expected to be sufficient for observing such events is 
$\Delta\eta\Delta\phi =1$.
Further considerations include the restricted area around the beampipe due to 
the main ring and the floor, and the lack of funding for the experiment.
In order to achieve ``minimal maximum acceptance'' (hence the name MiniMax),
the detector was designed as a telescope of multi-wire proportional chambers
(MWPC's) along the beampipe, with converter inside the telescope.
Thus, charged particles can be observed in the chambers before
and after the converter, and photon conversions in the chambers behind the 
converter.

In the absence of a magnetic field, the energy of the particles can only be
determined using calorimetry.  For this reason, and to observe photons which
do not convert inside the telescope, an electromagnetic calorimeter
was placed behind the MWPC telescope.

If Centauros are diffractive and are related to DCC,
then DCC might be more likely to be produced in diffractive interactions
than in non-diffractive ones.  In order to test this conjecture, 
diffractive events must be identified.  Scintillator and hadronic calorimetry
were used to detect leading anti-protons and anti-neutrons from diffractive
events.

A picture of the original design for the MiniMax detector is shown in Fig.
\ref{f:oldplan}.

A brief note on the similarity between MiniMax and UA5 is in order, 
although the details of the UA5 experiment did not play a role in the design
of the MiniMax detector.  UA5 was an experiment at the CERN proton--anti-proton
collider (Sp$\bar{\mbox{p}}$S) and ran from 1981-1982 and in 1985,
during which time 
collisions at $\sqrt{s}=546$, $200$, and $900\:$GeV were studied.  
The detector consisted of two streamer chambers, 
$600\times 125\times 50\:\mbox{cm}^3$, on opposite sides of the beampipe, 
in which charged particles left tracks which were photographed for analysis by 
well developed techniques.  For the runs at $\sqrt{s}=546\:$GeV, photons were 
detected through conversions in lead-glass plates, approximately $1\:$X$_0$ 
thick, located inside the streamer chambers near the sides.  In later runs,
a lead converter plate ($2\:$mm of lead supported by an aluminum box with walls 
$1\:$mm thick), placed between the beam pipe and the upper streamer chamber,
was used instead.  The beampipe was elliptical with mean
dimensions $6\times 15.2\:\mbox{cm}^2$.  Combining this transverse distance
of the chambers from the beam axis with the length of the chambers, UA5
was able to observe particles at pseudorapidities up to $|\,\eta\,|\approx 5$.
UA5 published results on many important studies, including
multiplicity and pseudorapidity distributions and 
correlations for charged particles and photons, strangeness production (e.g.
K's, $\Lambda$'s, and $\Xi$'s, observed through decays), diffractive 
dissociation, and a search for Centauros.  Reference \cite{UA5rep} is a 
comprehensive report on these studies at $\sqrt{s}=546\:$GeV.  
The Monte Carlo simulations created and used by UA5, which include 
non-diffractive and diffractive event generators, along with a Centauro 
generator, are described in Ref. \cite{UA5sim}.

\vfill\newpage
\begin{figure}[h] \vspace{-2.5in}\hspace*{-0.75in}
\psfig{file=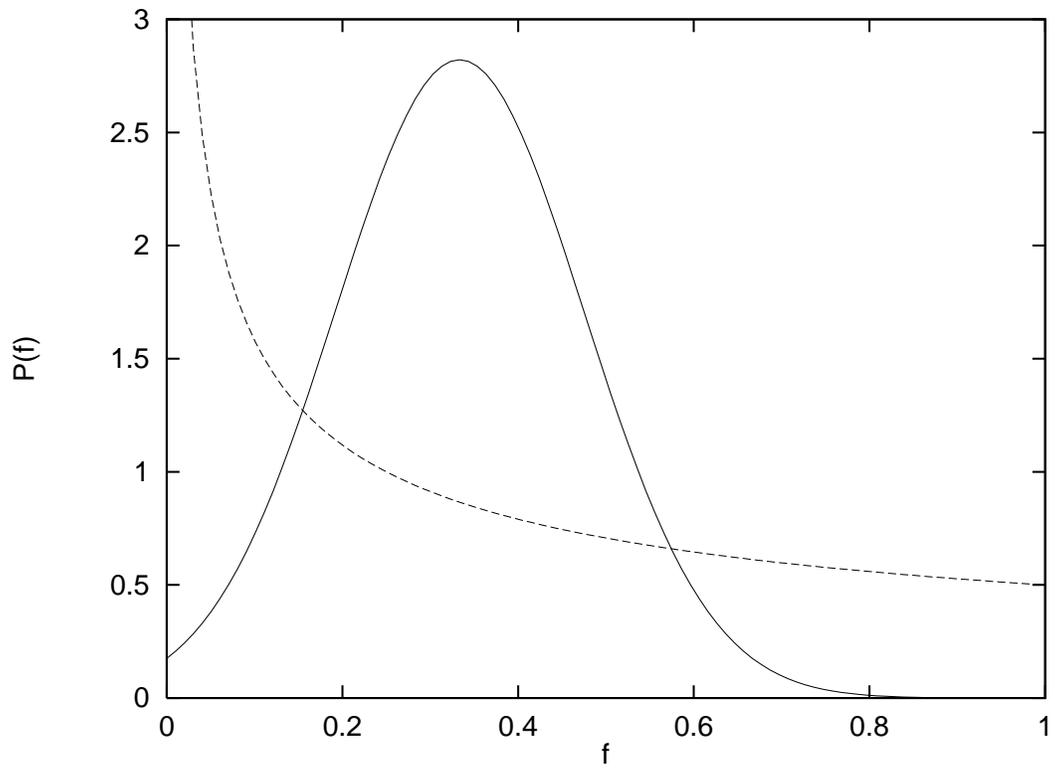,width=6.5in}
\caption{Binomial and DCC neutral fraction distributions.}
\label{f:sqrtf} \end{figure}
\clearpage
\vfill\newpage
\begin{figure}[h]
\ 
\caption{JACEE event}
\label{f:jacee} \end{figure}
\vfill\newpage
\begin{figure}[h] \vspace{-0.5in}\hspace*{-0.3in}
\psfig{file=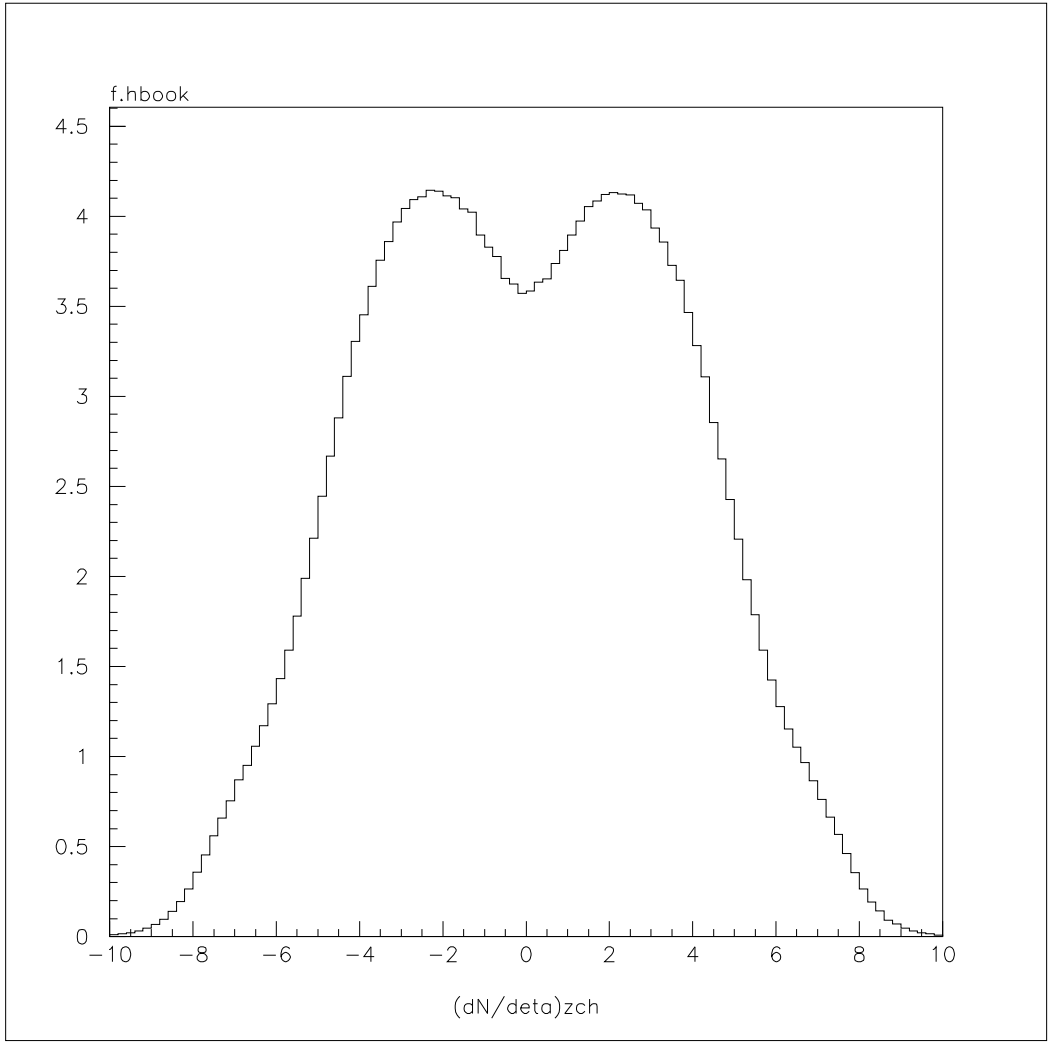,width=6in}
\vspace{-1in}
\caption{Charged multiplicity distribution $dN_{ch}/d\eta$ for non-single
diffractive inelastic PYTHIA events at $\protect\sqrt{s}=1.8$ TeV.}
\label{f:dndetach} \end{figure}
\vfill\newpage
\begin{figure}[h] \vspace{-0.5in} 
\psfig{file=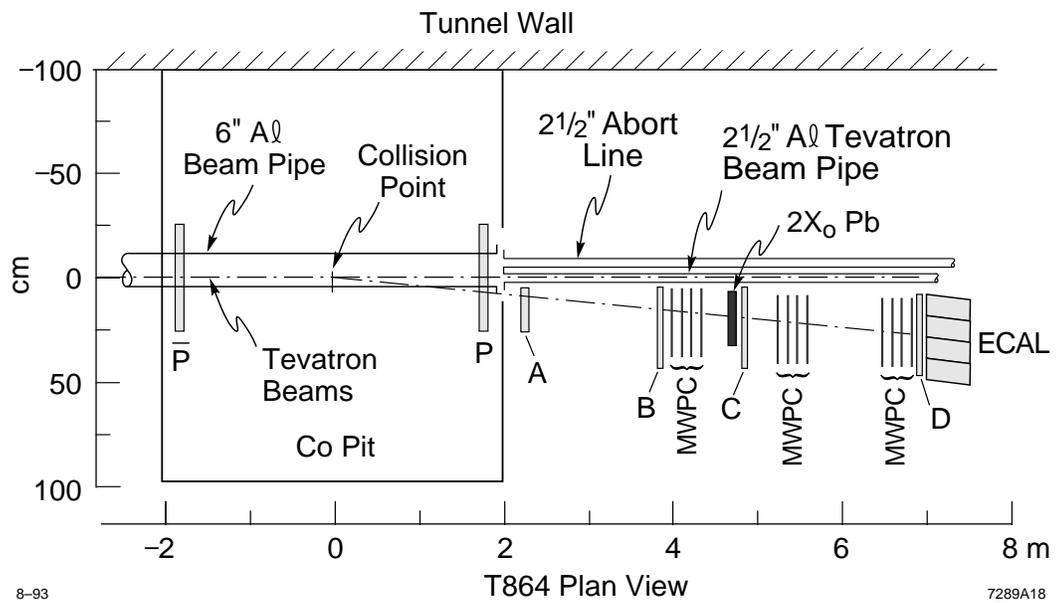,width=5.5in}
\caption{Original design of the MiniMax detector.}
\label{f:oldplan} \end{figure}

\clearpage
\chapter{MiniMax}

\section{The MiniMax environment}
The MiniMax detector was located at the C0 collision region of the Fermilab
Tevatron, which collides protons and anti-protons at cm energy
$\sqrt{s}=1.8$ TeV.  A sketch of the detector is shown in Fig. \ref{f:detect}.
The coordinate system was taken to be left-handed with positive $z$ in the 
downstream proton direction and positive $y$ upwards.
The region included a pit which was $18\:$in deep, $60\:$in wide, 
and $170\:$in long, with the bottom centered at 
$(x,y,z)=(0,-29\:\mbox{in},1\:\mbox{in})$
relative to the nominal collision point defined as $(0,0,0)$.
(Note that the actual collision point as determined by the mean distance of
closest approach between tracks is found to be at $z\approx 7\:$in, and within
$0.5\:$in of $x=y=0$.  The pointing of the tracks will be discussed in more
detail in Sec. \ref{sec:vertexer}.)

The Tevatron beampipe in the neighborhood of the C0 collision point 
during early stages of the experiment (from late 1993 to early 1995) was 
a $6\:$in-diameter, $0.0625\:$in-thick Al pipe, 
with a $2\:$in-thick Al flange at the transition to a 
$2.5\:$in-diameter, $0.035\:$in-thick Al pipe 
and an abort pipe of the same dimensions.
The detailed dimensions and locations of the pipe segments are given in 
Table \ref{t:oldpipe}.
Secondaries from interactions of collision primaries in the flange dominated 
the particles observed in the forward region.  

A new beampipe was designed
to minimize the number of interactions in the pipe which produced background
hits in the detector.  This was accomplished by steps of increasing diameter,
with flared transitions rather than vertical plates wherever possible between 
two sections of pipe (the pipe bent out at a $30^\circ$ angle to meet the 
adjoining segment), and by a thin window at $z=120\:$in through which all 
collision primaries entering the detector would pass.  
The steps in the Al pipe are given in Table \ref{t:newpipe}, and the beampipe
and abort are pictured in Fig. \ref{f:pipedraw}.
The new pipe was installed in February 1995, and was in place for the majority
of the data collected.

The main ring, which was used for accelerating protons which were then used 
in production of the anti-protons, and a main ring abort pipe 
were present in the C0 collision region, both approximately $20\:$in above 
the Tevatron pipe.  These pipes were steel, on the order of $1/4\:$in thick.
The main ring and abort were about $6\:$in and $4\:$in in diameter, 
respectively.
The accelerating protons in the main ring interacted with particles present in
gas inside the pipe, producing blinding flashes of background in the detector,
so that the triggers had to be gated off when the main ring protons passed
through the collision region.  In events which were triggered on, such 
beam-gas interactions also created hits in the upstream scintillator
used for diffractive tags
(which will be discussed in Sec. \ref{sec:detect}).
Interactions of collision primaries with the main ring pipes produced 
background tracks in the detector.
And accelerator studies using the main ring produced large bursts of radiation 
which badly damaged some of the electronics.  This was the greatest source
of radiation damage because the spread in momentum of beam particles in the
(older) main ring was much greater than that for the Tevatron beams, and those
particles with less-than-beam momentum often interacted in the pipe or escaped 
the pipe and interacted elsewhere.

\section{The MiniMax detector}
\label{sec:detect}
\subsection{MWPC telescope}
Particles were tracked through a telescope of 24 multi-wire proportional 
chambers on the downstream proton side of the collision point.  
The acceptance of these chambers was approximately a circle of radius
0.6 units in $\eta-\phi$ space centered on $\eta\approx 4$ 
(see Fig. \ref{f:lego}).  
The chambers were designed for past experiments at CWRU \cite{Bev}.
Figures \ref{f:mwpc_pln} and \ref{f:mwpc_elv} show diagrams of the wire chamber 
design.  Each chamber had an active volume of 
$12.8\:\mbox{in}\times 12.8\:\mbox{in}\times 0.375\:\mbox{in}$ 
containing 128 parallel wires with 
a spacing of $0.1\:$in.  The anode wires were gold-plated tungsten with a 
$0.0008\:$in diameter, and the cathode plane consisted of an evaporated Al
film on one side of $0.001\:$in-thick mylar.  The gap was $3/16\:$in.
Copper clad surfaces $1/16\:$in from the signal plane served as guard rings,
and $3/16\:$in from the signal plane as the contact for the high-voltage plane.
The 80\% Ar, 20\% CO$_2$ gas used was circulated freely on both sides of the
aluminized mylar.  The outer faces were sealed with $0.005\:$in-thick mylar.
After tests in early 1995 determined that some of the chambers were sensitive 
to light, $0.002\:$in-thick black plastic film (kevlar, or similar material)
was added to the front and back faces of the chambers.

The chambers were held together by $1/8\:$in- and $1/16\:$in-thick frames made 
of G-10 which were epoxied together.
Grooves $0.004\:$in-wide were etched in the printed circuit board face of the
signal plane to keep the signal wires aligned during the epoxy process.
The $1/16\:$in-thick G-10 circuit board extended $1.85\:$in below the frame, 
which was $1\:$in in width, so that 
the completed chambers had a width of $14.8\:$in, and a height of $16.65\:$in.

Two types of readout electronics were used.  Half of the chambers used
nanometrics supplied by Fermilab, which did not give pulse-height information, 
but were very reliable.  The other half were equipped with cards made by
the University of Michigan.  These had 12 bits to digitize the signal from 
each wire, and gave pulse-height information about the amount of charge
deposited on the wire.  A certain chip in the amplifier cards was had to be
replaced frequently due to radiation damage, which often led to runs which
did not have all MWPC's functioning.\footnote{For a brief period in Dec. 1995, 
the Tevatron ran at $\sqrt{s}=630\:$GeV.  However, all of the data collected by 
MiniMax at that energy did not have information from the wire chambers which 
were read out using the Michigan electronics because these were under repair at 
the time.}

The chambers were held in position by aluminum stands which allowed them to be
rotated at various angles.  The original stands each held four chambers.
New stands were made in September of 1995 to hold eight chambers in order to 
compress them closer together.  An end-on view of a chamber stand is
pictured in Fig. \ref{f:apparatus}, along with other pieces of apparatus.

Chamber orientations for various periods of running are given in Tables
\ref{t:cham310}-\ref{t:cham0324}.
The alignments during the first year or so of running utilized large-angle 
stereo for three-dimensional resolution.  Each chamber was rotated by a
different angle in order to reduce the number of potential reconstructed
tracks as described below. 
This philosophy was reconsidered \cite{HERA}, and for reasons
discussed below, the chambers were reconfigured in February 1995.

The final orientation of the chambers was defined in a coordinate system where,
looking in the positive $z$ direction, the $(u,v)$-axes were a 45$^\circ$ 
counter-clockwise rotation of the $(x,y)$-axes about the $z$-axis.  
[We also redefine the azimuthal angle as $\phi=\tan^{-1}{(v/u)}$.]
In the chamber configuration used for the majority of the data 
taking, three of the front chambers and eight of the rear had their wires
aligned perpendicular to the $u$-axis (``$u$ chambers''), three of the 
front chambers had wires aligned within 15$^\circ$ of normal to the $v$-axis 
(``$v$ chambers''), and the remainder were rotated by small angles 
(4-15$^\circ$) from the $u$-chamber orientation (``$u'$ chambers'').

A configuration with wires in half the chambers aligned in
a given direction and the other half perpendicular gives equal resolution in 
both directions.  However, the number of potential tracks considered by a 
track-finding algorithm goes like the square of the number of real tracks.
For example, two real tracks will produce two hits in a front chamber and
two in a back chamber, leading to four potential track candidates if such
candidates are found using only the two chambers.  In a situation like that
of MiniMax, where there are many random hits due to pipe shower, confirming
hits in other chambers can make these extra track candidates look like actual
tracks.
The opposite extreme is a chamber configuration with all wires parallel.
This does not give any information about the location of a track in
the direction along the wires, but greatly reduces the number of potential
fake tracks.  
The other great advantage to parallel wires, as opposed to large-angle stereo 
at random angles, is that tracks can be reconstructed by eye in a 
two-dimensional display of the hit wires, which can be quite useful for
studying the properties of, e.g., tracks coming from interactions in the 
beampipe.

The MiniMax Collaboration determined that the configuration discussed above, 
employing small-angle stereo, was a good compromise 
between the two extremes because it allowed for some resolution in both 
directions, the ability to view wire hits in about half the chambers,
and reduced probability of finding fake tracks.  Note also
that the front chambers
had three $v$ chambers for increased resolution of charged tracks\footnote{A
charged track is defined as the reconstructed track from a charged particle
which appears to originate in the collision. \hfill\newline} and of their
intersection with the $z$-axis (i.e. the collision point); the number of wire
hits, especially of those due to pipe shower, was much smaller in the front
chambers than in the back, so that the reduction of fakes by small-angle
stereo would not have been as much of an advantage as it was in the rear
chambers.

Another major reconfiguration was performed in the fall of 1995 in which
the chambers were moved closer to the collision point.  The reason was that
pipe shower and other sources of background increased towards the rear
of the detector.  This will be discussed further in Sec. \ref{sec:nhits}.

\subsection{Converter}
In order to detect photons, converter was inserted behind
the eighth chamber.  During early running (through 1994), the converter was a 
stationary plane of lead, 
$6\:\mbox{in}\times 4\:\mbox{in}\times 1\:\mbox{X}_0$,\footnote{One radiation 
length ($1\:$X$_0$) is defined as the mean distance over which a
high-energy electron loses all but $1/e$ of its energy by bremsstrahlung,
and varies with atomic number and mass of a medium.  Also, for very high-energy
photons, the e$^+$e$^-$ pair-production cross section is given by
$\sigma\approx{7\over 9}(A/\mbox{X}_0N_A)$, where $A$ is the atomic 
mass and $N_A$ is Avagadro's number \cite{PDG}.}
centered on $(5.25\:\mbox{in},6.20\:\mbox{in},198\:\mbox{in})$, level with
the floor.
For runs after Feb. 1995, the movable converter was an
$8\:\mbox{in}\times 8\:\mbox{in}$ square, rotated by 45$^\circ$ relative to 
the floor, so that its edges were parallel to the $(u,v)$ axes.
Various thicknesses and materials were used to study systematics in detecting
photon conversions; these were:
1, 2, $0.5\:$X$_0$ lead, $1\:$X$_0$ iron, and no converter 
(converter in the ``out'' position).
For runs before chamber compression, the lead was located at 
$(7.7\:\mbox{in},0.1\:\mbox{in},184.8\:\mbox{in})$ for ``lead-in'' running,
and approximately $20\:$in farther from the beampipe in $x$ for ``lead-out''
runs.
For the production runs, $1\:$X$_0$ lead was used about equally often in the
lead-in position $(5.13\:\mbox{in},5.13\:\mbox{in},150\:\mbox{in})$, and in the
lead-out position.

\subsection{Trigger scintillator}
\label{sec:trig}
Scintillator counters were used
to trigger on particles passing through the detector, signaling a collision.
The counters had both ADC (analog to digital converter) readout, which gave 
information about the energy deposited by particles passing through
the scintillator, and TDC (time to digital converter) information about the
time-of flight of the particles.  
The signature of beam-beam (rather than beam-gas) collisions 
is correctly-timed hits in the scintillator arrays.

The scintillator counters which were interspersed among the chambers were
referred to as A-E, or collectively as ``alphabet counters''.
The B and D counters which played a role in the usual trigger were 
located one directly behind the lead and one directly behind the last chamber.  
Each was comprised of two 
$8\:\mbox{in}\times 16\:\mbox{in}\times 0.5\:\mbox{in}$ pieces which together 
formed a square, which was rotated into the $(u,v)$ frame in a similar
manner as the lead.
The C counter was of the same type as B and D.  In early runs, B and C were
located on either side of the converter, and signals from both were
required for the trigger, significantly reducing the probability of triggering 
on an event with a photon conversion but no charged tracks.  The smaller A 
counter, located in front of the MWPC telescope, and the larger E counter
behind the calorimeter were also sometimes used in the trigger.
The history of configurations is given in 
Tables \ref{t:scint0}-\ref{t:scintcomp}.

The trigger also required a hit in scintillator counters on the downstream
anti-proton side of the collision (``pbar counters'').  This array contained
four $12.5\:\mbox{in}\times 2\:\mbox{in}\times 1\:\mbox{in}$ pieces and four 
$12.5\:\mbox{in}\times 4\:\mbox{in}\times 1\:\mbox{in}$
pieces arranged to form an $18.5\:$in square with a $6.5\:$in square hole in 
the middle which was centered around the $6\:$in-diameter beampipe at 
$z=-81\:$in, as shown in Fig \ref{f:apparatus}.  
Early running required a hit in an identical array on the
opposite side of the collision  at $z=83\:$in (``p counters''), but the 
information provided by these and by the alphabet counters was redundant, 
and the p counters were removed when the new beampipe was installed.

Histograms of the ADC values for some of the counters are shown in Fig.
\ref{f:adc}.  The trigger was designed to record events in which 
minimum-ionizing particles passed through the B, D, and pbar counters.
[Any moderately-relativistic charged particle other than an electron is
a minimum-ionizing particle (mip), and loses energy in a medium mainly
through ionization.  The amount of energy lost depends on the thickness
of the material and the velocity of the particle; the mean energy lost over
a given distance is described by the Bethe-Bloch equation \cite{PDG}.]
Plots of the pbar ADC's show these mip peaks, which indicate the number of 
minimum-ionizing particles which passed through.  These peaks are not seen
clearly in the alphabet ADC's because of degradation of the scintillator,
causing poorer correlation between the amount of light created by the mip
track and that collected by the photomultiplier tubes; however, by increasing
the voltage on the counters, triggering on minimum-ionizing particles was
still possible.  An entry in the log book from January 7, 1996 notes this,
along with particular mention of the poor performance of the top D counter.

A signal from the accelerator division reported the occurrence of a beam 
crossing at C0.
At this signal, counting began in the TDC's, and the time when a charged
particle was detected in the scintillator was recorded. 
If no charged particle was detected after $2^{11}$ counts, a hit was recorded
in the last bin.
Plots of the TDC values for the B, D, and Pbar counters (summed over
all pieces of scintillator) are shown in Fig. \ref{f:tdc}.  
One TDC count is equivalent to $0.25\:$ns for all counters.  However,
the offsets for each counter are different due to differing lengths of cable,
so that a particular TDC value does not correspond to the same amount of real
time for all counters.  
Also shown in Fig. \ref{f:tdc} is the TDC for the bottom D counter plotted
against the TDC for one of the pbar counters.  The dark region indicates
the timing for beam-beam collisions.  Hits which occur earlier or later in
one of the counters are due to something other than collision primaries.
For example, a beam-gas interaction involving a proton and occurring before the
proton bunch reaches the collision region could send particles into the
pbar counters at times earlier than those for beam-beam collisions.
Also, secondary interactions in the detector could cause hits at later times.

Since the counters on the detector side were behind the lead
converter, we were able to trigger on events with a photon conversion
regardless of the presence of charged tracks.  In fact, the
large frequency of interactions with the beampipe which sent particles
into the detector resulted in many triggered collisions which produced
no primary charged particles or photons in the acceptance.

The trigger rate during normal running conditions (D0 luminosity of 
$10^{30}$ cm$^{-2}$s$^{-1}$) was around 300 Hz.  
Only a fraction of the events which were triggered on were recorded; 
events were written at a rate of approximately $30\:$Hz.
The fraction of triggers not due to beam-beam collisions was less than about 
5\%, and was mainly due to proton-induced beam-gas interactions.

Mean trigger rates and Tevatron running conditions for some of the January 1996 
runs are given in Table \ref{t:runlog}.
The BDpbar delay is an indirect measure of the time difference of signals from 
pbar and alphabet counters, and different values lead to triggers on beam-beam 
or beam-gas collisions.
The p and pbar currents are the numbers of protons
and anti-protons circulating in the ring.
The raw trigger rate requires ADC counts from the B, D, and pbar
counters which are high enough to indicate the presence of collision-produced
particles, and the beam trigger rate additionally requires coincidental signals 
from the TDC's of these counters. (The alphabet rate is not given because it
was not recorded correctly 
in these runs.)  The ratio of beam trigger rate to D0 luminosity seems to vary
more in the low luminosity runs than in the runs with higher luminosity.
The ratio of beam to raw rates is a measure of the amount of raw 
triggers due to beam-gas interactions.  For the last few runs in Table
\ref{t:runlog}, the beam rate is reported as being higher than the raw rate,
which is clearly impossible.  The log book notes a ``double pulsing'' in the
trigger rate beginning in run 1128, and during run 1137, ``beam trig / raw trig
is ramping to 2.0 then back down over $\sim 5$ min period -- has occurred
4 times over last 200k events''.  The source of the problem is unclear.

The cross section seen by the MiniMax detector 
was estimated using the cross sections determined by CDF.
These are \cite{CDF} $80.03\pm 2.24\:$mb total
cross section, $19.70\pm 0.85\:$mb elastic, and $9.46\pm 0.44\:$mb single 
diffractive (this is the sum of cross sections for diffractive dissociation of 
the proton and of the anti-proton).  The non-single diffractive inelastic
cross section is therefore $50.87\:$mb. 
Some single diffractive events (most likely with high diffractive mass) may 
have been triggered on, and not all double diffractive events pass the trigger.
The cross section of events which actually passed the trigger was determined
from the trigger rate and luminosity.  The luminosity at C0 was taken as
that at D0 corrected for differences in the magnetic architecture in the two 
regions and the fact that the bunches which collide at C0 are not the same 
pairs as those which collide at D0.  
The parameter $\beta$ is a measure of how tightly focussed the beams are at a 
given point.  The ratio of the $\beta$'s at C0 and at D0 is approximately
$0.35/72$, and is further corrected by a factor of $1/0.7$ to take into
account the variation of $\beta$ over the luminous region at D0.
The ratio of trigger rate to D0 luminosity was typically around 
$0.30\:$mb.  
Therefore we estimate that the observed cross section was approximately 
$43\:$mb.

\subsection{Electromagnetic calorimeter}
A 28-module electromagnetic calorimeter was located behind the MWPC's, with the
face at $z=282\:$in, centered on $x=8.49\:$in, $y=7.71\:$in.
Twelve of the cells were $3.8\times 3.8\:\mbox{in}^2$ with 30 layers each of 
$4.88\:$mm-thick lead and $6\:$mm-thick scintillator, which combined for a 
total length of about $27\:$X$_0$.  These were made by the University of 
Michigan \cite{Michcal} and were used in Fermilab E-756.
(Before 1995, the calorimeter consisted of 16 Michigan cells with the same 
total cross sectional area, and with the face at $z=288\:$in.)
Four $10\:\mbox{cm}\times 10\:\mbox{cm}\times 30\:\mbox{cm}$ cells each had
4 photomultiplier tubes, which effectively made 16 
$5\:\mbox{cm}\times 5\:\mbox{cm}\times 30\:\mbox{cm}$ modules.
The cells, which were cut from longer hadronic calorimeter cells made by
Wayne State University, were lead with $47\times 47$ scintillating fibers of 
diameter $1\:$mm spaced $0.213\:$cm apart and running the length of the module 
\cite{wsucal}.  The effective radiation length was $0.78\:$cm, so that the
modules were about $38.5\:$X$_0$ long.
The calorimeter was placed on a stand so that it also was rotated by 
$44.5^\circ$.
The smaller cells occupied positions close to the beampipe in order to
cover roughly the same area in $\eta-\phi$ space as the larger cells,
as seen in Fig. \ref{f:ecallego} (recall that larger values of $\eta$ are 
closer to the beampipe at a given $z$).  This is desirable since the density of 
particles is roughly uniform in lego space.

\subsection{Upstream tags}
In order to tag diffractive events, scintillator and hadron calorimetry
was placed upstream of the collision point.  A view of the upstream region is
shown in Fig. \ref{f:upstream}.
At $z\sim -25\:$m, two hadron calorimeter modules, similar to the 
smaller cells of the electromagnetic calorimeter except having 
dimensions of $10\:\mbox{cm}\times 10\:\mbox{cm}\times 117\:\mbox{cm}$,
were positioned to detect leading anti-protons and anti-neutrons.  The machine
magnets bent the $\bar{\mbox{p}}$'s with less-than-beam momentum by greater
angles than the beam, and the $\bar{\mbox{n}}$'s not at all.  The 
anti-protons detected here had a Feynman $x$ of $x_F\simeq 0.5$, i.e. their
momentum is about half that of the beam particles.
Farther upstream at about $z=-60\:$m, four scintillator counters were 
placed to detect showers from anti-protons with $x_F\sim 0.9$ that
interact in what are called the kicker magnets.  If at least one of these 
scintillator is hit, the event has a ``ktag''.  Note that these counters 
did not only see diffractive anti-protons, but also products from beam-gas
interactions in the upstream region.

\section{Run history}
The concept of a detector with good coverage in the forward region 
which ultimately led to the MiniMax experiment originated in
an initiative for a full-acceptance detector (FAD) \cite{FAD} at the SSC.
A proposal was later submitted for a maximum-acceptance experiment at Fermilab 
(MAX) \cite{MAX} which was designed to investigate some of the physics goals of 
FAD.  A small test/experiment with a detector covering only the forward region 
(MiniMax, Fermilab T-864) was approved in the spring of 1993.
From late 1993 to early 1995, the MiniMax detector was installed, and collider
data was taken with 8, 12, 16, and finally 24 chambers in large-angle stereo.
My work on the experiment began in September 1994.

The new beampipe was installed in February of 1995, and when the chambers were
re-installed, they were given the new small-angle-stereo orientation with
wires in 11 of the 24 chambers aligned parallel to each other for reasons
discussed in the previous section.  
During the following five months, short opportunistic runs were taken with 
various thicknesses of lead and iron converter, and two different window 
thicknesses, 
in which a total of about $3.5\times 10^6$ events were collected.
The diffractive tags were commissioned in May and were available in more 
than $10^6$ events.
The chambers were again removed and reconfigured in the fall of 1995 in order 
to compress them closer to the interaction point, away from some of the 
blinding pipe showers farther downstream, as will be discussed in 
Sec. \ref{sec:nhits}.

Production running in January of 1996 yielded $4.2\times 10^6$ events,
$2.7\times 10^6$ of which had $1\:\mbox{X}_0$ lead in, and the remaining
$1.5\times 10^6$ events were run with the lead out.  The 
luminosity in these runs was 
lower than in most earlier runs by about 1-2 orders of magnitude.
This also led to fewer beam-gas interactions which, among other effects,
meant that the sample
of events with a ktag had a much higher fraction of diffractive events.
Some information about the runs used in the analysis described in this work
is given in Table \ref{t:runlog}.

The MiniMax detector was decommissioned the following spring.

\vfill\newpage
\begin{figure}[h] \vspace{-1in}\hspace*{-0.3in}
\psfig{file=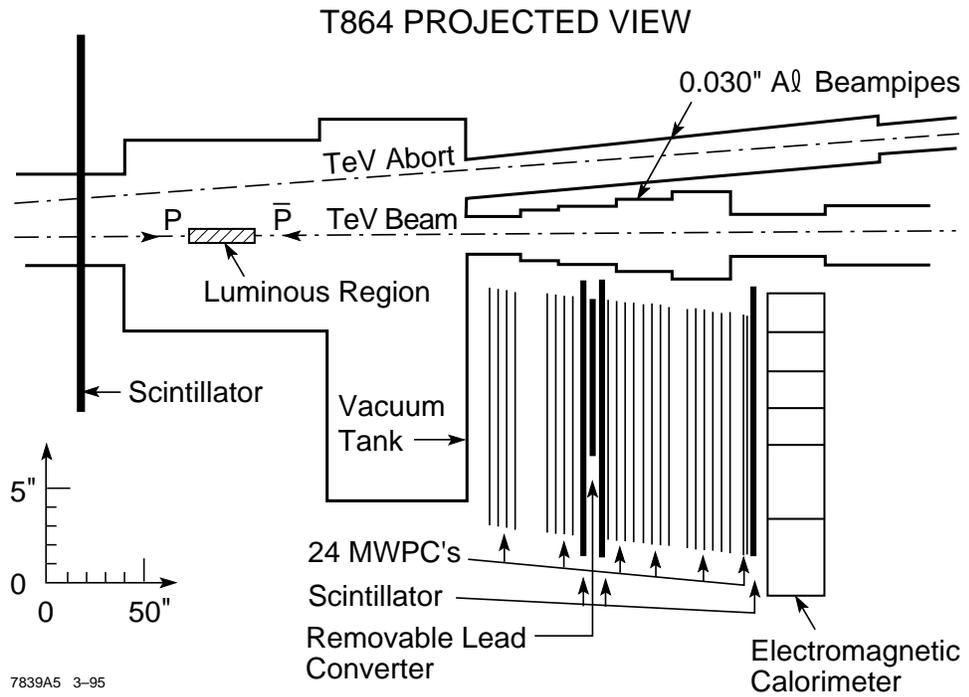,width=6.5in}
\vspace{-2in}
\caption{The MiniMax detector before compression of the MWPC telescope.}
\label{f:detect} \end{figure}
\vfill\newpage
\begin{figure}[h] \vspace{-1in}\hspace*{-0.3in}
\psfig{file=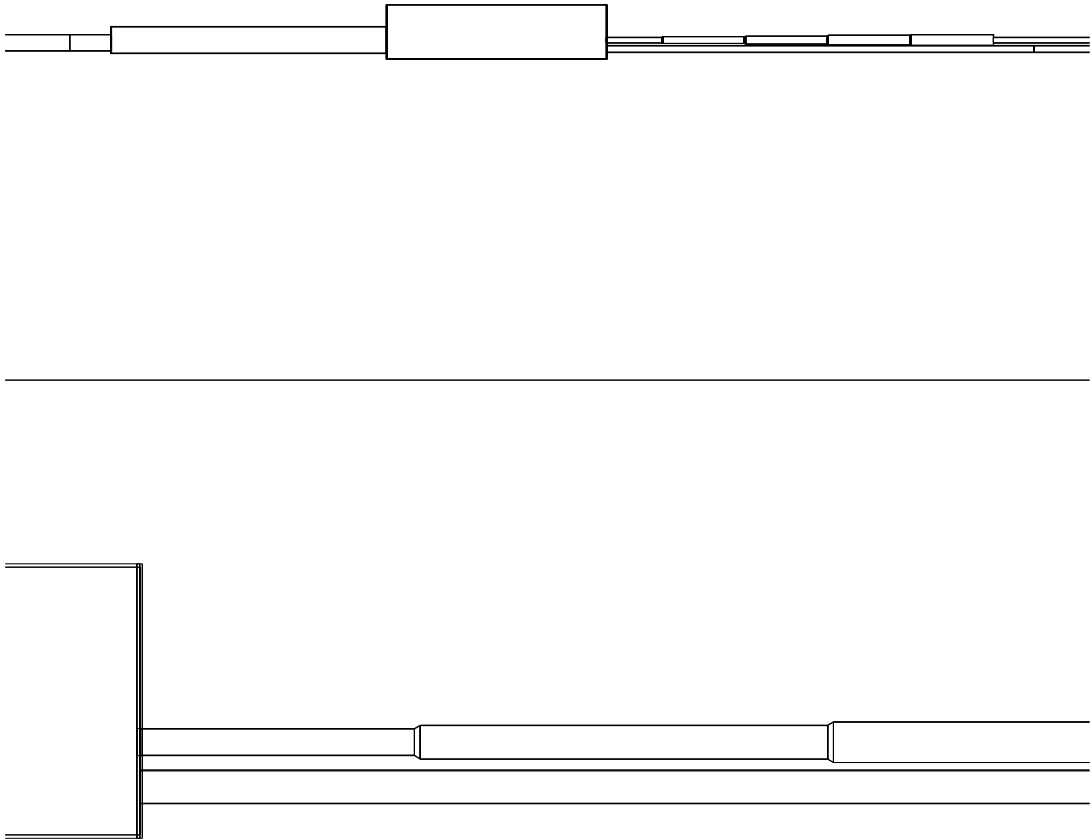,width=6.25in}
\vspace{-1in}
\caption{New Tevatron beampipe and abort at C0 and enlarged view of the region 
from $z\approx 110-190\:$in.}
\label{f:pipedraw} \end{figure}
\clearpage
\vfill\newpage
\begin{figure}[h]
\ 
\caption{Acceptance in lego space.}
\label{f:lego} \end{figure}
\clearpage
\vfill\newpage
\begin{figure}[h]
\ 
\caption{MWPC front view.}
\label{f:mwpc_pln} \end{figure}
\clearpage
\vfill\newpage
\begin{figure}[h]
\ 
\caption{MWPC top view.}
\label{f:mwpc_elv} \end{figure}
\clearpage
\vfill\newpage
\begin{figure}[h] \vspace{-0.25in}
\psfig{file=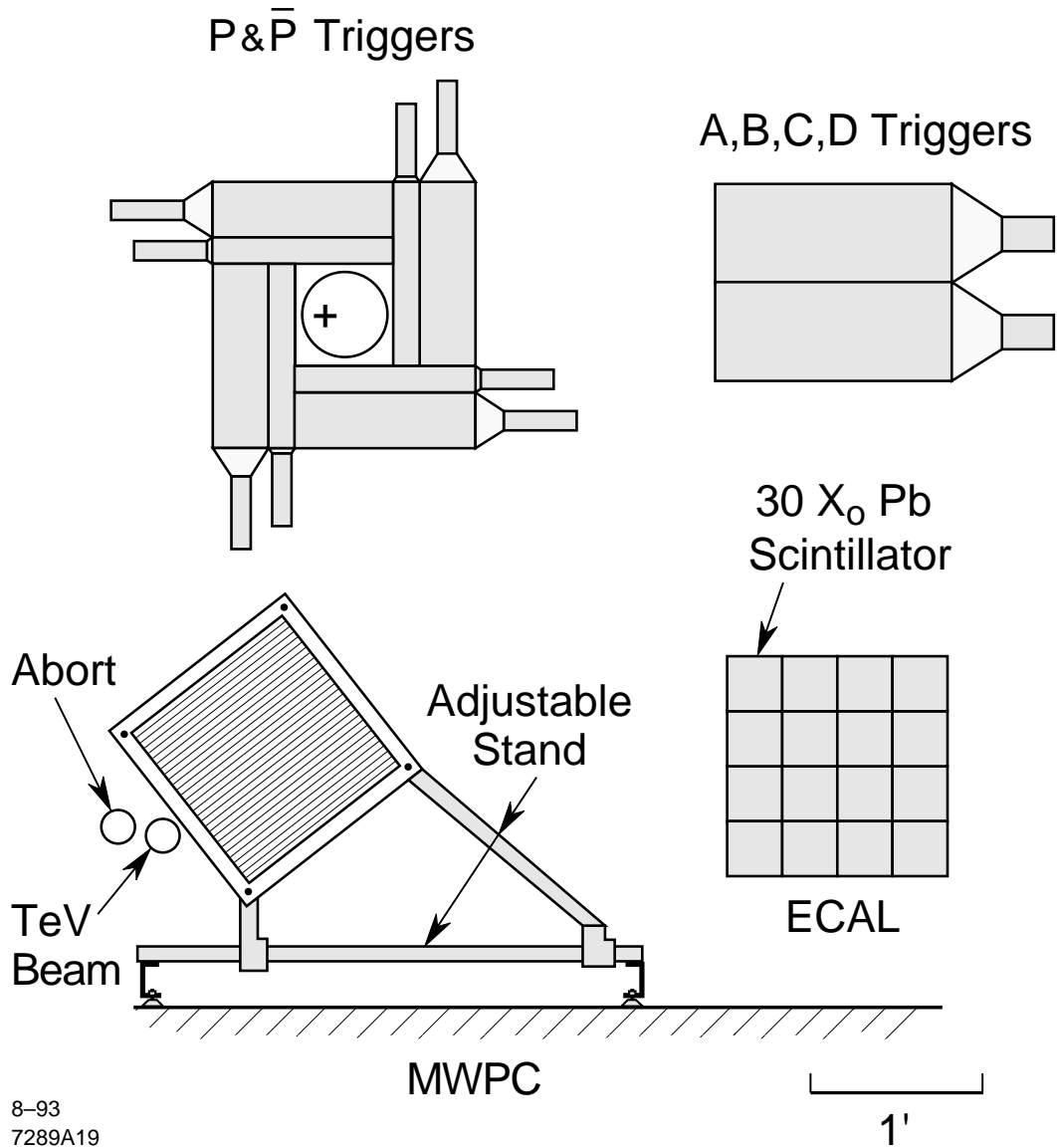,width=5.5in}
\caption{Pieces of apparatus including pbar and alphabet scintillator,
chamber stands, and the original calorimeter.}
\label{f:apparatus} \end{figure}
\vfill\newpage
\begin{figure}[h] \vspace{-1in}\hspace*{-0.3in}
\psfig{file=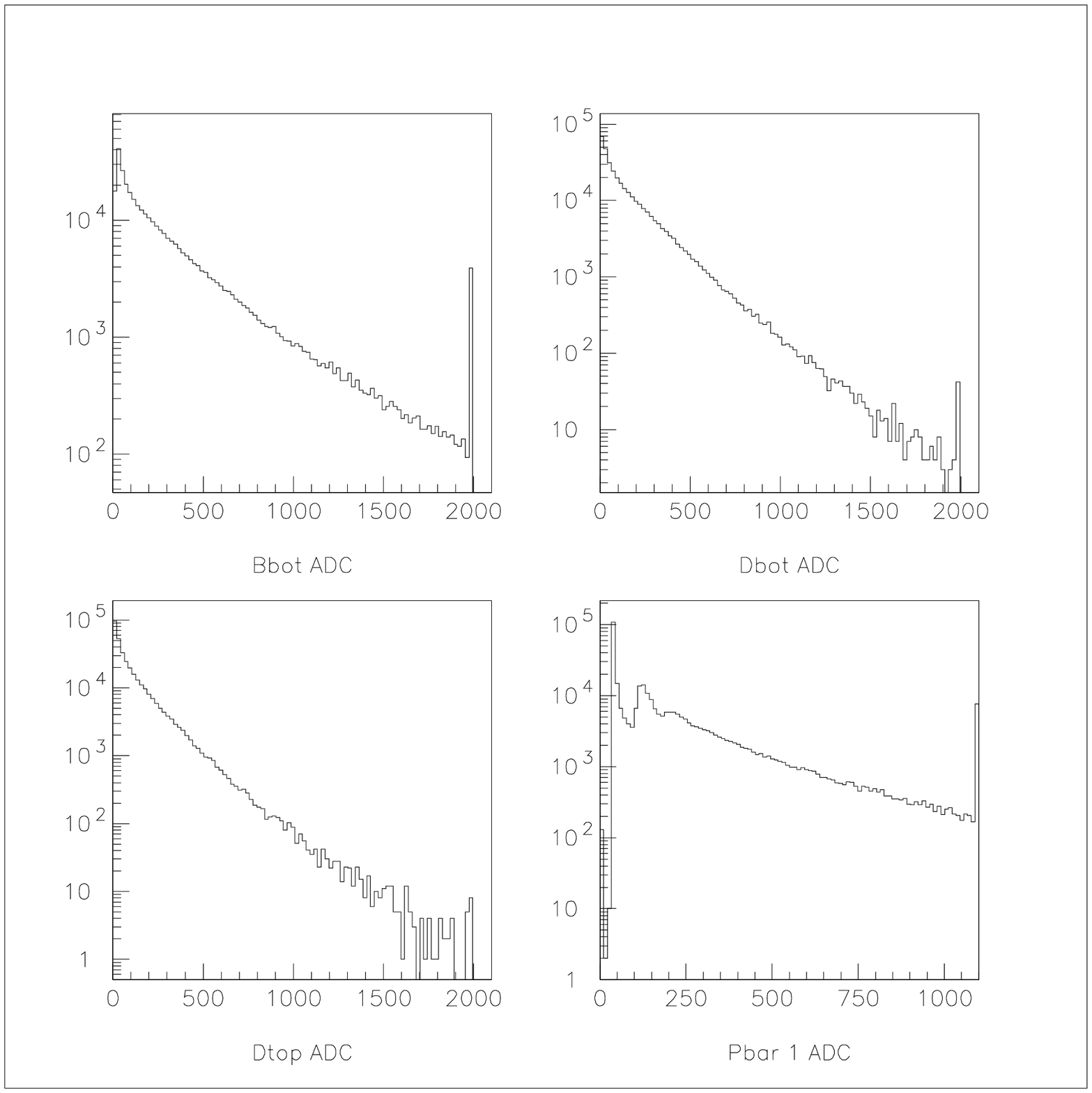,width=6.5in}
\vspace{-1.5in}
\caption{ADC spectra from some of the trigger counters.}
\label{f:adc} \end{figure}
\vfill\newpage
\begin{figure}[h] \vspace{-1in}\hspace*{-0.3in}
\psfig{file=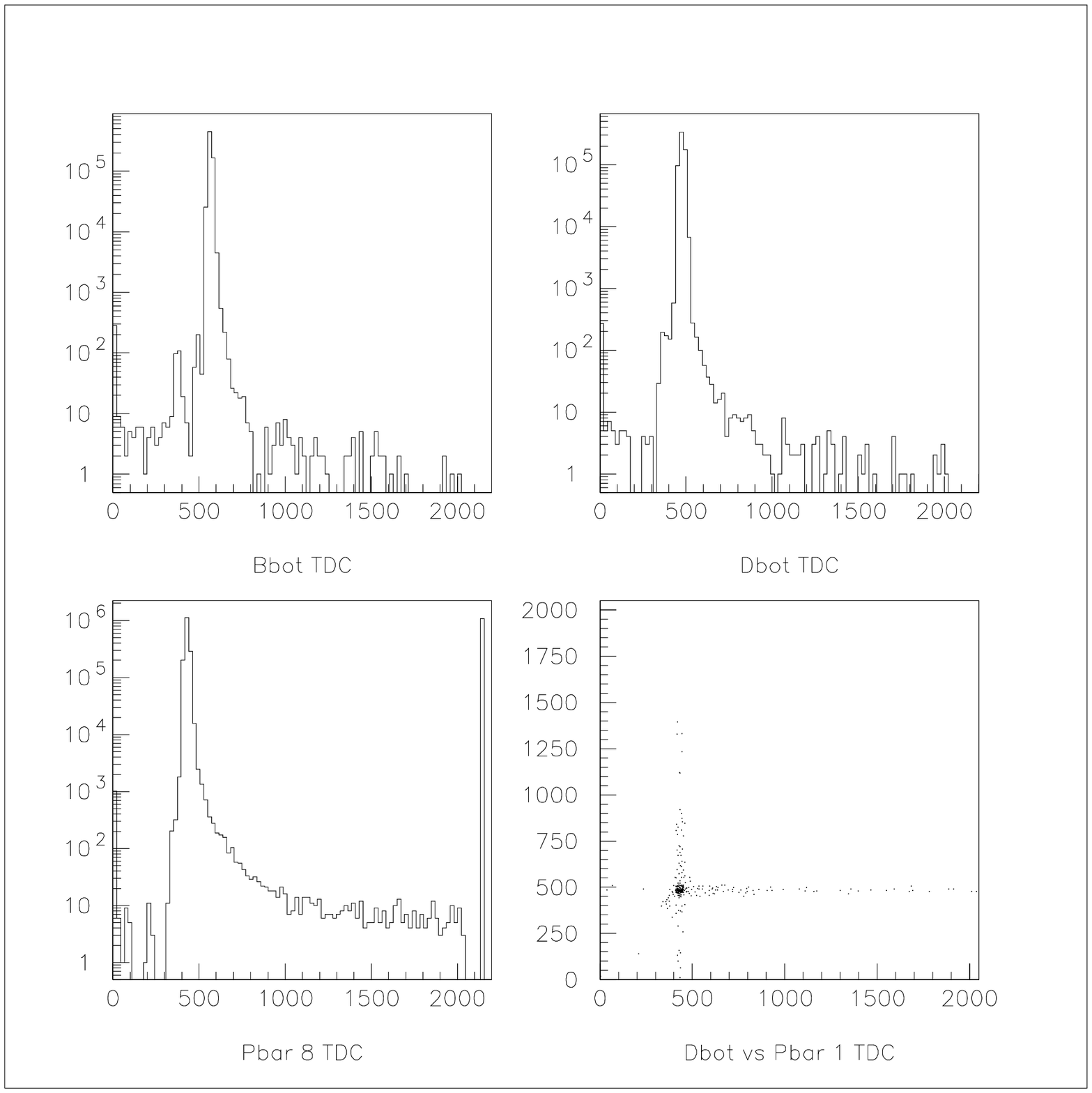,width=6.5in}
\vspace{-1.5in}
\caption{TDC spectra from some of the trigger counters.}
\label{f:tdc} \end{figure}
\vfill\newpage
\begin{figure}[h] \vspace{-0.25in}\hspace*{-0.3in}
\psfig{file=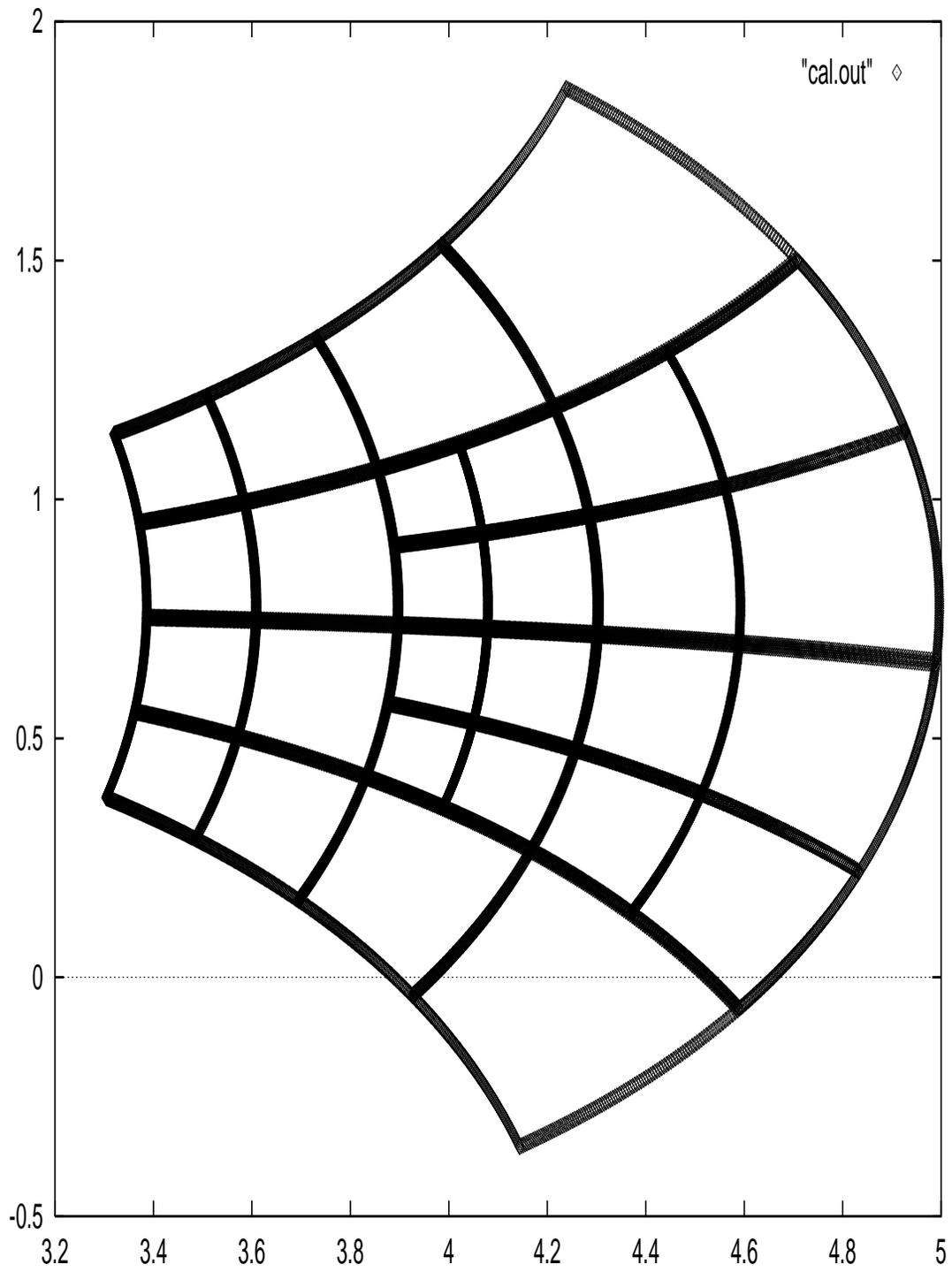,width=6in,height=7.5in,angle=270}
\caption[Electromagnetic calorimeter cells in lego space.]{Electromagnetic 
calorimeter cells in lego space; the vertical axis is $\phi$ and the horizontal 
axis is $\eta$.}
\label{f:ecallego} \end{figure}
\vfill\newpage
\begin{figure}[h] \vspace{-1in}\hspace*{-1.3in}
\psfig{file=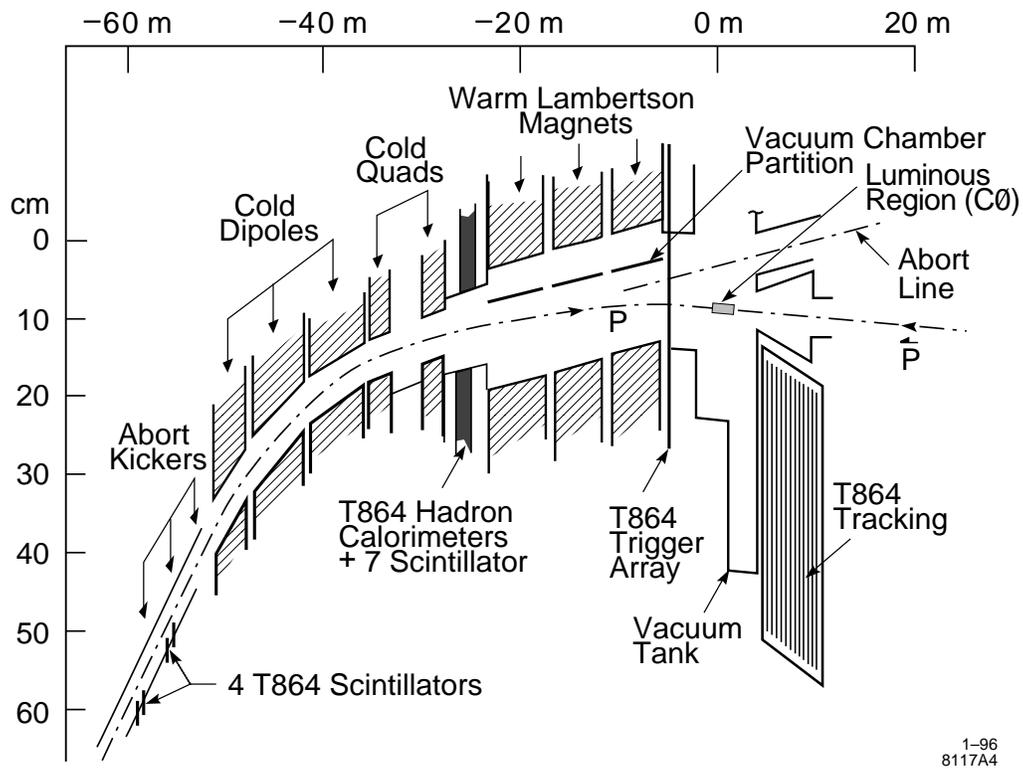,width=8in} 
\vspace{-3in}
\caption{A view of the detector including the upstream region.}
\label{f:upstream} \end{figure}

\clearpage

\vfill\newpage
\begin{table}
\begin{center}
\begin{tabular}{|c|c|c|c|c|c|c|c|}
\hline
pipe&   $z_1$   &  $z_2$  & diameter  & $x$ & $y$  & thickness & material \\
 &     &   & (outer)  &  &  & (radial) &  \\
\hline
Tev & -199.75 & -98. & 6.  & -1.5 & 0.5 & 0.0625 & steel \\
Tev & -98.    &  78. & 6.  & -1.5 & 0.5 & 0.0625 & Al \\
Tev &  78.    & 300. & 2.5 &  0.  & 0.  & 0.035  & Al \\
\hline
abort &  78. & 130. & 2.5 & -3.0 & 1.0 & 0.035 & Al \\
abort & 130. & 300. & 2.5 & -3.0 & 1.0 & 0.065 & steel \\
\hline
\hline
 &  $z$   & inner  & outer  & $x$  & $y$ & thickness & material \\
    & (center) & diameter  & diameter  &  &  & (in $z$) &  \\
\hline
flange    & 78. & 0. & 8. & 2.0 & -1.5 & 0.5 & Al \\
\hline
\end{tabular}
\end{center}
\caption{\label{t:oldpipe} Old Tevatron beampipe and abort at C0.}
\end{table}

\vfill\newpage
\begin{table}[h]
\begin{center}
\begin{tabular}{|c|c|c|c|c|c|c|c|}
\hline
pipe &   $z_1$   &  $z_2$    & diameter  & $x$ & $y$  & thickness & material \\
 &     &   & (outer)  &  &  & (radial) &  \\
\hline
Tev &  -199.75  &  -75.    &  6.  &  -1. &    1.   &  0.0625 & steel \\
Tev &   -75.    &  -60.    &  6.  &  -1. &    1.   &  0.0625 & Al \\
Tev &   -60.    &  40.     & 10.  &   0. &    0.   &  0.25   & Al \\
Tev &    40.    &  120.    & 20.  &   3. &    3.   &  0.25   & Al \\
Tev &   120.    &  139.942 &  2.  &   0. &    0.   &  0.03   & Al \\
Tev &   140.375 &  169.942 &  2.5 &   0. &    0.   &  0.03   & Al \\
Tev &   170.375 &  199.942 &  3.  &   0. &    0.   &  0.03   & Al \\
Tev &   200.375 &  229.942 &  3.5 &   0. &    0.   &  0.03   & Al \\
Tev &   230.375 &  260.375 &  4.  &   0. &    0.   &  0.03   & Al \\
Tev &   260.375 &  307.    &  2.  &   0. &    0.   &  0.03   & Al \\
\hline
abort &   120. &    275. &    2.5 &     -3.264 &  1.088 &   0.0625 & Al \\
abort &   275. &    307. &    2.5 &     -3.264 &  1.088 &   0.0625 & steel \\
\hline
\hline
 &  $z$   & inner  & outer  & $x$  & $y$ & thickness & material \\
    &   & diameter  & diameter  & offset & offset & (in $z$) &  \\
\hline
flange      & -60.     & 0.  &  10.  &   0.     &  0.     & 0.25  & Al \\ 
face plate  &  40.     & 0.  &  20.  &   3.     &  3.     & 0.375 & Al \\ 
face plate  & 120.     & 0.  &  20.  &   3.     &  3.     & 0.375 & Al \\ 
``window''  & 120.     & 0.  &   9.  &   3.8891 &  3.8891 & 0.25  & Al \\ 
flare       & 140.1585 & 2.  &   2.5 &   0.     &  0.     & 0.03  & Al \\
flare       & 170.1585 & 2.5 &   3.  &   0.     &  0.     & 0.03  & Al \\
flare       & 200.1585 & 3.  &   3.5 &   0.     &  0.     & 0.03  & Al \\
flare       & 230.1585 & 3.5 &   4.  &   0.     &  0.     & 0.03  & Al \\
flare       & 260.375  & 2.  &   4.  &   0.     &  0.     & 0.03  & Al \\
\hline
\end{tabular}
\end{center}
\caption{\label{t:newpipe} New Tevatron beampipe and abort at C0.}
\end{table}

\vfill\newpage
%
%
%
%

\vskip5ex
\begin{table}[h]
\begin{tabular}{|c|c|c|c|c|c|}
\hline
chamber & angle & $x_0$   &  $y_0$    &  $z_0$    & readout \\
\hline
 1  &  1.70504  & 9.38306 & -1.96032  & 173.993 & nanometrics \\
 2  &  3.01962  & 15.9741 &  3.55498  & 178.014 & nanometrics \\
 3  &  1.18197  & 4.19295 &  1.63239  & 181.974 & nanometrics \\
 4  &  2.49308  & 11.5763 &  2.44270  & 185.996 & nanometrics \\
 5  & -2.23082  & 13.6522 &  9.03440  & 203.816 & Michigan \\
 6  &  1.96328  & 10.8733 & -0.442493 & 208.952 & Michigan \\
 7  & -3.03528  & 14.9853 &  6.86178  & 212.912 & Michigan \\
 8  &  1.41509  & 7.64913 &  1.50400  & 216.934 & Michigan \\
 9  & -0.400473 & 4.92871 & 10.5747   & 247.129 & Michigan \\
 10 & -2.21307  & 14.7361 & 13.0676   & 250.841 & Michigan \\
 11 &  2.23920  & 14.9585 &  3.01641  & 256.658 & Michigan \\
 12 & -2.74165  & 16.9554 & 10.4961   & 260.679 & Michigan \\
\hline
\end{tabular}
\caption[Chamber alignment from 2/16/94.]
{ \label{t:cham310} Chamber alignment from 2/16/94.
      The location and orientation of each chamber is
       defined by the $(x_0,y_0,z_0)$ coordinates of the
       midpoint of wire number zero, together with the
       angle with respect to the $x$-axis of the vector
       pointing in the direction of increasing wire
       number.}
\end{table}

\vfill\newpage
\begin{table}
\begin{tabular}{|c|c|c|c|c|c|}
\hline
chamber & angle & $x_0$   &  $y_0$    &  $z_0$   & readout \\
\hline
 101 & -0.5990 &   1.017 &  10.103 &  122.83  & Michigan \\
 102 & -2.1730 &  10.135 &  12.038 &  126.70  & Michigan \\
 103 & -0.9782 &   2.762 &  11.960 &  130.83  & Michigan \\
 104 & -2.5529 &  11.810 &  10.092 &  134.70  & Michigan \\
 105 &  2.9449 &  14.174 &   6.474 &  155.33  & nanometrics \\
 106 & -1.7676 &   9.055 &  13.945 &  157.77  & nanometrics \\
 107 & -1.3747 &   6.559 &  14.038 &  161.52  & nanometrics \\
 108 &  0.2038 &   1.371 &   6.457 &  165.52  & nanometrics \\
 109 & -0.5331 &   1.466 &   9.924 &  184.70  & nanometrics \\
 110 & -2.1038 &  10.046 &  12.062 &  188.70  & nanometrics \\
 111 & -0.8795 &   2.601 &  11.817 &  192.70  & nanometrics \\
 112 &  0.6992 &  1.246  &   2.655 &  196.70  & nanometrics \\
 113 &  2.8384 &  13.312 &   5.131 &  203.27  & Michigan \\
 114 & -1.8662 &   9.191 &  13.135 &  207.02  & Michigan \\
 115 &  2.0755 &  10.282 &   1.746 &  212.83  & Michigan \\
 116 &  0.5036 &   1.530 &   4.230 &  216.83  & Michigan \\
 117 &  2.9981 &  15.017 &   7.409 &  228.02  & nanometrics \\
 118 & -1.6949 &   9.255 &  14.488 &  232.02  & nanometrics \\
 119 &  0.2893 &   2.138 &   6.258 &  236.02  & nanometrics \\
 120 & -1.2743 &   6.283 &  14.190 &  241.58  & nanometrics \\
 121 & -1.4826 &   7.788 &  14.272 &  246.58  & Michigan \\
 122 & -3.0389 &  14.772 &   8.508 &  251.89  & Michigan \\
 123 & -0.7090 &   3.398 &  12.185 &  254.58  & Michigan \\
 124 & -2.2739 &  12.464 &  12.906 &  256.14  & Michigan \\
\hline
\end{tabular}
\caption{\label{t:cham24} Chamber alignment from 11/23/94.}
\end{table}

\vfill\newpage

\begin{table}
\begin{tabular}{|c|c|c|c|c|c|}
\hline
chamber & angle & $x_0$   &  $y_0$    &  $z_0$    & readout \\
\hline
 101 & -1.0477 &   3.230 &  11.849 &  132.65  & Michigan \\
 102 & -2.6107 &  11.972 &   9.215 &  136.53  & Michigan \\
 103 & -0.7886 &   1.675 &  11.045 &  140.65  & Michigan \\
 104 & -2.3572 &  10.755 &  10.889 &  144.53  & Michigan \\
 105 & -2.3554 &  10.506 &  11.239 &  163.59  & nanometrics \\
 106 & -0.5289 &   1.512 &   9.405 &  167.59  & nanometrics \\
 107 & -2.0917 &  10.552 &  11.886 &  171.59  & nanometrics \\
 108 &  0.7801 &   1.752 &   2.136 &  175.59  & nanometrics \\
 109 &  0.7874 &   1.680 &   2.368 &  194.53  & Michigan \\
 110 & -2.1169 &  10.376 &  12.193 &  198.41  & Michigan \\
 111 & -2.3591 &  11.156 &  11.490 &  202.28  & Michigan \\
 112 &  0.5364 &   1.687 &   3.299 &  206.53  & Michigan \\
 113 & -2.3558 &  11.167 &  11.409 &  212.84  & nanometrics \\
 114 & -2.5271 &  12.242 &  10.461 &  216.84  & nanometrics \\
 115 & -2.3580 &  12.240 &  11.516 &  221.09  & nanometrics \\
 116 &  0.9668 &   3.447 &   1.569 &  226.66  & nanometrics \\
 117 & -2.3592 &  11.679 &  11.749 &  237.72  & nanometrics \\
 118 & -2.2361 &  11.216 &  12.239 &  241.72  & nanometrics \\
 119 &  0.7847 &   2.616 &   2.686 &  245.84  & nanometrics \\
 120 & -2.4771 &  12.387 &  11.188 &  249.72  & nanometrics \\
 121 & -2.2556 &  11.497 &  12.241 &  256.15  & Michigan \\
 122 &  0.7104 &   2.377 &   3.154 &  260.15  & Michigan \\
 123 &  0.7801 &   2.779 &   2.820 &  264.40  & Michigan \\
 124 &  0.7818 &   2.775 &   2.816 &  265.97  & Michigan \\
\hline
\end{tabular}
\caption{\label{t:cham24uv} Chamber alignment with 11 chambers parallel
from 3/19/95.}
\end{table}

\vfill\newpage
\begin{table}
\begin{tabular}{|c|c|c|c|c|c|}
\hline
chamber & angle & $x_0$   &  $y_0$    &  $z_0$   & readout \\
\hline
101 & -1.0315 &   2.806 &  11.653 &  123.69  & Michigan \\ 
102 & -2.3422 &  10.117 &  11.032 &  126.57  & Michigan \\ 
103 & -0.7575 &   0.853 &  10.870 &  129.69  & Michigan \\ 
104 & -2.5953 &  11.792 &   9.255 &  132.57  & Michigan \\ 
105 & -2.3318 &   9.997 &  10.910 &  135.69  & nanometrics \\ 
106 & -0.5044 &   0.818 &   9.193 &  138.69  & nanometrics \\ 
107 & -2.0857 &  10.076 &  11.700 &  141.69  & nanometrics \\ 
108 &  0.8029 &   1.252 &   1.813 &  144.69  & nanometrics \\ 
109 &  0.7959 &   1.692 &   2.568 &  167.09  & Michigan \\ 
110 & -2.1084 &  10.090 &  12.305 &  169.89  & Michigan \\ 
111 & -2.3370 &  10.970 &  11.702 &  172.84  & Michigan \\ 
112 &  0.5794 &   1.535 &   3.508 &  176.09  & Michigan \\ 
113 & -2.3370 &  11.199 &  11.781 &  178.84  & nanometrics \\ 
114 & -2.5063 &  12.360 &  10.771 &  181.84  & nanometrics \\ 
115 & -2.3405 &  11.424 &  11.812 &  185.09  & nanometrics \\ 
116 &  0.9669 &   3.366 &   1.695 &  189.65  & nanometrics \\ 
117 & -2.3370 &  11.602 &  12.036 &  194.29  & nanometrics \\ 
118 & -2.2201 &  11.327 &  12.524 &  197.29  & nanometrics \\ 
119 &  0.8029 &   2.814 &   2.963 &  200.29  & nanometrics \\ 
120 & -2.4574 &  12.648 &  11.415 &  203.29  & nanometrics \\ 
121 & -2.3527 &  12.056 &  12.114 &  206.16  & Michigan \\ 
122 &  1.0420 &   3.560 &   1.613 &  209.16  & Michigan \\ 
123 &  0.7837 &   3.140 &   3.048 &  212.41  & Michigan \\ 
124 &  0.6266 &   2.681 &   3.748 &  215.41  & Michigan \\ 
\hline
\end{tabular}
\caption{\label{t:cham0324} Chamber alignment after compression from 3/24/96.}
\end{table}

\vfill\newpage
\begin{table}
\begin{tabular}{|c|c|c|c|c|c|c|}
\hline
counter & $x$ & $y$  & $z$ & height & width & angle \\ 
\hline
 A    & -2.8  & -2.2 & 108. &  5.5 &  5.5 & 0.26 \\ 
 B1   & -5.9  & -3.7 & 155. &  8.0 & 16.0 & 0.26 \\ 
 B2   & -3.8  &  3.9 & 155. &  8.0 & 16.0 & 0.26 \\ 
 C1   & -5.9  & -3.7 & 223. &  8.0 & 16.0 & 0.26 \\ 
 C2   & -3.8  &  3.9 & 223. &  8.0 & 16.0 & 0.26 \\ 
 D1   & -5.9  & -3.7 & 269. &  8.0 & 16.0 & 0.26 \\ 
 D2   & -3.8  &  3.9 & 269. &  8.0 & 16.0 & 0.26 \\ 
 E    & -12.2 & -4.0 & 322. & 13.0 & 13.0 & 1.04 \\ 
\hline
\end{tabular}
\caption{\label{t:scint0}Configuration of alphabet counters for runs with the
old beampipe, through 1994.}
\end{table}

\begin{table}
\begin{tabular}{|c|c|c|c|c|}
\hline
counter & $x$ & $y$ & $z$ & angle \\
\hline
 A & & & & \\
 B1 & 5.1   & 10.8 & 189. & 45. \\
 B2 & 10.8  & 5.1  & 189. & 45. \\
 C1 & 5.0   & 10.7 & 180. & 45. \\
 C2 & 10.7  & 5.0  & 180. & 45. \\
 D1 & 6.7   & 11.3 & 280. & 41.7 \\
 D2 & 12.0  & 5.3  & 280. & 41.7 \\
 E  & -12.2 & -4.0 & 322. & 0. \\
\hline
\end{tabular}
\caption{\label{t:scintnew}Configuration of alphabet counters during running
with the new beampipe from 2/95-7/95.}
\end{table}

\begin{table}
\begin{tabular}{|c|c|c|c|c|}
\hline
counter & $x$ & $y$ & $z$ & angle \\
\hline
 A  &   2.8 &  2.8 & 264. &  45.  \\ 
 B1 &   6.1 & 11.8 & 279. &  45.  \\ 
 B2 &  11.8 &  6.1 & 279. &  45.  \\ 
 C1 &   6.0 & 11.7 & 270. &  45.  \\ 
 C2 &  11.7 &  6.0 & 270. &  45.  \\ 
 D1 &   7.2 &  7.6 & 157. &  27.4 \\ 
 D2 &  10.9 &  0.5 & 157. &  27.4 \\ 
 E  & -12.2 & -4.0 & 322. &  0.0  \\
\hline
\end{tabular}
\caption{\label{t:scintcomp}Configuration of alphabet counters during runs
with compressed MWPC telescope.}
\end{table}


\clearpage
\vfill\newpage
\begin{table}\hspace*{-1.2in}
\begin{tabular}{|c|c|c|r|c|c|c|c|r|r|c|c|}
\hline
run & date & lead & number & BDpbar & p & pbar & D0 & raw  & 
beam  & beam/  &   beam/  \\
number & &  & events &  delay & current &  current &  luminosity &  rate & 
    rate & D0lum & raw \\ 
 & & & & (ns) & ($10^{9}$) & ($10^{9}$) & ($10^{30}$cm$^{-2}$s$^{-1}$) 
  & (Hz) & (Hz) & (mb) & \\
\hline
1089 & 1/18 & out &  50112 & 8 & 105. & 83. & 0.91 & 269. & 
 258. & 0.28  & 0.96 \\
1093 & 1/18 & out & 103061 & 8 &  95. & 79. & 0.63 & 192. & 
 190. & 0.30  & 0.99 \\
1096 & 1/18 & in  & 249986 & 8 &  38. & 30. & 0.14 &  43. & 
  42. & 0.30  & 0.97 \\
1099 & 1/18 & in  & 174775 & 8 &  37. & 29. & 0.09 &  32. & 
  32. & 0.33  & 0.99 \\
1103 & 1/19 & in  &  67117 & 8 &  34. & 27. & 0.05 &  18. & 
  18. & 0.35  & 0.98 \\
1104 & 1/19 & out &  50461 & 8 &  34. & 26. & 0.04 &  13. & 
  13. & 0.36  & 0.98 \\
1108 & 1/19 & out & 249990 & 8 &  37. & 30. & 0.15 &  45. & 
  45. & 0.31  & 0.99 \\
1109 & 1/19 & in  & 250081 & 8 &  36. & 29. & 0.11 &  34. & 
  34. & 0.32  & 0.99 \\
1110 & 1/20 & in  & 249967 & 8 &  35. & 28. & 0.08 &  27. & 
  26. & 0.32  & 0.98 \\
1111 & 1/20 & out & 234361 & 8 &  34. & 27. & 0.06 &  20. & 
  19. & 0.33  & 0.98 \\
1123 & 1/21 & out &  37126 & 8 &  61. & 74. & 0.45 & 128. & 
 128. & 0.29  & 1.00 \\
1124 & 1/21 & out &  91958 & 8 &  21. & 14. & 0.05 &  13. & 
  13. & 0.28 & 1.00 \\
1125 & 1/21 & in  & 251434 & 8 &  20. & 13. & 0.03 &   9. & 
   9. & 0.36 & 1.00 \\
1126 & 1/22 & in  &  47944 & 8 &  19. & 13. & 0.02 &   8. & 
   7. & 0.35 & 1.00 \\
1127 & 1/22 & in  &  28933 & 6 &  19. & 13. & 0.01 &   3. & 
   3. & 0.38 & 0.99 \\
1129 & 1/22 & out &  53767 & 10 & 70. & 85. & 0.55 & 195. & 
 204. & 0.37 & 1.05 \\
1132 & 1/22 & out &  52402 & 4 &  68. & 80. & 0.41 & 138. & 
 146. & 0.36 & 1.06 \\
1137 & 1/22 & in  & 249977 & 8 &  65. & 74. & 0.26 & 128. & 
 137. & 0.52 & 1.07 \\
1139 & 1/23 & in  & 153014 & 8 &  60. & 67. & 0.16 &  75. & 
  74. & 0.46 & 1.00 \\
\hline
\end{tabular}
\caption{\label{t:runlog}Running conditions from some January 1996 runs.}
\end{table}
\clearpage
\chapter{Simulations}
\section{PYTHIA}
Minimum-bias collisions at $\sqrt{s}=1.8\:$TeV are simulated using
PYTHIA version 5.702 and JETSET 7.401 \cite{pyth,pythman}.
The combination of PYTHIA and JETSET provide a commonly-used event generator
for high-energy collisions of elementary particles based on the parton model
\cite{parton}, and will be hereafter referred to as ``PYTHIA''.
The physics of the collisions that PYTHIA deals with includes

\vspace{0.10in}
the parton distribution functions of the beam particles, i.e., the flavor of 

the quarks and the fraction of energy carried by the constituent quarks 

and gluons of the incident proton and anti-proton,

\vspace{0.10in}
any radiation from the beam particles, 
such as the emission of a gluon 

from a quark, 
which leads to an initial-state shower,

\vspace{0.10in}
the hard process between the incoming partons from each shower which 

produces outgoing particles,

\vspace{0.10in}
any final-state showers produced by outgoing partons,

\vspace{0.10in}
the beam remnants, which must return to color-neutral states after 

losing the interacting partons,

\vspace{0.10in}
the hadronization of outgoing partons -- confinement requires that they

form color-neutral hadrons,

\vspace{0.10in}
and the decay of any unstable outgoing particles.

For the hadronization, PYTHIA uses a string fragmentation model, specifically
what is called the Lund model \cite{lund}.  
Its basis is a string connecting a quark and
an anti-quark which stretches as the partons move away from each other until
it breaks by creating a new quark--anti-quark pair.  This continues until
only on-mass-shell hadrons remain,
where a hadron is a color-singlet pair connected by a small piece of string.

Default values are taken for
all parameters except that particles with a mean invariant lifetime $c\tau$
greater than $1\:$cm are not decayed.  This allows the decays of K$^0_s$
and $\Lambda^0$ particles to be studied later. 

For non-single diffractive inelastic p$\bar{\mbox{p}}$ collisions at 
$1.8\:$TeV, PYTHIA gives mean numbers of particles into the acceptance per 
event of 0.61 $\pi^\pm$'s, 0.15 other charged particles, and 0.70 $\gamma$'s.  
The inclusive pseudorapidity
distributions, $dN/d\eta$, for charged pions, all charged particles, and 
photons are shown in Fig. \ref{f:dndeta}.  (An interesting note is that
the larger fraction of particles with $6<\eta<10$ in the charged distribution 
relative to the charged-pion distribution
is due to leading protons and anti-protons from double diffractive processes.)

The cross sections for various types of events included in the minimum-bias
events generated by PYTHIA are given in Table \ref{table:trig}.
Note that compared to the cross sections reported by CDF 
(Sec. \ref{sec:detect}),
PYTHIA underestimates the total and elastic cross sections, and overestimates
the single diffractive.

\section{GEANT}
The particles generated in a collision are then taken as input into a GEANT 
simulation (GEANT version 3.21 \cite{geant}) 
where they are propagated through the material of the detector.
The initial position of the collision primary particles is taken to be
$u=v=0$, and $z$ given by a Gaussian distribution with a mean of 
$7\:$in and standard deviation $16.2\:$in, in order to reproduce the
variance of the collision point seen in the data.
Cross sections and simulations of processes such as hadronic interactions, 
electromagnetic processes, ionization by charged particles,
multiple scattering, and decays in flight are included in GEANT.  
Default routines are used for all such processes, as well as the default 
energies below which particles are not tracked ($1\:$MeV for photons and 
electrons, $10\:$MeV for hadrons and muons).  
The energy deposited in material due to interactions is given at each step, 
and the secondary particles produced are added to the list of particles to be
tracked through the detector.

The energy deposited in the detector elements is used to model the
response to particle interactions seen in the real data.  At least 
$4.4(2.2)\:$MeV is required to be deposited in the 
$1\:$in\,($0.5\:$in)-thick scintillator trigger counters 
in order to keep the event.  
Figure \ref{f:scintcut}\,a shows the energy deposited in the $1\:$in-thick
scintillator counters for minimum bias events, and Fig. \ref{f:scintcut2}\,a
shows the energy deposited by single charged pions (minimum-ionizing tracks).
In Fig. \ref{f:scintcut}\,a, the peak for one minimum-ionizing particle
passing through the scintillator can be seen around $4.5\:$MeV, and the
smaller peak for two mip's at $9-10\:$MeV.
The first minimum-ionizing peak starts at about $4\:$MeV, and has a maximum at
$4.5\:$MeV.  The low energy background is fairly insignificant.  Therefore, any
cutoff energy between these values is reasonable.
The value $4.4\:$MeV is chosen to be as large as possible in order to increase 
the mean number of wires hit in the MWPC's (``NHITS'') for triggered 
events without missing a significant part of the mip peak.
(The reason for this will be given in Sec. \ref{sec:nhits}.)

For the GEANT simulation, the gas in each MWPC is segmented into 128 pieces 
corresponding to the volume which surrounds each wire in the actual chambers.  
The wires are not included in the simulation.  
Plots similar to those for the scintillator energy are given for the
chamber pulse heights (energy deposited in a segment) in 
Fig. \ref{f:scintcut}\,b and Fig \ref{f:scintcut2}\,b.
For single charged pions, the low energy background which falls off around
$0.3-0.4\:$keV where the mip peak begins is presumably due to tracks which pass
through more than one segment and deposit some fraction of the minimum-ionizing 
energy in each segment.  For minimum-bias events, the low energy background
is much higher and extends into the minimum-ionizing peak.
The cutoff value chosen to represent a hit wire is $0.4\:$keV, where the 
background falls to a level comparable to the mip contribution.
Note that in the actual MWPC's, ionized electrons created by the charged tracks 
drift to the positively charged wires and the amount of charge collected on a 
wire determines the pulse height.  This may not be linearly related to the 
energy deposited in the gas; however, direct comparisons of pulse heights in 
the GEANT and in the data discussed in Sec. \ref{sec:nhits} show that this is
not an unreasonable assumption. 

The energy deposited in each cell of the lead-scintillator
electromagnetic calorimeter is recorded, 
so that it can be used to calibrate the signals from the actual calorimeter.

The other crucial detector element which is included is converter of 
various thicknesses.

Because of the large observed background of particles due to interactions 
in material in the forward (detector) region, many objects not related to 
the detector are included in the simulation.  These include the Tevatron
beam pipe and abort pipe, the main ring and its abort, support stands, 
vacuum pumps, the concrete floor, etc.  In spite of all these additions,
the mean number of wires hit in each MWPC is lower than that in the real
data by about a factor of two (see Figs. \ref{f:support24}, \ref{f:support}).

\subsection{GEANT NHITS study}
\label{sec:nhits}

The NHITS distribution is dependent on the pulse-height and trigger cuts.
A lower pulse-height cut allows lower-energy hits to be counted as hits by
actual tracks, thus increasing the NHITS in each event.  A higher cut on
energy deposited in the scintillator counters in order to trigger (i.e. to keep
that event) tends to cut out events with less energy or number of tracks going
into the detector, so that the mean NHITS of the remaining events is higher
than that for all events.  The cuts made on the GEANT data are intended
to match those in the real data, namely to make cuts based on the signatures
of real charged tracks going through the chambers and scintillator.
However, we tried varying these cuts through all reasonable (and even
unreasonable) values to see if the NHITS distribution of the data could
be reproduced by the GEANT simulation.  Only 12 of the 24 MWPC's read out 
pulse-height information, so we use only those chambers in the following 
analysis.
The pulse height is calibrated in terms of charge deposited on the wires,
not energy deposited in the gas as it is in GEANT, and is not the same in all 
chambers.  The
first step, therefore, is to adjust the energy scales of the chambers
in the GEANT so that the mean of a Gaussian fit to the peaks of
both the real and GEANT pulse-height distributions are equivalent.
The result is shown in Figs. \ref{f:michplshgt}, \ref{f:michplshgt2}.  
The mean NHITS in these
12 chambers is 93 for run 867.  It is clear in Table \ref{table:plshgt} 
that there is no combination of pulse-height and scintillator cuts which
gives a mean NHITS this high in the GEANT data.  For the lowest possible
pulse-height cut ($\mbox{pulse height}>0$), and the highest shown cut
on energy deposited in the scintillator ($8\:$MeV, which practically requires
two charged tracks rather than one), the mean NHITS in those 12 chambers is 
only 79.

The next attempts to find an explanation for the lower NHITS in GEANT 
include the following:
\begin{enumerate}
\item Increasing the thickness of the beampipe from $0.03\:$in to $0.035\:$in 
to check for GEANT problems with the boundary;
\item lowering the energy thresholds of particles for GEANT to track
to 1/2 the default value;
\item changing the defaults of GEANT so that delta rays are produced;
\item supposing that the collision point is not where we think it is 
relative to the pipe, and therefore changing the $z$ of the collision point 
to $10\:$in closer to the pipe, or changing $x$ and $y$ by $1\:$in each towards 
the pipe.
\end{enumerate}
None of these have much effect on the NHITS.  Next we tried adding the 
main ring and correcting the description of the abort pipe to include the 
change of material to steel at $z>275\:$in,
which increased the mean NHITS by about 20\%.
Encouraged by this, we added more of the material nearby the detector:
$2\:$ft-thick concrete floor, tunnel wall, chamber stands,
Tevatron-beampipe and main-ring support stands, and vacuum ion pumps.

The effect on NHITS of material close to the wire chambers was studied
in more detail by removing that material.  The beampipe is the largest 
source of non-collision-point tracks hitting the chambers (Fig. \ref{f:nopipe}).
The abort pipe and the G-10 frames of the wire chambers 
were also studied (Figs. \ref{f:noabort}, \ref{f:nog10}).
The rear chambers are hit by tracks from these sources more than the front; 
in fact, when all three of the sources are removed, the mean number of hits 
per event in each chamber is approximately a constant between 3.5 and 4
(Fig. \ref{f:noall}), as opposed to the roughly linear increase in hits 
going from about 4 in the front chamber to more than 8 in the last, as in
Fig. \ref{f:support24}.
(This is, of course, for lead-out runs.  For lead-in runs, the 
number of hits in the chambers behind the lead is much greater than that in the 
front due to conversions of primary photons.) In the real data, the number of 
hits per chamber increases faster than 
linearly towards the rear of the detector.  The addition of the beampipe 
supports in the GEANT simulation reproduces this effect fairly well.
These observations are what led to the compression of the wire chambers
which moved them closer to the collision point before the production running.

It has been noted \cite{Pumplin} that the multiplicity distributions of 
low-$p_T$ charged particles in PYTHIA were incorrectly extrapolated from 
collider data at high $p_T$.  Since the MiniMax detector is more sensitive 
than conventional collider detectors to low-$p_T$ particles, these can
potentially contribute significantly to the NHITS.
We tried using HERWIG, another commonly-used event generator (especially for 
jet studies) to see if the NHITS distribution might be better reproduced. 
Comparisons of $dN/d\eta$ from non-diffractive PYTHIA events and HERWIG events
show that the mean number of particles produced is higher for PYTHIA 
(see Fig. \ref{f:ndpythhw} for $dN_{ch}/d\eta$).  
The NHITS distribution obtained using HERWIG was not significantly different.

Another attempt to increase the mean NHITS in the simulations was
to change the defaults for multiple interactions in PYTHIA.
The occurrence of hard interactions of more than one parton pair in a hadronic
collision 
is not well understood, and PYTHIA provides several models.
The default is that the probability of multiple interactions is equal for all
events, with a  sharp $p_{\perp min}$ cut-off.
It was suggested to us that the option with ``multiple interactions assuming a 
varying impact parameter and a hadronic matter overlap consistent with a double 
Gaussian matter distribution \dots 
with a continuous turn-off of the cross section at $p_{\perp 0}$'' 
(Ref. \cite{pythman}, p. 222) would give larger 
multiplicity fluctuations, and therefore possibly larger mean NHITS.
The effect of using this model is also negligible. 

A study was done by the ALICE Collaboration on MWPC's which are intended to be
used in a muon spectrometer at the LHC \cite{alice}.  The chambers are 
remarkably similar to the ones used by MiniMax, and therefore the results
obtained in the study may be relevant to the performance of the MiniMax
chambers.  A pion beam was aimed at a lead absorber, and background particles
from interactions in the lead were detected in a wire chamber located at the
position which was predicted by simulations as the location of the shower 
maximum.  A GEANT simulation proved to underestimate the measured number of 
charged particles into the chamber by a factor of 2-4.  A simulation using
stand-alone FLUKA (as opposed to the option of using FLUKA subroutines inside 
GEANT) produced a distribution of particles much closer to what was actually
seen.  Further analysis suggested that a significant contribution to the
hits in the chambers was due to neutrons interacting in the mylar
face of the chamber and knocking out protons which were then detected.
Such interactions of low-energy neutrons are included in the stand-alone FLUKA.
We have not been able to obtain the stand-alone FLUKA simulation code.

\subsection{GEANT trigger rates}
The trigger rates and cross sections for events in the GEANT simulation
from the minimum-bias PYTHIA input are shown in Table \ref{table:trig}.
The total fraction of events which pass the trigger in GEANT is slightly
more than 50\%, and is higher for lead-in runs than for lead-out by about
1\% for all types of events.  Single diffractive events in which the proton
is dissociated are triggered on about 15\% of the time, and about 10\% of
those with a fragmenting anti-proton pass the trigger.  For double 
diffraction, the fraction is about 20\%. 

Note that if we add the fraction of triggered single diffractive events (26\%)
times the single diffractive cross section from CDF (9.46 mb) to the
non-single diffractive fraction (79\%) times the corresponding
CDF cross section (50.87 mb),
the result is that we should trigger on 42.7 mb, which agrees very well with
the estimate in Sec. \ref{sec:trig}.

\section{DCC generator}
\label{sec:dccgen}
The operational definition of DCC used by the MiniMax collaboration is that
DCC is a cluster of pions (domain) with neutral fraction $f$ given by the
$1/(2\sqrt{f})$ distribution and with near-identical momenta which is
non-relativistic in the DCC rest frame.
For the studies presented here, the domain size is taken to be on the order of 
the detector acceptance, and the momentum is such that the DCC is aimed
at the center of the acceptance with a reasonably large $p_T$ in the lab frame.

We assume that the pions are non-relativistic in the cm frame of the DCC 
  domain (and therefore refer to the domain as a ``snowball''). 
  We take the momentum distribution to be Gaussian with mean 
  $\left<\vec{p}\right> =0$ and
  variance $\left<\vec{p}\cdot\vec{p}\right> =3\sigma_p^2$.
  Since the DCC domain must be large enough to contain physical pions, 
  i.e. have dimensions on the order of a few fm, 
  uncertainty principle arguments require a small momentum, 
  $\sigma_p\sim 50 - 100\:$MeV.
  We use the value $\sigma_p=100\:$MeV, which gives the pions a relatively
  large momentum in the lab frame, making them easier to detect.
If the pions are not too relativistic in the cm frame, then the boosted DCC 
  domain is approximately a circular disk of radius in lego space 
  $R_{DCC}\sim\sigma_p/p_T$.  
  For the MiniMax detector, with acceptance approximately a circle with radius
  0.6,  a domain with $R_{DCC}\sim 0.7-1.0$ would be easiest to find, 
  which implies $p_T=140\:$MeV.
  This size seems reasonable since the typical radius of a jet is 0.7.

We also assume that the number of pions in the DCC domain is approximately 
  independent of 
  the pseudorapidity of the center of the boosted domain.  
  A result of this is that the ratio $\psi$ of the mean energy density of 
  a DCC pion to that of a generically-produced pion is approximately constant.
  Then the mean number of pions is given by
  $$ \left< N_\pi\right> \sim \psi {p_T^{gen}\over p_T} 
     \left( {d^2N\over d\eta d\phi}\right)_{gen} \delta\eta\,\delta\phi,$$
or, substituting in the geometric values,
  $$ \left< N_\pi\right> \sim \psi 
     {p_T^{gen}\over p_T} 
  {1\over 2\pi}\left( {dN\over d\eta}\right)_{gen}
  \pi R_{DCC}^2.$$
We take $(dN/d\eta)_{gen}=6$ in the region of the acceptance, 
$p_T^{gen}=500\:$MeV, and $\psi=1$, which gives a mean number of pions 
$\left< N_\pi\right>=5.0$.
For a given event, the number of pions is taken from
a Poisson distribution about this mean.

Next the charge of the pions is determined according to the $1/(2\sqrt{f})$ 
distribution.  The neutral fraction in a given event is generated using the 
transformation method, where, if $x$ is a uniform deviate, then $f=x^2$ is 
distributed
according to $1/(2\sqrt{f})$.  A uniform deviate $y_i$ is then generated for 
each of the pions ($i=1,N_\pi$); if $y_i<f$, the pion is defined to be neutral, 
otherwise it is defined to be charged.

%
%

Figures \ref{f:dccp}-\ref{f:dccn} show for the DCC pions:
the momentum distributions, the distributions in $\eta$ and $\phi$ for a
domain aimed at $\eta=4.1$ and $\phi=0.75$, the neutral-fraction distribution
and the number distribution.  Histograms of the number of events with 
given numbers of charged and neutral pions, for both total numbers and 
the numbers of those that enter the acceptance are shown in 
Fig. \ref{f:dccnchn}, and can be compared with that from generically-produced
pions from PYTHIA in Fig. \ref{f:dccnchn_pyth}.

Unless noted otherwise, all further mention of the DCC generator refers to the 
use of the parameters given here.

\vfill\newpage
\begin{figure}[h] \vspace{-0.5in}\hspace*{-0.3in}
\psfig{file=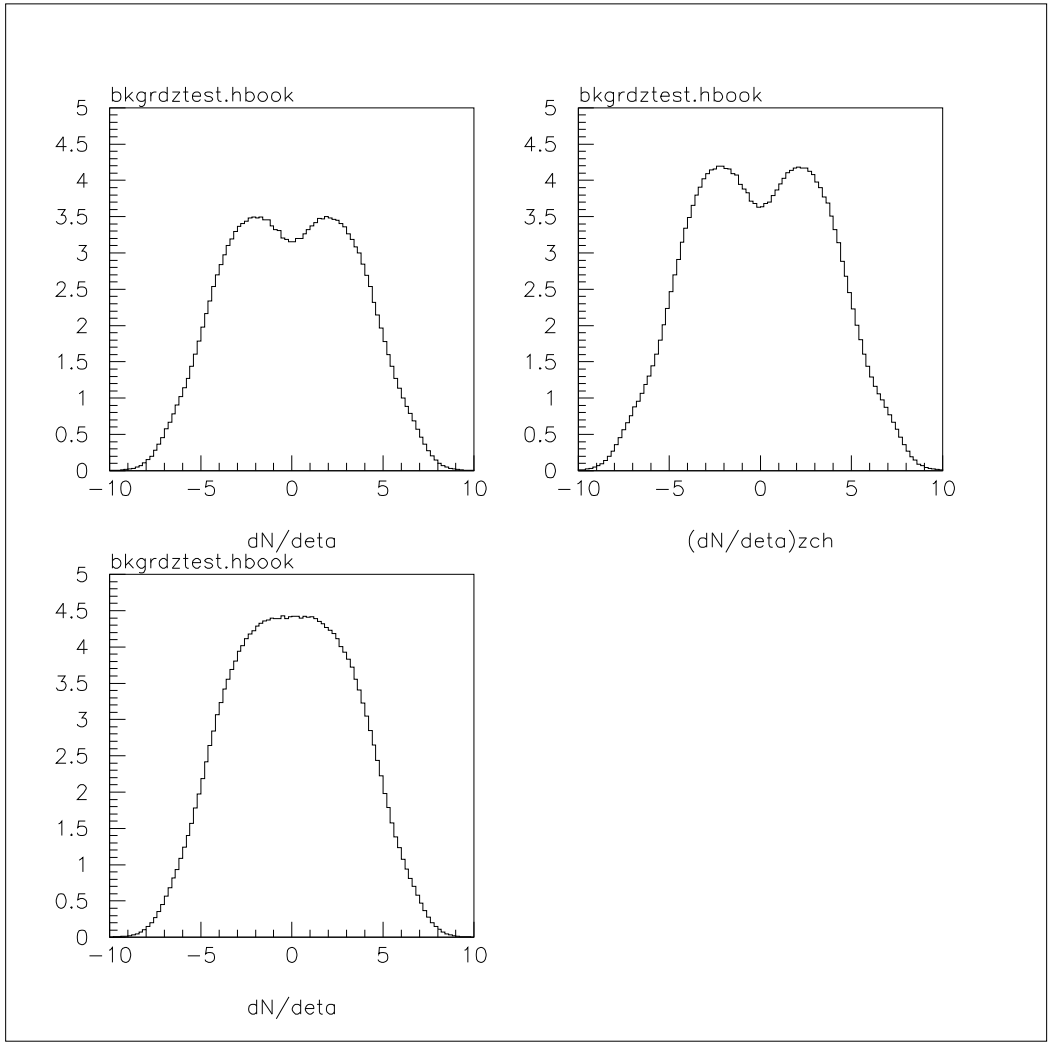,width=6in}
\vspace{-1in}
\caption{Pseudorapidity distribution $dN/d\eta$ for charged pions, all charged
particles, and photons.}
\label{f:dndeta} \end{figure}
\vfill\newpage
\begin{figure}[h] \vspace{-0.5in}\hspace*{-0.3in}
\psfig{file=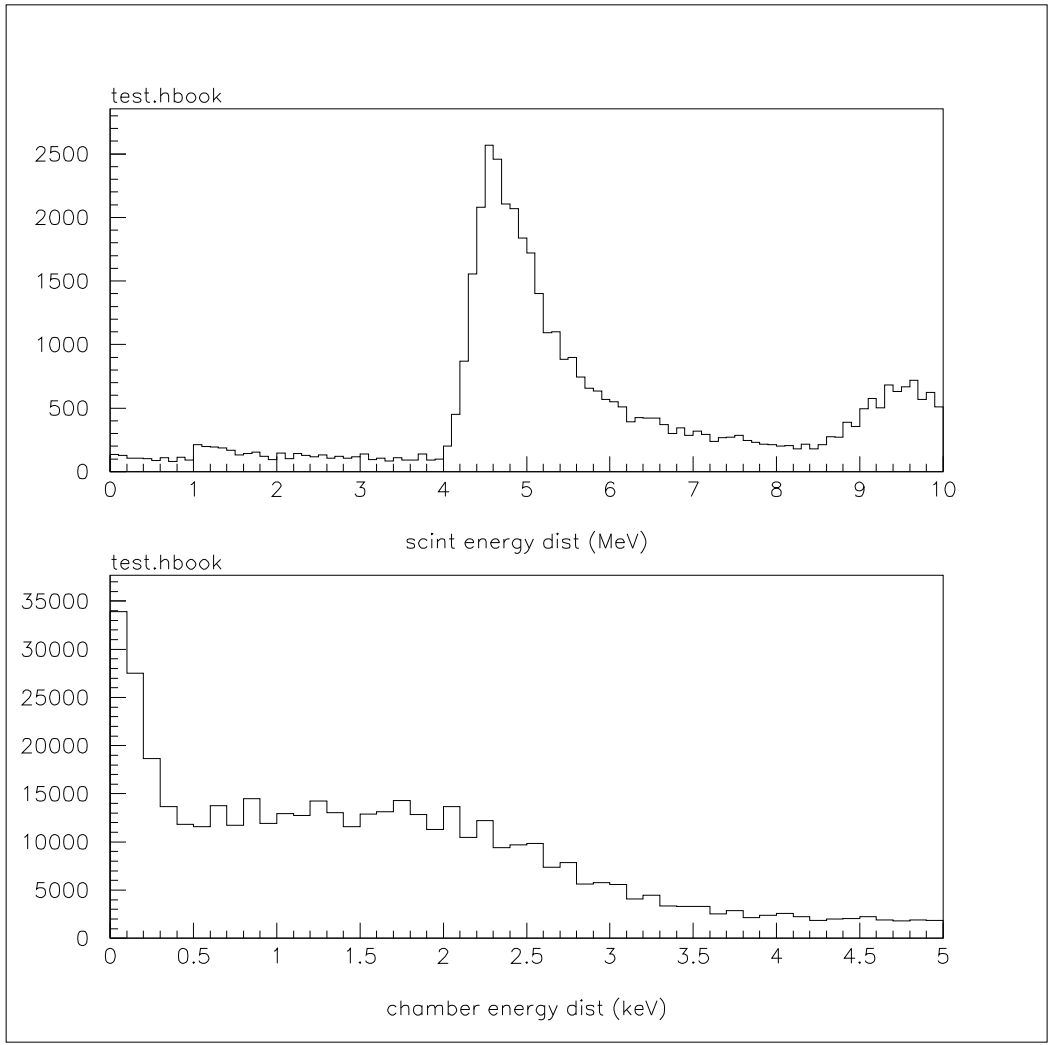,width=6in}
\vspace{-1in}
\caption[Energy deposited in scintillator and chambers in minimum bias events.]
{Energy deposited in (a) scintillator and (b) chambers in minimum bias events.}
\label{f:scintcut} \end{figure}
\vfill\newpage
\begin{figure}[h] \vspace{-0.5in}\hspace*{-0.3in}
\psfig{file=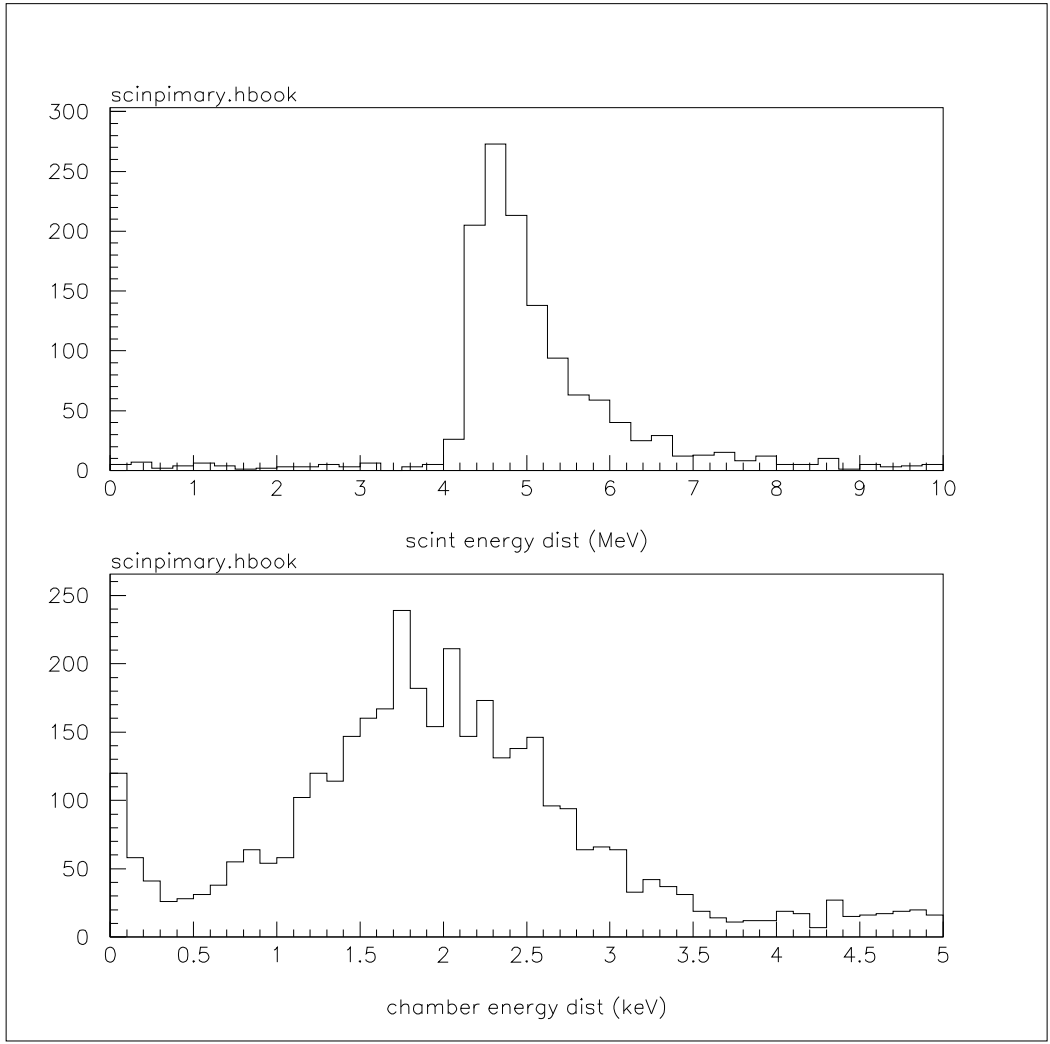,width=6in}
\vspace{-1in}
\caption[Energy deposited in scintillator and chambers by single charged pions.]
{Energy deposited in (a) scintillator and (b) chambers by single 
charged pions.}
\label{f:scintcut2} \end{figure}
\begin{figure}[h] \vspace{-0.5in}\hspace*{-0.3in}
\psfig{file=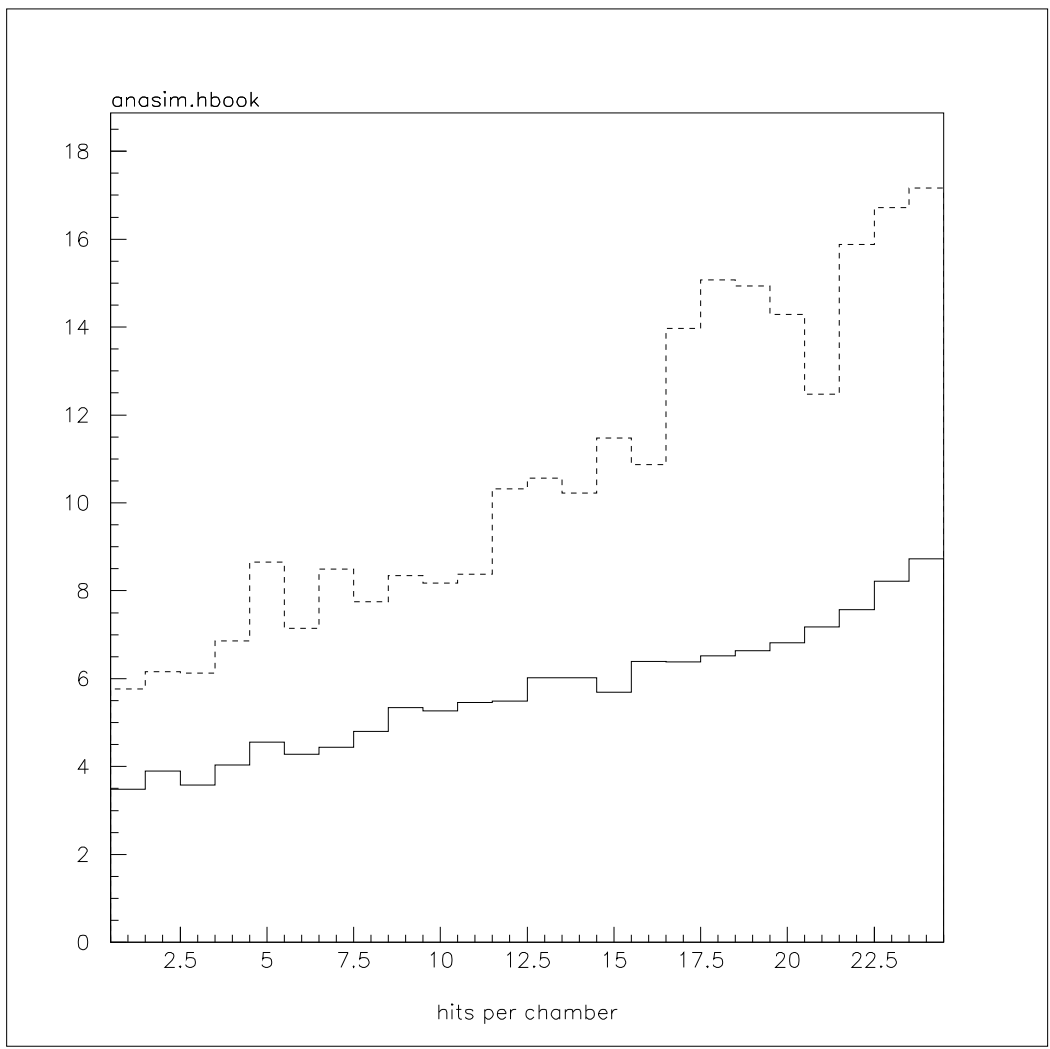,width=6in}
\vspace{-1in}
\caption[Mean number of wire hits per chamber: real and GEANT with
beam supports, etc.]{Mean number of wire hits per chamber: real (dashed) and 
GEANT with beam supports, etc (solid).}
\label{f:support24} \end{figure}
\begin{figure}[h] \vspace{-0.5in}\hspace*{-0.3in}
\psfig{file=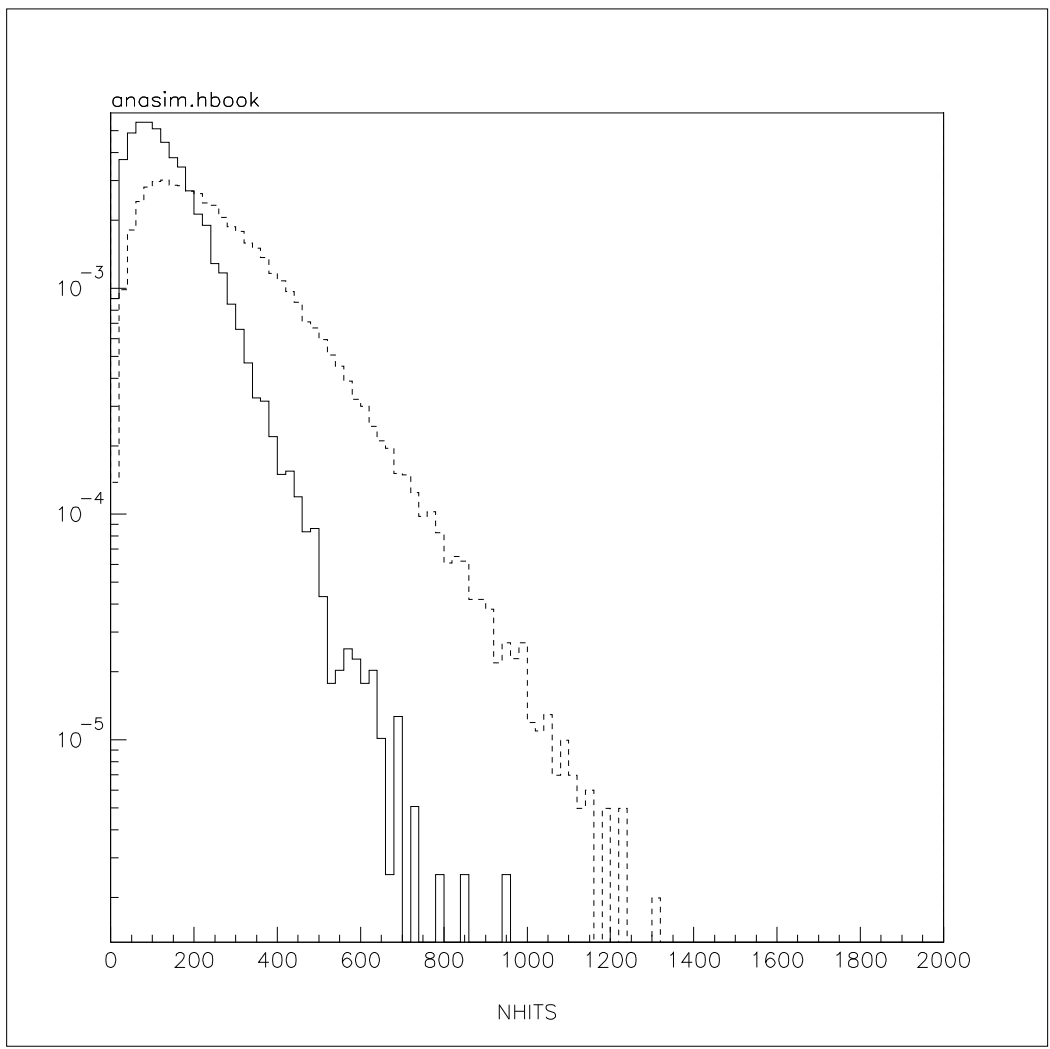,width=6in}
\vspace{-1in}
\caption[Distribution of NHITS: real and GEANT with beam supports, etc.]
{Distribution of NHITS: real (dashed) and GEANT with beam supports, etc
(solid).} 
\label{f:support} \end{figure}
\begin{figure}[h] \vspace{-0.5in}\hspace*{-0.3in}
\psfig{file=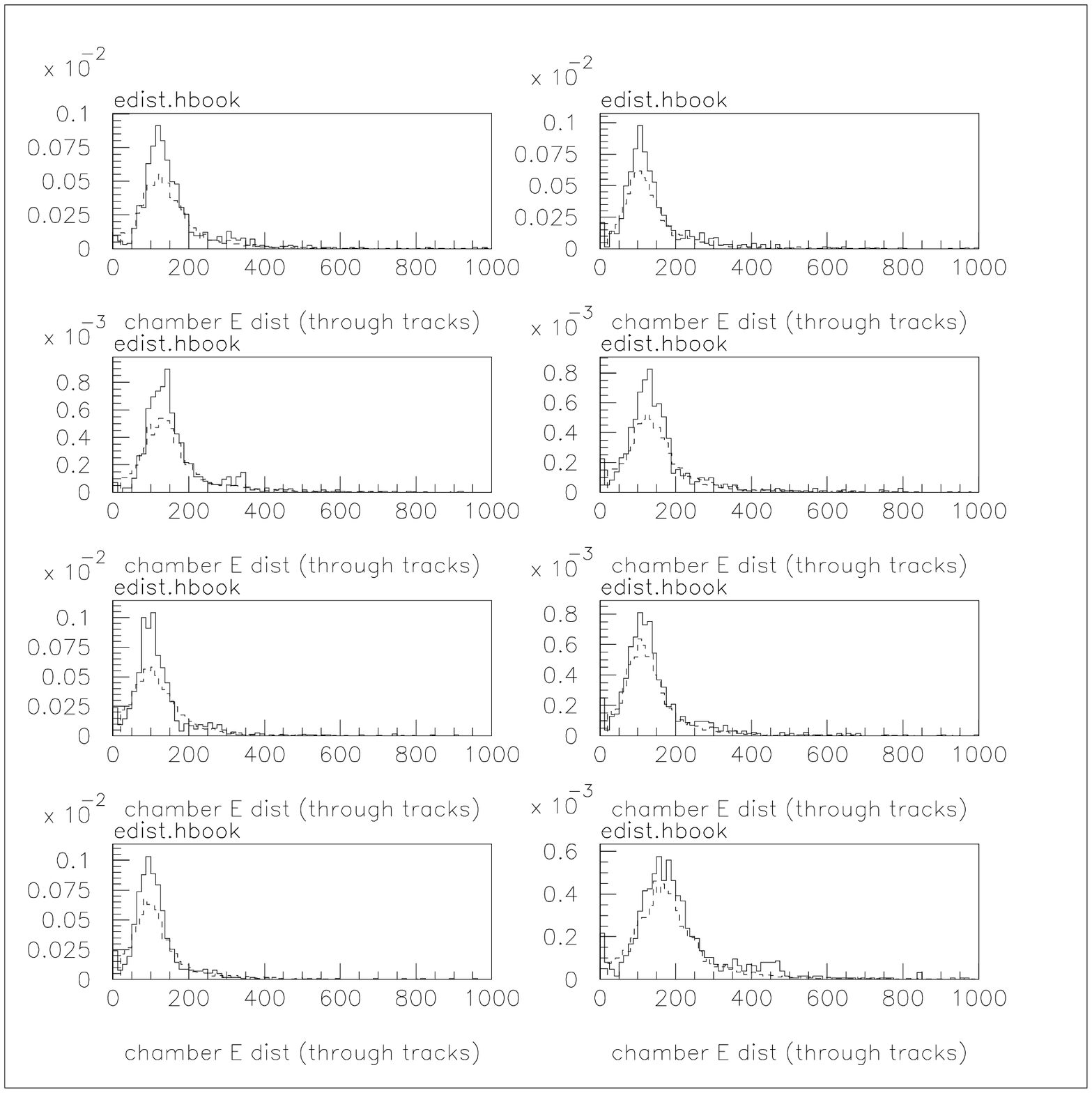,width=6in}
\vspace{-1in}
\caption[Pulse height distributions for chambers 1-4, 9-12, real and GEANT.]
{Pulse height distributions for chambers 1-4, 9-12, real (dashed)
and GEANT (solid).} 
\label{f:michplshgt} \end{figure}
\begin{figure}[h] \vspace{-0.5in}\hspace*{-0.3in}
\psfig{file=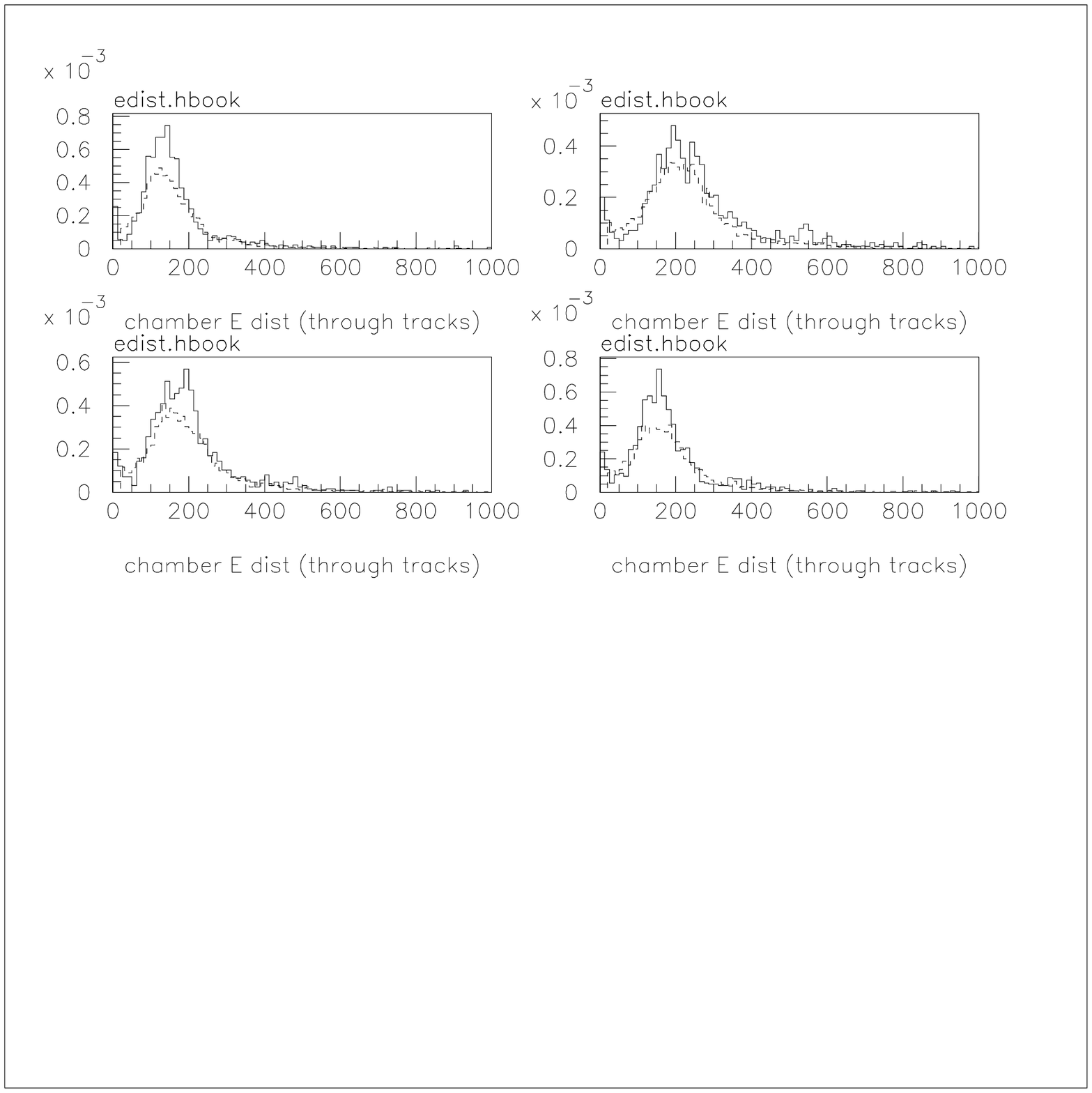,width=6in}
\vspace{-1in}
\caption[Pulse height distributions for chambers 21-24: real and GEANT.]
{Pulse height distributions for chambers 21-24: real (dashed) 
and GEANT (solid).} 
\label{f:michplshgt2}\end{figure}
\begin{figure}[h] \vspace{-0.5in}\hspace*{-0.3in}
\psfig{file=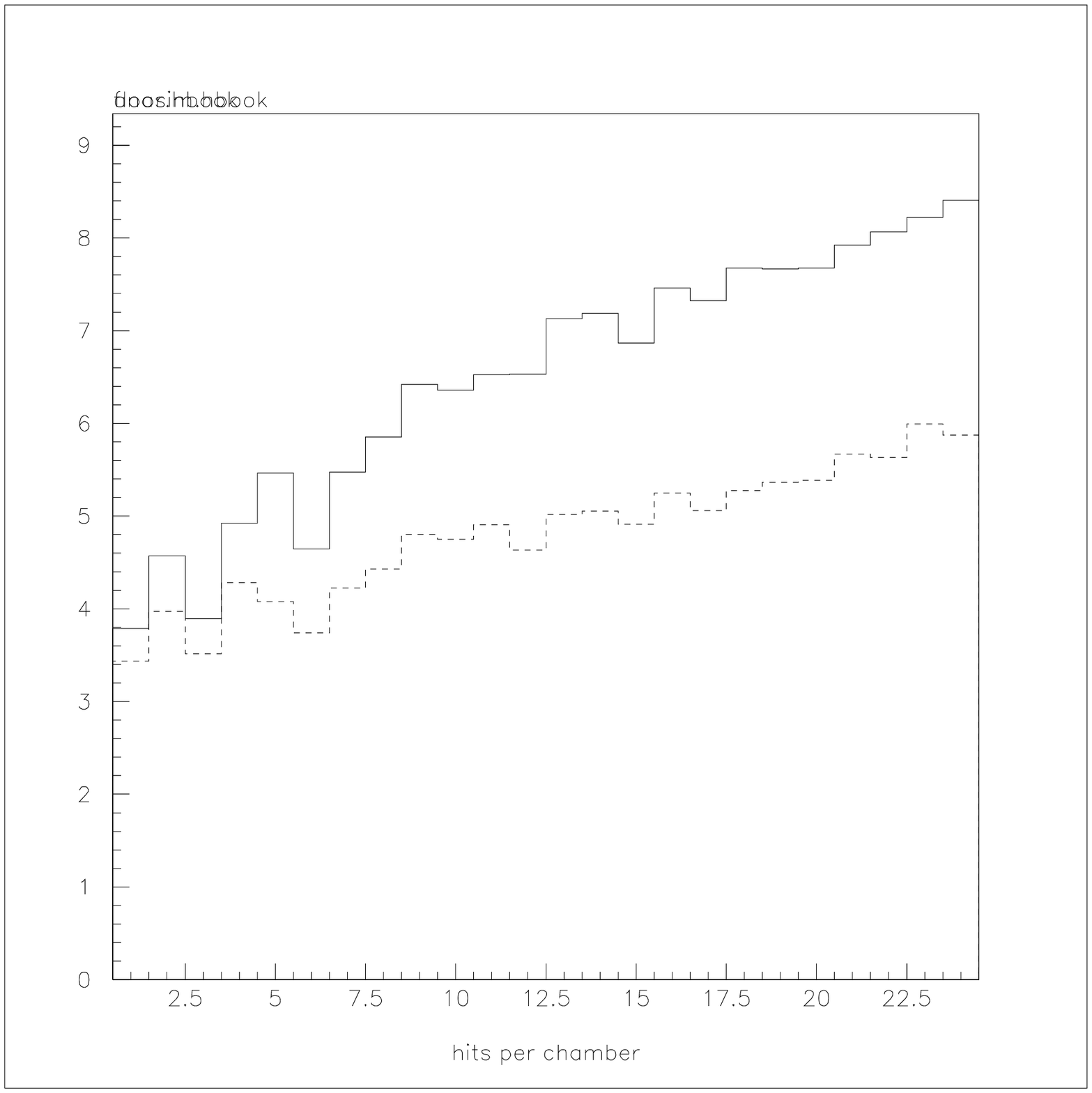,width=6in}
\vspace{-1in}
\caption[Mean number of wire hits in each chamber and with the beampipe 
removed.]{Mean number of wire hits in each chamber (solid) and with the 
beampipe removed (dashed).}
\label{f:nopipe} \end{figure}
\begin{figure}[h] \vspace{-0.5in}\hspace*{-0.3in}
\psfig{file=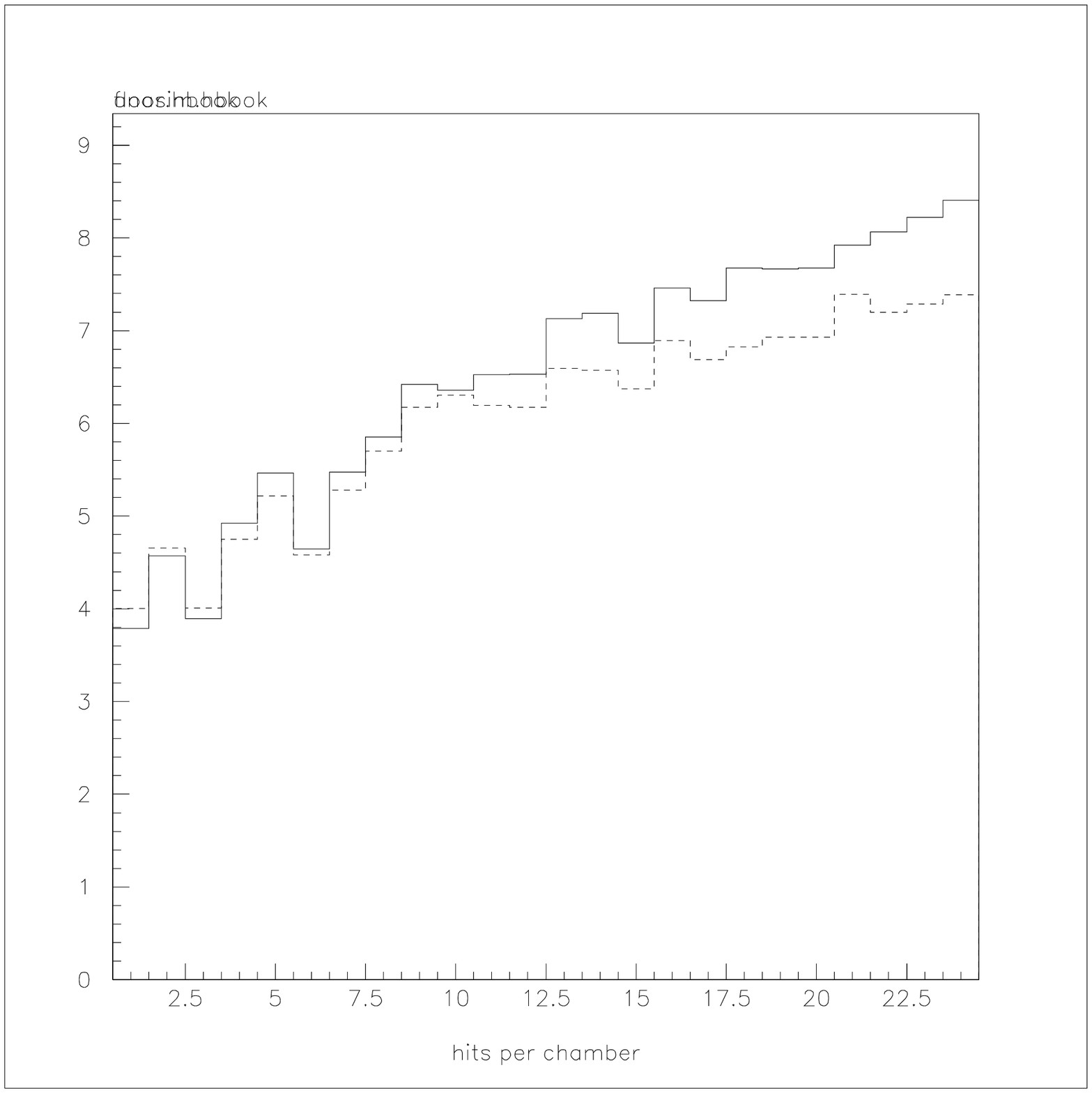,width=6in}
\vspace{-1in}
\caption[Mean number of wire hits in each chamber and with the abort pipe 
removed.]{Mean number of wire hits in each chamber (solid) and with the abort 
pipe removed (dashed).}
\label{f:noabort} \end{figure}
\begin{figure}[h] \vspace{-0.5in}\hspace*{-0.3in}
\psfig{file=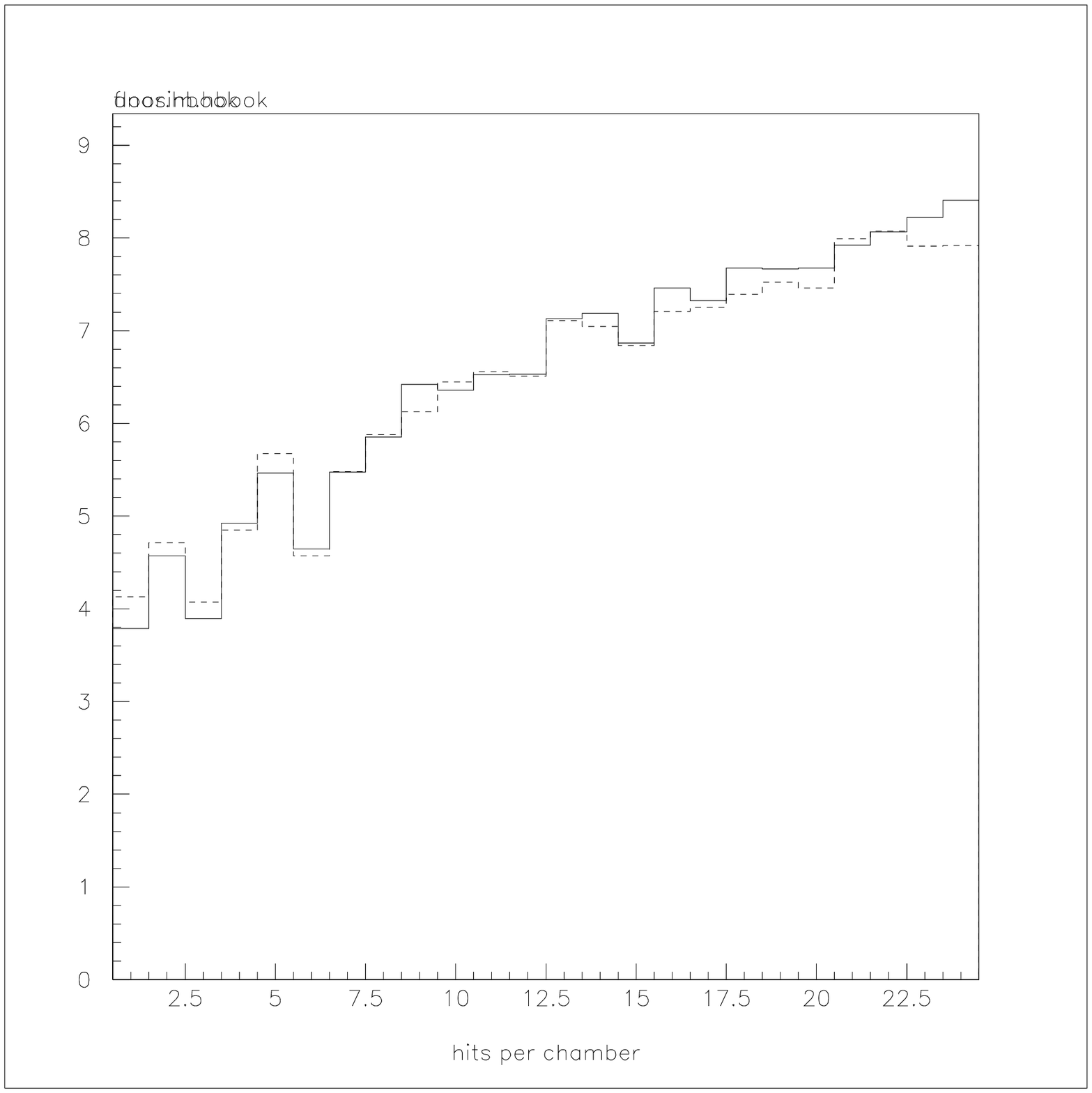,width=6in}
\vspace{-1in}
\caption[Mean number of wire hits in each chamber and with the chamber frames 
removed.]{Mean number of wire hits in each chamber (solid) and with the chamber 
frames removed (dashed).}
\label{f:nog10} \end{figure}
\begin{figure}[h] \vspace{-0.5in}\hspace*{-0.3in}
\psfig{file=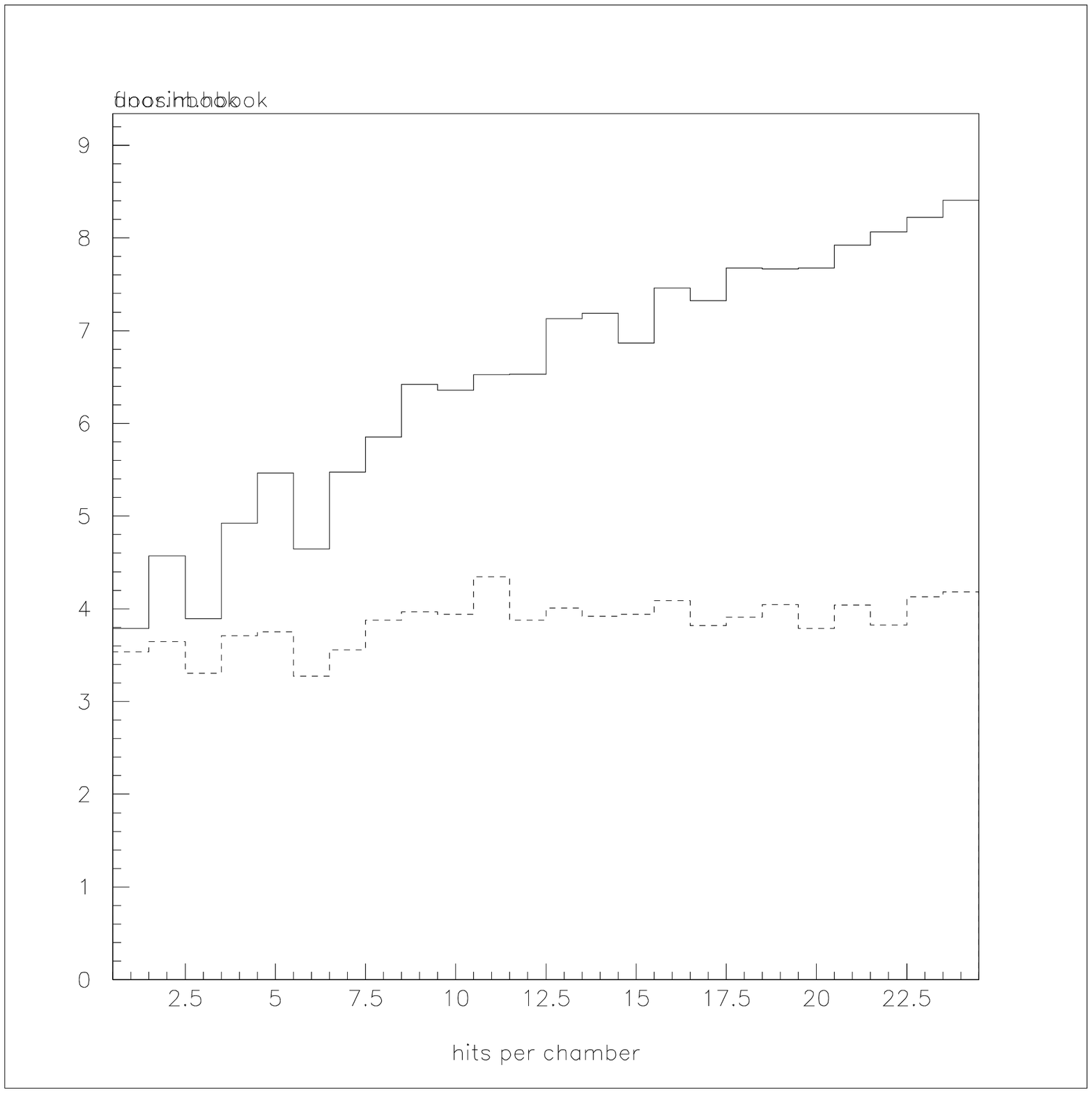,width=6in}
\vspace{-1in}
\caption[Mean number of wire hits in each chamber and with the beampipe, abort 
pipe, and chamber frames removed.]{Mean number of wire hits in each chamber 
(solid) and with the beampipe, abort pipe, and chamber frames removed (dashed).}
\label{f:noall} \end{figure}
\begin{figure}[h] \vspace{-0.5in}\hspace*{-0.3in}
\psfig{file=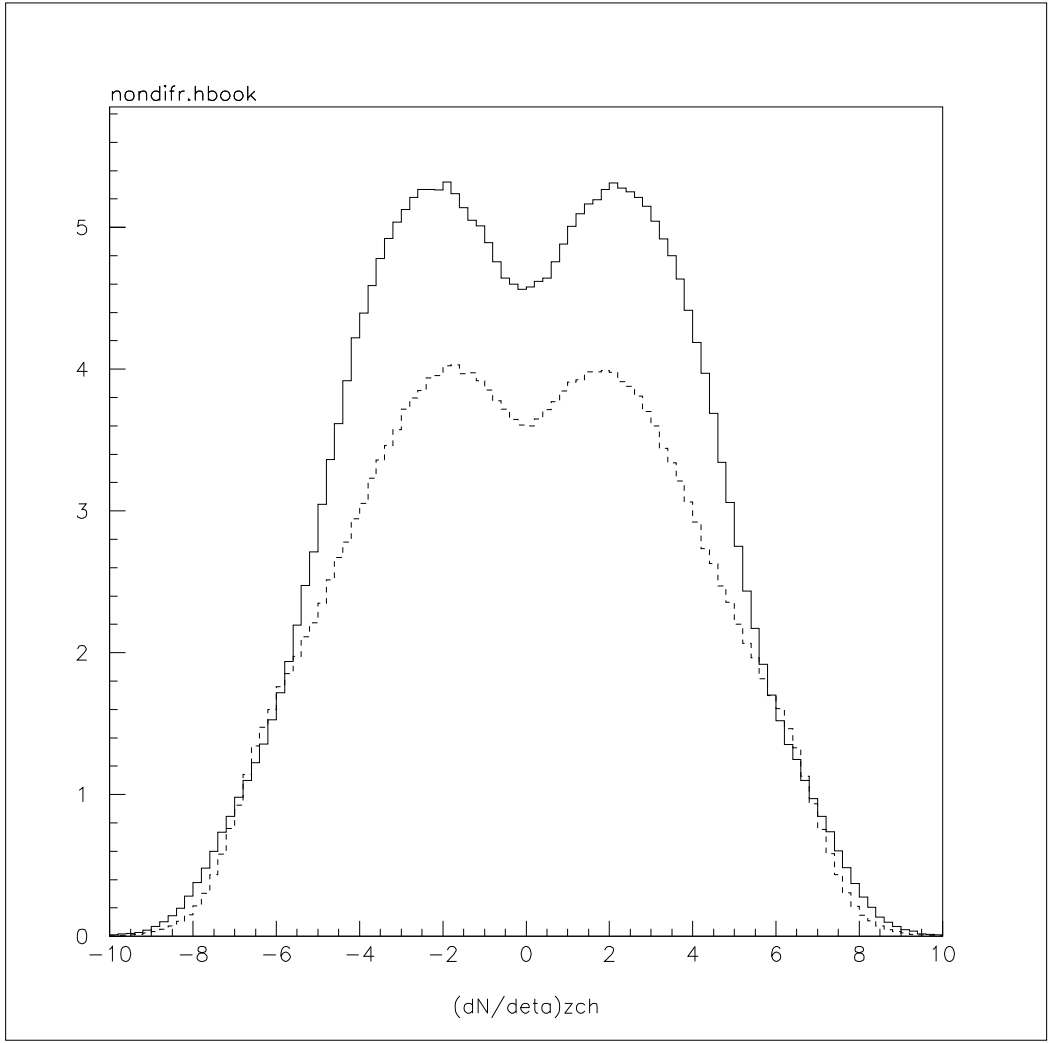,width=6in}
\vspace{-1in}
\caption[Charged-particle multiplicity $dN_{ch}/d\eta$ for non-diffractive
PYTHIA and HERWIG.]{Charged-particle multiplicity $dN_{ch}/d\eta$ for 
non-diffractive PYTHIA (solid) and HERWIG (dashed).}
\label{f:ndpythhw} \end{figure}
\begin{figure}[h] \vspace{-0.5in}\hspace*{-0.3in}
\psfig{file=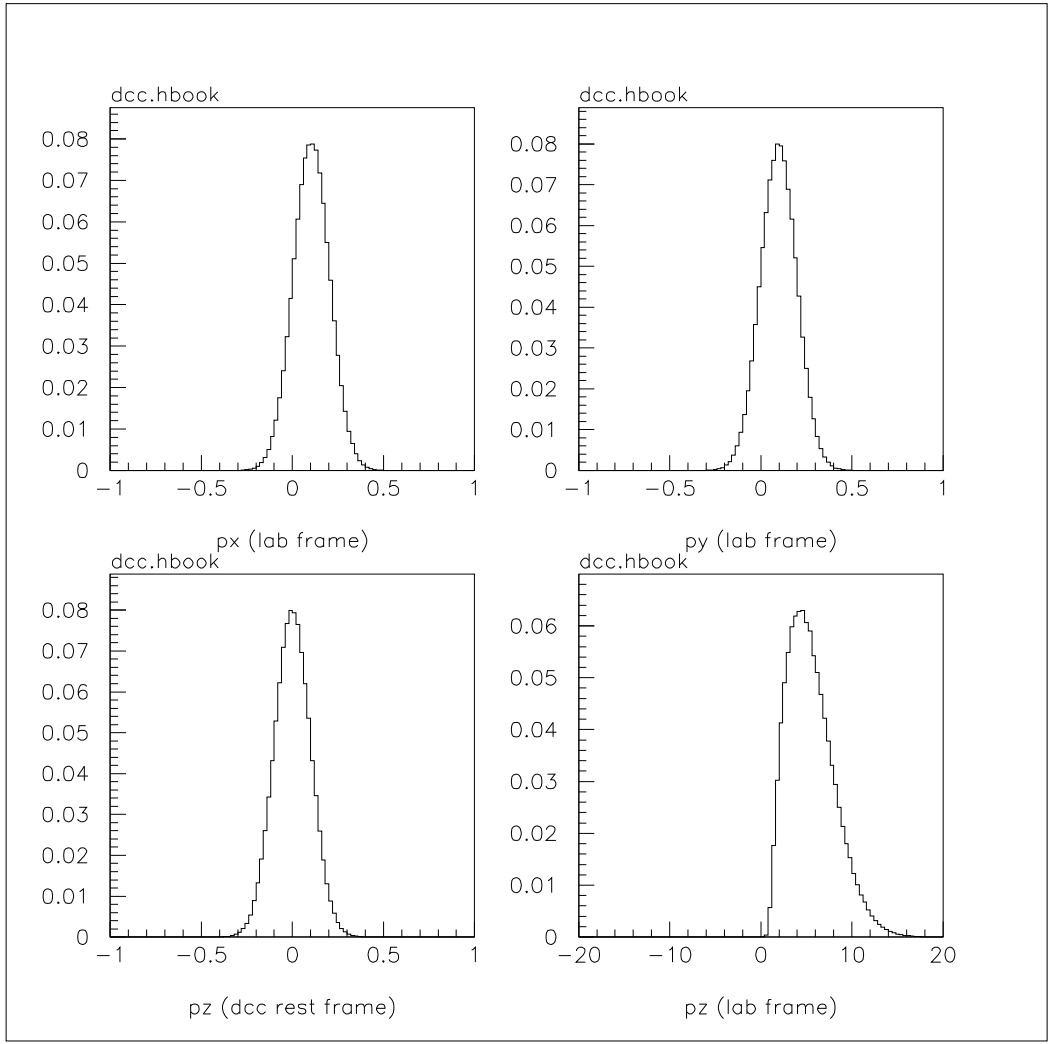,width=6in}
\vspace{-1in}
\caption{Momentum (GeV) of DCC pions: $p_x$ and $p_y$ in lab frame, $p_z$
in DCC rest frame, and $p_z$ in lab frame.}
\label{f:dccp} \end{figure}
\begin{figure}[h] \vspace{-0.5in}\hspace*{-0.3in}
\psfig{file=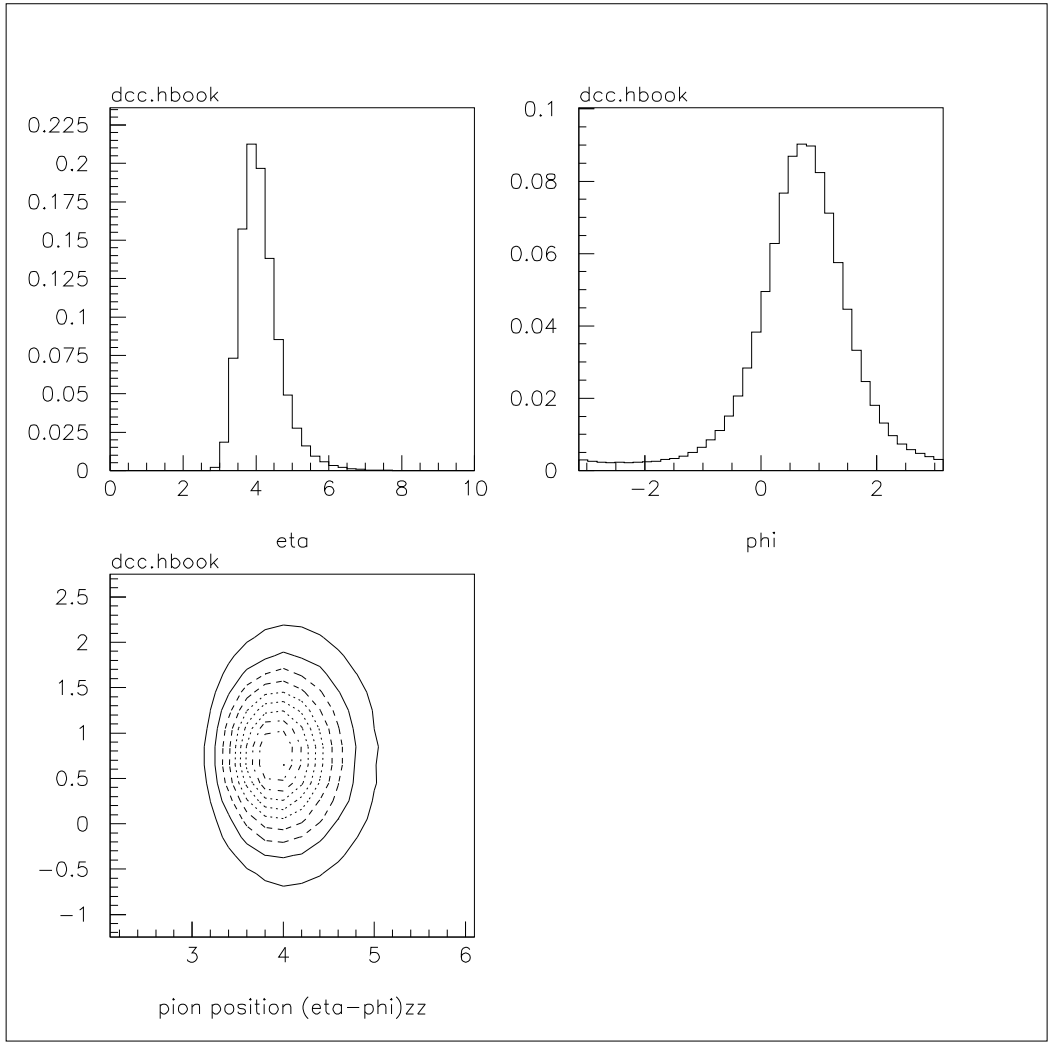,width=6in}
\vspace{-1in}
\caption{Location of the DCC pions in lego space.}
\label{f:dccetaphi} \end{figure}
\begin{figure}[h] \vspace{-0.5in}\hspace*{-0.3in}
\psfig{file=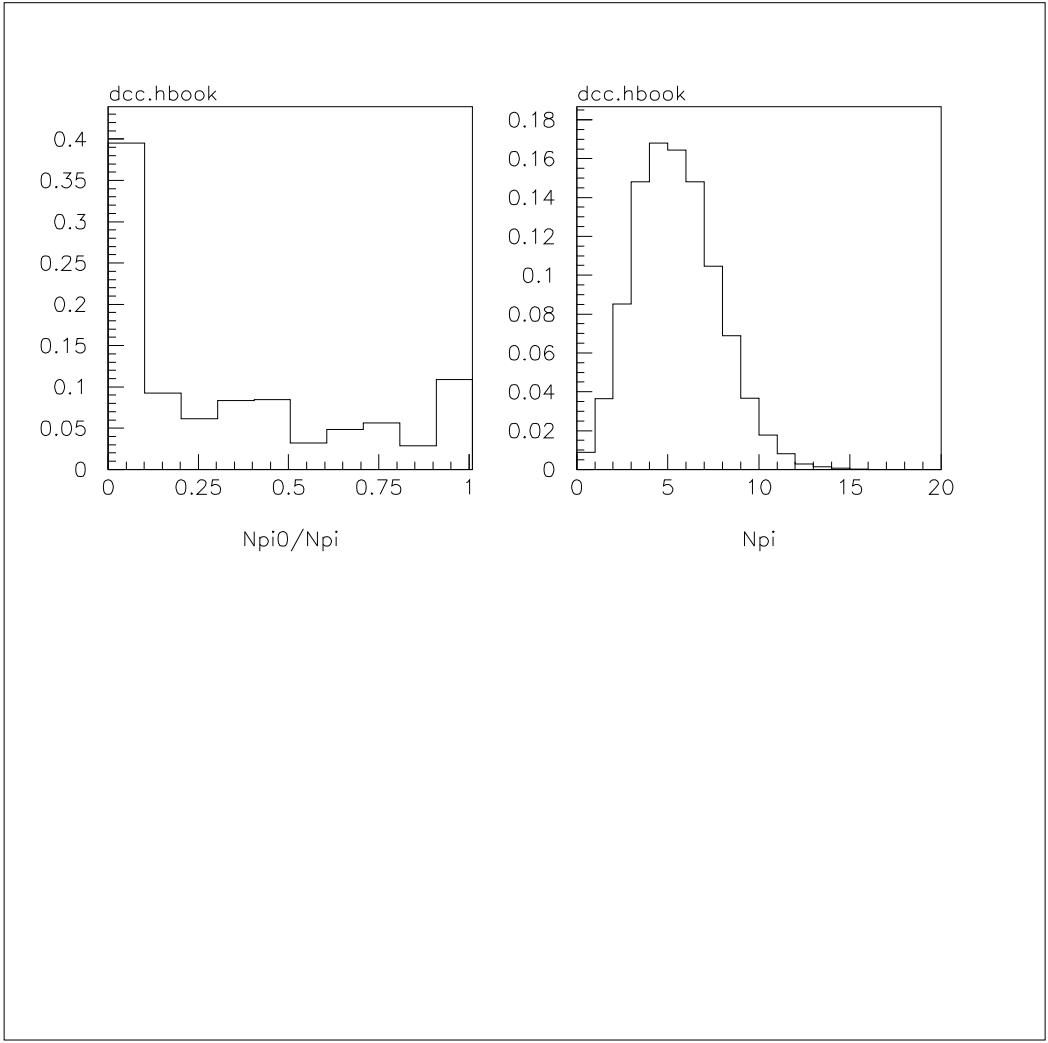,width=6in}
\vspace{-1in}
\caption{Distribution of fraction of DCC pions which are neutral 
($N_{\pi^0}/N_{\pi}$) and total number of DCC pions ($N_{\pi}$).} 
\label{f:dccn} \end{figure}
\begin{figure}[h] \vspace{-0.5in}\hspace*{-0.3in}
\psfig{file=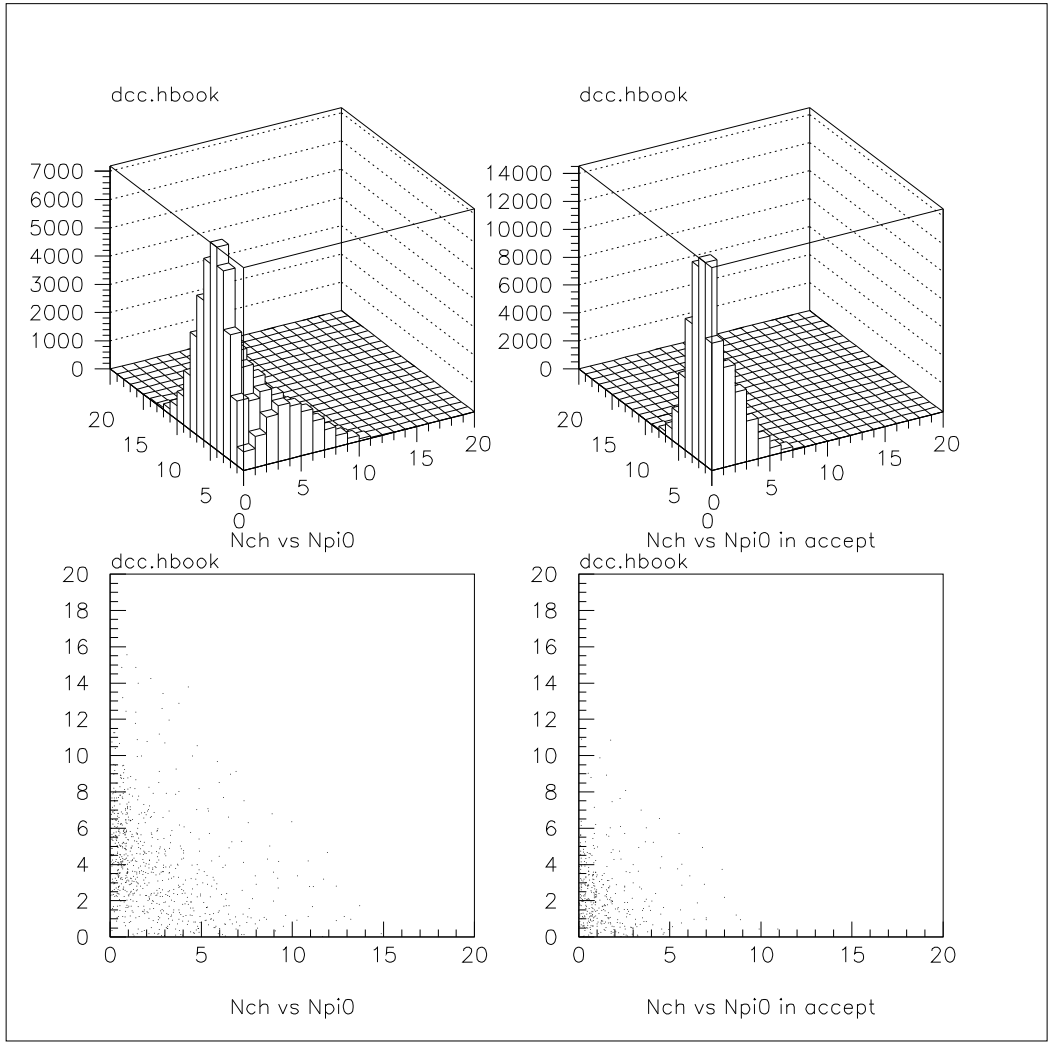,width=6in}
\vspace{-1in}
\caption{Number of charged vs number of neutral DCC pions: total number
produced, and the number which enter the MiniMax acceptance.}
\label{f:dccnchn} \end{figure}
\begin{figure}[h] \vspace{-0.5in}\hspace*{-0.3in}
\psfig{file=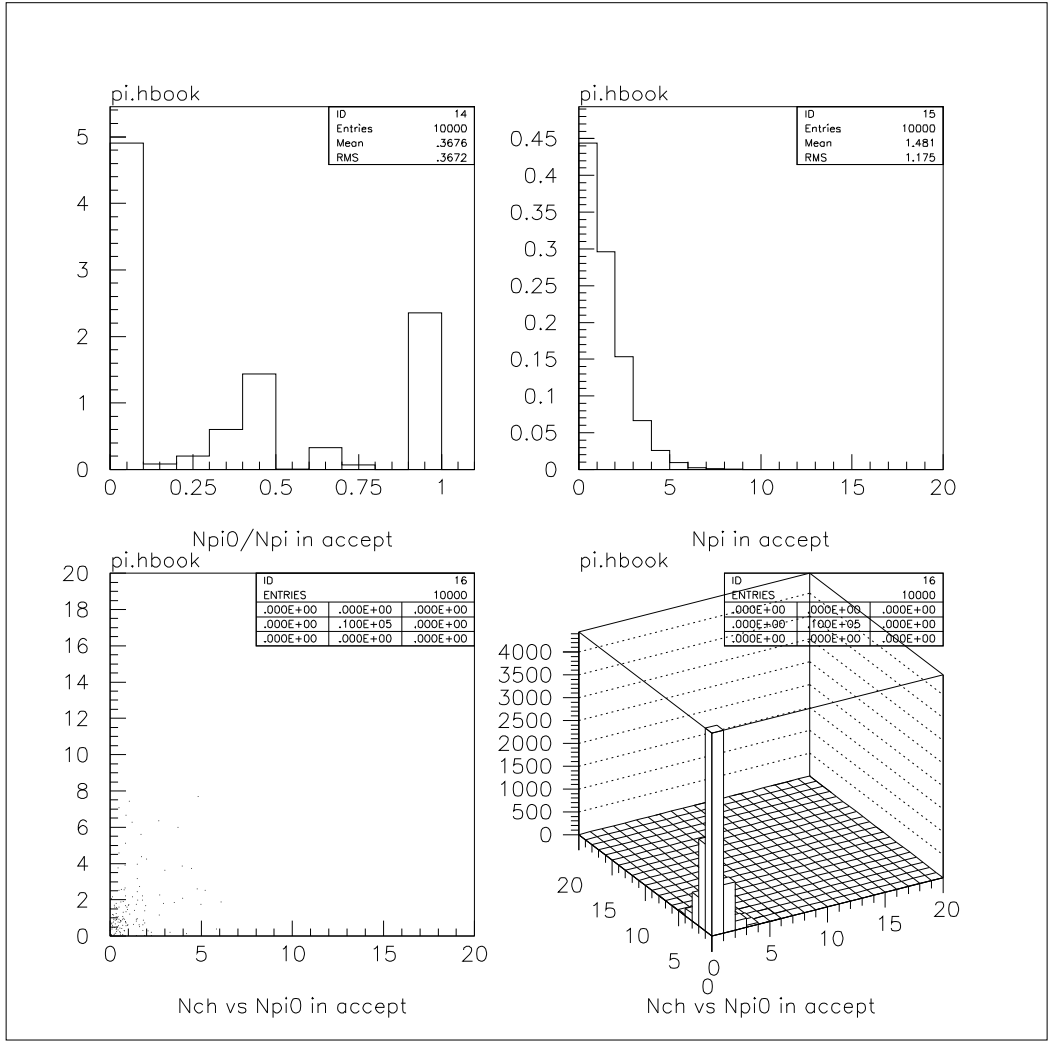,width=6in}
\vspace{-1in}
\caption{Number distributions of pions produced generically (by PYTHIA).}
\label{f:dccnchn_pyth} \end{figure}

\clearpage
\vfill\newpage
\begin{table}[p]
\vspace{.5in}\hspace*{-0.5in}
\begin{tabular}{|l|l|r|r|r|r|r|}
\hline
    subprocess & cross &
number & GEANT & GEANT & GEANT & GEANT\\
     & section & of & trigger & trigger & trigger & trigger \\
     & (mb) & events & 1 X$_0$ & fraction & 0 X$_0$ & fraction \\
\hline
    All included subprocesses  &    72.98 &
      300000  &   155709 & 0.5190 &  152069  & 0.5069 \\
    f + f' $\rightarrow$ f + f' (QCD)     &  \ 2.217 &
        9012  &     7494 & 0.8316 &    7289  & 0.8088 \\
    f + f~ $\rightarrow$ f' + f~'         &  \ 0.03198 &
         130  &       98 & 0.7538 &      88  & 0.6769 \\
    f + f~ $\rightarrow$ g + g            &  \ 0.03813 &
         155  &      122 & 0.7871 &     121  & 0.7806 \\
    f + g $\rightarrow$ f + g             &    14.67 &
       59642  &    52456 & 0.8795 &   51412  & 0.8620 \\
    g + g $\rightarrow$ f + f~            &  \ 0.4667 &
        1897  &     1719 & 0.9062 &    1692 &  0.8919 \\
    g + g $\rightarrow$ g + g             &    21.42 &
       87072  &    79857 & 0.9171 &   78618  & 0.9029 \\
    Elastic scattering         &    14.75 &
       60825   &       0 & 0.0000 &       0  & 0.0000 \\
    Single diffractive (XB)    &  \ 6.194 &
       25525   &    4057 & 0.1589 &    3815  & 0.1495 \\ 
    Single diffractive (AX)    &  \ 6.194 &
       25367   &    2706 & 0.1067 &    2452  & 0.0967 \\
    Double  diffractive        &  \ 6.839 &
       28271   &    5920 & 0.2094 &    5336  & 0.1887 \\
    Low-pT scattering          &  \ 0.1508 &
        2104   &    1280 & 0.6084 &    1246  & 0.5922 \\
\hline
\end{tabular}
\vspace{0.25in}
\caption{\label{table:trig}PYTHIA cross sections and GEANT trigger rates}
\end{table}
\clearpage
\begin{table}[h]
\setbox0=\hbox{scint cut/(4 MeV)}
\rotl{0}
\vspace{-1.5in}
\begin{tabular}[pos]{|cc|cccccccccccc|}
\hline
  & & \multicolumn{11}{c}{pls hgt cut/(0.4 keV)} &  \\
  & &  0.00 &  0.10 &  0.20 &  0.30 &  0.40 &  0.50 &  0.60 &  0.70 &
  0.80 &  0.90 &  1.00 &  \\
\hline
 & 0.0 & 65.24 & 63.58 & 62.11 & 60.82 & 59.50 & 58.39 & 57.45 & 56.58 &
 55.79 & 55.16 & 54.46 &  \\
 & 0.2 & 65.24 & 63.58 & 62.11 & 60.82 & 59.50 & 58.39 & 57.45 & 56.58 &
 55.79 & 55.16 & 54.46 &  \\
 & 0.4 & 65.24 & 63.58 & 62.11 & 60.82 & 59.50 & 58.39 & 57.45 & 56.58 &
 55.79 & 55.16 & 54.46 &  \\
 & 0.6 & 65.24 & 63.58 & 62.11 & 60.82 & 59.50 & 58.39 & 57.45 & 56.58 &
 55.79 & 55.16 & 54.46 &  \\
 & 0.8 & 65.24 & 63.58 & 62.11 & 60.82 & 59.50 & 58.39 & 57.45 & 56.58 &
 55.79 & 55.16 & 54.46 &  \\
 & 1.0 & 65.24 & 63.58 & 62.11 & 60.82 & 59.50 & 58.39 & 57.45 & 56.58 &
 55.79 & 55.16 & 54.46 &  \\
 & 1.2 & 71.32 & 69.50 & 67.88 & 66.47 & 65.00 & 63.78 & 62.74 & 61.77 &
 60.91 & 60.21 & 59.45 &  \\
 & 1.4 & 75.45 & 73.53 & 71.81 & 70.31 & 68.77 & 67.47 & 66.36 & 65.33 &
 64.42 & 63.68 & 62.87 &  \\
 & 1.6 & 77.16 & 75.21 & 73.46 & 71.93 & 70.36 & 69.04 & 67.92 & 66.87 &
 65.94 & 65.19 & 64.36 &  \\
 & 1.8 & 78.35 & 76.39 & 74.61 & 73.05 & 71.47 & 70.13 & 68.99 & 67.93 &
 66.99 & 66.22 & 65.39 &  \\
 & 2.0 & 79.37 & 77.39 & 75.58 & 74.00 & 72.40 & 71.04 & 69.88 & 68.82 &
 67.86 & 67.08 & 66.24 &  \\
\hline
\end{tabular}
\vspace{0.25in}
\caption{\label{table:plshgt}NHITS of ``Michgan chambers" in GEANT
(threshold = 20) for varying pulse-height and scintillator cuts.}
\end{table}
\vskip2ex
\hspace{1.5in}
\vfill\newpage

\chapter{Data Analysis Tools}

\section{Tracker}
At least three different track-finding programs were written and used by the 
MiniMax collaboration \cite{trackers}.
Each has a different algorithm for reconstructing tracks from hit wires,
but all give similar results.

The work described here was done using the combinatorial tracker.
That algorithm is constructed to find all possible combinations of the hit 
wires in four ``crosshair chambers''.  At least three of these chambers have
different orientations, so that a unique straight line can be drawn through 
any such combination of hits.  The line is considered a potential track, and 
the non-crosshair chambers are searched for hit wires within 3.5 wire spacings
of the line.
If enough wire hits are found, a straight line is fit to the hits.
Then, if the fitted track passes quality cuts such as
cuts on the $\chi^2$ of the fit, the track is recorded.
Several sets of crosshair chambers are used in order to
increase the probability that a track and all the associated hit wires are 
found.

Tracks with a greater number of wires are always looked for first because 
they are more
likely to be the correct track than a similar one with fewer wires; when
two tracks are similar (have a large fraction of wires in common) the
track found first has more weight in determining the track parameters.
Also, the tracker code is written to search for tracks which are pointed such 
that they appear to originate near the collision point before allowing the 
tracks to point in any direction.
Tracks are searched for in the following order.  First the charged 
tracks that go straight through the detector are found and recorded.
Then the tracking is run on the back 16 chambers and tracks 
which do not share too many wires (the exact numbers are given in Table 
\ref{t:tracker})
with previously-found charged tracks are recorded.  These tracks include 
photon conversion tracks and segments of charged tracks which bend in the lead 
and are therefore not found as through-going charged.  Tracks of the same 
type are then looked for in the eight chambers directly behind the lead 
(chambers 9-16) in case the tracks leave the acceptance before hitting enough 
chambers.  Interactions in the beam pipe also produce background tracks which 
can be found in this region.  
Finally, the front eight chambers are searched for segments of
charged tracks that are not found as going straight through all chambers.
Figures \ref{f:cleanch1}-\ref{f:ugly1} show event displays of 
various types of tracks.

In the process of searching for each type of track, track candidates are
compared to previously-found tracks to make sure that the same real track
is not being recorded as many similar tracks.  In the charged tracking,
for example, tracks with hit wires in at least 22 chambers are searched for
first.  
After a track is found, the wires hit by the track are stored in a list of
used wires.  As each new track candidate is found, its wires are checked
against the used wires.  If the candidate does not have at least 17 unique 
wires hit, it is dropped.  This significantly
reduces the number of fake tracks recorded by the tracker.  
The candidate track is also compared to the other tracks
individually.  The hits in each chamber are compared, and if the tracks are 
within two wires of each other in at least 16 chambers, the tracker determines 
that the same track has been found and the hits of the current
candidate are added to the list of hits in the previously-found track.
We refer to this as ``grouping''.  Wires from the grouped track are added
to the used-wire list.
Next, tracks which go through only 21 chambers are considered.
If such a track candidate has more than 17 wires which are in the
used-wire list, the candidate is dropped, and if it has 16 wires within two
wire spacings of hits in another track, the two tracks are grouped.  
Table \ref{t:tracker} shows the various cuts on number of chambers hit or
wires in common for each type of track.

The tracks are broken into segments in front of and behind 
the converter plane, referred to as ``heads'' and ``tails'', respectively.
A dst is written, which, though traditionally stands for data 
summary tape, is just 
a file which contains, for each event, the event number, number of heads and 
number of tails, 
and then for each 
track segment, the number of wires hit followed by the list of wires.
The wire number is given by 
$(\mbox{chamber}-1)\times 128+\mbox{wire}$, for chambers 1-24 and wires 1-128 
in each chamber.
Earlier dst formats included the NHITS of the event (which can still be
retrieved from the run data file)
and the track parameters determined by a fit (which are now determined using
separate code, as discussed in the next section).
Sample dst entries are shown in Figs. \ref{f:dst1}, \ref{f:dst2}.
The tracker is run once for each actual or simulated run, and the dst
is used in all further analysis.

Only events with an NHITS less than 600 have been analyzed by the tracker.
Higher-NHITS events tend to have large hit densities in the rear chambers,
which leads to a huge number of potential tails, most of which are not real
charged or photon-conversion tracks, and the tracker takes a longer time
to analyze these events.
About 6\% of the events in lead-in runs
have NHITS higher than this value.  The NHITS is
correlated with multiplicity, so that the sample of events which were
analyzed by the tracker is somewhat biased towards lower-multiplicity events.

\section{Track fitter}
\label{sec:fit}
A separate program is necessary to fit tracks from the wires written in a dst 
since the fit parameters from the combinatorial tracker are calculated in the 
$(x,y)$ coordinate system instead of in $(u,v)$ (this is important because the
resolution in $u$ is much better than in $v$ in the chambers behind the lead)
and uncertainties in the track parameters are not reported.  The code for
the track fitter is given in Appendix B.  With the track
fitter, straight lines are fit to the hit wires in a track, parameterized by
$u(z)=a_u+b_u z$ and $v(z)=a_v+b_v z$, and the covariance matrix (which 
gives the correlated uncertainties of parameters \cite{Bevstat}) is generated.
Uncertainties in the position of the track at a chamber are taken to be
$1/\sqrt{12}$ \footnote{For a hit to be recorded in a single wire in a chamber, 
a charged track must pass within 1 wire spacing of that wire.  The
uncertainty in position is then given by 
$(\Delta x)^2=\left< x^2\right>-\left< x\right>^2
=\int_0^1 d\! x x^2 - (\int_0^1 d\! x x)^2 = {1/12}$, where $x$ is the
distance in units of wire spacings.}  times the wire spacing for all hits.


Another subroutine (see Appendix B)
is used to correct a small problem with tracks found by the tracker.
Occasionally (in about 10\% of the events), a tail is recorded which, 
in a single chamber,
includes two wires as part of the same track which are separated by several 
wires; these double hits are present in most of the non-$u$ chambers of
the track.  

The cause turned out to be that two track candidates have common
wires in all of the $u$-chambers, and enough close wires in other chambers
for the tracker to group
the track candidates into one track.  In some cases, the candidates which are
grouped in this way are two real tracks, but more often one of the candidates
is a type of fake track which will be referred to as a ``ghost''.
(A ghost is a track which borrows hits in the $u$ chambers from a
single real track, and finishes up the track with random hits in the non-$u$ 
chambers such as from pipe-shower secondaries.)
The track which is recorded is halfway between the two candidates.

Since the occurrence of these tracks with double hits is fairly rare,
especially for those made of two real tracks, we do not try to separate
out two distinct tracks, but rather to find one good track.
The code used to correct this problem tries all possible combinations of
hit wires such that the non-$u$ chambers each have only one wire included 
in the track.  Only two of these combinations are considered as potential
correct tracks: the one which best points (in $v$) towards the collision point,
and the one which has the best $\chi^2$ when fit to a line in the $v$-$z$ 
plane.  Of course, a single combination can satisfy both these requirements.

The pointing in $v$ is somewhat complicated by the fact that tracks going
through the acceptance are at very small angles from the $v=0$ plane.
Therefore, the uncertainty of the $z$ for which $v=0$ is much larger than
that of $z(u=0)$.  In other words, the uncertainty in the angle between the
track and the $v=0$ plane and that angle are both comparably small.
Instead, we 
use the following measure of pointing in $v$.  We find the $z$ for which 
$u=0$ for the track, defined as $z_0$.  Then, working in the $v$-$z$ plane, we 
define $\theta_0$ as the angle between the $z$-axis and a line drawn 
from $(v=0,z=z_0)$ to the point where the track being considered intersects
the lead.  The angle between the track itself and the $z$-axis in the $v$-$z$
plane is defined as $\theta$, and the uncertainty in that angle as 
$\sigma_\theta$.  The measure of pointing is taken to be 
$\left|\theta_0-\theta\right|/\sigma_\theta$,
which is small for tracks which point (to within uncertainty) to the same
$z$ as is pointed to in $u$.

Figure \ref{f:refit} shows histograms of $\chi^2$ for tracks with the best 
$\chi^2$, $\left|\theta_0-\theta\right|/\sigma_\theta$ for tracks with the best 
pointing, and then these values for good tails (those which did not have the 
double-hit problem).
The plot of $\chi^2$ for good tails has an obvious separation of what we 
believe are real tracks and background, which we use to choose the cut 
$\chi^2<7$.  For the pointing in $v$, we require 
$\left|\theta_0-\theta\right|/\sigma_\theta<8$.
We are most interested in tails which point back to the collision point.
(Further discussion of pointing cuts is given in Sec. \ref{sec:vertexer}.)
Therefore, if the track with the best pointing also passes the $\chi^2$
cut, it is kept as a good track.  If this is not the case, the
track with the best $\chi^2$ is considered, and kept if it passes the
pointing cut.  In most cases, 
none of the combinations of wires produce
a track which passes the pointing cuts, and the track is dropped.

\section{Vertexer}
\label{sec:vertexer}
Charged tracks from the collision point and photon conversion tracks
are then reconstructed from the track segments.  The vertexer code,
given in Appendix C, is used to
determine the probability that track segments meet at a common point
at the plane of the lead.  A charged track is then defined as a head
which either ``matches'' at least one tail or 
passes cuts discussed below, and a photon as any group of tails (or a single 
tail) not matched to anything in front of the lead.  
In terms of the notation (number of heads in match, number of tails in match),
a charged track is a $(>0,\geq 0)$ and a photon is a $(0,>0)$.
The location of the vertex is taken as the mean position of the included
track segments at the lead, weighted by the uncertainties.  In order to 
avoid effects from the edge of the lead, the vertex position is required
to be within the region defined by $4.25\:\mbox{in}<u<10.25\:\mbox{in}$, 
$-3\:\mbox{in}<v<3\:\mbox{in}$, and 
$\sqrt{(u-7.25\:\mbox{in})^2+v^2}<4\:\mbox{in}$, which is roughly the area 
$1\:$in from all sides of the lead.

The uncertainty in the position of the tail at the lead is apparently 
underestimated by propagating the uncertainties in track parameters
calculated by the fitter.  
The most likely cause of this is multiple scattering in the lead.  
The mean variance in the $u$ position from this error propagation is 
${\sigma_u}^2=5.5\times 10^{-4}\:\mbox{in}^2$ and in $v$ is 
${\sigma_v}^2=0.030\:\mbox{in}^2$.  A new estimate 
was found by histogramming the distance between two conversion tracks
from a single photon in GEANT at the lead, and taking the standard deviation 
of the histogram as the uncertainty in position.  Figure \ref{f:dudv}
shows these histograms, which give the new values 
${\sigma_u}^2=0.007\:\mbox{in}^2$ and ${\sigma_v}^2=0.092\:\mbox{in}^2$.  
(The order of magnitude is really more important than the specific value,
since cuts on other variables used in the vertexer can compensate for
small differences in these values.)  
Calculated uncertainties for a track segment which are 
less than the new values are increased to the new values in the vertexer.

Tracks are required to point to within some distance of the mean collision 
point in order to remove background such as tracks from beampipe shower, and 
also combinatorial fakes.
Figures \ref{f:pointout} and \ref{f:pointin} show the $z$ for which $u=0$ of 
heads and tails of (1,1) charged tracks, and for single (0,1) photon conversion 
tails.
Heads of charged tracks are required to point to the region $-50<z<60$, 
which includes almost all primary charged tracks.  
Most of the tracks outside this region are fakes, although
some are decay products of neutral particles such as the K$_s^0$.
Tails of charged tracks are not required to point so that tracks which
multiple scatter in the lead will not be dropped.  For lead-out runs
(Fig. \ref{f:pointout}), most (0,1)'s which point to 
$z\,\raisebox{-.75ex}{$\stackrel{\textstyle{>}}{\sim}$}\, 50$
are tracks from interactions in the 
beampipe, while for lead-in runs (Fig. \ref{f:pointin}), many are from
photon conversions in the lead.  In order to cut out the fakes from pipe shower,
single photon conversion tracks are required to point to $-40<z<50$.
For photon conversions which produce more than one track, at least one track
in the vertex must point to $z<50$.
The pointing cut in $v$ is taken from the study in the previous section to be
$| (\theta_0-\theta)/\sigma_\theta | < 8$.

Parameters used in the vertexer code were determined
by finding those which best reconstruct the PYTHIA and GEANT tracks, 
which are recorded in separate files.
The pointing cuts, the increased uncertainties in the $u$ and $v$ position of
tails, and cuts on the $\chi^2$ for the fit of how well two tracks match at the 
lead (which will be discussed in Sec. \ref{sec:match}, see Fig. \ref{f:chisq})
are not independent; the cuts chosen are all self consistent.
The code seems to work sufficiently well for both simulated and real data.
As an example, Fig. \ref{f:uglyv} shows the tracks which survived the vertexer
cuts for the large NHITS event of Fig. \ref{f:ugly1}.

Plots of efficiencies for finding charged tracks and photons are shown in
Figs. \ref{f:eff_chE}-\ref{f:eff_phi}.  Efficiencies are given as a function
of energy, transverse momentum, multiplicity, NHITS, and position in $\eta$ 
and $\phi$.  The numbers of charged tracks and of photons in the 
acceptance as a function of these parameters are also shown, both because these 
distributions are interesting, and also to give an indication of the 
statistical significance of bins in the efficiency plots.
The plots also include the mean number of fakes as a function of multiplicity 
and NHITS, and the mean number of actual charged tracks and photons as a 
function of NHITS.

\subsection{Matching tracks at the lead}
\label{sec:match}
Tracks are ``matched'' or ``vertexed'' in the following way.
Consider two tracks intersecting the lead at points
${\mbox{\boldmath $\xi_1$}}=({\xi_1}^1,{\xi_1}^2)=(u_1,v_1)$,
${\mbox{\boldmath $\xi_2$}}=({\xi_2}^1,{\xi_2}^2)=(u_2,v_2)$, 
with covariance matrices ${\bf C_1}$ 
and ${\bf C_2}$.
Assume that the two tracks meet at exactly the same point 
[$\mbox{\boldmath $\bar{\xi}$}=(\bar{\xi}^{\:1},\bar{\xi}^{\:2})$]
at the lead.  Also assume that the \mbox{\boldmath $\xi_i$} are
Gaussian distributed about mean \mbox{\boldmath $\bar{\xi}$}.

The $\chi^2$ for the fit of the two tracks to \mbox{\boldmath $\bar{\xi}$} is
\begin{equation}
\chi^2=({\mbox{\boldmath $\xi_1$}}-{\mbox{\boldmath $\bar{\xi}$}}) 
	{\bf C_1}^{-1} 
	({\mbox{\boldmath $\xi_1$}}-{\mbox{\boldmath $\bar{\xi}$}})+
        ({\mbox{\boldmath $\xi_2$}}-{\mbox{\boldmath $\bar{\xi}$}})
	{\bf C_2}^{-1} 
	({\mbox{\boldmath $\xi_2$}}-{\mbox{\boldmath $\bar{\xi}$}}).
\label{eq:chisq}
\end{equation}
Minimizing $\chi^2$ with respect to $\bar{\xi}^{\:i}$ gives 
\begin{equation}
\bar{\xi}^{\:i}=\left[ ({{\bf C}_1}^{-1}+{{\bf C}_2}^{-1})^{-1}\right]^{ij}
\left[({{\bf C}_1}^{-1})^{jk}{\xi_1}^k+({{\bf C}_2}^{-1})^{jk}{\xi_2}^k\right].
\end{equation}
Putting this \mbox{\boldmath $\bar{\xi}$} back into Eq. \ref{eq:chisq} yields 
$\chi^2$.
There are four known parameters (\mbox{\boldmath $\xi_1$} and 
\mbox{\boldmath $\xi_2$}), and two which
are determined (\mbox{\boldmath $\bar{\xi}$}), leaving two degrees of freedom.
The reduced $\chi^2$ is therefore $\chi^2/2$.

All track segments in an event were matched together in order to find
the cutoff in $\chi^2/2$ for which matches with a lower $\chi^2$ are most
likely real tracks, while those with a higher $\chi^2$ are most likely
unrelated track segments.  Plots of matches between pairs of heads, pairs
of tails, and head-tail pairs are shown in Fig.  \ref{f:chisq}.
Head-head matches which have a low $\chi^2$ are due only to the heads being
coincidentally close together at the lead.  The peak at low $\chi^2/2$ for
head-tail matches is due to charged tracks, and is fairly cleanly separated
from false matches.  The tail-tail plot does not show a clear division,
but a slight division is present at about the same value of $\chi^2/2$
as the division for head-tail matches.  (Note that the cutoff values
could have been different for the head-tail and tail-tail matches since
the uncertainties in position of the tails are much larger than those
for heads.)  A reasonable cut appears to be $\log{(\chi^2/2)}<0.7$, or
$\chi^2/2 < 5$.

\subsection{Tracks which may be grouped together}

The original vertexer algorithm grouped any track segments which matched any 
other track segments in a vertex.  The uncertainty in position of the tails
is large enough that this occasionally produced vertices which contained more 
than one real track.  For instance, a (1,1) charged track and a (0,n) photon
could be grouped into a (1,n+1), or two charged tracks could become a (2,2).

By demanding that all tails in a vertex match the head (which has a much 
smaller uncertainty), charged tracks are less likely to be grouped with other 
tracks.  The vertexer code was therefore changed to look for charged tracks 
first and remove those tails which were matched to heads before matching tails 
together.  The following is the result of this change for a sample of  GEANT
events.

Of the original 135 (2,2) vertices, 104 have heads which do not 
match each other
and were therefore separated.  About 94\% of these are really two charged 
tracks, and 6\% are fake heads grouped with real tracks.  
The 31 vertices which have two heads that match each other consist 
of 48\% fake heads along with other tracks, 13\% e$^+$e$^-$ pairs from photon 
conversions in the window (``window conversions''), and 39\% vertices with two 
charged tracks which are very close together at the lead.

The majority of (1,n)'s which were separated into more than one vertex are
charged tracks combined with tails from photons, fake tracks, or pipe 
shower tracks.  The latter two types of tails are likely to be removed by 
pointing cuts when separated from the charged track.
About 10\% result from window conversions where a conversion secondary
showers in the lead, and approximately 5-10\% are 
single charged tracks with multiple tails (from interactions in the lead).  
Of those which remained (1,n)'s,
approximately 20\% are pions which interact in the lead, 30\% are pairs of a 
charged track and a fake tail, 25\% result from window conversions, and
20\% are pairs of a close charged track and a photon.  The rest involve
missing heads,
decays in GEANT, etc.  Only about half of the vertices which contained both a
charged track and a photon were reclassified as such.

The overall improvement in vertexer performance due to these changes is small
since these types of vertices are not very common.  Efficiencies are improved
for high multiplicity events, without introducing many fakes.  

Studies were
also done with the real data by adding a single charged track or photon 
conversion track from a clean event to every other event in the run.  This
was done at the level of the dst, so that the behavior of the tracker for
the combined events is ignored.  At this level, the efficiencies for finding
all tracks of the individual events in the combined event
are not compromised unless the added track is within some distance of a
track in the other event.  The mean distance between an added charged track
and another charged track for a track to be missed is about $0.5\:$in,
and for an added charged track and a photon conversion about $0.7\:$in.

\subsection{Tail-less heads as charged tracks}

Often the tail of a low-energy charged track will not be detected due to
multiple scattering in the lead which may cause it to stop or to bend out of 
the acceptance or to lose so much energy that it does not travel in a straight
path through the rear chambers.
In GEANT, most of the (1,0)'s are real (collision-point) charged tracks which 
stopped in the lead or had very soft tails; some are from neutral particle 
decays. Those tail-less heads which are fakes are distinguished by the fact 
that they do not have hits in all eight of the front chambers.
However, in the real data, the (1,0)'s seem to include a large fraction
of fakes. The ratio of the number of (1,0)'s to the total number of identified
charged tracks [including (1,0)'s] is much 
higher for the real data (run 1125) than for GEANT, 17\% vs 7\%, whereas if
the energy distributions of charged tracks are similar for real data and GEANT,
this fraction should be about the same.   Eliminating those (1,0)'s
which do not have hits from all eight chambers reduces the fractions to 
12\% and 6\%. 
Other types of (1,0)'s which appear to be fakes either share wires in all 
three $u$ or all three $v$ chambers with another track (ghosts), or point
to somewhere other than 
the intersection with the $z$-axis (in $u$) of
other charged tracks in the event which go all the way through the detector. 
 
Based on these observations, tight cuts are made which are intended to
eliminate most fakes at the expense of losing some real charged (1,0)'s.
Any (1,0)'s which share wires in all $u$ or all $v$ chambers are dropped.
The pointing cut is defined by other charged tracks in the event; if there
are none, it is not used.
The collision point $z_{cp}$ is determined as the weighted mean of the point 
of intersection with the $z$-axis of any other heads of (1,$>0$) charged 
tracks in the event, using the parameterization in $u$ of the heads and the
associated uncertainties.  
Plots of the distance between the $z(u=0)$ of the (1,0) and the
mean $z_{cp}$ divided by $\sigma$, the root mean square of the uncertainty 
in these $z$, for both GEANT and real-data (1,0)'s are used to choose the 
condition that the $z(u=0)$ of the (1,0) must be within $2\sigma$
of $z_{cp}$.  This cuts out many real charged (1,0)'s. 
These cuts reduce the ratio of ``charged'' (1,0)'s to total charged tracks to 
5.0\% for GEANT and 9.6\% for run 1125.  Although the fraction for the data
is almost twice as large as that for the GEANT, no apparent qualities of the 
remaining (1,0)'s suggest that they are fake. 

\subsection{Middle-eight tracks}

Originally, tails which do not go through at least 14 of the rear
chambers, but do go through the first eight chambers behind the lead
were not used by the vertexer.  The reason is that the rear chambers
are often flooded with pipe shower, which results in a large number of
``middle-eight'' tracks which are pipe shower tracks or combinatorial fakes 
which can be found as photon tracks by the vertexer.  However, these tracks
can be used carefully in certain circumstances.
Allowing the middle-eight tracks to be vertexed as tails of charged 
tracks saves some real charged tracks that would otherwise be (1,0)'s 
and might have be thrown away by the tight cuts.  The vertexer also allows
middle-eight tracks as photon conversions as long as they are vertexed with at 
least one other tail. 
This is important because the tight pointing cuts on (0,1) photons are often
failed by a single tail of a photon conversion which produces other tracks
that leave the acceptance before reaching the final chambers.  
If a middle-eight tail is vertexed to such a 
tail then the vertex is not subject to the tight cuts.  Also, since fake
tracks are unlikely to vertex at the lead, a vertex made only of at least
two middle-eight tracks is counted as a photon.

It turns out that very few events have middle-eight tracks which can be
vertexed with other track segments at the lead,
so that the effect on overall efficiencies is almost negligible. 
This may be related to the difficulty of getting a good fit for middle-eight
tracks due to the lack of $v$-resolution.
Fake charged tracks are occasionally created when a fake (1,0) 
is coincidentally matched with a middle-eight tail.
The middle-eight tracks were nevertheless included in the vertexing, since even 
a tiny improvement in the photon-finding efficiencies is welcome.

\subsection{Origin of fakes}
\label{sec:uthetau}

A charged track from the vertexer is classified as a fake if there are no 
charged tracks from PYTHIA inside the acceptance within the specified distances
in $u$ and $v$ of $0.5\:$in and $1\:$in, respectively, of the identified track 
at the lead.
A sample of events which have one fake charged track was taken from 
$1.5\times 10^5$ GEANT events.  
In order to avoid effects due to the presence of photons, the events were
required to have no photons observed or known to convert in the acceptance.
The sample is divided into the following categories: events in which
0, 1, or 2 charged tracks are sent into the acceptance and one extra charged
track is found (the total sample of events had 3402, 1475, 348 events,
respectively, in these classes, and 39, 38, and 38 events were used in this 
study), and those in which 0 charged tracks are sent into the acceptance and 
2 are found (217 total events, 37 events used, accounting for 74 fake tracks).
Of the 188 fake charged tracks,

150 are decay product(s) of K$_s$'s or $\Lambda$'s, 

24 are secondary charged tracks from other decays or interactions in the 

$\quad$ material surrounding the detector 

1 is from the conversion of a photon in the window, 

12 are real charged tracks just outside the acceptance, and 

1 is an actual combinatorial fake. \\ 
Decays of single K$_s$'s or $\Lambda$'s are responsible for about 70\% of the 
events with two fake charged tracks.  The important outcome of this study is
that almost all of these ``fakes'' are caused by interesting physics processes
or edge effects.

Also studied were 20 events (from a total of 416) where a photon conversion 
is present in GEANT\footnote{GEANT conversions are defined as e$^\pm$ tracks 
originating in the region containing the lead and scintillator between chambers 
8 and 9, and within $0.1\:$in at the lead of a photon from PYTHIA.} 
and that photon is not found, 
and no PYTHIA charged tracks are sent into the acceptance, but one is found. 
In 11 of these events, the photon converts in the window, and a resulting
e$^\pm$ showers in the lead, i.e., the missed photon and fake charged track
are correlated. 
The remaining events involve photons which are not found by the vertexer
or are found just outside the acceptance, with uncorrelated fake charged
tracks which fall into the categories mentioned above.

A photon found by the vertexer is defined as fake if no PYTHIA photon is
aimed such that it will
hit the lead within $0.5\:$in in $u$ and $1\:$in in $v$ of the identified
conversion.  In 38 events (from a total of 1800) where no charged tracks are 
sent into the acceptance by PYTHIA and no photon conversions from GEANT are 
present in the acceptance, and one photon is found,

19 are conversions of secondary photons, roughly half of which appear 

$\quad$ to be due to decays, and the other half due to interactions in the 

$\quad$ detector region,

14 are real photons just outside the acceptance, 

2 are charged decay products of K$_s$'s where the heads are lost, 

1 is the product of an interaction of a neutron in the lead, and 

2 are from events which are too complicated to determine the source 

$\quad$ of the fake. 

\vskip1ex
Of 20 events (from a total of 319) where a charged particle is lost and a fake 
photon found, 17 involve a charged track which multiple scatters in the lead so 
that the tail is found as a photon conversion, but the head is dropped. 
In the remaining 3 events, the missed and fake tracks are uncorrelated. 

When a pipe shower produces a large density of hits in the rear chambers,
the probability is greatly increased for finding fake photons.
Combinations of these hits, possibly together with hits from real tracks from 
the collision, form tracks which point to the collision region and 
therefore should be found by the tracker.  Ghosts (described in 
Sec. \ref{sec:fit}) are a common example of such a fake track.

If pipe showers were uncorrelated with important observables such as the number
of charged tracks and photons entering the acceptance in an event, then
vetoing events which have pipe shower would reduce the number of fake photons
found without biasing the remaining sample of events, say towards those with 
low multiplicity.  Unfortunately, this is not the case.
In order to study this, we first identified events 
with pipe-shower tracks.  Primary tracks which are produced in the collision
and pass all the way through the MWPC telescope occupy a very limited region
of the phase space defined by the position and angle in $u$ at some $z$.
(Recall that the acceptance is roughly defined by the coverage of the
lead converter, which extends from $u=3.25\:$in to $11.25\:$in and $v=-4.0\:$in
to $4.0\:$in at $z_{lead}=150\:$in.)
Figure \ref{f:uthetau} shows the location of tracks in $(u,\theta_u)$ space
where $u=u(z_{lead})$ and $tan{\theta_u}=u(z_{lead})/z_{lead}$, averaged over
many events.  The dense band near the center of the plot is due to 
primary charged tracks.  We defined rough boundaries around this region
and vetoed events with any tails in the regions $\theta_u > 0.009u+0.016$ or
$\theta_u < 0.006u-0.028$, excluding tails which the vertexer found as 
photon conversion tracks.  (Note that this is a very tight cut.)

Because of the higher hit density in the rear chambers created by pipe shower,
the events which remain after the pipe-shower veto have a much lower mean 
NHITS than that for all events.  In GEANT, the mean NHITS for lead-in runs
dropped from 142 to 90, and in run 1125 (lead in) from 229 to 113.
Also, surviving events with higher NHITS tend to have more real tracks,
as is shown for GEANT in Fig. \ref{f:gpipeveto}.  This figure also shows the 
distributions of actual charged-track and photon multiplicities (from PYTHIA), 
which appear to be biased towards lower multiplicities.  This seems to be true
to a greater extent for the real data; Figure \ref{f:pipeveto} shows that the 
probability distribution of the observed number of photons before and after the 
pipe-shower veto for real data is more biased towards lower multiplicities 
than that for GEANT.  The mean number of charged tracks found decreased from 
$\left< n_{ch}\right>=0.53$ for all GEANT events to $0.47$ for this sample, 
and of photons from $\left< n_\gamma\right>=0.21$ to $0.14$.  For the
data (run 1125), the cut is even harder: $\left< n_{ch}\right>=0.48$ falls to
$0.32$ and $\left< n_\gamma\right>=0.19$ drops to $0.06$, practically no
photons.
We expect the veto to have a greater effect on the data because we believe 
that the real events contain more pipe shower than the GEANT, but since
the vetoed events are apparently correlated with high multiplicity events,
the remaining events are a clean, but not unbiased, sample.

This sample of non-pipe-shower events is useful for studying fake photons
which were not caused by pipe shower, although the events are also less
likely to contain converted photons.
Fake photons in two sets of the remaining events were classified.
The first set was taken from about $10^4$ minimum-bias
(triggered) events, and contains 104 events which pass the pipe-shower veto 
and have a fake photon.  The second contains 71 such events which also have at 
least four primary charged particles and/or photons (from PYTHIA) in the 
acceptance; this sample includes all events of that type in the approximately
$1.5\times 10^5$ triggered events.

Of the fakes in the first set,

5 are ghosts,

11 are tails of charged tracks which are not vertexed with the heads, 

9 are associated with photon conversions but are not vertexed with 

the other conversion tracks,

31 are other secondary photons such as those produced by interactions 

in the beampipe,

2 are secondaries from the interactions of neutrons in the lead,

11 are photons from the decays of $\pi^0$'s produced by 
K$_s\rightarrow\pi^0\pi^0$, 

18 are real photons just outside the acceptance, 
\newline
and the remaining 17 either are not able to be classified, or fell 
into unique categories. 
The first few types of fakes are the most serious because they are correlated
with the presence of real tracks.  In almost all cases, when the tail of
a charged track is not vertexed to the head, the tail-less head does not
pass the cuts necessary to be counted as a charged track.  Therefore
a charged track is lost and is replaced by a fake photon.
(This complicates the generating function analysis described in Chapter 5,
because it introduces a correlation between charged-track and photon 
efficiencies.)
The classification of photons which are outside the acceptance but
are found inside the acceptance as fakes is really just due to details
in the code for finding fakes.

Due to the higher multiplicities in the second set, we expect more fakes
associated with real tracks, such as ghosts.  We find that this is true,
but not to an extent which would seriously invalidate the assumptions made
in the next chapter.  The fake photons include

8 ghosts,

25 tails of charged tracks,

12 tracks associated with a photon conversion,

11 other secondary photons, 

2 photons from K$_s\rightarrow\pi^0\pi^0$,

9 photons just outside the acceptance,
\newline
and 4 unclassified fakes.

\begin{figure}[h] \vspace{-0.5in}\hspace*{-0.3in}
\psfig{file=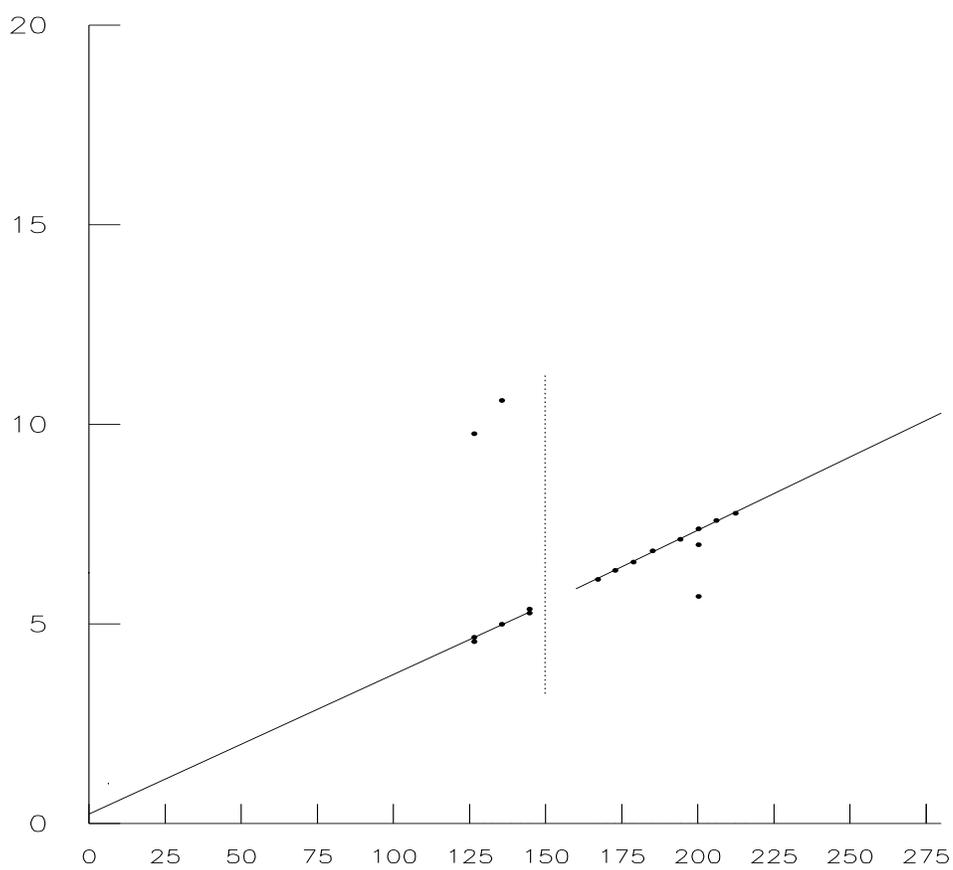,width=6in,height=6in}
\caption{Event display of a clean charged track.}
\label{f:cleanch1} \end{figure}
\begin{figure}[h] \vspace{-0.5in}\hspace*{-0.3in}
\psfig{file=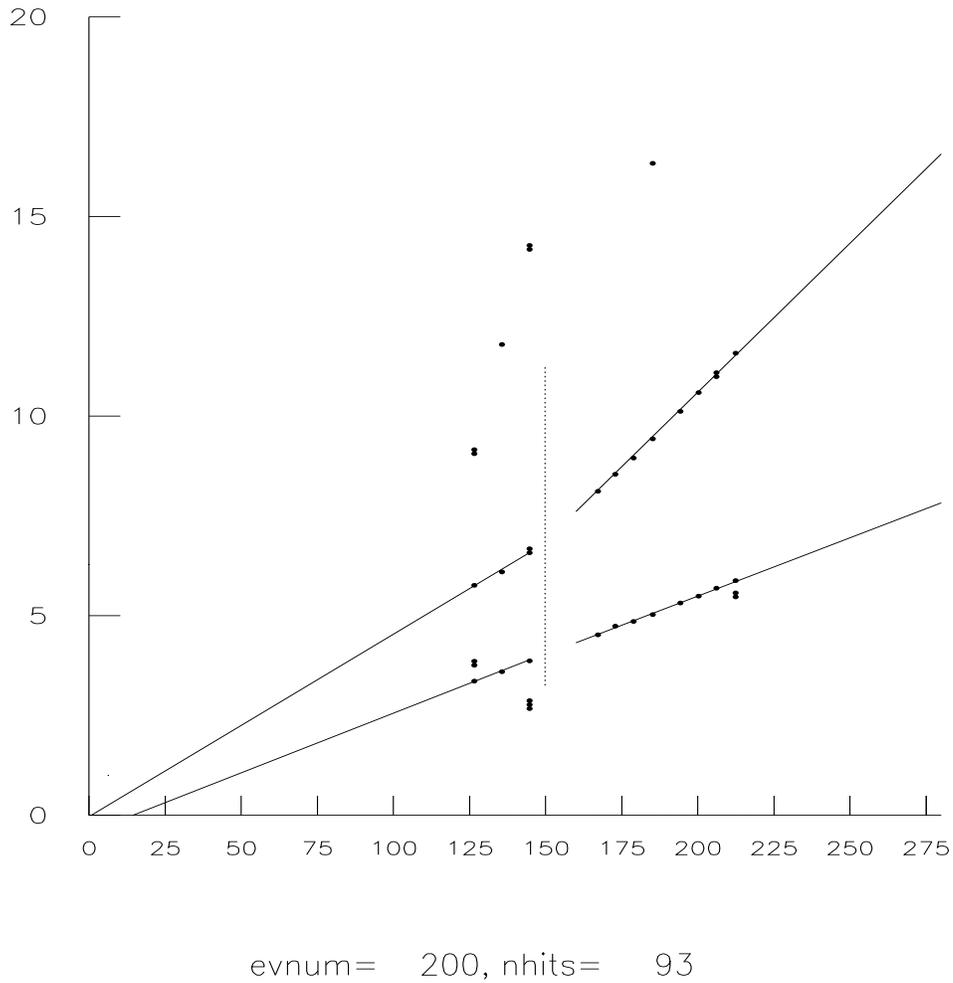,width=6in,height=6in}
\caption{Event display of a straight-through charged track and a charged
track which bends in the lead.}
\label{f:cleanch2} \end{figure}
\begin{figure}[h] \vspace{-0.5in}\hspace*{-0.3in}
\psfig{file=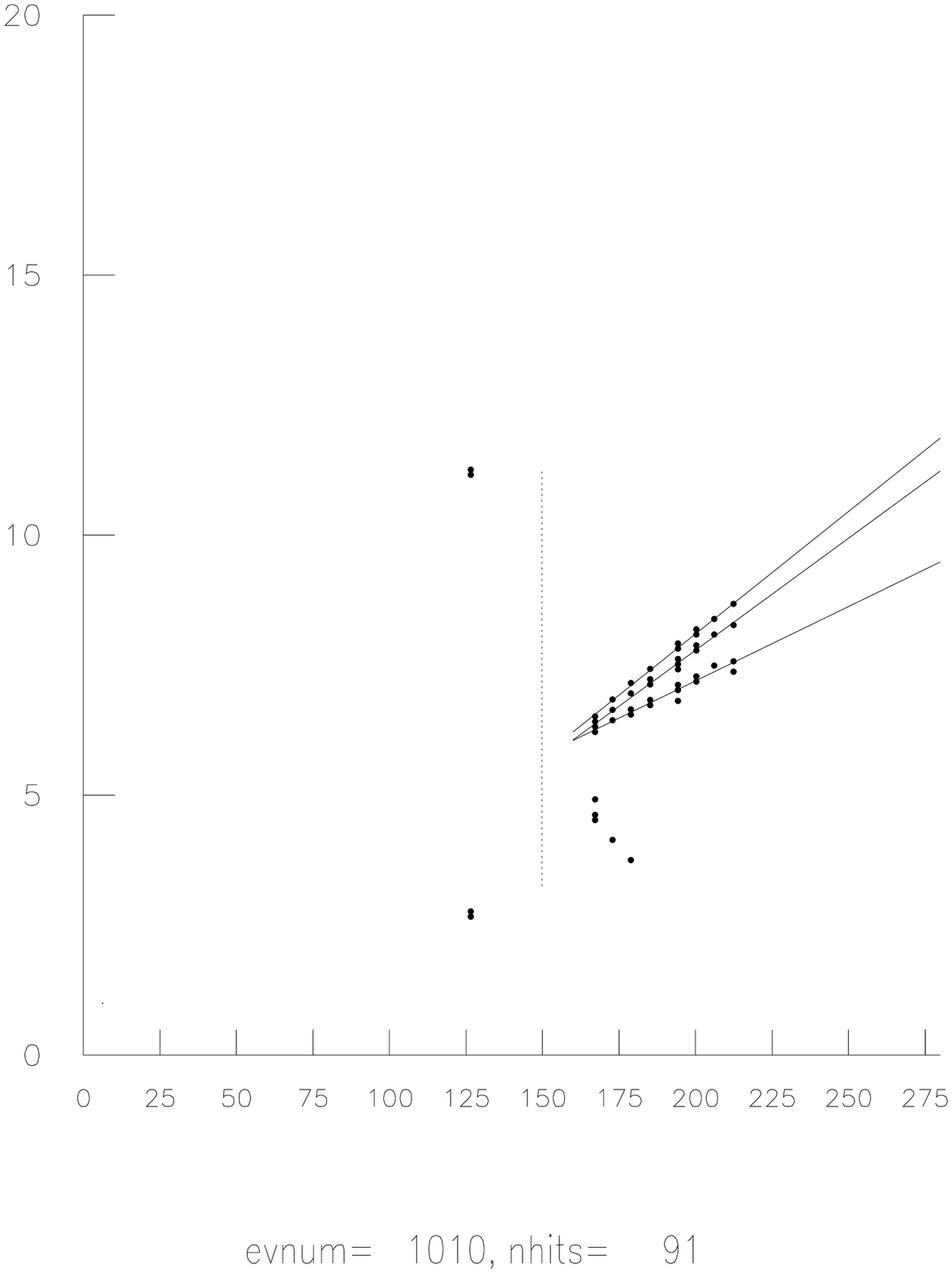,width=6in,height=6in}
\caption{Event display of a photon with three conversion tracks.}
\label{f:cleangam1} \end{figure}
\begin{figure}[h] \vspace{-0.5in}\hspace*{-0.3in}
\psfig{file=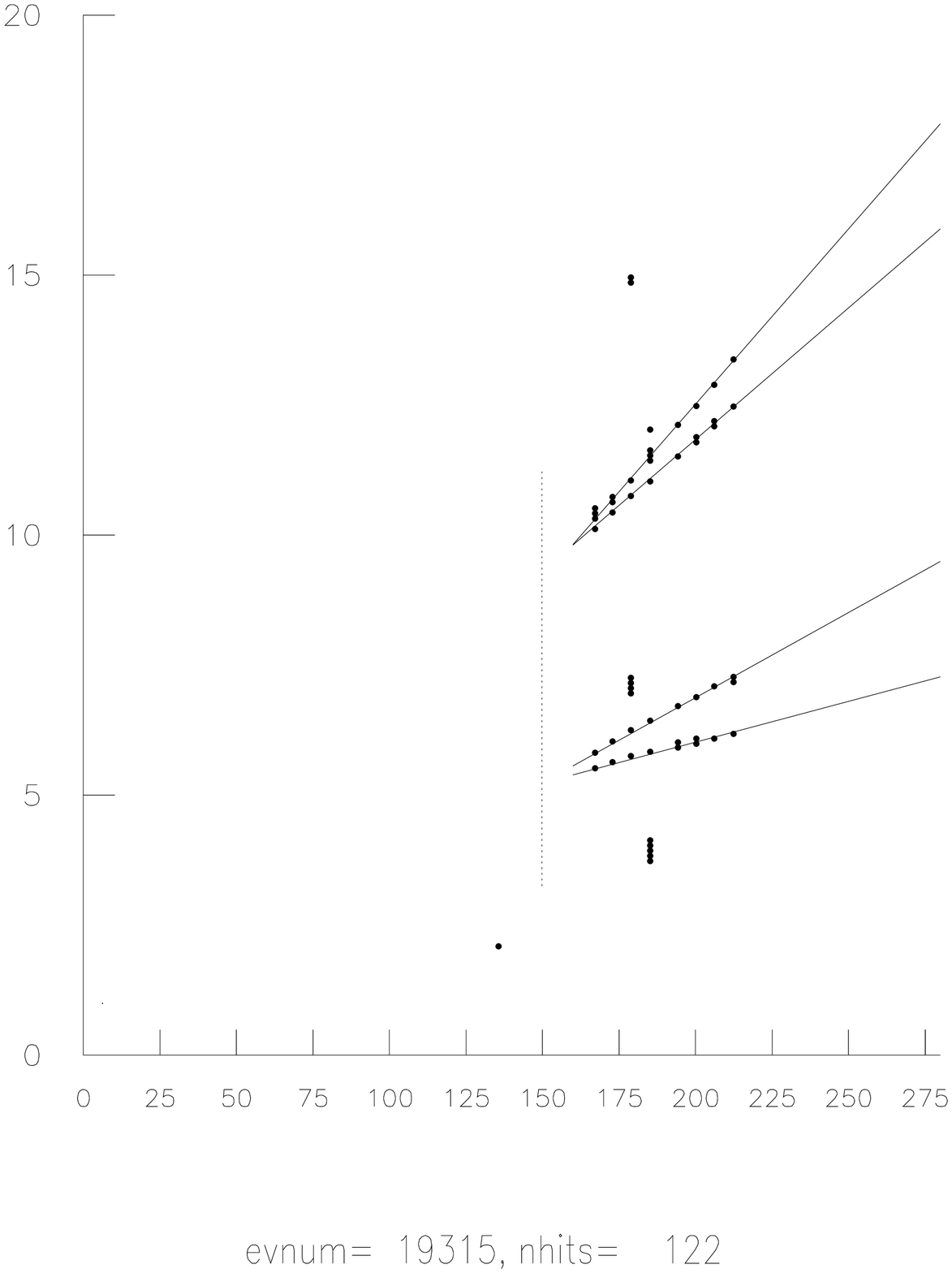,width=6in,height=6in}
\caption{Event display of two clean photon conversions.}
\label{f:cleangam2} \end{figure}
\begin{figure}[h] \vspace{-0.5in}\hspace*{-0.3in}
\psfig{file=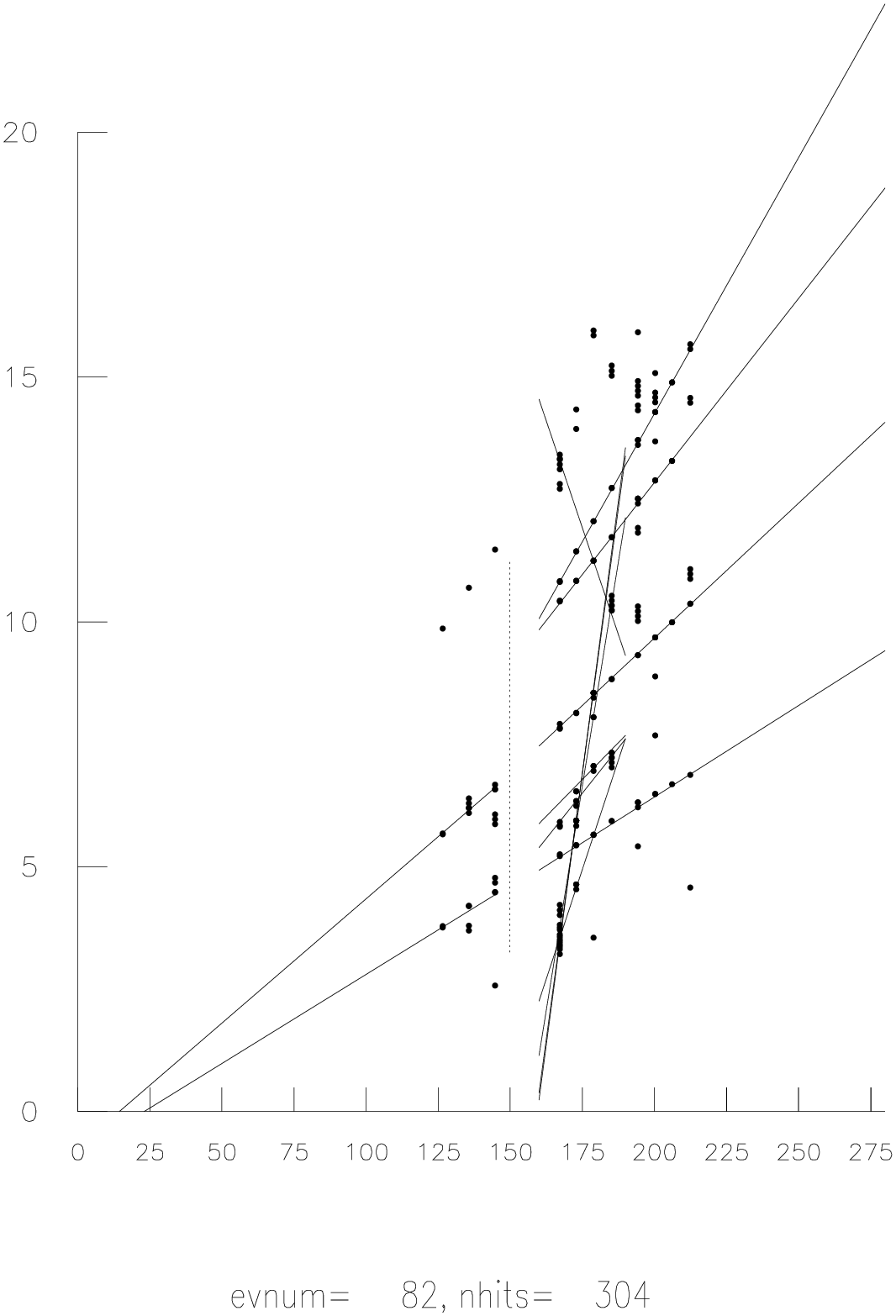,width=6in,height=6in}
\caption{Display of an event with two charged tracks and a photon,
and typical NHITS.}
\label{f:typical1} \end{figure}
\begin{figure}[h] \vspace{-0.5in}\hspace*{-0.3in}
\psfig{file=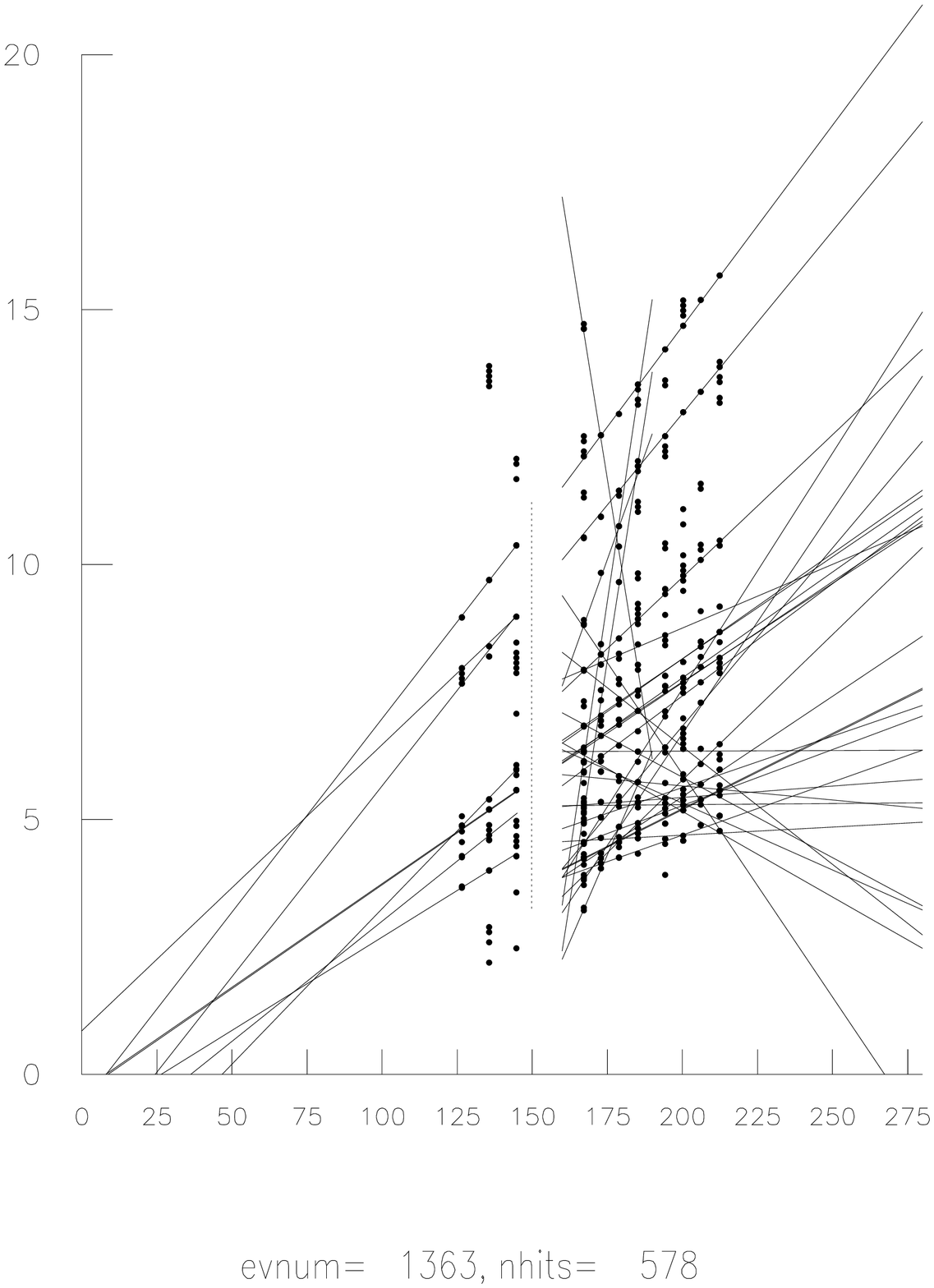,width=6in,height=6in}
\caption{Display of an event with a large NHITS.}
\label{f:ugly1} \end{figure}
\begin{figure}[h] 
{\small
\begin{verbatim}
   164       1       1
          7
  77   328   485   610   694   870   927
         15
1055  1249  1377  1440  1633  1761  1888  1955  2144  2273  2337  2655
2730  2850  2979



  1010       0       3
         15
1058  1241  1374  1436  1629  1763  1884  1965  2140  2264  2341  2650
2745  2855  2977
         19
1057  1056  1241  1242  1376  1433  1632  1766  1888  1960  2144  2269
2335  2533  2656  2737  2846  2848  2969
         15
1059  1239  1372  1439  1627  1758  1882  1965  2136  2264  2344  2647
2741  2859  2986
\end{verbatim}}
\caption{Entries in the dst for the events shown in Figs. 4.1 and 4.3.}
\label{f:dst1} \end{figure}
\begin{figure}
{\small
\begin{verbatim}
    82       2      11
          8
  52   240   298   501   618   659   870   919
          8
  82   221   336   472   598   705   860   940
         14
1046  1248  1386  1421  1642  1778  1897  1955  2152  2275  2328  2543
2735  2960
         15
1072  1234  1359  1613  1740  1868  1974  2122  2251  2360  2502  2631
2751  2876  3006
         15
1098  1200  1332  1478  1586  1719  1839  2011  2090  2215  2392  2475
2598  2795  3033
         14
1102  1195  1326  1485  1578  1709  1829  2020  2078  2203  2406  2459
2582  3054
          7
1258  1375  1454  1628  1748  1883  1952
          7
1127  1178  1485  1586  1726  1854  1984
          7
1028  1268  1381  1453  1613  1726  2011
          7
1027  1279  1381  1459  1613  1723  2014
          7
1031  1256  1381  1618  1742  1853  2011
          7
1029  1279  1394  1425  1642  1771  1974
          7
1052  1268  1377  1456  1748  1884  1950
\end{verbatim}}
\caption{Entry in the dst for the event shown in Fig. 4.5.}
\label{f:dst2} \end{figure}

\begin{figure}[h] \vspace{-0.5in}\hspace*{-0.3in}
\psfig{file=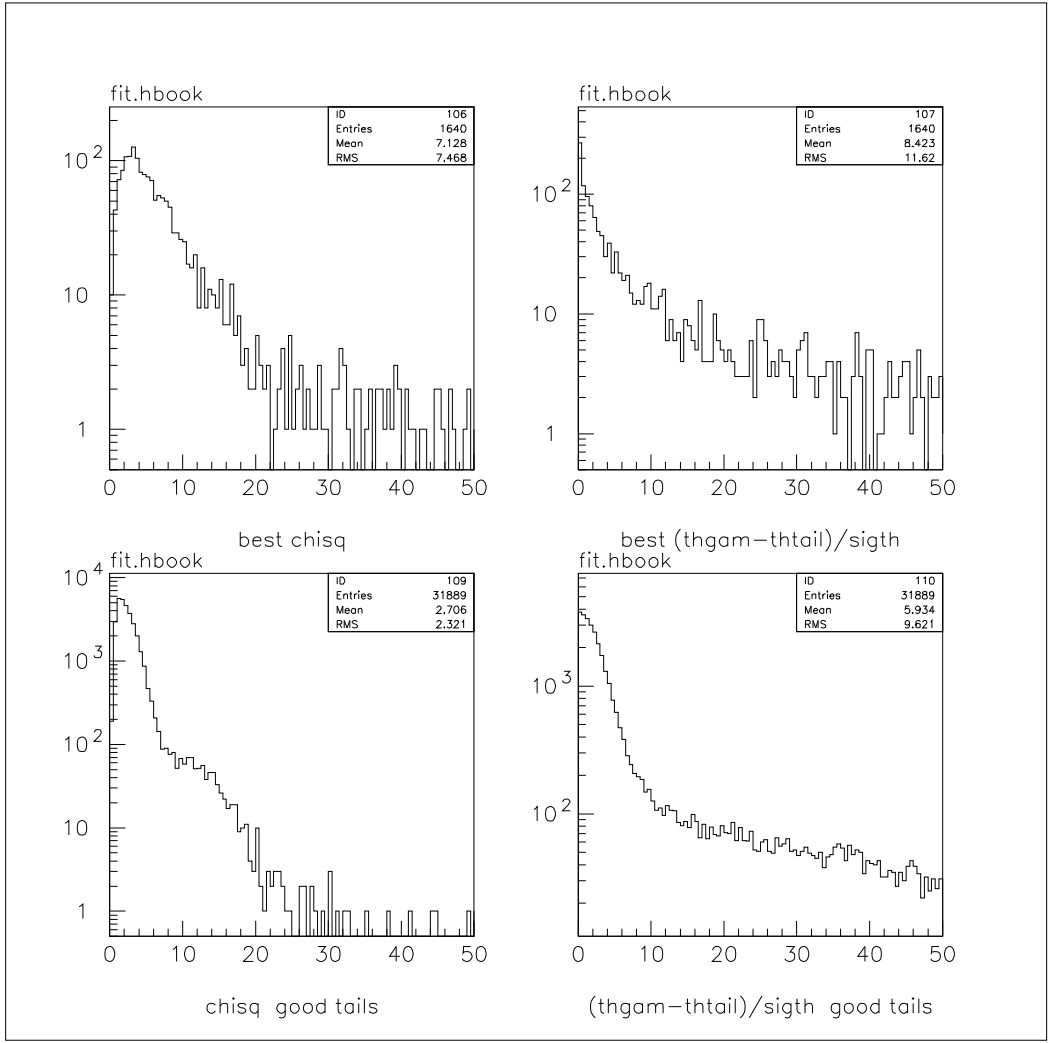,width=6in}
\vspace{-1in}
\caption[Histograms of $\chi^2$ for the fit in $v$ and of 
$\left|\theta_0-\theta\right|/\sigma_\theta$ for re-fitted tails and good 
tails.]{Histograms of $\chi^2$ for the fit in $v$ and of 
$\left|\theta_0-\theta\right|/\sigma_\theta$ for re-fitted tails and good 
tails.  Cuts were made of $\chi^2<7$ and 
$\left|\theta_0-\theta\right|/\sigma_\theta <8$.}
\label{f:refit} \end{figure}

\begin{figure}[h] \vspace{-0.5in}\hspace*{-0.3in}
\psfig{file=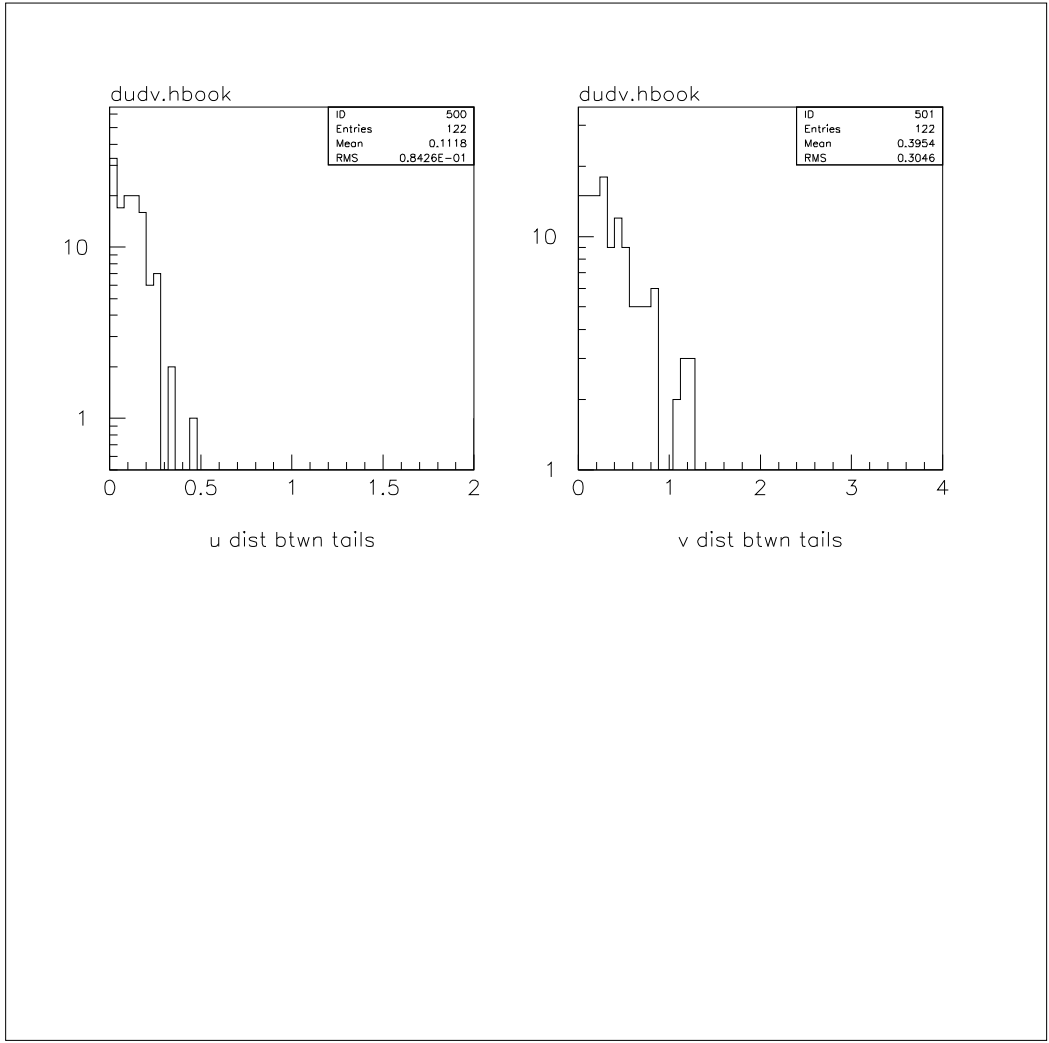,width=6in}
\vspace{-1in}
\caption{Distance in $u$ and $v$ between photon conversion tracks at the lead
when there are two conversion tracks in GEANT.}
\label{f:dudv} \end{figure}
\begin{figure}[h] \vspace{-0.5in}\hspace*{-0.3in}
\psfig{file=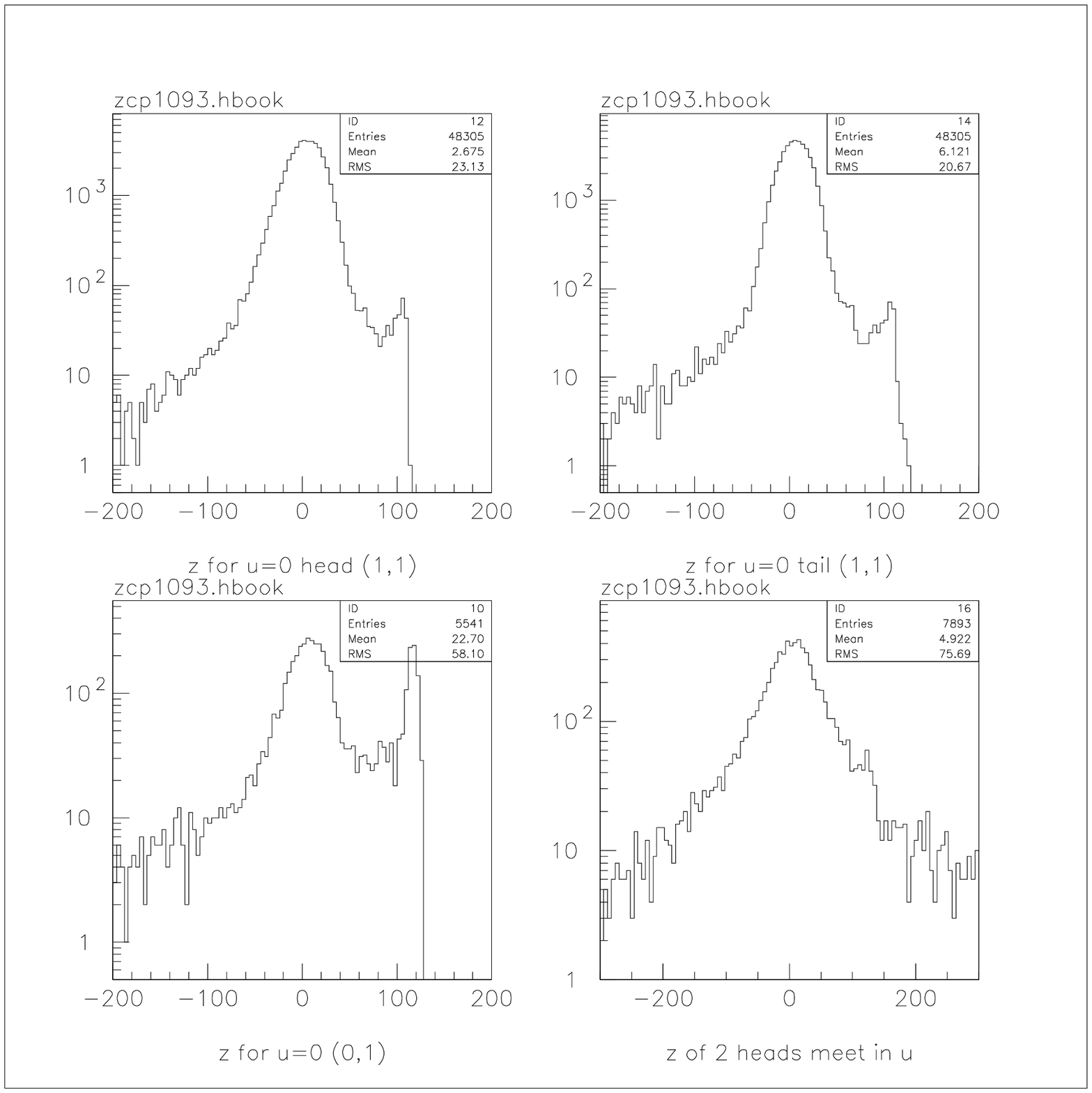,width=6in}
\vspace{-1in}
\caption[Pointing in $u$ for heads and tails of charged tracks, and for single 
photon conversion tracks (lead-out run 1093).]
{Pointing, defined as the $z$ where the track goes through $u=0$, for
heads and tails of charged tracks, single photon conversion tracks, and
the $z$ where any two heads have the same $u$ position, for lead-out run 1093.}
\label{f:pointout} \end{figure}
\begin{figure}[h] \vspace{-0.5in}\hspace*{-0.3in}
\psfig{file=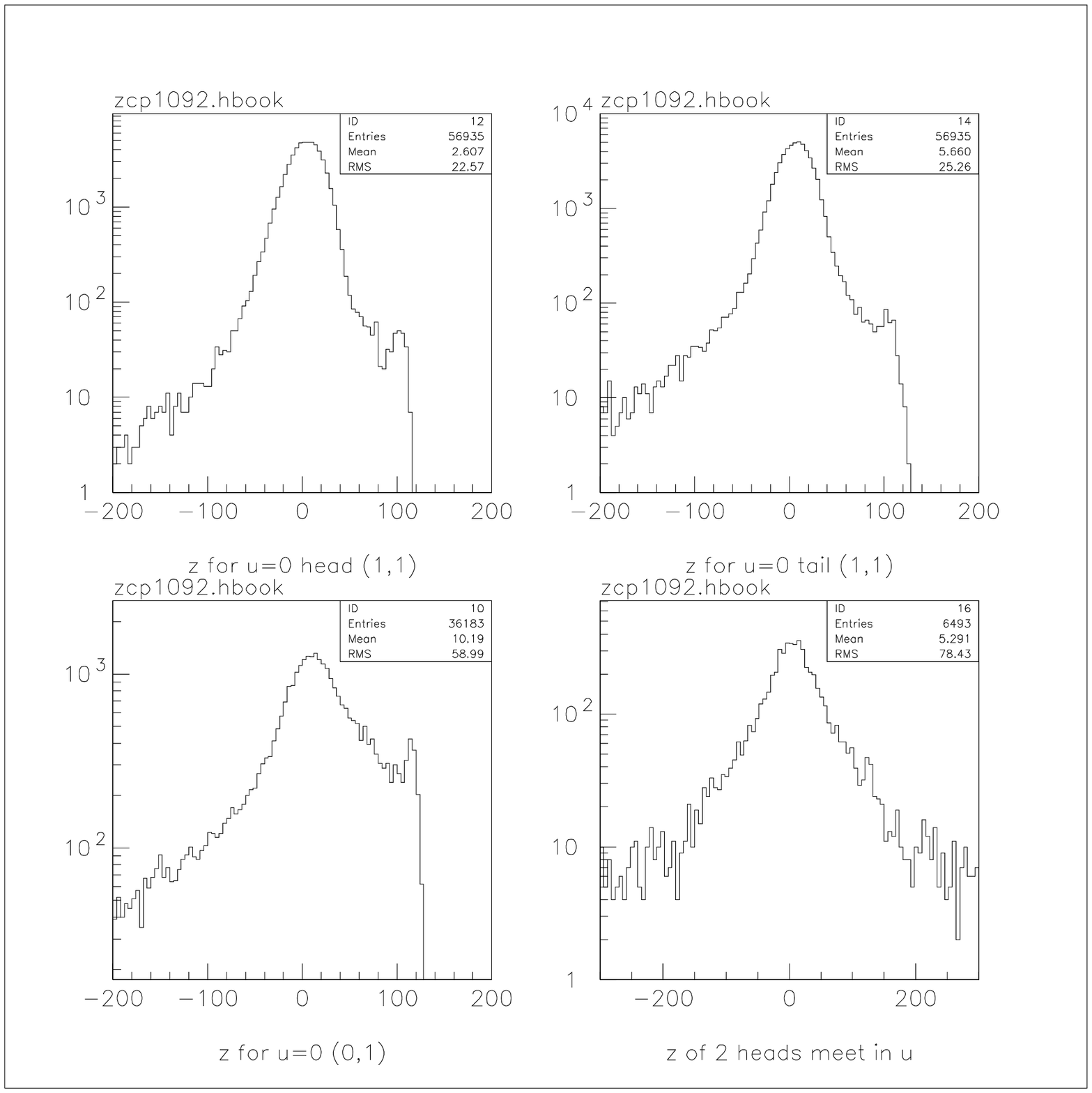,width=6in}
\vspace{-1in}
\caption[Pointing in $u$ for heads and tails of charged tracks, and for single 
photon conversion tracks (lead-in run 1092).]
{Pointing, defined as the $z$ where the track goes through $u=0$, for
heads and tails of charged tracks, single photon conversion tracks, and
the $z$ where any two heads have the same $u$ position, for lead-in run 1092.}
\label{f:pointin} \end{figure}
\begin{figure}[h] \vspace{-0.5in}\hspace*{-0.3in}
\psfig{file=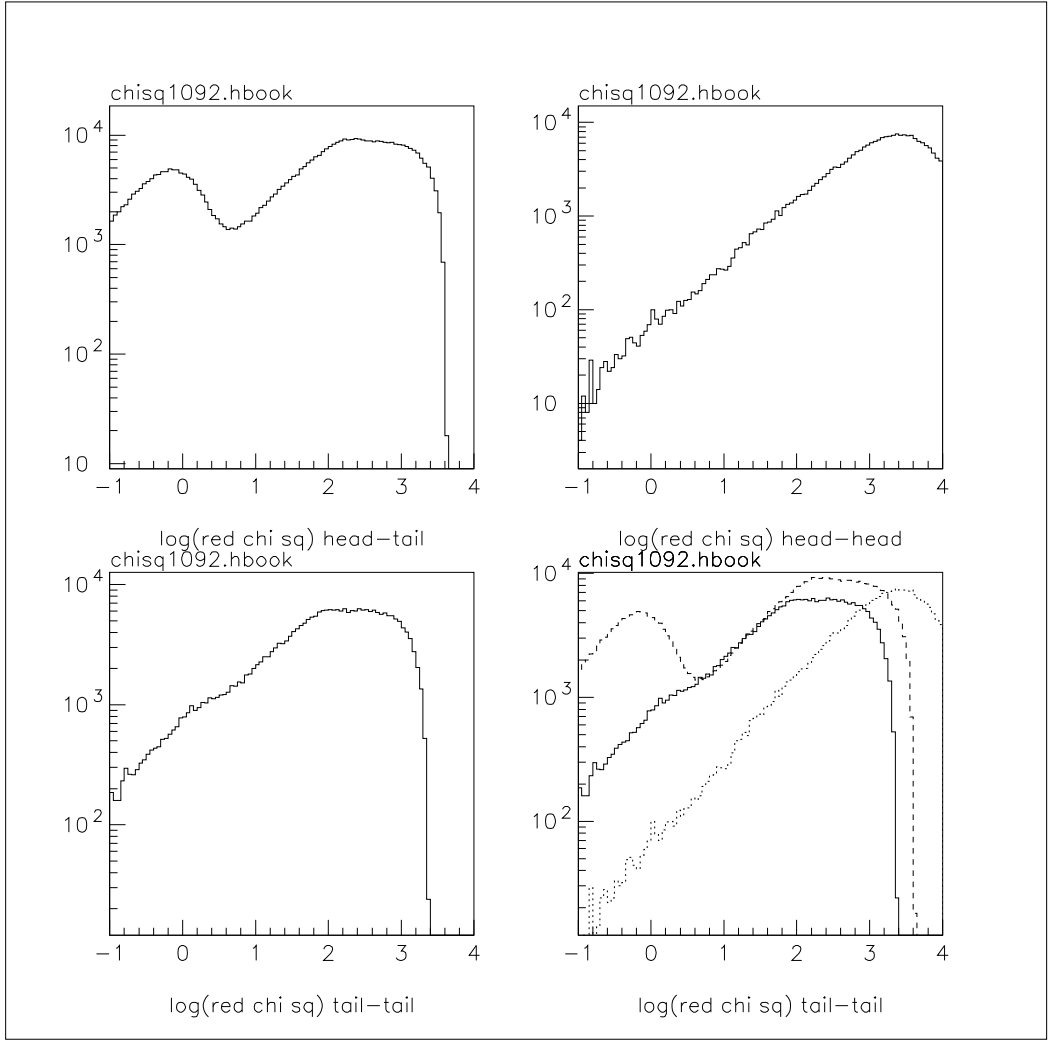,width=6in}
\vspace{-1in}
\caption[Values of $\log{\chi^2\over 2}$ for head-tail, head-head, and
tail-tail matches at the lead.]
{Values of $\log{\chi^2\over 2}$ for head-tail, head-head, and tail-tail
matches at the lead.  A superposition of these plots shows that
$\log{\chi^2\over 2}=0.7$, or ${\chi^2\over 2}=5$ is a reasonable cutoff 
between matching real tracks together and track segments which are 
coincidentally close at the lead.}
\label{f:chisq} \end{figure}

\begin{figure}[h] \vspace{-0.5in}\hspace*{-0.3in}
\psfig{file=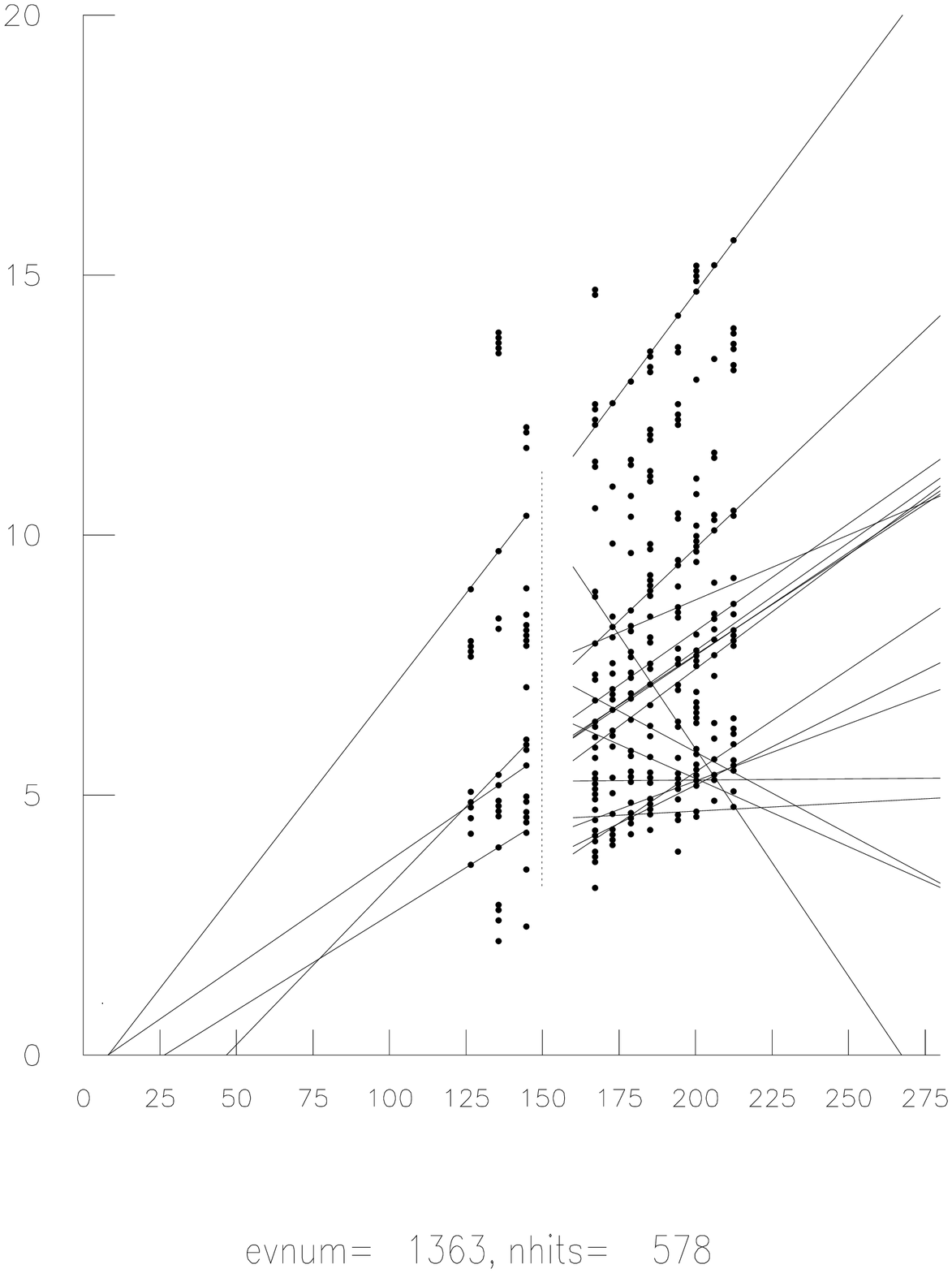,width=6in,height=6in}
\caption{Display of the event shown in Fig. 4.6 for tracks kept by the
vertexer.}
\label{f:uglyv} \end{figure}

\begin{figure}[h] \vspace{-0.5in}\hspace*{-0.3in}
\psfig{file=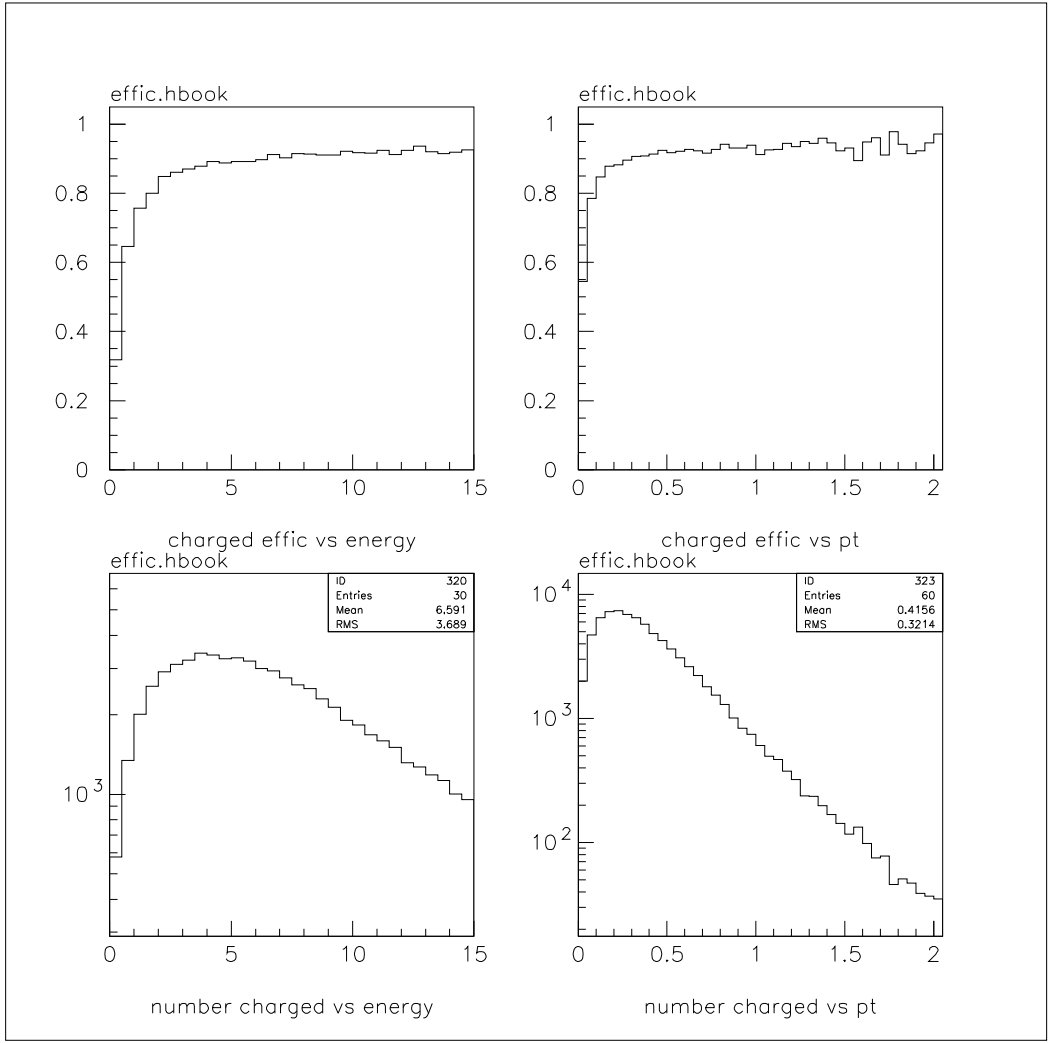,width=6in}
\vspace{-1in}
\caption[Efficiencies for finding charged tracks as a function of
energy and transverse momentum.]{Efficiencies for finding charged tracks as a 
function of energy and transverse momentum of the track, number of charged 
tracks vs energy, transverse momentum.}
\label{f:eff_chE} \end{figure}
\begin{figure}[h] \vspace{-0.5in}\hspace*{-0.3in}
\psfig{file=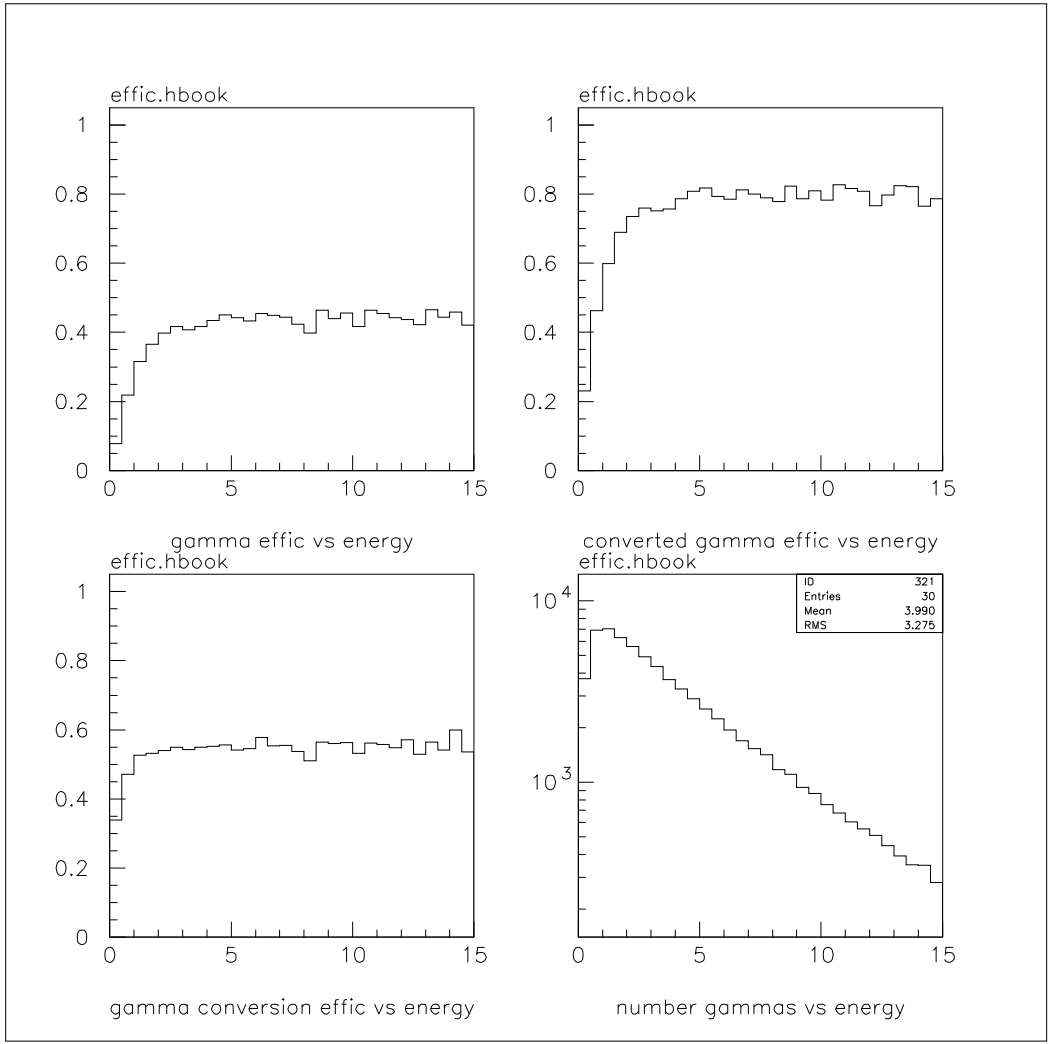,width=6in}
\vspace{-1in}
\caption[Efficiency for finding photons as a function of photon energy.]
{Efficiency for finding photons as a function of photon energy, 
efficiency for finding photons know to convert vs energy, probability
of conversion of a photon vs energy, number of photons vs energy.}
\label{f:eff_gE} \end{figure}
\begin{figure}[h] \vspace{-0.5in}\hspace*{-0.3in}
\psfig{file=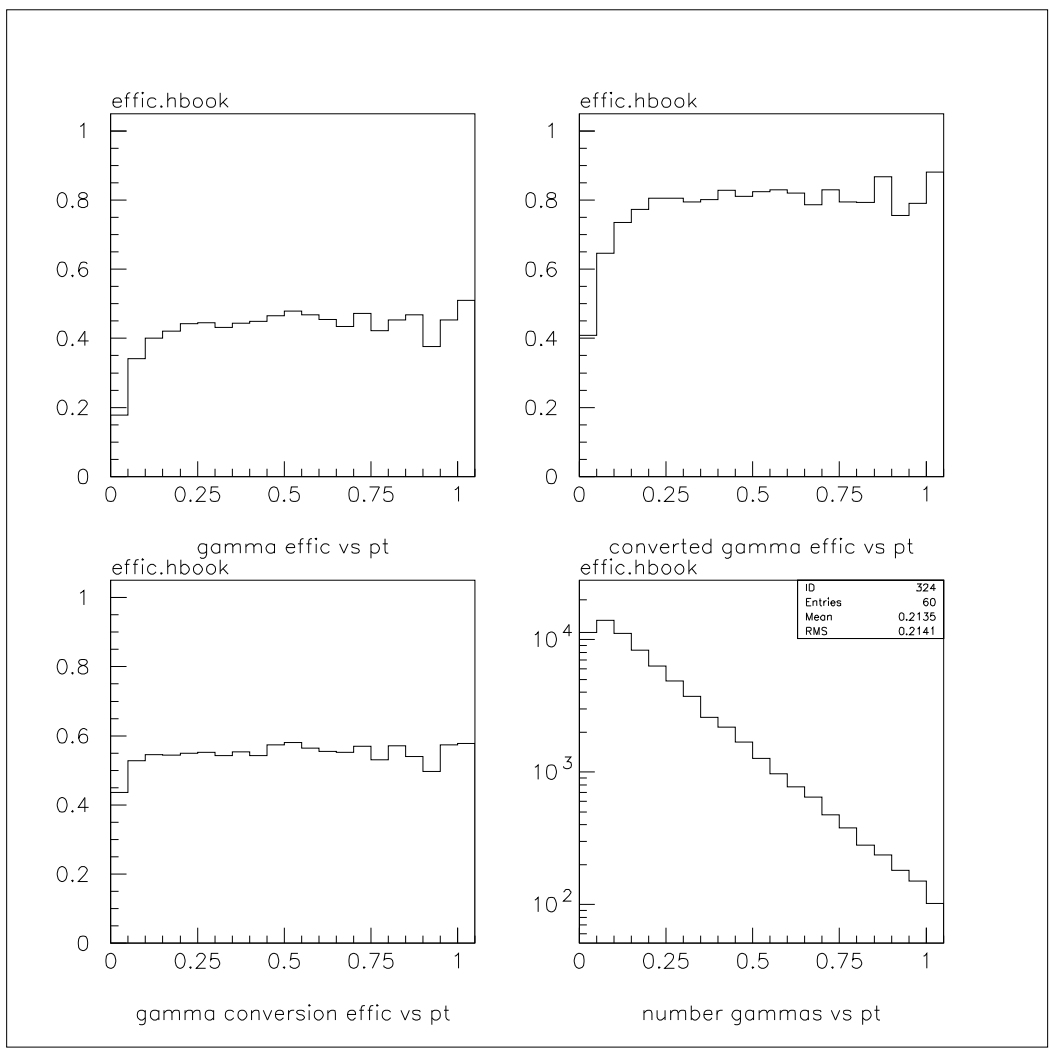,width=6in}
\vspace{-1in}
\caption[Efficiency for finding photons as a function of photon transverse
momentum.]{Efficiency for finding photons as a function of photon transverse
momentum, efficiency for finding photons know to convert vs $p_T$, probability
of conversion of a photon vs $p_T$, number of photons vs $p_T$.}
\label{f:eff_gpt} \end{figure}
\begin{figure}[h] \vspace{-0.5in}\hspace*{-0.3in}
\psfig{file=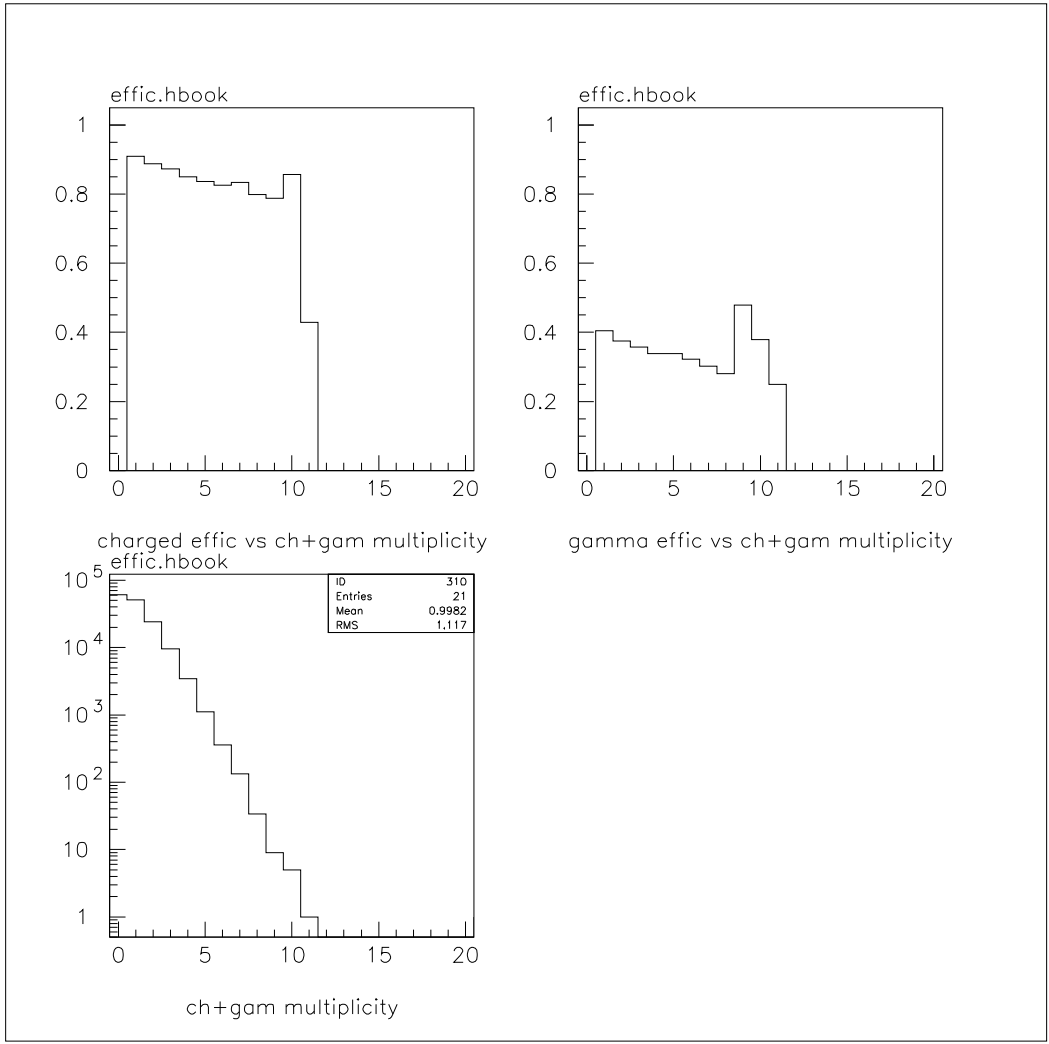,width=6in}
\vspace{-1in}
\caption[Efficiency for finding charged tracks and photons as a function of 
total multiplicity into the acceptance.]
{Efficiency for finding charged tracks as a function of total
multiplicity into the acceptance, efficiency for finding photons vs total
multiplicity, frequency of observing an event with a given total multiplicity.}
\label{f:eff_mult} \end{figure}
\clearpage
\begin{figure}[h] \vspace{-0.5in}\hspace*{-0.3in}
\psfig{file=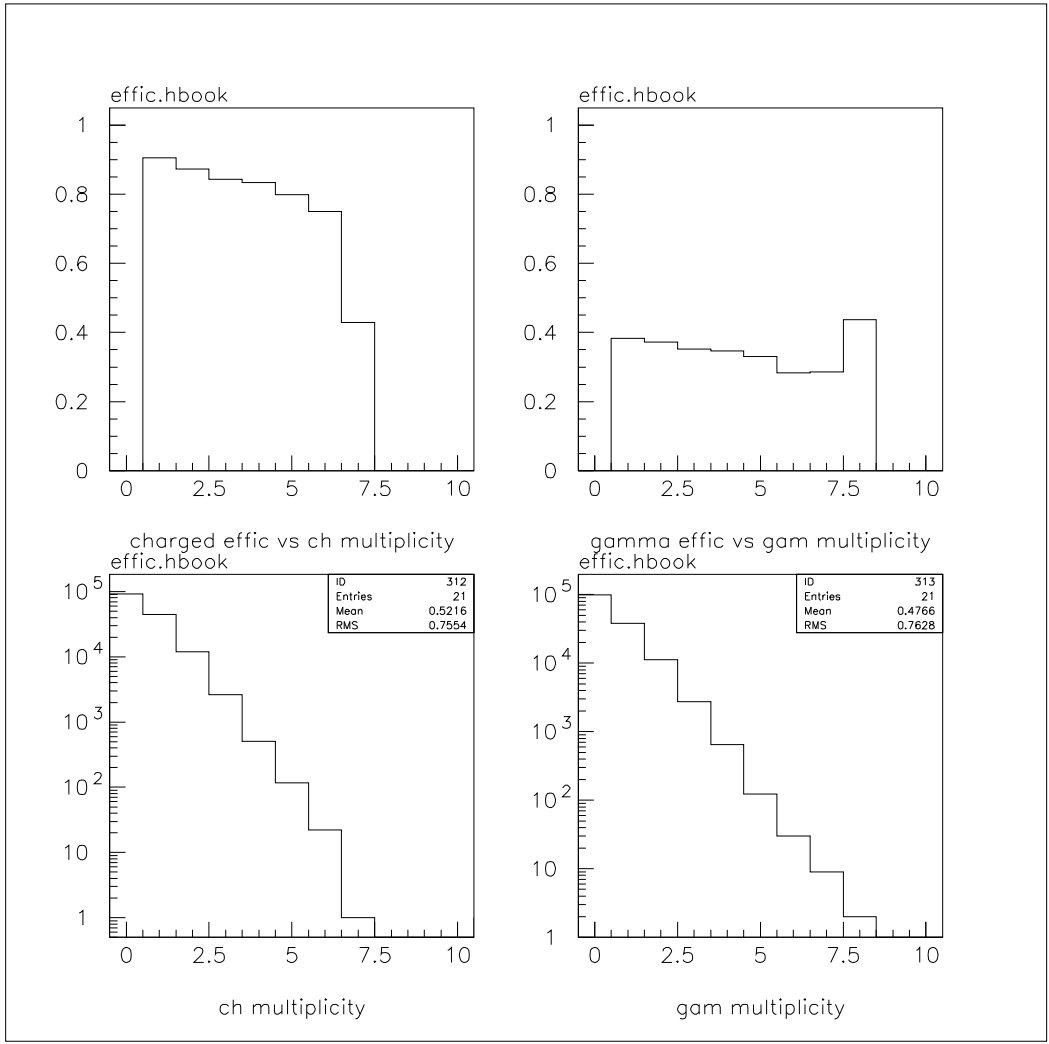,width=6in}
\vspace{-1in}
\caption[Efficiency for finding charged tracks as a function of charged
multiplicity into the acceptance, efficiency for finding photons vs photon
multiplicity.]{Efficiency for finding charged tracks as a function of charged
multiplicity into the acceptance, efficiency for finding photons vs photon
multiplicity, frequency of observing an event with a given charged multiplicity,
frequency vs photon multiplicity.}
\label{f:eff_mult2} \end{figure}
\begin{figure}[h] \vspace{-0.5in}\hspace*{-0.3in}
\psfig{file=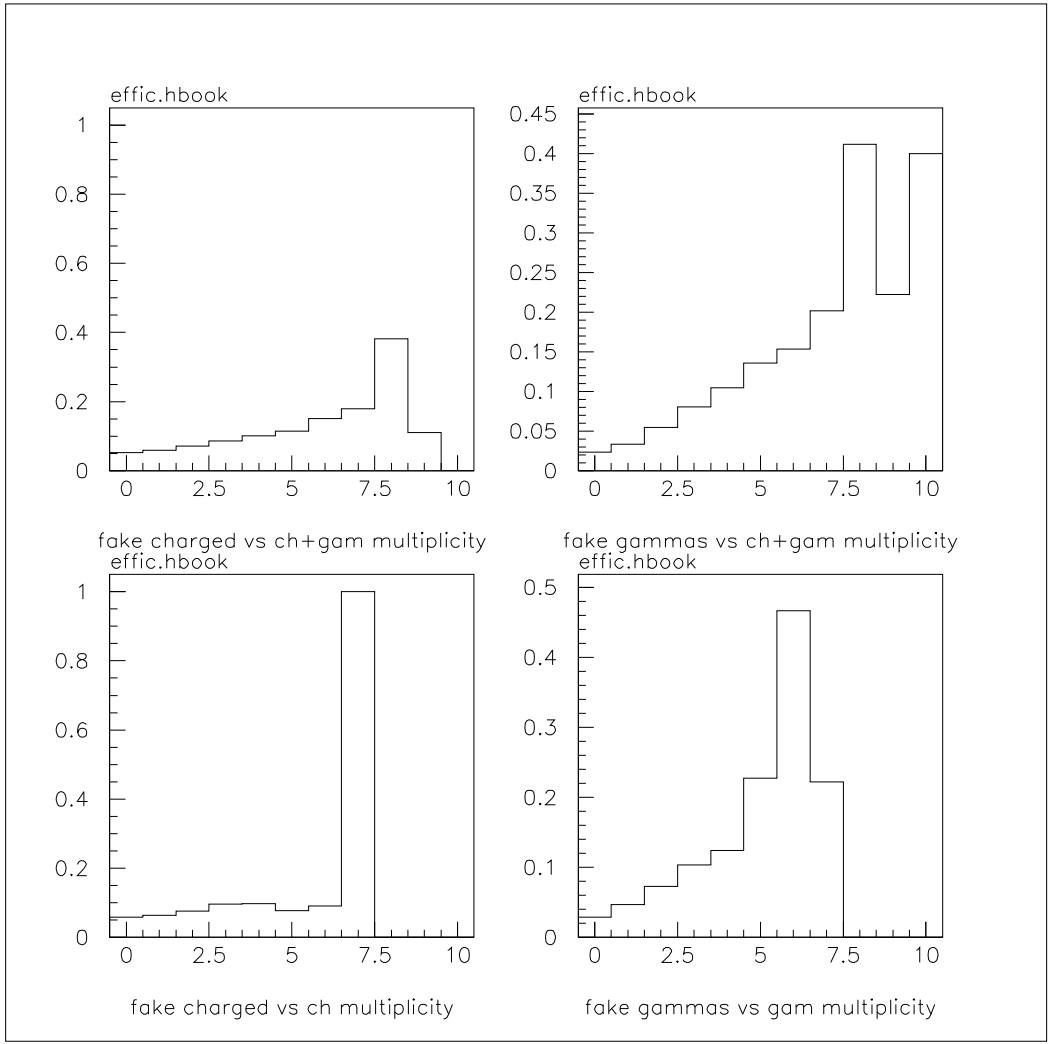,width=6in}
\vspace{-1in}
\caption[Mean number of fake charged tracks and photons found per event with a 
given multiplicity.] {Mean number of fake charged tracks found per event with a 
given total multiplicity, mean number of fake photons found per event with a 
given photon multiplicity, mean number of fake charged tracks vs charged 
multiplicity, mean number of fake photons vs photon multiplicity.}
\label{f:eff_fmult} \end{figure}
\clearpage
\begin{figure}[h] \vspace{-0.5in}\hspace*{-0.3in}
\psfig{file=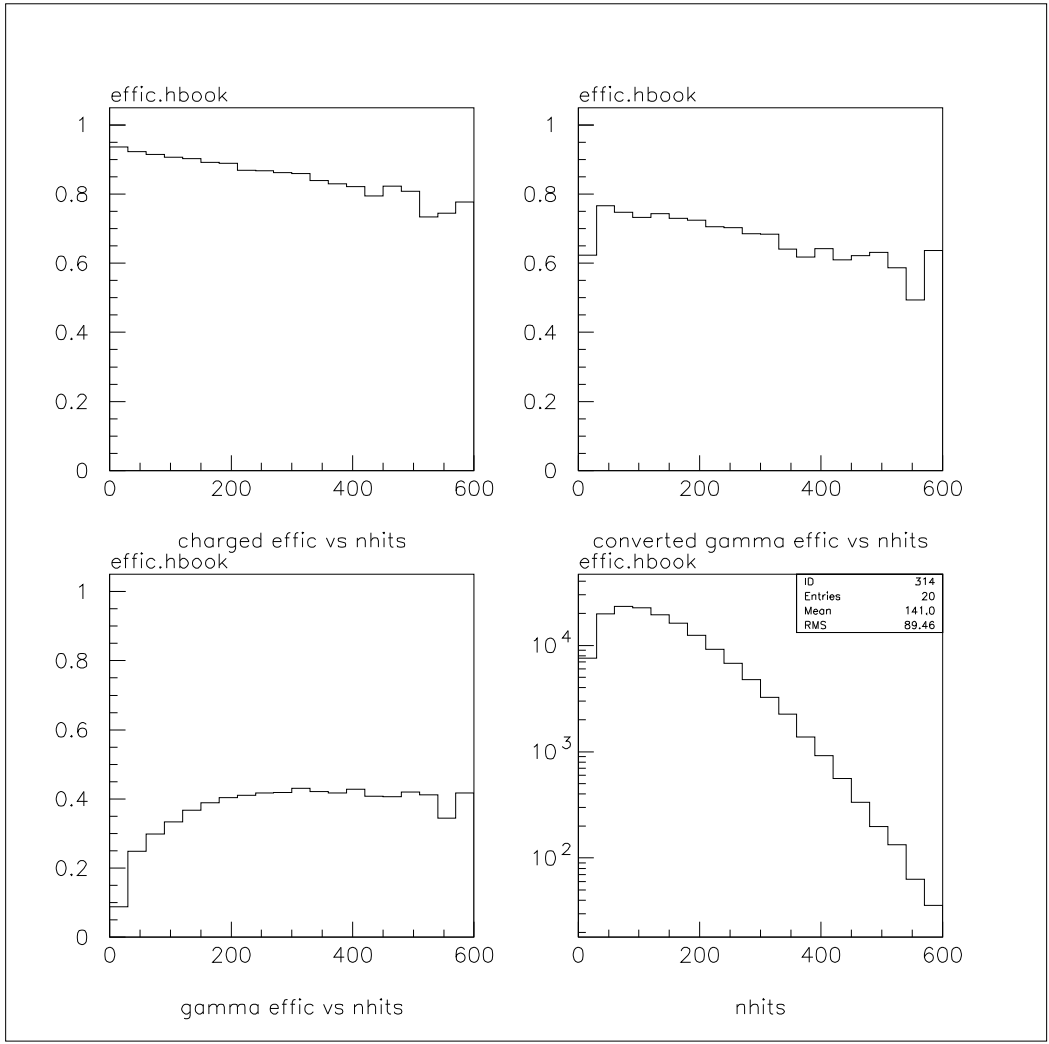,width=6in}
\vspace{-1in}
\caption[Efficiency for finding charged tracks and photons as a function of 
NHITS of the event.]{Efficiency for finding charged tracks as a function of 
NHITS of the event, efficiencies for finding converted photons and all photons vs 
NHITS, frequency of observing an event with a given NHITS.}
\label{f:eff_nhit} \end{figure}
\begin{figure}[h] \vspace{-0.5in}\hspace*{-0.3in}
\psfig{file=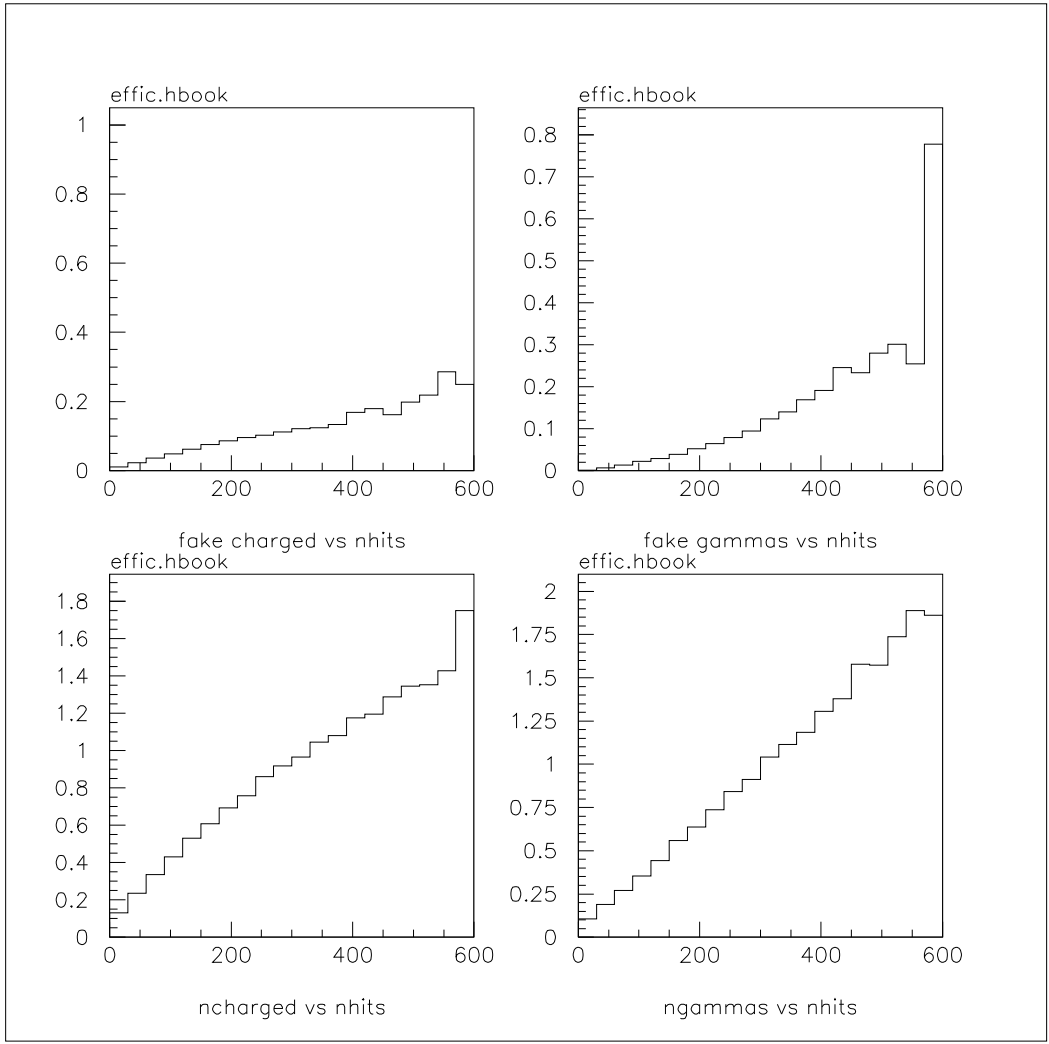,width=6in}
\vspace{-1in}
\caption[Mean number of fake charged tracks and photons found per event with a 
given NHITS.]{Mean number of fake charged tracks found per event with a given
NHITS, mean number of fake photons found per event with a given NHITS,
mean numbers of real charged tracks, real photons given NHITS.}
\label{f:eff_fnhit} \end{figure}
\begin{figure}[h] \vspace{-0.5in}\hspace*{-0.3in}
\psfig{file=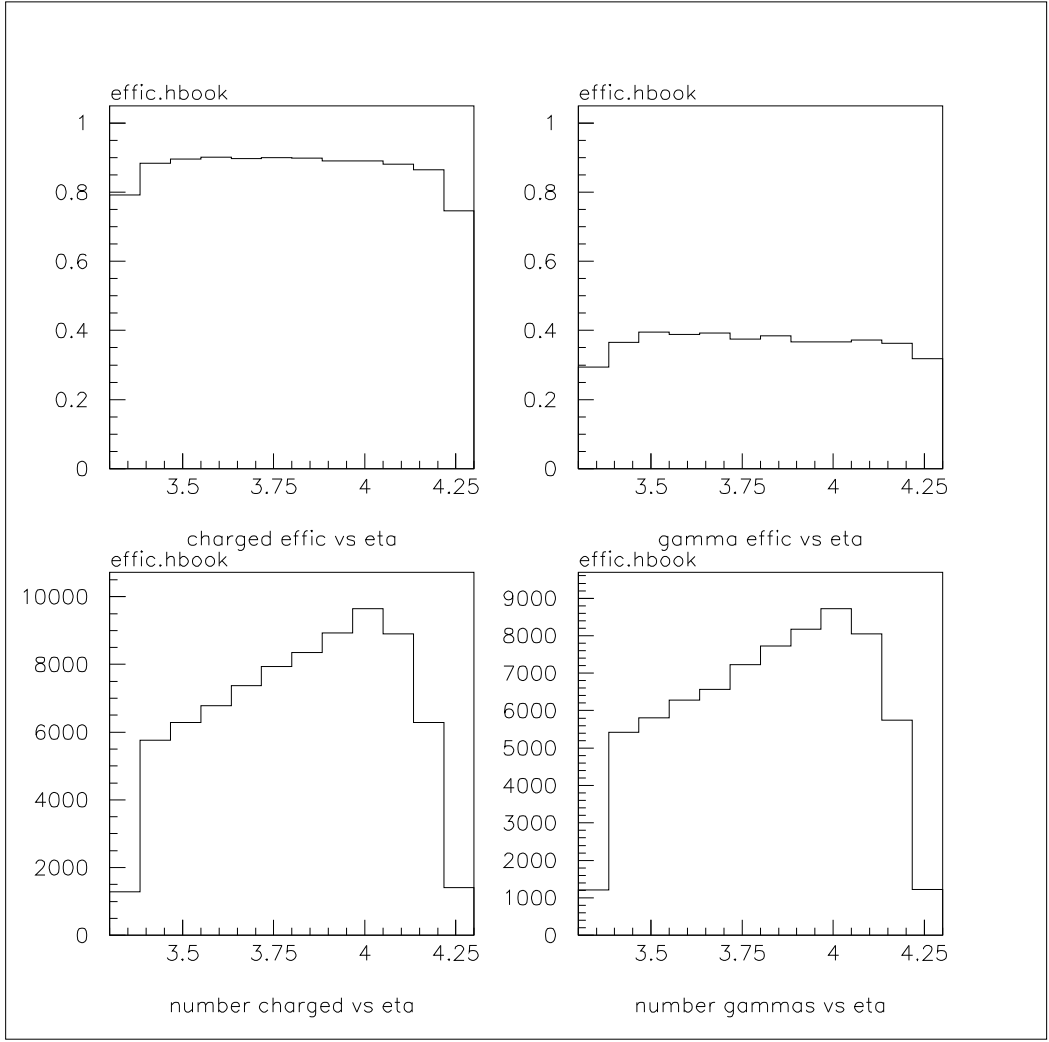,width=6in}
\vspace{-1in}
\caption[Efficiencies for finding charged tracks, photons at a given $\eta$.]
{Efficiencies for finding charged tracks, photons at a given $\eta$, 
number of charged tracks, photons vs $\eta$.}
\label{f:eff_eta} \end{figure}
\begin{figure}[h] \vspace{-0.5in}\hspace*{-0.3in}
\psfig{file=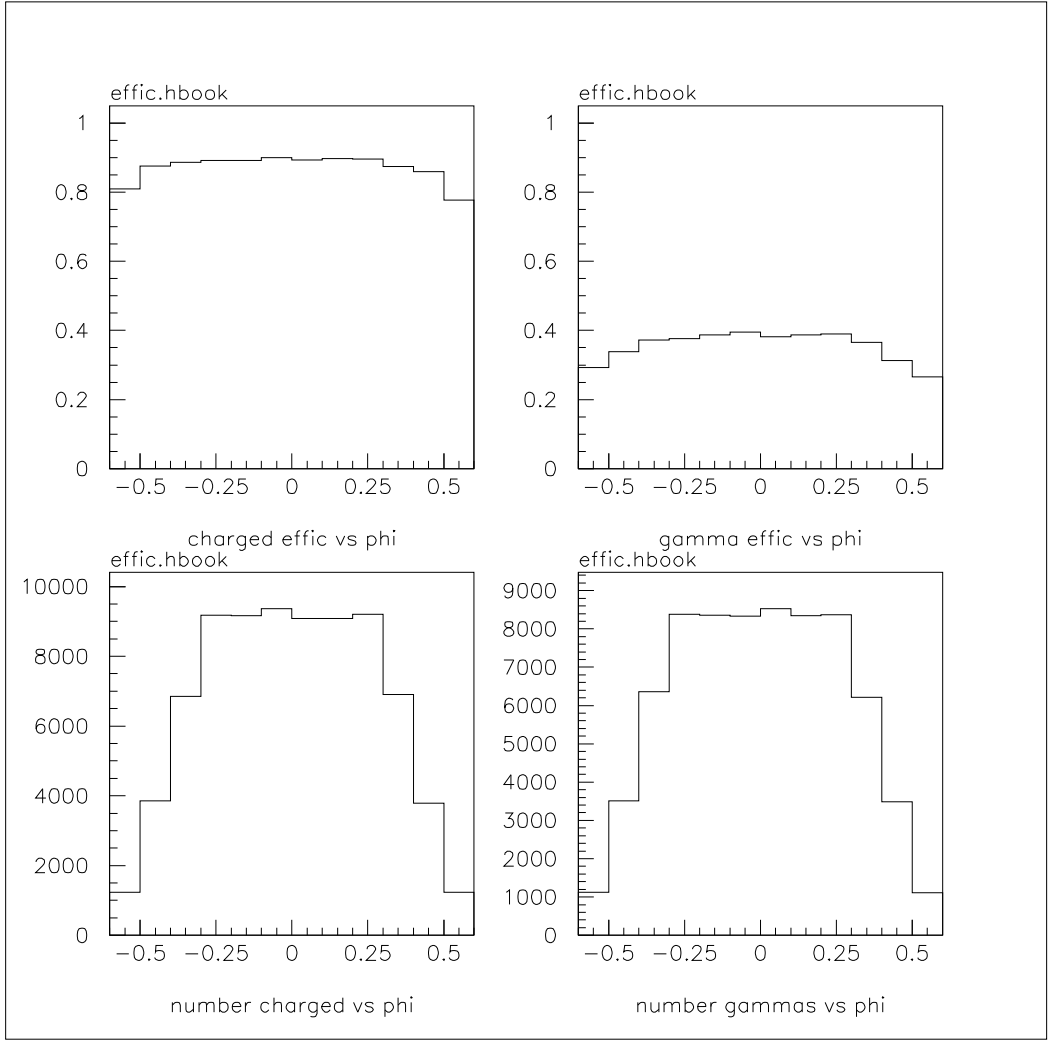,width=6in}
\vspace{-1in}
\caption[Efficiencies for finding charged tracks, photons at a given $\phi$.]
{Efficiencies for finding charged tracks, photons at a given $\phi$, 
number of charged tracks, photons vs $\phi$.}
\label{f:eff_phi} \end{figure}

\begin{figure}[h] \vspace{-0.5in}\hspace*{-0.3in}
\psfig{file=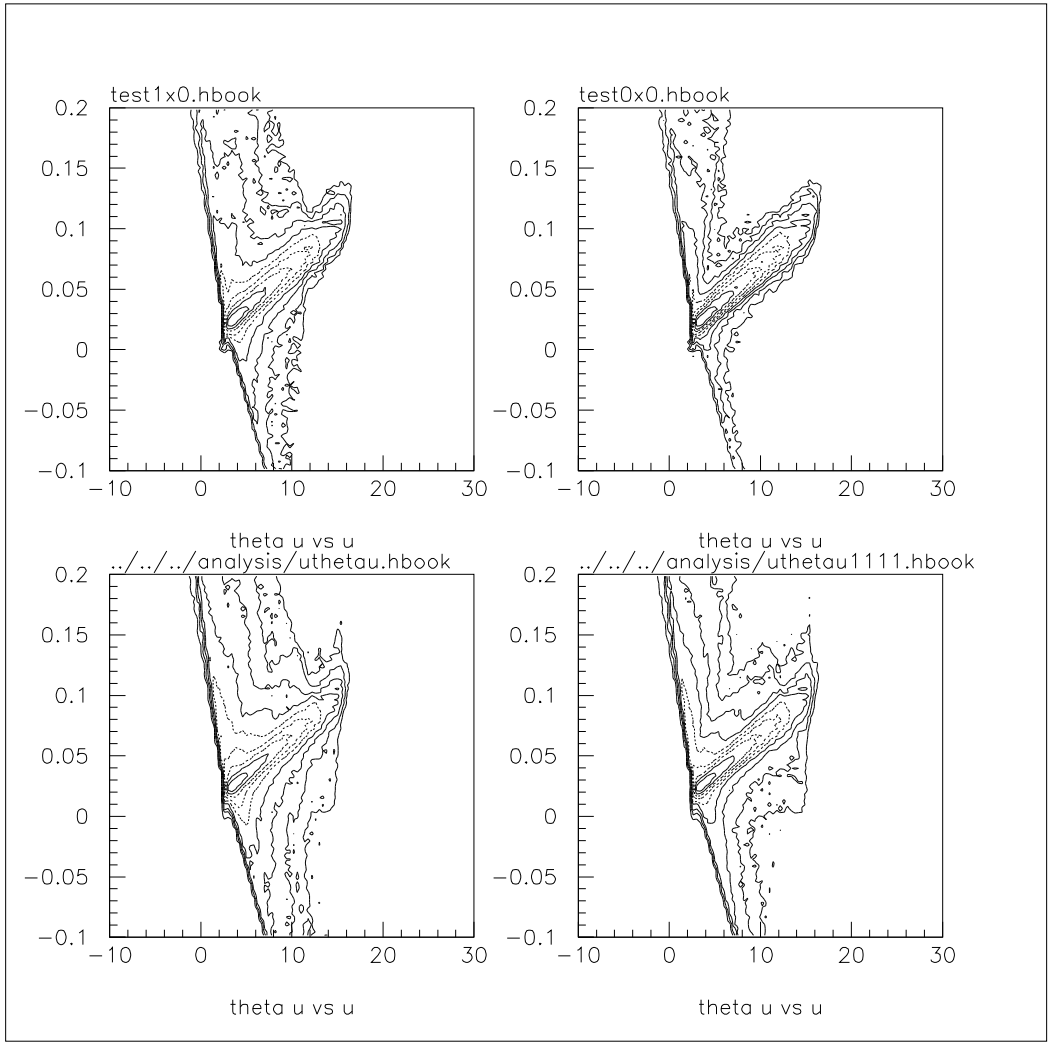,width=6in}
\vspace{-1in}
\caption[Location of tracks at the lead in phase space $(u,\theta_u)$.]
{Location of tracks at the lead in phase space $(u,\theta_u)$ for
lead-in GEANT, lead-out GEANT, lead-in run 1110, lead-out run 1111.}
\label{f:uthetau} \end{figure}
\begin{figure}[h] \vspace{-0.5in}\hspace*{-0.3in}
\psfig{file=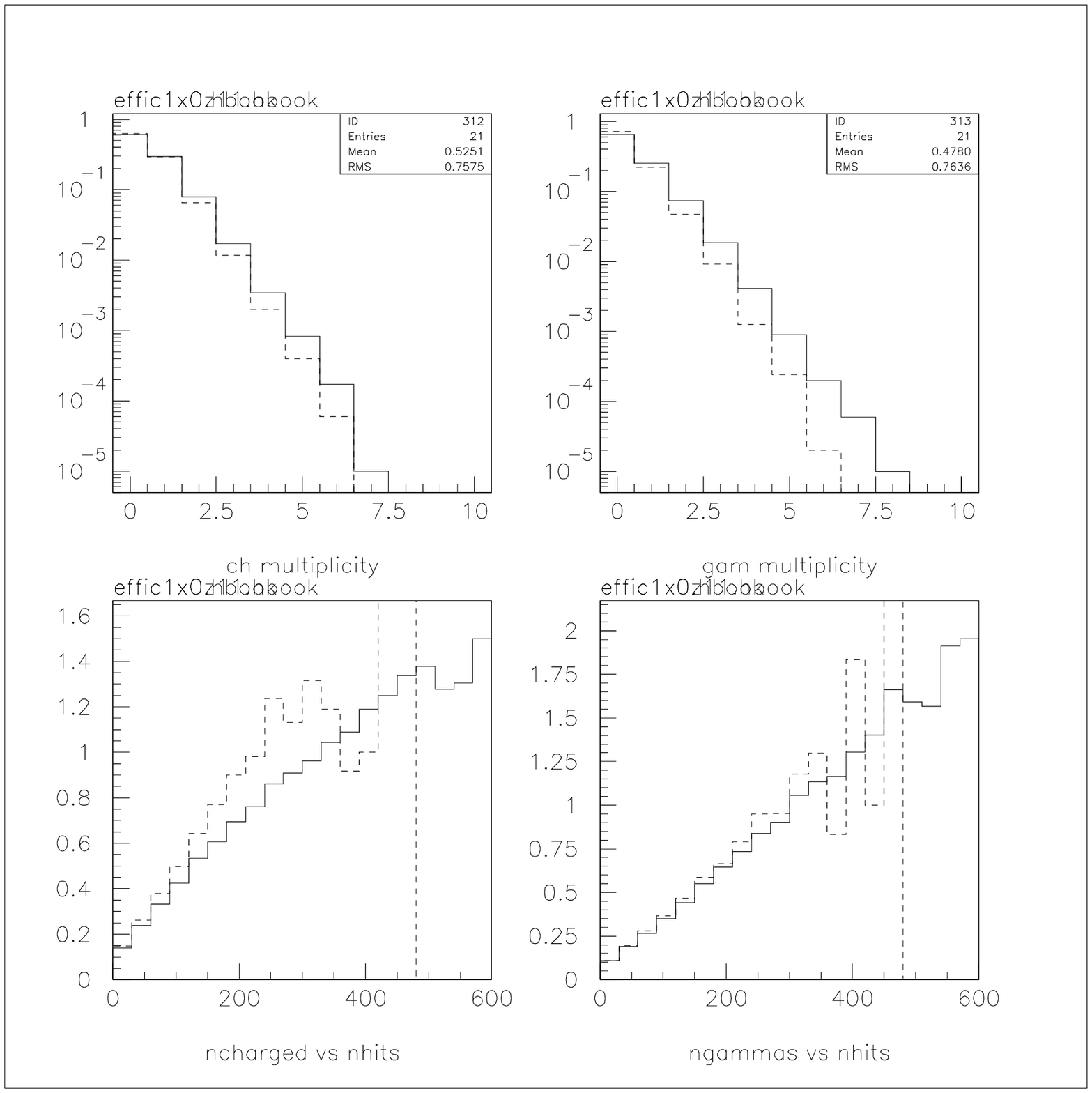,width=6in}
\vspace{-1in}
\caption[PYTHIA charged-track and photon multiplicity distributions,
mean multiplicities vs NHITS for all lead-in GEANT events and for
non-vetoed events.]{PYTHIA charged-track and photon multiplicity distributions, 
mean multiplicities vs NHITS for all lead-in GEANT events (solid) and 
for non-vetoed events (dashed).}
\label{f:gpipeveto} \end{figure}
\begin{figure}[h] \vspace{-0.5in}\hspace*{-0.3in}
\psfig{file=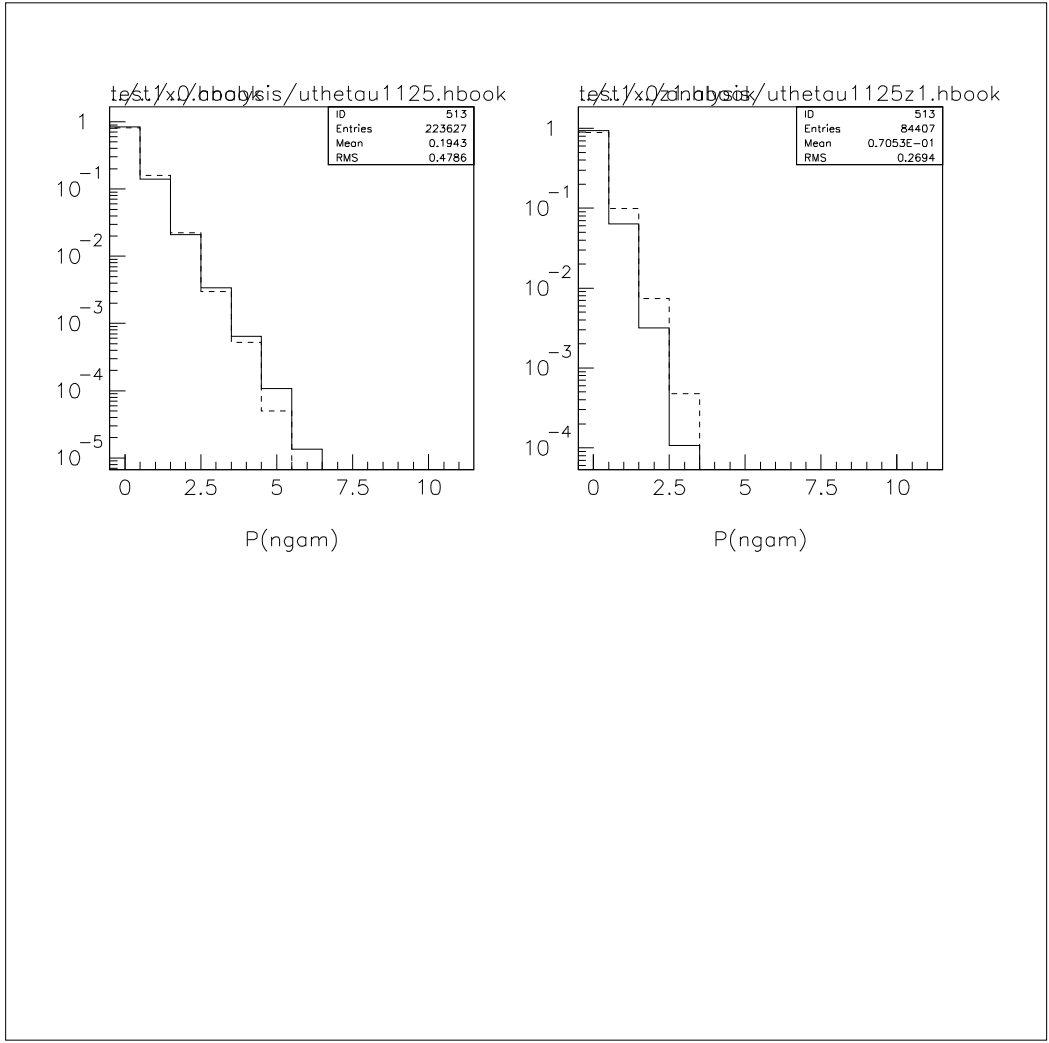,width=6in}
\vspace{-1in}
\caption[Observed photon multiplicity distribution for all events and for
non-vetoed events.]{Observed photon multiplicity distribution for all events 
and for non-vetoed events from lead-in run 1125 (solid) and lead-in GEANT 
(dashed).}
\label{f:pipeveto} \end{figure}

\clearpage

\begin{table}[e]
\begin{center}
\begin{tabular}{|c|c|c|c|c|}
\hline
track & chambers & number of & used-wire & grouping \\
type &  searched & chambers hit & cut & cut \\
\hline
charged & 1-24 & $\geq 22$ & 17 & 16 \\
 &  & \ \ \ 21 & 17 & 16 \\
\hline
photon/ & 9-24 & $\geq 15$ & 13 & 11 \\
ch tail &  & \ \ \ 14 & 13 & 11 \\
\hline
mid-8 & 9-16 & $\geq 7$ & 5 & 6 \\
\hline
front-8 & 1-8 & $\geq 7$ & 5 & 6 \\
\hline
\end{tabular}
\end{center}
\caption{\label{t:tracker}Cuts used by the tracker.}
\end{table}

\clearpage
\chapter{Generating Function Formalism and Robust Observables}

In order to determine the distribution of the neutral fraction ($f$)
of pions in the MiniMax detector, we would like to count the number of 
charged and neutral pions from a collision entering our acceptance.
(Even if only a piece of a DCC domain enters the acceptance, we would
expect on average to observe the same $f$ as for the entire domain.)
This approach is complicated by many things, including the fact that
$\pi^0$'s decay almost immediately into two photons which we identify only 
if they convert in the lead or scintillator within the MWPC telescope.
To precisely determine the number of $\pi^0$'s, we would need to identify
both $\gamma$'s from a decay and reconstruct the pion mass using the
electromagnetic calorimeter.

We are unable to do this due to the fact that the probability of both photons
from a decay entering the small acceptance of the calorimeter is only about 
15\%.  However, we have found a set of observables which sample the
charged-neutral distribution and are independent of many of the 
detector-related complications and (uncorrelated) efficiencies, and
take very different values for pure DCC and for generic particle production.
We make many bold assumptions about the production and detection of particles
going into the detector in order to establish the robustness of these
observables; however, simulations indicate that the assumptions are
not unreasonable (see Sec. \ref{sec:rgeant}).
We assume that particles other than pions can be ignored, 
that charged particles and photons are not misidentified, 
that the production process can be modeled as a two-step process, with
the total number of pions given by a parent distribution and the fraction
of pions that are neutral given by, e.g., a binomial or DCC 
distribution function, and that detection efficiencies for finding
a charged particle or photon do not depend on the nature of the rest of the 
event.

These observables and their properties are best understood using generating 
functions and their factorial moments to describe probability distributions
for the production of some species of particle.  
The generating function formalism has been widely used in multiparticle
analysis \cite{gf1}-\cite{gf5}
and was extended to two variables \cite{gf6}-\cite{gf9}, \cite{UA5mult},
e.g., charged and neutral pions
by the MiniMax Collaboration in Ref. \cite{robust}. 
The ideas in this section are described in detail in that paper.

Some of the difficulties in measuring the charged-neutral distribution
are listed below:
\begin{enumerate}
\item The MiniMax acceptance is small, so that it is improbable that 
both $\gamma$'s from a $\pi^{0}$ enter the detector acceptance;
\item the conversion efficiency per $\gamma$ is only about 50\%.;
\item not all $\gamma$'s come from $\pi^{0}$'s;
\item not all charged tracks come from $\pi^{\pm}$'s; 
\item because of the small acceptance, the multiplicities are rather low, 
so that statistical fluctuations are very important; 
\item detection efficiencies are not the same for charged tracks and $\gamma$'s 
and are momentum-dependent;
\item efficiency functions may depend on the observed multiplicity 
or other parameters;
\item the efficiency for triggering when no charged track 
or converted $\gamma$ is produced within the acceptance is relatively low and 
different from that for events in which at least one charged particle or 
converted $\gamma$ is detected. 
\end{enumerate}


\section{Generating functions for charged-pion--neutral-pion distributions}

If the set of (normalized) probabilities for producing $N$ particles in a 
given region of 
phase space is $\{P(N)\}$, then the generating function can be defined as
\begin{equation}
G(z) =\sum\limits_{N=0}^{\infty}z^{N}P(N),
\end{equation}
and contains all the information of the $\{P(N)\}$:
\begin{equation} P(N)={1\over N!}\left({d^N G\over dz^N}\right)_{z=0}.
\end{equation}
Information can likewise be extracted from the factorial moments, defined as
\begin{equation}
f_{i} \equiv \left({{d^{i}G(z)} \over {dz^{i} }}\right)_{z=1} 
       =  \langle N(N-1) \cdots (N -i+1) \rangle,
\end{equation}
where $\left<{\cal O}\right>=\sum_N {\cal O} P(N)$.

Now if $p(n_{ch}, n_{0})$ is the probability distribution for producing
$n_{ch}$ charged and $n_{0}$ neutral pions, the generating function is
\begin{equation}   
G(z_{ch}, z_{0}) = \sum_{n_{ch}=0}^\infty
\sum_{n_0=0}^{\infty} p(n_{ch}, n_{0}) z_{ch}^{n_{ch}} 
z_{0}^{n_{0}},
\label{eq:gch0}
\end{equation} 
and the factorial moments for charged ($ch$) and neutral (0) pions are
\begin{equation}
f_{i,j}(ch, 0) \equiv \left({{\partial^{i+j}G(z_{ch}, z_{0})} \over 
{\partial z_{ch}\,^{i} \partial z_{0}\,^{j}}}\right)_{z_{ch}=z_{0}=1}.
\end{equation}

We assume that $p(n_{ch},n_0)$ can be written as the product of a parent
distribution $P(N)$ for producing $N$ total pions, and 
$\hat{p}(n_{ch},n_0;N)$,
which gives the charged-neutral distribution of pions for a given 
$N=n_{ch}+n_0$:
\begin{equation}
p(n_{ch}, n_{0}) =  P(N)\hat{p}(n_{ch}, n_{0}; N), 
\label{eq:pnchn0}
\end{equation}
where 
\begin{equation}            
\sum\limits_{N=0}^{\infty}P(N)=1, 
\end{equation} 
\begin{equation}           
\sum_{n_{ch}=0}^{\infty}\sum_{n_{0}=0}^{\infty} 
\delta_{N , n_{ch} + n_{0}}
\,\hat{p}(n_{ch}, n_{0}; N) =  1. 
\end{equation} 

For generic production of pions, the charge is distributed according to a 
binomial ($bin$) with mean neutral fraction $\hat{f}=1/3$, so that
\begin{equation}
\hat{p}_{bin}(n_{ch}, n_{0}; N) = {N!\over n_0!(N-n_0)!}
\hat{f}^{n_{0}}(1-\hat{f})^{n_{ch}}.
\label{eq:pbin}
\end{equation}
The corresponding generating function [from Eqs. (\ref{eq:pbin}),
(\ref{eq:pnchn0}) and (\ref{eq:gch0})] is
\begin{equation}   
G_{bin}(z_{ch}, z_{0};\hat{f}) = \sum\limits_{N} P(N) [\hat{f}z_{0} + 
(1-\hat{f}) z_{ch}]^{N}.
\label{eq:gbin}
\end{equation} 
Note that $G_{bin}$ depends only on the linear combination
\begin{equation} 
\zeta \equiv \hat{f}z_{0} + (1-\hat{f}) z_{ch}.
\end{equation} 

Much of the simplicity of the generic case is also realized for
 what can be called the binomial transform,
\begin{equation}   
\hat{p}(n_{ch}, n_{0}; N) = {N!\over n_0!(N-n_0)!}
\int_0^1 {{df}}p(f)f^{n_{0}}(1-f)^{n_{ch}}, 
\end{equation} 
of the arbitrary normalized distribution $p(f)$, such as the DCC distribution
$p(f)=1/(2\sqrt{f})$.
This leads to a wide class of possible pion factorial-moment generating 
functions, namely
\begin{equation}   
G(z_{ch}, z_{0}) =  \int_0^1 {{df}}p(f) G_{bin}(z_{ch},z_{0};f), 
\label{eq:gwide}
\end{equation}
where $G_{bin}(z_{ch}, z_{0};f)$ is given by (\ref{eq:gbin}) with 
$\hat{f}$ replaced by an arbitrary $f$, $0 \leq f \leq 1$. 

If $P(N)$ is a Poisson distribution, [$P(N)={\mu^N\over N!}e^{-\mu}$ with
$\mu=\left< N\right>$] then
$G_{bin}(z_{ch},z_0;\hat{f})=e^{-\mu+\mu\zeta}$, so that
$\ln G_{bin}(z_{ch}, z_{0}; \hat{f})$ 
is linear in $\zeta$. 
The PYTHIA simulations yield generating functions that, 
to good approximation, depend only on a fixed 
linear combination of $z_{ch}$ and $z_{0}$ (Fig. \ref{f:genfp}); the full 
detector simulation with GEANT is found to alter this linear behavior slightly.
Compare this to the generating function for the DCC distribution 
which depends on both $z_{ch}$ and $z_0$ (Fig. \ref{f:genfdcc}).

\section{Generating functions for charged-pion--photon distributions}

Next we take into account the probability $\epsilon_{ch}$ for observing a 
given primary charged pion in the detector and a probability 
$(1-\epsilon_{ch})$ 
for not observing it, and the probabilities $\epsilon_{m}$, $m =0, 1, 2$, 
for observing $m$ photons from a $\pi^{0}$ decay, with
$\epsilon_{0} + \epsilon_{1} + \epsilon_{2} = 1$.
We assume that these efficiencies are uncorrelated.
Then the generating function for the distribution of observed particles, 
including efficiencies, is obtained from $G(z_{ch}, z_{0})$  \cite{Pumplin} 
by replacing $z_{ch}$ by the generating function
\begin{equation}
g_{ch}(z_{ch})=(1-\epsilon_{ch})+\epsilon_{ch}z_{ch},
\end{equation} 
 and $z_{0}$ by the generating function
\begin{equation}
g_{0}(z_{\gamma})=\epsilon_{0} +\epsilon_{1} z_{\gamma} +\epsilon_{2}
{z_\gamma}^{2}.
\end{equation}

For the class of production models characterized by (\ref{eq:gwide}), this
leads to the following factorial-moment generating 
function for the distribution of observed charged pions and photons:
\begin{equation}
  G_{obs}(z_{ch}, z_{\gamma}) = \int_0^1 {{df}}p(f)G_{bin}
(g_{ch}(z_{ch}), 
g_{0}(z_{\gamma});f).
\label{eq:gobs}
\end{equation}
The charged-pion--photon factorial moments are
\begin{equation}
f_{i,j} \equiv \left({{\partial^{i,j}G(z_{ch}, z_{\gamma})} 
\over {\partial z_{ch}\,^{i} \partial 
z_{\gamma}\,^{j}}}\right)_{z_{ch}= 
z_{\gamma}=1}, 
\end{equation}
which introduces the indexing $(i,j)$ with respect to charged 
particles and photons which will be used in the remainder of this section.
The two lowest orders of factorial moments are
\begin{equation}
\begin{array}{rll}
\!\!f_{1,0}= & \!\langle n_{ch}\rangle & 
\!\!=\langle1- f\rangle \,\epsilon_{ch} \,\langle N\rangle, \\
\!\!f_{0,1}= & \!\langle n_{\gamma}\rangle & 
\!\!=\langle f\rangle\,(\epsilon_{1} +2\epsilon_{2})\,\langle N\rangle, \\
\!\!f_{2,0}= & \!\langle n_{ch}(n_{ch}-1)\rangle & 
\!\! = \langle(1-f)^{2} \rangle \,{\epsilon_{ch}}^2\,\langle 
N(N-1)\rangle, \\
\!\!f_{1,1}= & \!\langle n_{ch}n_{\gamma}\rangle & 
\!\! = \langle f(1-f)\rangle\,\epsilon_{ch}(\epsilon_{1}+2\epsilon_{2})\,
\langle N(N-1)\rangle, \\
\!\!f_{0,2}= & \!\langle n_{\gamma}(n_{\gamma}-1) 
\rangle & \!\!= \langle f^{2}\rangle\,(\epsilon_{1} +2\epsilon_{2})^{2}\,
\langle N(N-1)\rangle + \langle f\rangle\,2\epsilon_{2}\,\langle N\rangle,
\end{array}
\label{eq:factmom}
\end{equation}
where the overall statistical averages for the charged, photon, and 
charged-photon factorial moments are expressed, in an obvious notation, in 
terms of the independent moments taken with respect to the $P(N)$ and $p(f)$ 
distributions.
The second-order factorial moments represent the lowest-order
correlative effects among charged pions and photons. 

\section{Robust observables}
We would like to construct a measure from the moments in the form of a ratio 
in order to cancel out as many effects as possible, apart from the $p(f)$ 
averages which give information about the charged-neutral distribution.
The gamma-gamma correlation ($f_{0,2}$) involves a term proportional to
$\epsilon_2\left< N\right>$ that cannot be cancelled by any other
moments of this order.
However, the ratio
\begin{equation}
  r_{1,1}= 
{f_{1,1}f_{1,0}\over f_{2,0}f_{0,1}}
={{\langle n_{ch}n_{\gamma}\rangle\langle n_{ch}\rangle } \over 
{\langle n_{ch}(n_{ch}-1)\rangle\langle n_{\gamma}\rangle}}\ .
 \end{equation}
involving the other four moments has complete cancellation of
all reference to the background distribution $P(N)$ and the efficiencies
$\epsilon_{1}$, $\epsilon_{2}$, and $\epsilon_{ch}$ 
for generating functions of the form (\ref{eq:gobs}):
\begin{equation}
  r_{1,1}= {{\langle f(1-f)\rangle\langle (1-f)\rangle } \over 
{\langle (1-f)^{2}\rangle\langle f\rangle}}\ .
\end{equation}

For generic pion production, $p(f)=\delta(f-\hat{f})$ and
\begin{equation}
  r_{1,1}(gen)= 1.
\label{eq:r11gen}
\end{equation}
For a DCC distribution, $p(f)=1/(2\sqrt {f})$, the ratio is
\begin{equation}
  r_{1,1}(DCC)= {1 \over 2}\ .
\label{eq:r11dcc}
\end{equation}
Therefore, the pure DCC and generic distributions should be easily 
distinguishable if the statistical uncertainties are not too large.

We can simplify the formulas slightly by introducing normalized factorial
moments.
For the production of $N$ particles, these are given by
\begin{equation}
F_{i}\equiv \frac{\langle N(N-1)\ldots (N-i+1)\rangle}{\langle N\rangle ^{i}}\ .
\label{eq:normF}
\end{equation}

A generalization of the $F_{i}$'s to normalized moments for charged track and 
photon production is
\begin{equation}
 F_{i,j}=
\frac{\left< n_{ch}(n_{ch}-1)\ldots (n_{ch}-i+1)~n_{\gamma}
(n_{\gamma}-1) \ldots (n_{\gamma}-j+1)\right>} 
{\left<n_{ch}\right>^{i}\left<n_{\gamma}\right>^{j}}\ .
\end{equation}
In particular, 
\begin{equation}
F_{i,0} = \frac{F_{i}\left<(1-f)^{i}\right>}{\left<(1-f)\right>^{i}}
\end{equation}
and
\begin{equation}
F_{i,1} = \frac{F_{i+1}\left<f(1-f)^{i}\right>} 
{\left<f\right>\left<(1-f)\right>^{i}}\ ,
\end{equation} 
where $F_{i}$ refers to the $i$th normalized factorial moment (\ref{eq:normF})
of the $P(N)$ distribution for the total multiplicity. 

We see that $r_{1,1}= F_{1,1}/F_{2,0}$.
A generalization of $r_{1,1}$ to a family of robust observables is
\begin{equation}
r_{i,1}=\frac{F_{i,1}}{F_{i+1,0}} 
= \frac{\left<(1-f)\right>\left<f(1-f)^{i}\right>}{\left<f\right> 
\left<(1-f)^{i+1}\right>}\ ,
\end{equation}
where again the dependence on the parent distribution and efficiencies has
dropped out.
For all $i \geq 1$, generic particle production yields
\begin{equation}
r_{i,1}(gen) = 1, 
\end{equation}
while for DCC,
\begin{equation}
r_{i,1}(DCC) = \frac{1}{i+1}\ .
\end{equation}
Thus, $r_{i,1}$ becomes more sensitive to the difference between DCC and 
generic production mechanisms with increasing order of the moments. 
This reflects the broadness of the DCC neutral-fraction distribution relative 
to the binomial distribution of the generic case.
If only a fraction of particle production is due to DCC, the signal will
be easier to see in the higher-order ratios, which are sensitive to the tail 
of the charged-photon distribution, where the ratio of DCC to generic 
production is relatively high.
The ratios 
\begin{equation}
r_{i,j}=\frac{F_{i,j}}{F_{i+j,0}} 
\end{equation} 
involving higher-order gamma correlations ($F_{i,j}$, $j>1$)
are not robust because the moments $F_{i,j}$ depend on
the photon detection efficiencies. However, the moments
can be expressed in terms of only one 
combination of these efficiencies along with the mean number of photons, namely
\begin{equation}
\xi=\frac{2\epsilon_{2}}{(\epsilon_{1}+2\epsilon_{2})\left<n_{\gamma}\right>}\ ,
\end{equation}
as 
\begin{equation}
F_{i,j} = \sum_{m=0}^{[j/2]}c_{j,m}\xi^{m}F_{i+j-m} \frac{\left<(1-f)^{i}
f^{j-m}\right>}{\left<(1-f)\right>^{i} \left<f\right>^{j-m}}\ .
\end{equation} 
The coefficients $c_{j,m}$ are obtained from the identity \cite{GR}, 
true for any differentiable function $D(z^{2})$,
\begin{equation}
\frac{d^{j}D(z^{2})}{(dz)^{j}} = \sum_{m=0}^{[j/2]}c_{j,m}2^{m}(2z)^{j-2m} 
  \frac{d^{j-m}D(z^{2})}{(dz^{2})^{j-m}}\ .
\end{equation}
The first few $c_{j,m}$ are
\begin{eqnarray}
c_{j,0}& = & 1, \nonumber \\
c_{j,1}& = & j(j-1)/2, \nonumber \\
c_{j,2}& = & 3~j!/4!(j-4)!\,. 
\end{eqnarray}
The ratios $r_{i,j}$ can be used in the analysis of experimental distributions, 
with the understanding that the parameter $\xi$ is to be 
determined from the data. 

Generally, we have the bounds and limiting values
\begin{equation}
r_{i,j}(gen) \geq 1,
\end{equation}
\begin{equation}
\left. r_{i,j}(gen)  \right|_{\xi=0}= 1,
\end{equation}
\begin{equation}
\left. r_{i,j}(DCC)  \right|_{\xi=0} ={{i!(2j-1)!!}\over (i+j)!}\ .
\end{equation}

Finally, we turn to the effect of the MiniMax trigger on the moments and ratios.
As discussed in Sec. \ref{sec:detect}, events in which no charged particle or 
converted $\gamma$ goes through the acceptance of the detector are triggered 
with lower efficiency, $\epsilon$, than other events.
Take $p^{obs}(n_{ch},n_{\gamma})$ to be the (normalized) probability 
for observing an event with $n_{ch}$ charged particles and $n_{\gamma}$ 
converted $\gamma$'s in the acceptance assuming perfect triggering, and
$p^{trig}(n_{ch},n_{\gamma})$ to be the probability including 
the effects of both the trigger and the particle detection efficiencies.
An effective model for the effect of the trigger on the probability
is given by 
\begin{equation}
p^{trig}(0,0) = \epsilon \alpha p^{obs}(0,0), \qquad\quad\ 
n_{ch}=n_{\gamma}=0,
\end{equation}
\begin{equation}
p^{trig}(n_{ch},n_{\gamma}) = \alpha p^{obs}
(n_{ch},n_{\gamma}), 
\quad n_{ch}+ n_{\gamma}>0,
\end{equation}
where $\alpha$ is a normalization factor,
\begin{equation}
\alpha = \left[ 1+ (1-\epsilon)\,p^{obs}(0,0)\right]^{-1}.
\end{equation}

When we incorporate these trigger efficiencies,
\begin{equation}
f_{i,j}\rightarrow \alpha f_{i,j},
\end{equation} 
and
\begin{equation}
F_{i,j}\rightarrow \alpha^{1-i-j} F_{i,j},
\end{equation} 
leaving the $r_{i,j}$ robust in the sense that they are also independent of 
$\epsilon$.

In summary,
\begin{enumerate}
\item the $r_{i,j}$ do not depend upon the form of the parent pion multiplicity 
distribution;
\item the $r_{i,j}$ are independent of the detection efficiencies for finding 
charged tracks, provided these efficiencies are not correlated with each other 
or with other variables such as total multiplicity or background level;
\item the $r_{i,1}$ are also independent of the $\gamma$ efficiencies 
in the same sense as above; the $r_{i,(j > 1)}$ depend only upon 
one parameter, $\xi$, which reflects the relative probability of both photons 
from a $\pi^{0}$ being detected in the same event;
\item in all cases the $r_{i,j}$ are independent of the magnitude of the null 
trigger efficiency;
\item the ratios $r_{i,j}$ possess definite and very different values for pure 
generic and pure DCC pion production.
\end{enumerate}

\vfill\newpage
\begin{figure}[h] \vspace{-0.5in}\hspace*{-0.3in}
\psfig{file=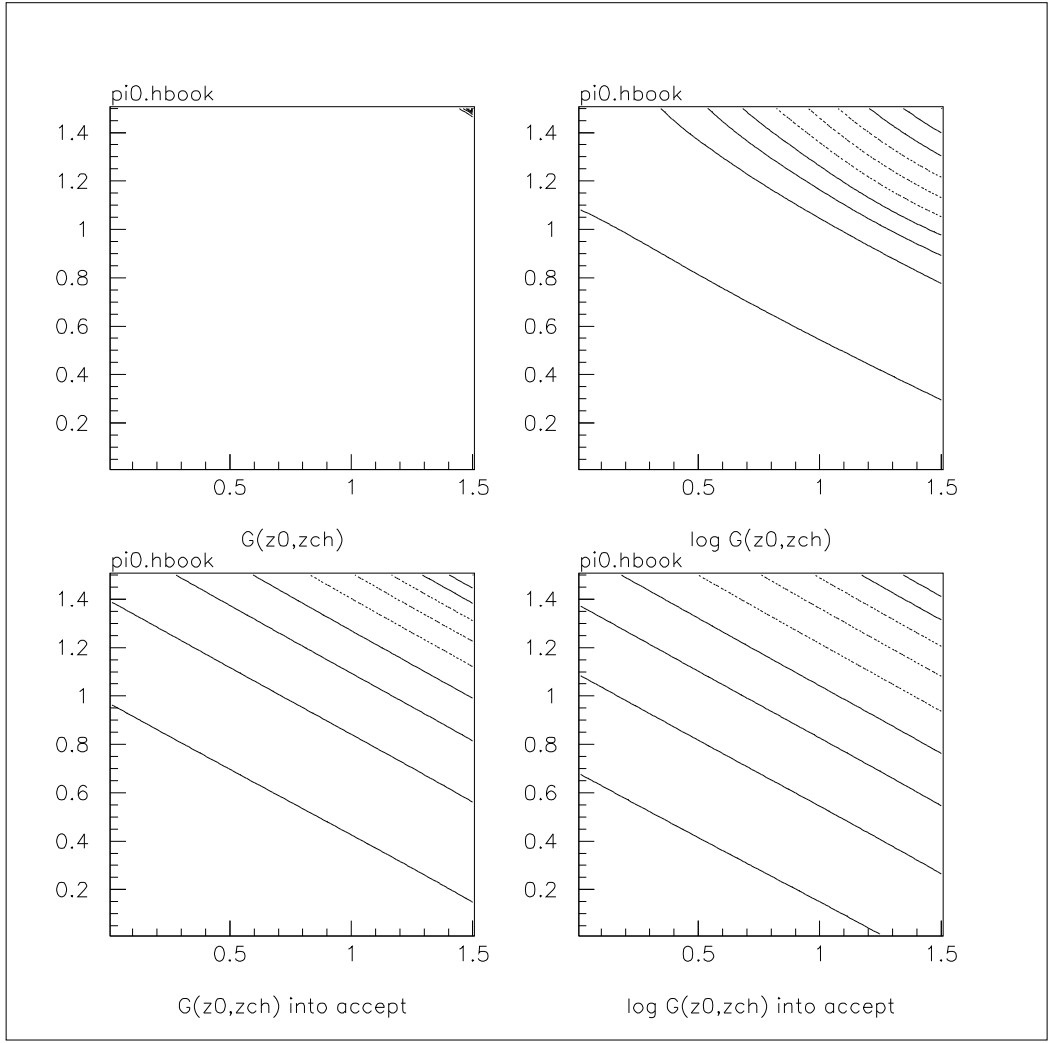,width=6in}
\vspace{-1in}
\caption[Contour plots of the generating function $G(z_{ch},z_0)$ and
$\log{G(z_{ch},z_0)}$ from all PYTHIA charged and neutral pions, and from 
those entering the MiniMax acceptance.]
{Contour plots of the generating function $G(z_{ch},z_0)$ and
$\log{G(z_{ch},z_0)}$ from all PYTHIA charged and neutral pions, and from 
those entering the MiniMax acceptance (the vertical axis is $z_{ch}$ and 
the horizontal is $z_0$).}
\label{f:genfp} \end{figure}
\vfill\newpage
\begin{figure}[h] \vspace{-0.5in}\hspace*{-0.3in}
\psfig{file=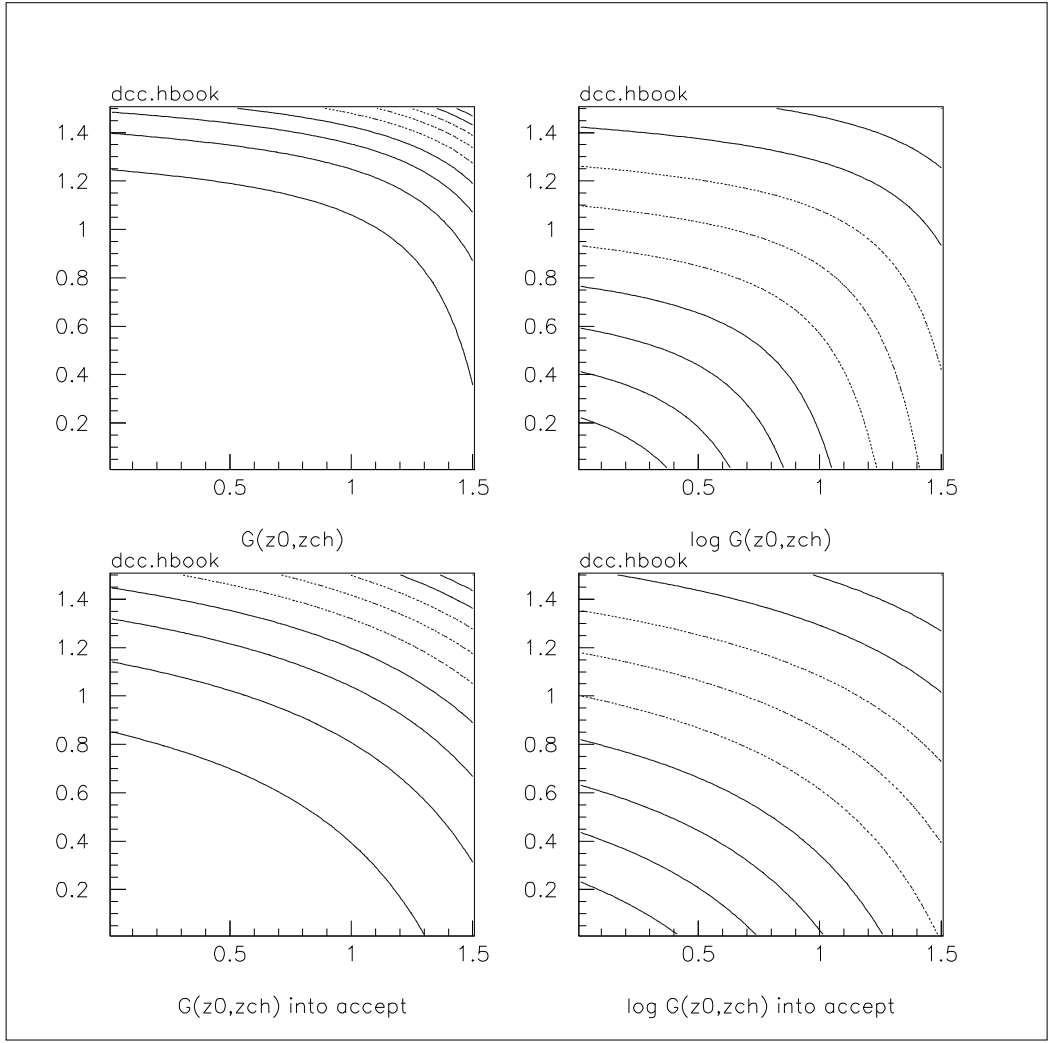,width=6in}
\vspace{-1in}
\caption[Contour plots of the generating function $G(z_{ch},z_0)$ and
$\log{G(z_{ch},z_0)}$ from all charged and neutral pions from the DCC 
generator, and from those entering the MiniMax acceptance.]
{Contour plots of the generating function $G(z_{ch},z_0)$ and
$\log{G(z_{ch},z_0)}$ from all charged and neutral pions from the DCC 
generator, and from those entering the MiniMax acceptance (the vertical axis is 
$z_{ch}$ 
and the horizontal is $z_0$).}
\label{f:genfdcc} \end{figure}
\clearpage
\chapter{Results}

\section{DCC: Calculated factorial moments and robust observables}
For each real or simulated event, the vertexer returns the number of
charged tracks and photons found.  Then the frequency distribution for
observing a given number of charged and photon-conversion tracks,
${\cal N}(n_{ch},n_\gamma)$, is used to
calculate the factorial moments, assuming that the probability of observing an 
event with a given number of charged tracks and converted photons is given by
the observed distribution:
\begin{equation}
P(n_{ch},n_\gamma)\approx{\cal N}(n_{ch},n_\gamma)/N,
\end{equation}
where $N$ is the total number of events
\begin{equation}
N=\sum_{n_{ch}=0}^\infty\sum_{n_\gamma=0}^\infty {\cal N}(n_{ch},n_\gamma).
\end{equation}

Statistical errors are estimated assuming Poisson fluctuations and 
standard propagation of errors formalism.  Appendix A shows the derivation
of some of the formulas used to calculate uncertainties in the moments
and $r_{i,j}$.
In order to check the accuracy of the calculated uncertainty in the $r_{i,j}$, 
the values of $r_{1,1}$ and $\sigma_{r_{1,1}}$ were determined for groups 
of 25000 events 
from runs 1096, 1099, 1103, 1109, 1110, 1125, 1126, 1127, 1137, and 1139.
Then a Gaussian was fit to the histogram of 
$(r(i)-\left< r\right>)/\sigma_r(i)$,
where $r(i)$ and $\sigma_r(i)$ are $r_{1,1}$ and $\sigma_{r_{1,1}}$, 
respectively, for the $i$th group of events, and $\left< r\right>=1.0228$ is 
the mean $r_{1,1}$ for all events.  The fitted Gaussian is shown in 
Fig. \ref{f:uncer}, and has mean $\mu=0.012\pm 0.144$ and standard deviation 
$\sigma=0.985\pm 0.160$.  Note that if the calculated uncertainty is
equivalent to the actual statistical uncertainty, then the standard deviation 
of the Gaussian should be $\sigma=1$.  Since this is true to within errors,
we believe that the calculated uncertainty accurately represents the
true statistical uncertainty.

\subsection{Simulations}
\label{sec:rgeant}
The moments and $r_{i,j}$ were calculated for approximately $1.5\times 10^5$ 
PYTHIA events which would be seen by the detector (pass trigger cuts) and 
$2\times 10^4$ pure DCC events, and the results are shown in Table \ref{t:r}. 
The PYTHIA results are given for perfect charged- and photon-finding 
efficiencies, along with the output of running these events through the
GEANT simulation.  The DCC events are also processed by the same GEANT
simulation, except that the trigger is not used since there are no 
particles generated in the $\bar{\mbox{p}}$ direction.
For purposes of comparison, the predicted values for idealized binomial and
$1/(2\sqrt{f})$ distributions are included.  [For $j>1$, these depend on
the parameter $\xi$, which was estimated using the relationship between
$f_{0,2}$, $f_{2,0}$ assuming a binomial distribution.  Then the observed
$\left< n_\gamma\right>$ from PYTHIA/GEANT was used to obtain
$2\epsilon_2/(\epsilon_1+2\epsilon_2)\approx 0.08\pm 0.01$.  The $F_i$
used in these predictions
were also determined from the PYTHIA/GEANT data.  We do not expect these
values to be correct for the DCC case; in particular, the simulated DCC pions
have significantly lower $\left< p_T\right>$ than the PYTHIA pions, so that 
the detection efficiencies $\epsilon_1$ and $\epsilon_2$ are not the same.
However, the values from PYTHIA are used here to illustrate the problems which
will arise in DCC searches using these non-robust ratios.]

The $r_{i,1}$ obtained by counting numbers of charged tracks and photons
aimed into the acceptance by PYTHIA are within two standard deviations
of 1.00.  This suggests that something close to a binomial charged-neutral 
distribution is used by PYTHIA.  Detector effects are included by running the
events through GEANT.  The fact that the lower-order (statistically significant)
$r_{i,1}$ did not differ by more than about 10\% 
(and are in fact within $2\sigma$) of the values
from the PYTHIA input justifies the claim that the robust observables
are indeed insensitive to detection efficiencies, and also that correlations
of efficiencies with multiplicities and momentum do not greatly alter the
robustness of these variables.
The ratios $r_{i,1}$ obtained from running the DCC events through GEANT are 
somewhat higher than the values predicted for the $1/(2\sqrt{f})$ distribution;
however, they are clearly distinguishable from the values for generic 
production.

The choice of parameters used in the DCC generator is somewhat optimistic;
the DCC domain is aimed directly at the center of the acceptance, and the
DCC pions have a rather large $\left< p_T\right>$.  As is shown in 
Fig. \ref{f:eff_gpt}, very low-$p_T$ photons are less likely to convert, and 
the efficiency for detecting those which do convert is only about 40\% for 
$p_T\,\raisebox{-.75ex}{$\stackrel{\textstyle{<}}{\sim}$}\, 50\:$MeV.  
Charged pions with low $p_T$ are also more difficult to find because they tend 
to stop or multiple scatter more in the lead; some scatter so much that the 
tail is not associated with the head by the vertexer, and may be found as a 
fake photon conversion.  Therefore, we varied the $\left< p_T\right>$ of
the pions in the DCC generator.  The resulting moments and $r_{i,j}$ are
given in Table \ref{t:rdcc}.  The parameters in model A are those mentioned
previously.  Models B and C have lower $\left< p_T\right>$, $50\:$MeV and
$25\:$MeV, respectively.  In these models, the ratio of mean energy density of
the DCC pions to that of generic pions is not changed ($\psi=1$), which
leads to larger numbers of pions in the domain since  
$\left< N_\pi\right>\propto \psi/p_T$.  In order to keep the mean number of
pions constant, we also varied $\psi$ with $p_T$  in models D and E.
For DCC domains with very low-$p_T$ pions, the values of the robust observables
are closer to what we expect for a binomial distribution, so that 
distinguishing these domains from generically-produced pions using these
observables becomes much more difficult.

Of course, we do not expect to observe events consisting of only a DCC domain
aimed into the acceptance.  Possible scenarios for mixing DCC and generic
multiparticle production are discussed in Ref. \cite{robust}.  For example,
any given event could be due to either DCC or generic production, but not
both (exclusive production).  
Perhaps more realistically, the occurrence of DCC in an event could be 
independent of the generically-produced pions (independent production), or
the amount of DCC production could depend on the amount of generic production 
(associated production).  An example of the latter type is the Baked Alaska 
model \cite{bj}, which has the number of DCC pions given by 
$N_{DCC}\sim{N_{gen}}^{3/2}$.

In order to study the effect of an admixture of DCC with generic events
where the amount of DCC produced is independent of the amount of generic
production, we added DCC domains from the DCC-generator/GEANT to various
fractions of (random) PYTHIA/GEANT events.  The effect on the $r_{i,1}$
is shown in Table \ref{t:rfract}.  (Note that a slightly older version of the 
vertexer was used for this study, which accounts for the discrepancy in 
$r_{1,1}$ here with no DCC added and in Table \ref{t:r}.)
The values for $r_{i,1}$ fall off faster than linearly with fraction of DCC.
For a fraction of 1, i.e. when a DCC domain is added to every event, 
the ratios are higher than for DCC alone, but are still easily
distinguishable from the generic values.

\subsection{Characteristics of events from lead-in runs}
\label{sec:rgen}
Ten of the lead-in runs 
(1096, 1099, 1103, 1109, 1110, 1125, 1126, 1127, 1137, and 1139), 
totaling about 1.5 million events, were used in the following analysis.
The running conditions were very clean (e.g. low luminosity) 
and therefore the diffractive tags had very little contamination
from beam-gas interactions.
The frequency distribution of events with given numbers of charged tracks and 
photons is determined (Table \ref{t:ncg}) and used to calculate the factorial
moments and $r_{i,j}$.  Table \ref{t:rreal} gives the values for some of these
variables.  The mean number of charged tracks found per event is about 0.5 and
of converted photons is about 0.2.  The lower-order
$r_{i,1}$ are close to what is expected for a binomial
distribution ($r_{i,1}=1$).  The values for $r_{1,1}$ and $r_{1,2}$
are within two standard deviations of the PYTHIA results.  
The higher-order ratios are weighted towards bins of 
${\cal N}(n_{ch},n_\gamma)$ which 
are statistically limited, and therefore the deviations from unity are not very
significant.  In any case, the ratios are not smaller than one as would be 
expected for a contribution from DCC.  Therefore the events appear to be 
consistent with production by only generic mechanisms.

\subsection{Characteristics of events with diffractive tags}

From the 10 runs used in Sec. \ref{sec:rgen}, 21412 events have a ktag from
the scintillator which detected showers from interactions of diffractive 
anti-protons in the kicker magnets.
The mean numbers of charged particles and photons are lower for ktag events,
as would be expected for diffractive events, where a large fraction of the
total energy is carried away by the beam remnant,
and the charged-charged and charged-gamma correlations are correspondingly
lower.
Table \ref{t:rreal} gives the values of the $r_{i,j}$ for the ktag events.
 
The upstream hadronic calorimeters at $z\approx -25\:$m are used to tag events
with diffractive anti-protons with $x_F\sim 0.5$ and anti-neutrons.
Differences related to isospin exchange in diffractive events might be apparent
in comparisons between events with an $\bar{\mbox{n}}$ and those with a 
$\bar{\mbox{p}}$.
Figure \ref{f:hcal} shows histograms of the ADC readout from the 
$\bar{\mbox{n}}$ and $\bar{\mbox{p}}$ calorimeters.  Events with 
$\mbox{ADC}>400$ (in order to cut out background from products of showers in 
the magnets) were used to calculate the
moments and $r_{i,j}$, given in Table \ref{t:rreal}.  
The mean number of particles found
is higher than that in events with the ktag, but lower than in the
total sample of events, and is lower for the tag on leading $\bar{\mbox{n}}$'s
than for $\bar{\mbox{p}}$'s with half the beam momentum,
consistent with energy conservation.

The $r_{i,j}$ for diffractive-tagged events do not differ by more than two 
standard deviations from the values for the total sample.  
Therefore, we conclude that there is no evidence
for more DCC production in events with diffractive tags, offering no support
to the conjecture that Centauros are related to DCC and are diffractive in 
nature.



\subsection{Characteristics of events with a pbar multiplicity tag}
Since DCC may be more often present in events with large multiplicity,
we would like to find a measure of the multiplicity independent of that
in the small acceptance.
The multiplicity in the scintillator on the downstream anti-proton side 
of the collision at $z=-81\:$in (``pbar counters'') may be correlated with the 
total multiplicity, and therefore could be such a measure.

In GEANT, where the number of charged tracks hitting the pbar counters is  
known, a plot of the total energy deposited in the pbar counters 
against the pbar multiplicity (Fig. \ref{f:geant_pbar}) shows the correlation
between energy and multiplicity which we expect from minimum-ionizing
particles.
The pbar multiplicity in the real data can thus be found from 
the mip peaks in the ADC readout of the counters (Fig. \ref{f:pbar}).

In this way, the pbar multiplicities of events in runs
1099, 1103, 1109, 1125, 1126, and 1127 were determined, and events
were grouped in bins of pbar multiplicity, such that each bin contained 10\%
of the events.
The multiplicity in the pbar counters is indeed correlated with 
that in the acceptance.  Table \ref{t:rpb} shows an increase in 
$\left< n_{ch}\right>$ and $\left< n_{\gamma}\right>$ with increasing
pbar multiplicity.
However, the $r_{i,1}$ do not appear to vary with multiplicity,
and show no sign of an increased presence of DCC for higher multiplicity
events.  The $r_{1,1}$'s are consistent with each other and 
with that for all events.

\section{Multiparticle analysis}
Plots of $dN_{ch}/d\eta$ and $dN_\gamma/d\eta$, uncorrected for detection
and trigger efficiencies,
are shown in Fig. \ref{f:dndeta1x0ch} and Fig. \ref{f:dndeta1x0gam}, 
respectively.  The solid line is the result from approximately $1.5\times 10^6$
events from the 10 runs used in previous analysis.  The dotted line is the
result from the PYTHIA/GEANT simulation, and the dashed line is from the
PYTHIA events which passed the GEANT trigger (about $1.5\times 10^5$ events).
The charged multiplicity found by the vertexer for GEANT events is higher than 
that from the PYTHIA input by about 0.15, or 3\%. 
The number of photons found for GEANT is about 42\% of the PYTHIA input.
The charged multiplicity from the data is lower than that from GEANT and
from PYTHIA (about 90\% of the GEANT value).  The photon multiplicity
is approximately 92\% of that in GEANT.

Since multiplicity is correlated with NHITS, and the contribution to the NHITS 
from background is greater in the real data than in GEANT, it is not unlikely 
that the cut of NHITS\,$<600$ would cut more high-multiplicity
events from the data sample than from the GEANT.  This conjecture is supported
by the results in Sec. \ref{sec:uthetau}, which showed a correlation between 
pipe shower and 
multiplicity.  However, we do not necessarily expect the data to agree with 
the simulations, since the PYTHIA input is not based on real measurements
at the cm energy and $\eta$ range of the MiniMax experiment.

\section{Low $p_T$ photons}
There has been controversial evidence suggesting that the number of low-$p_T$
photons produced is larger than what is expected from hadronic decays and
QED inner bremsstrahlung, in particular, that there is an excess of photons 
with $p_T\,\raisebox{-.75ex}{$\stackrel{\textstyle{<}}{\sim}$}\, 20$ MeV
\cite{spyr}.
We have attempted to study this by observing photon conversions in the 
$1/4\:$in ($0.07\:$X$_0$) -thick  Al window at $z=120\:$in, and measuring the 
momentum of the conversion products either by their bending in the scintillator 
($0.03\:$X$_0$-thick at $z=157\:$in)
due to multiple scattering or by tracking them into the calorimeter and
determining the energy deposited in the hit cells.  The lead-out runs
were used for this so that the conversion tracks would not stop or
scatter and lose energy in the lead.  Conversions in the window, rather than
in the scintillator, for example, were chosen because the increased resolution
in $v$ in the front chambers was desirable for determining the momentum of the
converted photon.

First, it was necessary to determine which tracks were due to window 
conversions.  To do this, we plotted the $z$ of closest approach between
pairs of tracks.
The GEANT lead-out data shows a wide peak at the collision point,
and a smaller peak at the window, $z\approx 120\:$in, as can be seen in 
Fig. \ref{f:lowptz}.
For the real data, the peak at the collision point and at the window are 
fairly clear, but there is also a smaller peak at intermediate $z$.
This is apparently due to fake tracks made from $u$ wires
borrowed from several real tracks, such that the resulting fakes tended
to vertex with the real tracks at a location in $z$ between the collision point 
and the first chamber (which is just behind the window).  These fake vertices 
are removed by requiring the heads to point such that they will hit the 
lead, and that they hit all eight of the front chambers.  The distribution
of the $z$ of closest approach for remaining tracks is shown in 
Fig. \ref{f:rlowptz}.

A window conversion is defined as a pair of tracks with a
$z$ at closest approach between 115 and $150\:$in, and with a reduced $\chi^2$
of the match between the tracks at the window (done in the same way as the 
usual vertexing at the lead) of $\chi^2/2 < 2.5$ (see Fig. \ref{f:lowptz}
for the GEANT distribution and Fig. \ref{f:rlowptz} for that of the data).  
Also, in order to be able to determine the
momentum of the conversion tracks by bending in the scintillator or by 
following them into the calorimeter, each head is required to have exactly
one tail which matched to the head with $\chi^2/2 < 5$.

Both conversion tracks are found by the tracker in about 26\% of the window 
conversions in GEANT which send both tracks into the acceptance.  Only
one track is recorded 70\% of the time, usually because the tracks are so
close together that they are grouped by the tracker into a single track.
This is apparent in Fig. \ref{f:openang}, which shows the opening angle
of window conversion tracks in GEANT and of the pairs of tracks found as
window conversions.  Conversions with smaller opening angles are much
less likely to be found.

Of the pairs of tracks found as window conversions, approximately 90\% have
one tail matched to each head.  As is also shown in Fig. \ref{f:openang},
the efficiency for finding vertices from the low-$p_T$ photons that
convert in the window is fairly high (because such conversions tend
to have larger opening angles).  Unfortunately, the low-$p_T$ conversion
tracks scatter more in the scintillator than high energy particles,
so that the efficiency for matching the tails to the heads is relatively low,
preventing a determination of the momentum of these tracks.

The momentum of the conversion tracks is determined from the angles between
the heads and tails at the scintillator using the relationship between
the momentum of a particle and the bending angle of the 
track due to multiple scattering in a material \cite{PDG}
\begin{equation}
\theta_0\approx {13.6\:\mbox{MeV}\over\beta c p} z \sqrt{x\over \mbox{X}_0},
\end{equation}
where $x/\mbox{X}_0$ is the thickness of the material in radiation lengths, 
$p$ is the momentum and $\beta c$ the velocity of the particle, $z$ is
its charged number, and
$\theta_0$ is the bending angle in the plane of the incident and bent track.
The momentum of the photons which converted in the window is taken to be the
sum of the momenta of the two conversion tracks.  The $p_T$ is determined from
the total momentum and the angle of the trajectory of the photon, which
must have traveled from the collision point to the vertex of the
conversion tracks.  For GEANT conversions, where the photon momentum is known, 
the $p_T$ as determined by bending in the scintillator tends to be lower than 
the actual $p_T$ (Fig. \ref{f:pt}).  Other attempts to find a measure of
the momentum, using, for example, the opening angle, were unsuccessful.

Following the conversion tracks into the calorimeter, their energy is
determined to be simply the energy deposited in the cell(s) into which the 
tracks are aimed, without using a clustering algorithm.  The $p_t$ found with
this method is surprisingly accurate (Fig. \ref{f:pt}).
Unfortunately, the actual calorimeter was never successfully calibrated,
and therefore this method can not be reliably used on the data.

The $p_T$ distributions of ``photons'' from pairs of tracks which are found
as window conversions in the data are shown in Fig. \ref{f:rpt}, although
these distributions are not believed to be very accurate.
Instead of comparing $p_T$ distributions from GEANT and from the data,
a comparison of the number of window conversions integrated over all $p_T$
was used to look for evidence for an excess of low-$p_T$ photons.

First, the origins of pairs of tracks found as GEANT window conversions are 
determined.  The relative contribution from sources other than actual window
conversions should be low, and hopefully from sources of similar magnitude
in the GEANT and in the data.
In a sample of about $10^5$ GEANT events, 225 pairs of tracks are identified
as window conversions. About half of these really are window conversions
(106 conversions in the window and 4 conversions in the first chamber at 
$z=123\:$in).  The next largest
contribution is 19\% from pairs of charged tracks from the collision point
which are close together at the window,
including 13 dalitz pairs.  The large uncertainty in the point of closest
approach due to the nearly identical slope of such tracks allows the 
possibility of their appearing to vertex at the window. 
Strong interactions in the window, neutral particle decays 
(e.g. K$_s^0$, $\Lambda^0$), 
pairs of a primary charged track from the collision and a secondary
charged track, and pairs of a charged track and a fake each contribute between
5 and 7\% of the pairs found as window conversions.  Five events have
conversions in the window of 
secondary photons.  The remaining few events involve
pairs of secondary tracks from interactions in other material.

Almost $10^6$ events from lead-out runs (1089, 1093, 1104, 1108, 1111, 1123,
1124, 1129, 1132) were used to search for window conversions in the data.
The ratio of the number of window conversions found in the data to that
in GEANT is $2121/339=6.3$.
The ratio of events in the data sample to that in the GEANT sample is
$923238/155737=5.9$, so that there appears to be an excess of window
conversion in the data of about 6\%.  The uncertainties involved in this
analysis do not allow a precise statement either supporting or contradicting
the claim of Ref. \cite{spyr}.

\vfill\newpage
\begin{figure}[h]\vspace{-0.5in}\hspace*{-0.3in}
\psfig{file=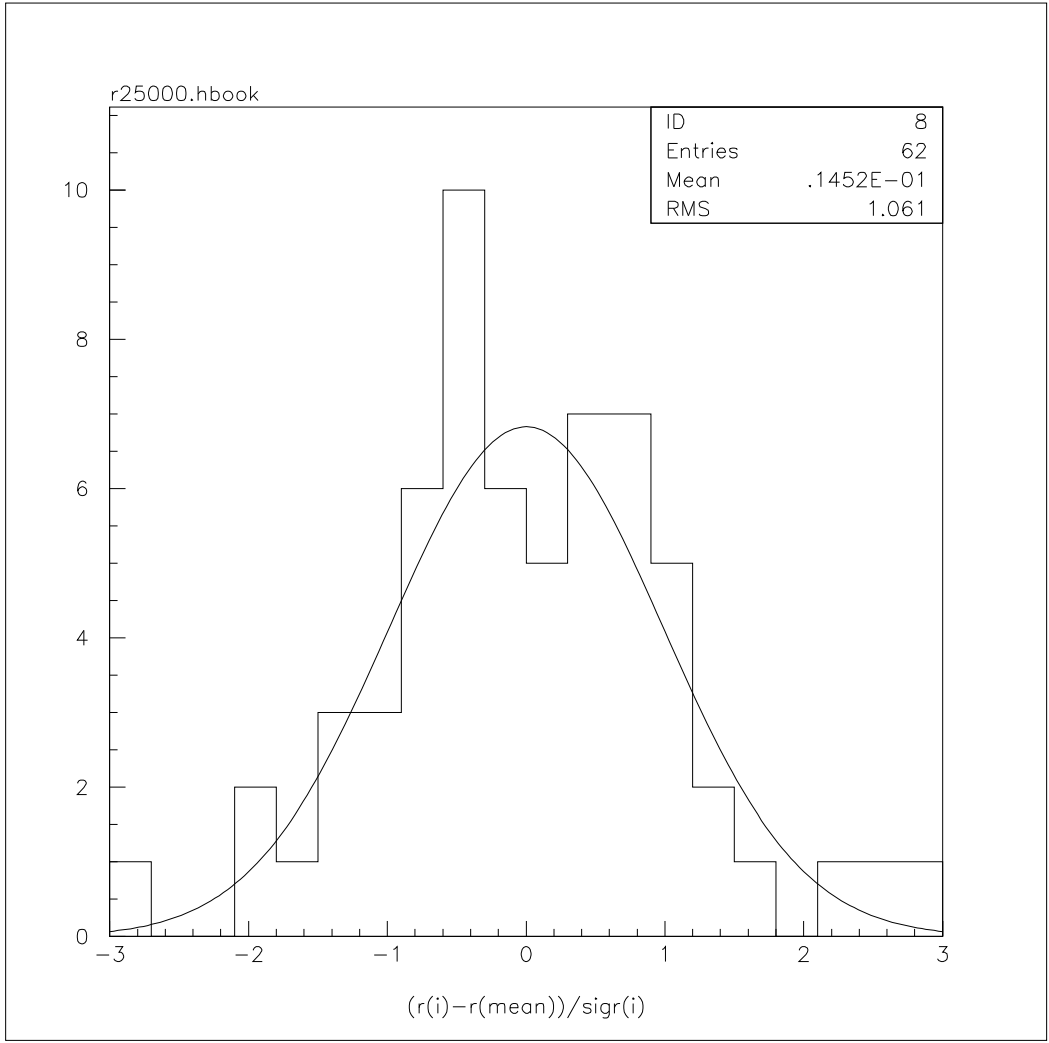,width=6in}
\vspace{-1in}
\caption{Histogram of $(r(i)-\left< r\right>)/\sigma_r(i)$ for groups ($i$) of 
25000 events and Gaussian fit.}
\label{f:uncer} \end{figure}
\vfill\newpage
\begin{figure}[h]\vspace{-0.5in}\hspace*{-0.3in}
\psfig{file=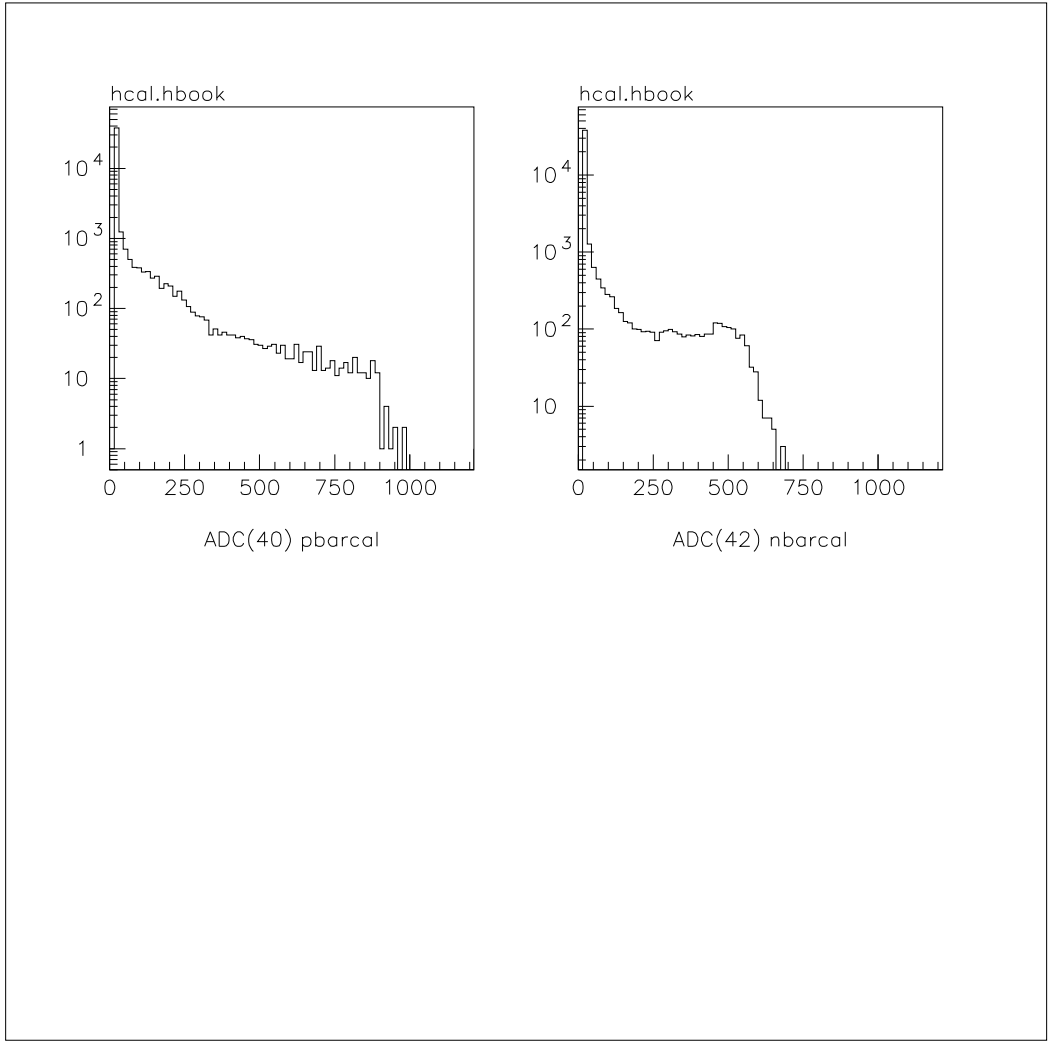,width=6in}
\vspace{-1in}
\caption[ADC values of the hadronic calorimeters which see anti-neutrons
and anti-protons.]{ADC values of the hadronic calorimeters which see 
anti-neutrons [ADC(40)] and anti-protons [ADC(42)].}
\label{f:hcal} \end{figure}
\vfill\newpage
\begin{figure}[h]\vspace{-0.5in}\hspace*{-0.3in}
\psfig{file=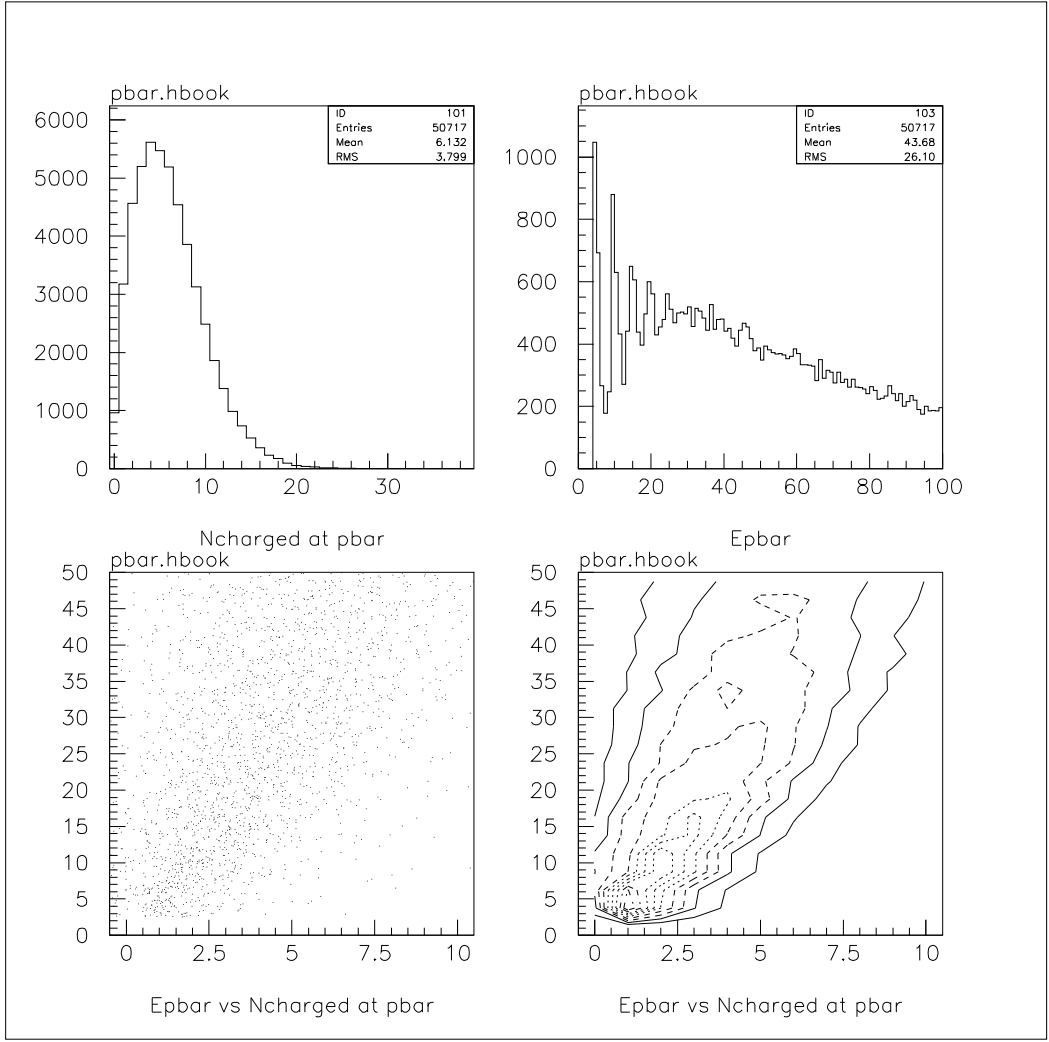,width=6in}
\vspace{-1in}
\caption{Energy (in MeV) deposited in the pbar counters vs number of charged
particles hitting these counters}
\label{f:geant_pbar} \end{figure}
\vfill\newpage
\begin{figure}[h]\vspace{-0.5in}\hspace*{-0.3in}
\psfig{file=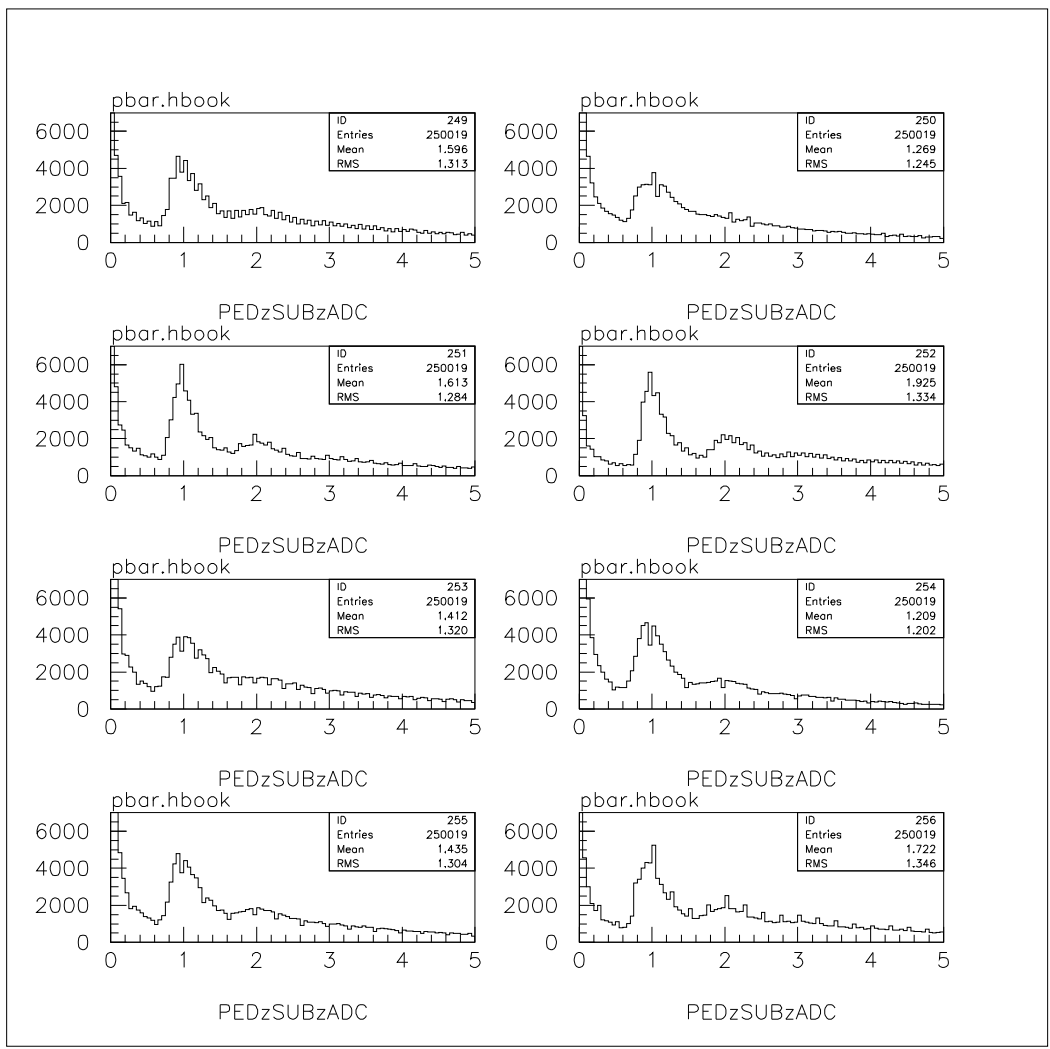,width=6in}
\vspace{-1in}
\caption{PED\_SUB\_ADC values (in units of mip energies or, equivalently, number
of charged particles) of the pbar counters}
\label{f:pbar} \end{figure}
\vfill\newpage
\begin{figure}[h]\vspace{-0.5in}\hspace*{-0.3in}
\psfig{file=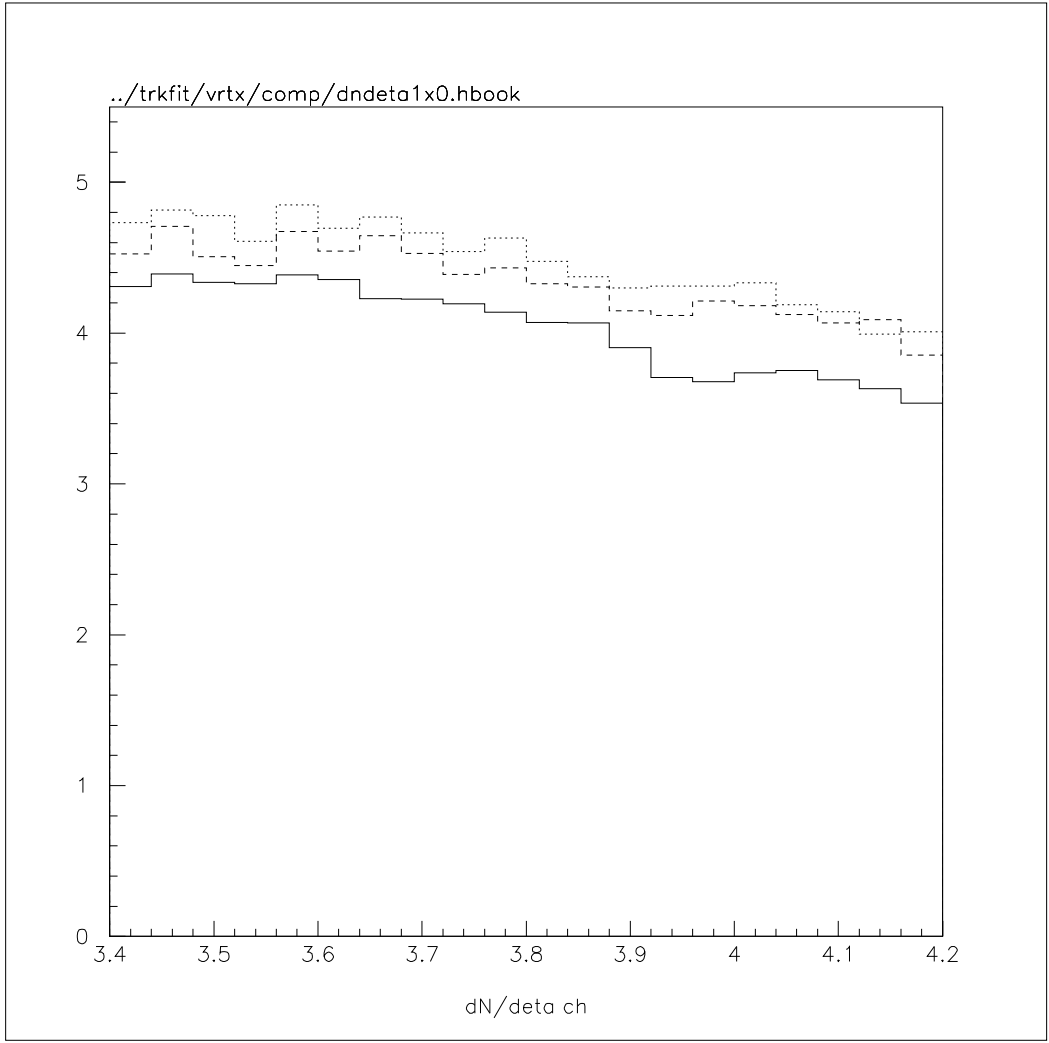,width=6in}
\vspace{-1in}
\caption[Pseudorapidity distribution $dN/d\eta$ vs $\eta$ for charged tracks in
the data, GEANT, and PYTHIA, uncorrected for efficiencies.]{Pseudorapidity 
distribution $dN/d\eta$ vs $\eta$ for charged tracks in the data (solid), 
GEANT (dotted), and PYTHIA (dashed), uncorrected for efficiencies.}
\label{f:dndeta1x0ch} \end{figure}
\vfill\newpage
\begin{figure}[h]\vspace{-0.5in}\hspace*{-0.3in}
\psfig{file=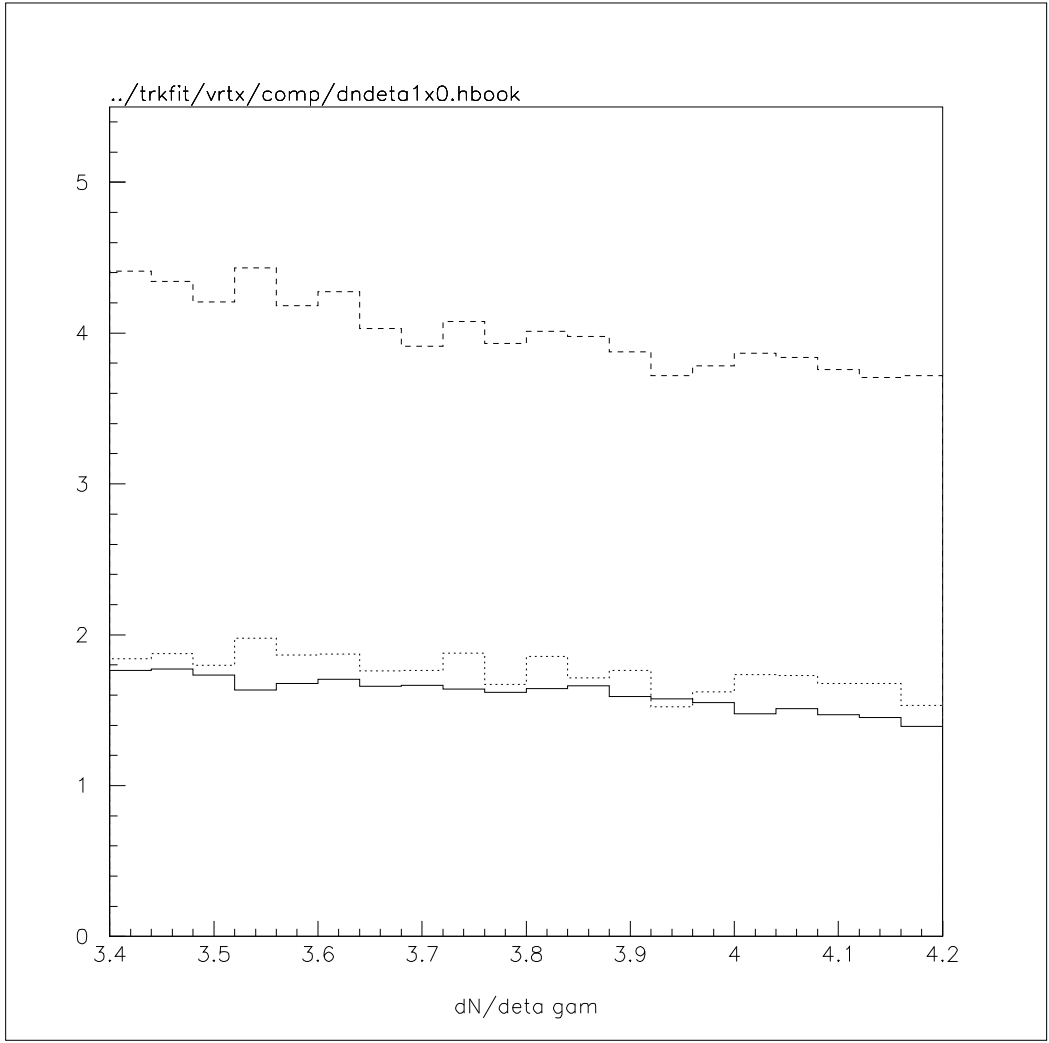,width=6in}
\vspace{-1in}
\caption[Pseudorapidity distribution $dN/d\eta$ vs $\eta$ for photons in
the data, GEANT, and PYTHIA, uncorrected for efficiencies.]{Pseudorapidity 
distribution $dN/d\eta$ vs $\eta$ for photons in the data (solid), 
GEANT (dotted), and PYTHIA (dashed), uncorrected for efficiencies.}
\label{f:dndeta1x0gam} \end{figure}
\vfill\newpage
\begin{figure}[h]\vspace{-0.5in}\hspace*{-0.3in}
\psfig{file=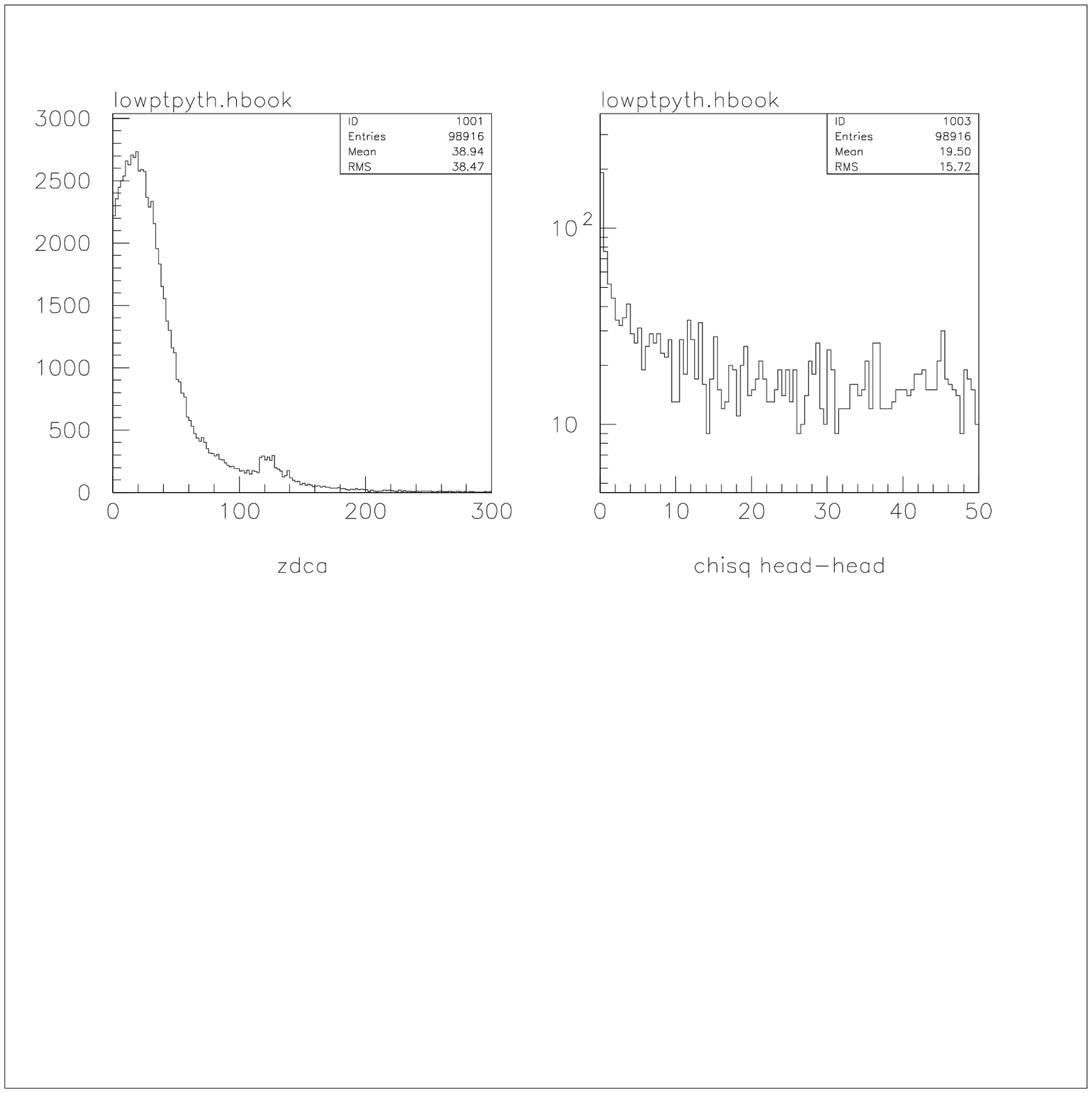,width=6in}
\vspace{-1in}
\caption{Histogram of the $z$ of closest approach between all heads,
and of the $\chi^2$ of the matching of those pairs in GEANT.}
\label{f:lowptz} \end{figure}
\vfill\newpage
\begin{figure}[h]\vspace{-0.5in}\hspace*{-0.3in}
\psfig{file=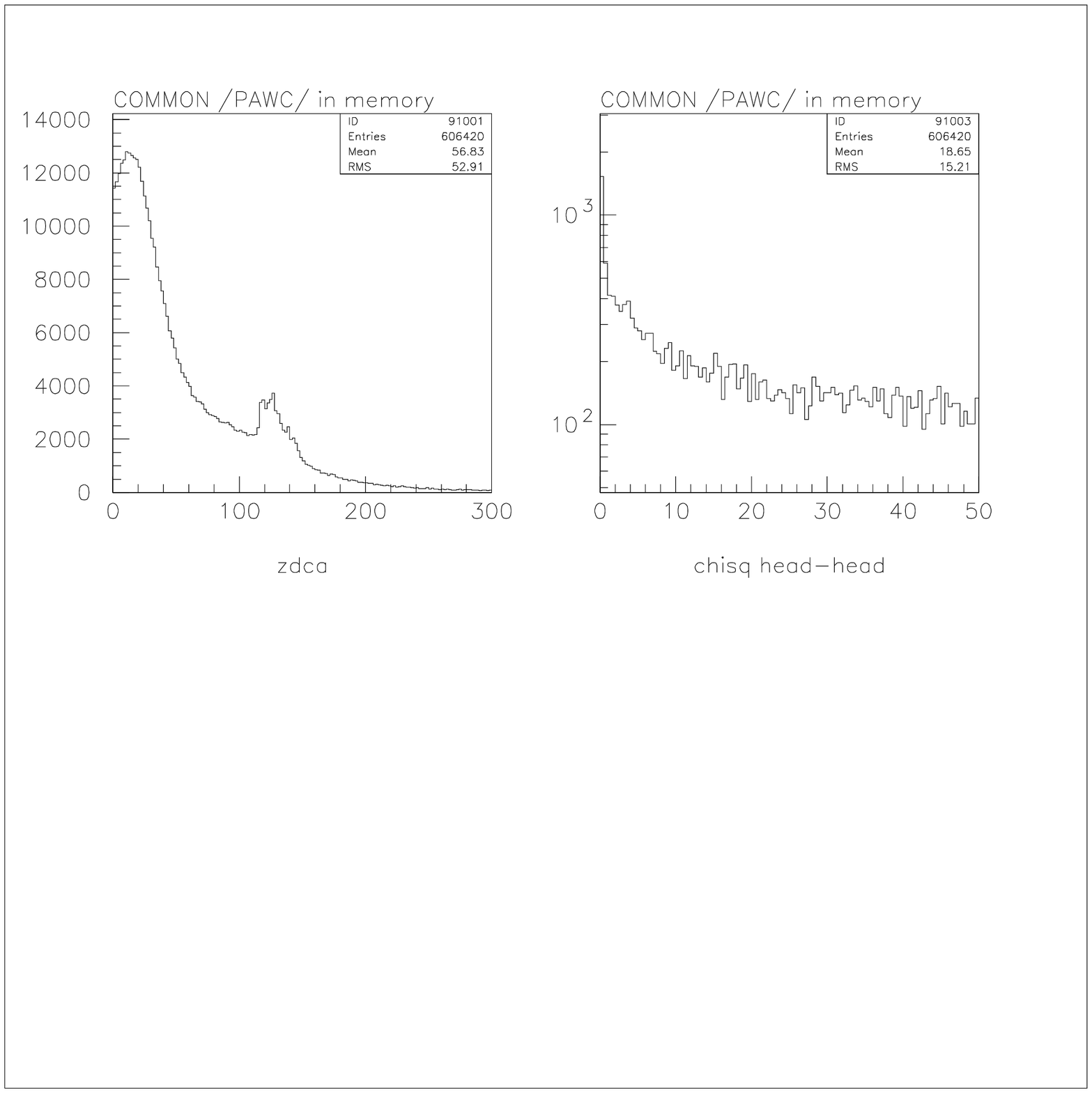,width=6in}
\vspace{-1in}
\caption{Histogram of the $z$ of closest approach between all heads,
and of the $\chi^2$ of the matching of those pairs in the data.}
\label{f:rlowptz} \end{figure}
\vfill\newpage
\begin{figure}[h]\vspace{-0.5in}\hspace*{-0.3in}
\psfig{file=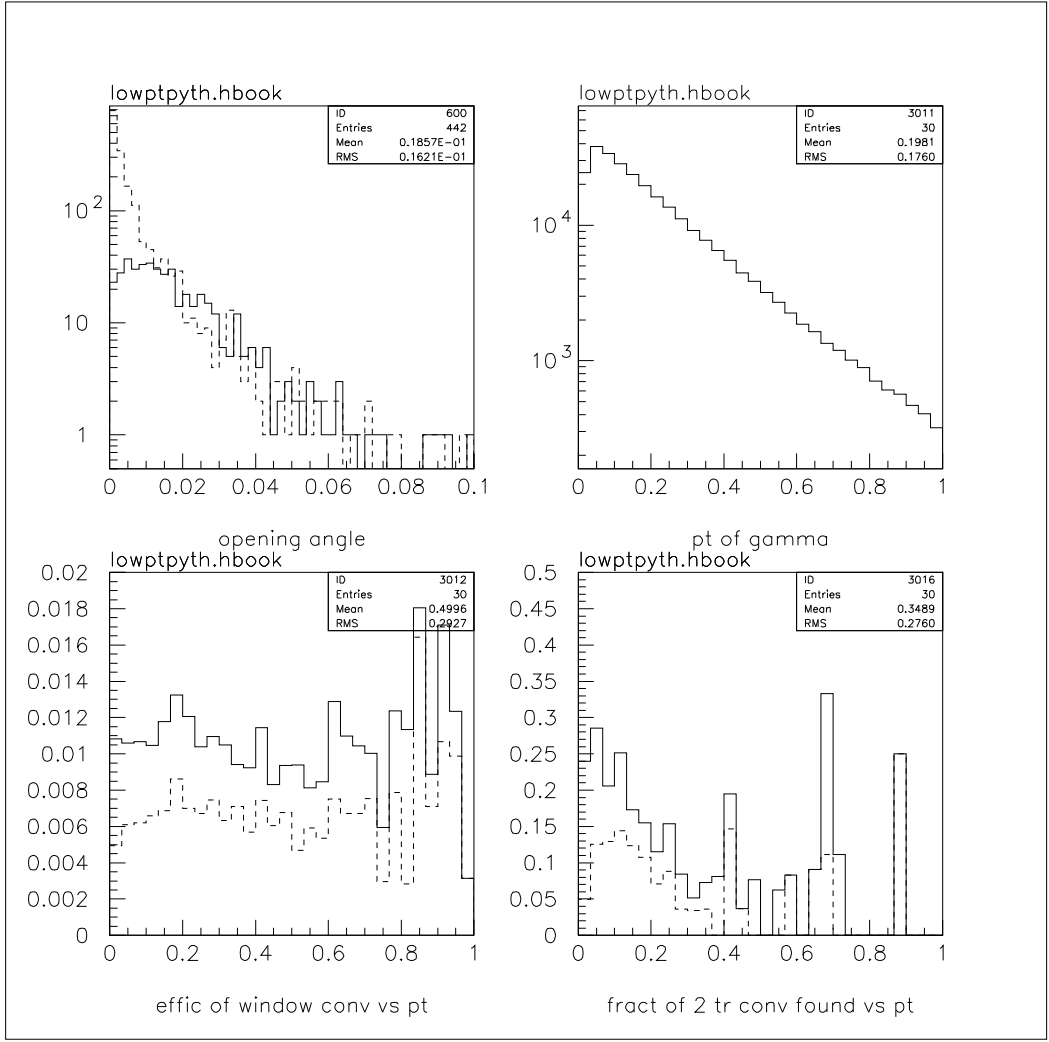,width=6in}
\vspace{-1.5in}
\caption[Opening angle between pairs found as window conversion tracks in GEANT,
$p_T$ of all primary photons in GEANT, and efficiencies for finding window
conversions as a function of $p_T$.]{Opening angle between pairs found as 
window conversion tracks in GEANT (solid) and between all GEANT conversion 
tracks (dashed),
$p_T$ of all primary photons in GEANT, and efficiencies for finding window
conversions as a function of $p_T$.  In the lower-left plot, the solid line
represents the fraction of primary photons which converted in the window, 
the dashed line is the fraction which also produced two conversion tracks
in the acceptance. 
In the lower right plot, the solid line
is the fraction of window conversions with both tracks in the acceptance
which were found as a pair of heads, and the dashed line is the fraction
which also had one tail per head.}
\label{f:openang} \end{figure}
\vfill\newpage
\begin{figure}[h]\vspace{-0.5in}\hspace*{-0.3in}
\psfig{file=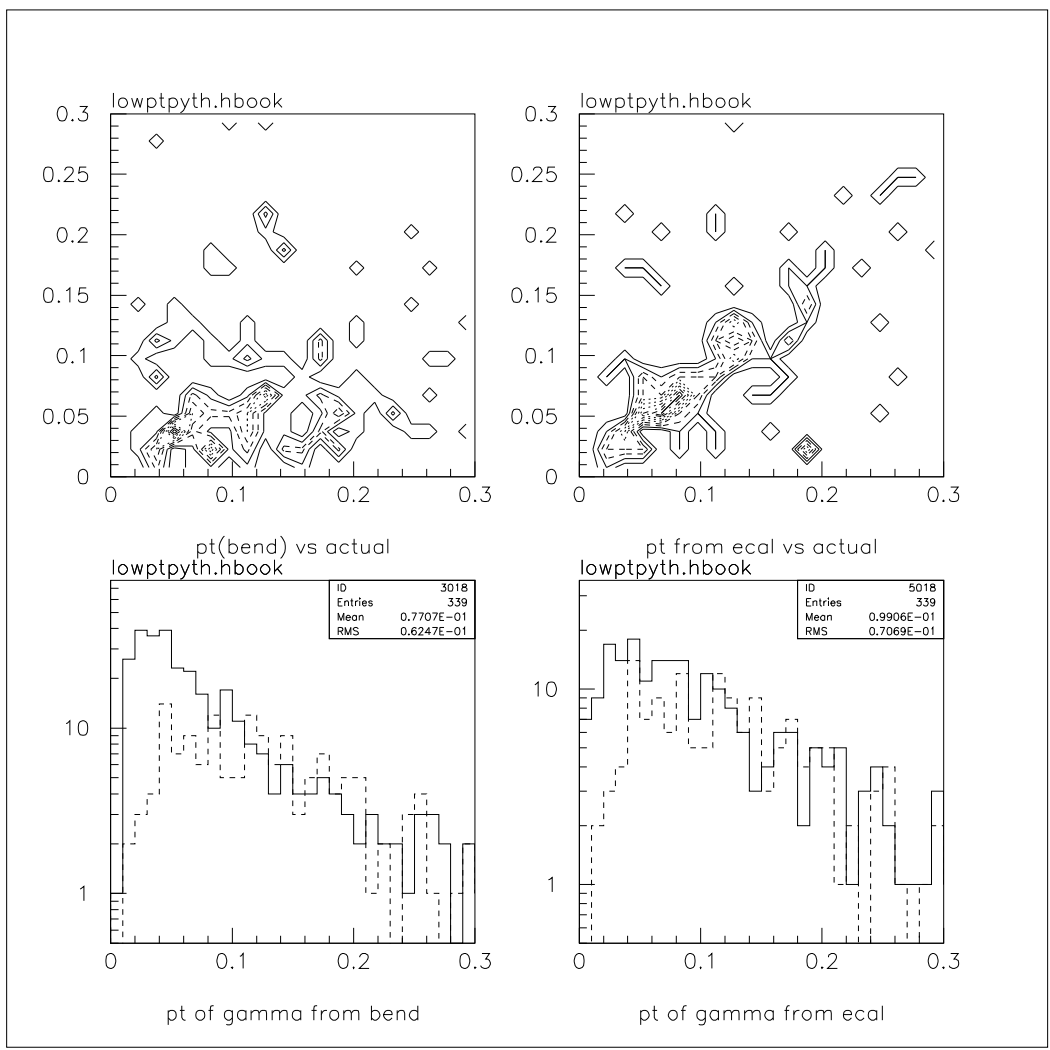,width=6in}
\vspace{-1in}
\caption[Contour plots of the $p_T$ found by bending in the scintillator and
by using the calorimeter vs the actual $p_T$ of the converted photon in GEANT,
and corresponding $p_T$ distributions.]{Contour plots of the $p_T$ found by 
bending in the scintillator and
by using the calorimeter vs the actual $p_T$ of the converted photon in GEANT,
and corresponding $p_T$ distributions (solid is calculated $p_T$, dashed is 
actual).}
\label{f:pt} \end{figure}
\vfill\newpage
\begin{figure}[h]\vspace{-0.5in}\hspace*{-0.3in}
\psfig{file=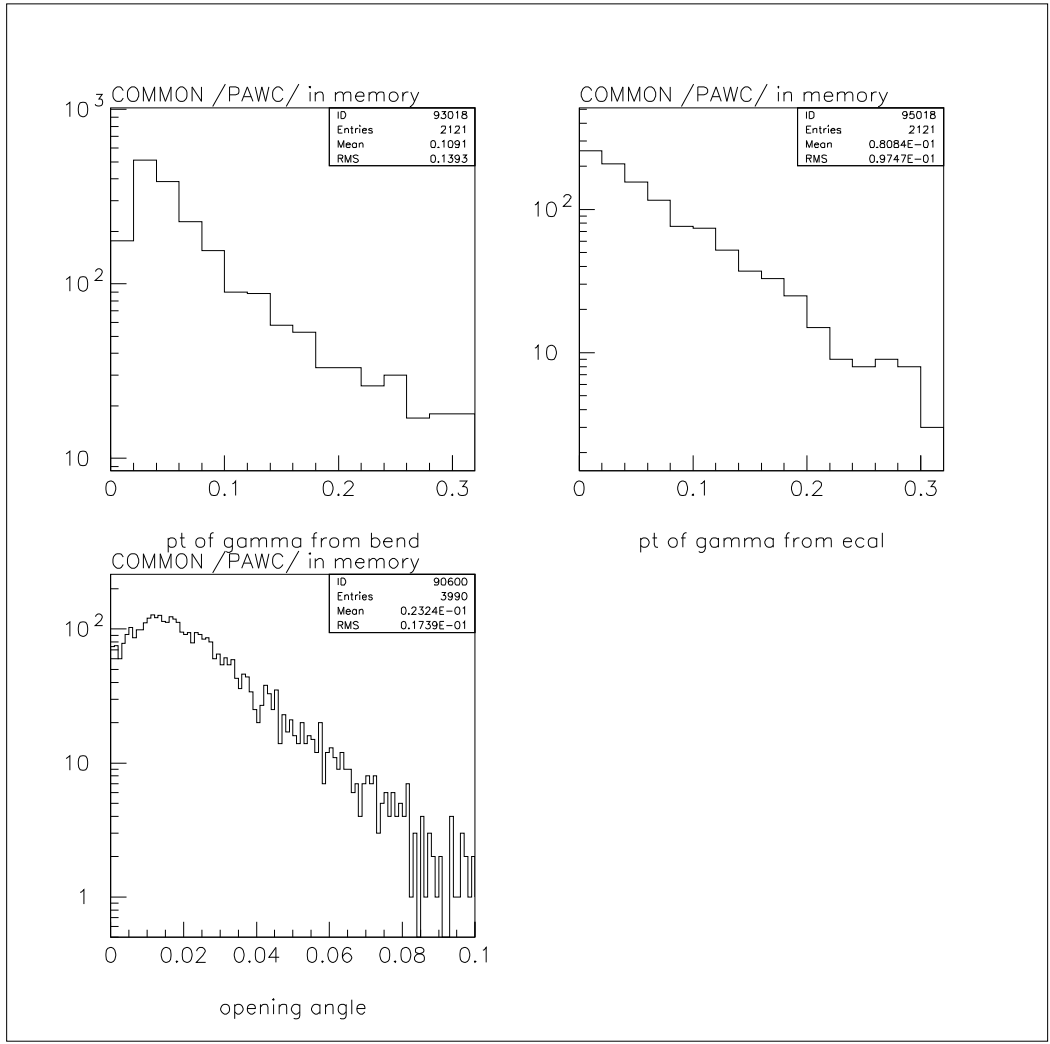,width=6in}
\vspace{-1in}
\caption{Distributions of $p_T$ found by bending in the scintillator and
by using the calorimeter for converted photons in the data.}
\label{f:rpt} \end{figure}

\clearpage

\vfill\newpage
\begin{table}[h]\vspace{0.5in}\hspace*{-0.5in}
\begin{tabular}{|cc|c|c|c|c|c|}
\hline
     &    & $\quad$PYTHIA$\quad$ & PYTHIA/GEANT &
  $\ $DCC/GEANT$\ \ $ & binomial & $1/(2\sqrt{f})$ \\
 $i$ & $j$ &  $r_{ij} \pm \sigma_{r_{ij}}$ &
$r_{ij} \pm \sigma_{r_{ij}}$ & $r_{ij} \pm \sigma_{r_{ij}}$ &
$r_{ij}$ & $r_{ij}$ \\
\hline
   1 & \ 1 \ &   $\quad$1.00 $\pm$ 0.01$\quad$ &
$\quad$1.00 $\pm$ 0.01$\quad$ &  $\quad$0.58 $\pm$ 0.01$\quad$ & 1.00 & 0.50 \\
   2 & \ 1 \ &    $\quad$0.99 $\pm$ 0.02$\quad$ &
$\quad$1.00 $\pm$ 0.03$\quad$ &  $\quad$0.44 $\pm$ 0.03$\quad$ & 1.00 & 0.33 \\
   3 & \ 1 \ &    $\quad$0.96 $\pm$ 0.06$\quad$ &
$\quad$1.03 $\pm$ 0.07$\quad$ &  $\quad$0.39 $\pm$ 0.05$\quad$ & 1.00 & 0.25 \\
   4 & \ 1 \ & $\quad 1.02\pm 0.11\quad$ & 
$\quad 1.29\pm 0.18\quad $ & $\quad 0.42\pm 0.11\quad$ & 1.00 & 0.20 \\
   5 & \ 1 \ & $\quad 1.32\pm 0.20\quad$ & 
$\quad 2.03\pm 0.56\quad $ & $\quad 0.51\pm 0.37\quad$ & 1.00 & 0.17 \\
   6 & \ 1 \ & $\quad 2.06\pm 2.52\quad$ & 
$\quad 3.61\pm 1.35\quad $ & $\quad 0.51\pm 0.73\quad$ & 1.00 & 0.14 \\
\hline
   0 & \ 2 \ &    $\quad$1.24 $\pm$ 0.01$\quad$ &
$\quad$1.36 $\pm$ 0.02$\quad$ &  $\quad$1.51 $\pm$ 0.05$\quad$ & 1.36 & 1.80 \\
   1 & \ 2 \ &    $\quad$1.20 $\pm$ 0.04$\quad$ &
$\quad$1.37 $\pm$ 0.05$\quad$ &  $\quad$0.67 $\pm$ 0.05$\quad$ & 1.30 & 0.62 \\
   2 & \ 2 \ &    $\quad$1.13 $\pm$ 0.09$\quad$ &
$\quad$1.45 $\pm$ 0.15$\quad$ &  $\quad$0.43 $\pm$ 0.08$\quad$ & 1.25 & 0.31 \\
   0 & \ 3 \ &    $\quad$1.60 $\pm$ 0.06$\quad$ &
$\quad$2.35 $\pm$ 0.15$\quad$ &  $\quad$2.77 $\pm$ 0.34$\quad$ & 1.89 & 3.54 \\
   1 & \ 3 \ &    $\quad$1.50 $\pm$ 0.15$\quad$ &
$\quad$2.47 $\pm$ 0.30$\quad$ &  $\quad$1.17 $\pm$ 0.27$\quad$ & 1.74 & 0.90 \\
   0 & \ 4 \ &    $\quad$2.09 $\pm$ 0.24$\quad$ &
$\quad$4.82 $\pm$ 0.86$\quad$ &  $\quad$6.04 $\pm$ 1.84$\quad$ & 2.70 & 7.34 \\
\hline
\end{tabular}
\caption{\label{t:r}Robust observables $r_{i,j}$ for generic events simulated
by PYTHIA and pure-DCC events simulated by the DCC generator, along
with predictions for binomial and $1/(2\protect\sqrt{f})$ distributions.}
\end{table}
\vfill\newpage
\begin{table}[h]\vspace{0.25in}\hspace*{-1.0in}
\begin{tabular}{|c|c|c|c|c|c|}
\hline
  model    &  A    &   B   &   C   &   D   &   E   \\
\hline
 \# events & 20000 & 10000 & 10000 & 10000 & 10000 \\
\hline
$\left< p_T\right>$ (MeV) & 100 & 50 & 25 & 50 & 25 \\
$\psi$ & 1 & 1 & 1 & 0.5 & 0.25 \\
$\left< N_\pi\right>$ & 5.0 & 10.0 & 19.9 & 5.0 & 5.0 \\
\hline
$\left< n_{ch}\right>$ & $\quad 0.97\pm 0.01\quad$ & $\quad 1.96\pm 0.02\quad$ 
  & $\quad 2.98\pm 0.02\quad$ & $\quad 0.99\pm 0.01\quad$ & 
  $\quad 0.93\pm 0.01\quad$ \\
$\left< n_{\gamma}\right>$ & $0.27\pm 0.01$ & $0.29\pm 0.01$ & 
  $0.33\pm 0.01$ & $0.14\pm 0.01$ & $0.06\pm 0.01$ \\
$\left< n_{ch}(n_{ch}-1)\right>$ & $1.03\pm 0.02$ & $4.22\pm 0.07$ & 
  $9.43\pm 0.11$ & $1.08\pm 0.03$ & $0.94\pm 0.02$ \\
$\left< n_{ch}n_{\gamma}\right>$ & $0.17\pm 0.01$ & $0.43\pm 0.01$ & 
  $1.03\pm 0.02$ & $0.09\pm 0.01$ & $0.05\pm 0.01$ \\
\hline
 $r(1,1)$ & $0.58\pm 0.01$ & $0.69\pm 0.01$ & $0.97\pm 0.01$ & 
   $0.62\pm 0.02$ & $0.83\pm 0.05$ \\
 $r(2,1)$ & $0.44\pm 0.03$ & $0.61\pm 0.03$ & $1.03\pm 0.02$ & 
   $0.48\pm 0.05$ & $0.92\pm 0.11$ \\
 $r(3,1)$ & $0.39\pm 0.05$ & $0.57\pm 0.04$ & $1.12\pm 0.04$ & 
   $0.39\pm 0.08$ & $1.07\pm 0.28$ \\
\hline
 $r(0,2)$ & $1.51\pm 0.05$ & $1.27\pm 0.06$ & $1.06\pm 0.04$ & 
   $1.32\pm 0.12$ & $1.52\pm 0.31$ \\
 $r(1,2)$ & $0.67\pm 0.05$ & $0.64\pm 0.05$ & $1.10\pm 0.06$ & 
   $0.61\pm 0.10$ & $0.79\pm 0.30$ \\
 $r(2,2)$ & $0.43\pm 0.08$ & $0.41\pm 0.06$ & $1.27\pm 0.09$ & 
   $0.34\pm 0.13$ & $0.75\pm 0.44$ \\
 $r(0,3)$ & $2.77\pm 0.34$ & $2.08\pm 0.34$ & $1.17\pm 0.21$ & 
   $2.52\pm 0.95$ & $3.06\pm 3.04$ \\
 $r(1,3)$ & $1.17\pm 0.27$ & $0.79\pm 0.16$ & $1.24\pm 0.24$ & 
   $0.75\pm 0.38$ & $0.00\pm 0.00$ \\
 $r(0,4)$ & $6.04\pm 1.84$ & $3.81\pm 1.49$ & $1.66\pm 0.98$ & 
   $5.41\pm 5.42$ & $0.00\pm 0.00$ \\
\hline
\end{tabular}
\caption{\label{t:rdcc}Results for varying parameters in the DCC generator.}
\end{table}
\vfill\newpage
\begin{table}[h]
\begin{center}
\begin{tabular}{|c|c|c|c|c|}
\hline
fraction & $r_{1,1}\pm\sigma_{r_{1,1}}$ &
 $r_{2,1}\pm\sigma_{r_{2,1}}$ & $r_{3,1}\pm\sigma_{r_{3,1}}$ & \# events \\
\hline
0.00 & $\quad 1.01\pm 0.02\quad$ & $\quad 1.02\pm 0.05\quad$ & 
  $\quad 1.09\pm 0.14\quad$ & 51471 \\
0.02 &  $1.00\pm 0.02$ &  $1.00\pm 0.05$ &  $1.01\pm 0.15$ &  51741 \\
0.05 &  $0.97\pm 0.02$ &  $0.93\pm 0.05$ &  $0.95\pm 0.10$ &  51741 \\
0.10 &  $0.95\pm 0.02$ &  $0.89\pm 0.04$ &  $0.89\pm 0.08$ &  51741 \\
0.20 &  $0.93\pm 0.02$ &  $0.83\pm 0.04$ &  $0.77\pm 0.07$ &  51741 \\
0.50 &  $0.84\pm 0.01$ &  $0.71\pm 0.03$ &  $0.68\pm 0.06$ &  40000 \\
1.00 &  $0.74\pm 0.01$ &  $0.60\pm 0.03$ &  $0.55\pm 0.06$ &  20000 \\
\hline
\end{tabular}
\end{center}
\caption[The effect on the $r_{i,1}$ of an admixture of DCC and
generic (PYTHIA) events.]
{\label{t:rfract} The effect on the $r_{i,1}$ of an admixture of DCC and
generic (PYTHIA) events.  DCC domains from the DCC generator/GEANT are added to 
various fractions of random PYTHIA/GEANT events.  The first column represents
the fraction of events in which a DCC is overlaying a generic event.
A DCC fraction of 1 means that DCC has been added to every event, not that 
the events are pure DCC as in Table 6.1.}
\end{table}
\vfill\newpage
\begin{table}[h]
\vskip2ex
\begin{tabular}{|cc|cccccccccl|}
\hline
 & & \multicolumn{9}{c}{$n_\gamma$}  & \\
 & & \ \ 0\ \ \  & \ \ 1\ \ \  & \ \ 2\ \ \  & \ \ 3\ \ \  & \ \ 4\ \ \  & 
\ \ 5\ \ \  &\ \ 6\ \ \  & \ \ 7\ \ \ &\ \ 8\ \ \ & \\
\hline
 & 0 & 834625 & 126160 & 16739  & 2515  &  454   &  67 & 10 & 8 & 1 & \\
 & 1 & 355999 &  62717 & 10398  & 1732  &  272   &  56 & 12 & 1 & 0 & \\
 & 2 &  87117 &  19140 &  3580  &  639  &  148   &  31 &  4 & 0 & 0 & \\
 & 3 &  17786 &   4771 &  1016  &  208  &   34   &   9 &  0 & 0 & 0 & \\
$n_{ch}$ & 4 &   3350 &    943 &   224  &   48  &   10   &   2 &  0 & 0 & 0 & \\
 & 5 &    516 &    182 &    49  &    7  &    4   &   0 &  0 & 0 & 0 & \\
 & 6 &     88 &     38 &     7  &    3  &    0   &   0 &  0 & 0 & 0 & \\
 & 7 &     10 &      6 &     1  &    0  &    0   &   0 &  0 & 0 & 0 & \\
 & 8 &      0 &      1 &     0  &    0  &    0   &   0 &  0 & 0 & 0 & \\
\hline
\end{tabular}
\caption{\label{t:ncg}Number of events with a given $n_{ch}$, $n_\gamma$.}
\end{table}
\vfill\newpage
\clearpage
\begin{table}[h]\hspace*{-0.8in}
\begin{tabular}{|c|c|c|c|c|}
\hline
 & all events & ktag & $\bar{\mbox{n}}$ hcal & $\bar{\mbox{p}}$ hcal \\
\hline
 \# events & 1551738 & 21412 & 27807 & 42449 \\
\hline
$\left< n_{ch}\right>$ & $\quad 0.4814\pm 0.0006\quad$ & 
  $\quad 0.407\pm 0.005\quad$ & $\quad 0.422 \pm 0.004\quad$ & 
  $\quad 0.437 \pm 0.003\quad$ \\
$\left< n_{\gamma}\right>$ & $0.1922\pm 0.0004$ & $0.167\pm 0.003$ &
 $0.173 \pm 0.003$ & $0.183 \pm 0.002$ \\
$\left< n_{ch}(n_{ch}-1)\right>$ & $0.2830\pm 0.0010$ & $0.195\pm 0.007$ &
 $0.213 \pm 0.006$ & $0.227 \pm 0.005$ \\
$\left< n_{ch}n_{\gamma}\right>$ & $0.1156\pm 0.0004$ & $0.084\pm 0.003$ &
 $0.092 \pm 0.003$ & $0.103 \pm 0.002$ \\
\hline
 $r(1,1)$ & $1.0228 \pm 0.0035$ & $1.052 \pm 0.036$
          & $1.051 \pm 0.031$ & $1.076 \pm 0.024$  \\
 $r(2,1)$ & $1.0320 \pm 0.0097$ & $1.134 \pm 0.107$
          & $1.042 \pm 0.083$ & $1.116 \pm 0.069$  \\
 $r(3,1)$ & $1.0563 \pm 0.0251$ & $1.302 \pm 0.281$
          & $1.220 \pm 0.211$ & $1.180 \pm 0.173$  \\
 $r(4,1)$ & $1.1183 \pm 0.0612$ & $1.802 \pm 0.637$ 
          & $2.102 \pm 0.599$ & $1.370 \pm 0.413$ \\
 $r(5,1)$ & $1.3382 \pm 0.1436$ & & & $1.721 \pm 0.911$ \\
 $r(6,1)$ & $2.0758 \pm 0.4159$ & & & \\
 $r(7,1)$ & $5.0105 \pm 5.6885$ & & & \\
\hline
 $r(0,2)$ & $1.5793 \pm 0.0092$ & $1.647 \pm 0.090$
          & $1.683 \pm 0.077$ & $1.658 \pm 0.059$  \\
 $r(1,2)$ & $1.5632 \pm 0.0204$ & $1.623 \pm 0.217$
          & $1.700 \pm 0.176$ & $1.675 \pm 0.152$  \\
 $r(2,2)$ & $1.5538 \pm 0.0497$ & $1.914 \pm 0.616$
          & $1.908 \pm 0.474$ & $1.933 \pm 0.441$  \\
 $r(0,3)$ & $3.3711 \pm 0.0759$ & $3.378 \pm 0.627$
          & $3.359 \pm 0.545$ & $3.333 \pm 0.390$  \\
 $r(1,3)$ & $3.1115 \pm 0.1210$ & $3.335 \pm 1.165$
          & $3.196 \pm 0.895$ & $3.367 \pm 0.958$  \\
 $r(0,4)$ & $8.9729 \pm 0.6201$ & $6.675 \pm 2.934$
          & $7.266 \pm 2.980$ & $6.236 \pm 1.780$  \\
\hline
\end{tabular}
\caption{\label{t:rreal}Values of $r_{i,j}$ for all lead-in events and for
those with diffractive tags.}
\end{table}
\vfill\newpage
\begin{table}[h]\vspace{0.75in}\hspace*{-0.8in}
\begin{tabular}{|c|c|c|c|c|c|}
\hline
 bin & $\left< n_{ch}\right>$ & $\left< n_\gamma\right>$ &  
 $r_{1,1}$   & $r_{2,1}$    & $r_{3,1}$ \\
\hline
1 & $\quad 0.364\pm 0.002\quad$  & $\quad 0.145\pm 0.002\quad$  & 
  $\quad 1.02\pm 0.02\quad$  & $\quad 0.97\pm 0.06\quad$  & 
  $\quad 0.87\pm 0.14\quad$  \\
2 & $0.401\pm 0.002$  & $0.165\pm 0.002$  & 
  $1.00\pm 0.02$  & $1.00\pm 0.05$  & $1.04\pm 0.14$  \\
3 & $0.426\pm 0.002$  & $0.173\pm 0.002$  & 
  $1.03\pm 0.02$  & $1.11\pm 0.05$  & $1.28\pm 0.13$  \\
4 & $0.449\pm 0.002$  & $0.183\pm 0.002$  & 
  $1.00\pm 0.02$  & $1.05\pm 0.05$  & $1.10\pm 0.13$  \\
5 & $0.468\pm 0.003$  & $0.190\pm 0.002$  & 
  $1.03\pm 0.02$  & $1.10\pm 0.05$  & $1.17\pm 0.14$  \\
6 & $0.493\pm 0.003$  & $0.195\pm 0.002$  & 
  $1.02\pm 0.02$  & $1.01\pm 0.04$  & $1.05\pm 0.11$  \\
7 & $0.503\pm 0.003$  & $0.202\pm 0.002$  & 
  $1.04\pm 0.02$  & $1.03\pm 0.05$  & $0.98\pm 0.12$  \\
8 & $0.531\pm 0.003$  & $0.215\pm 0.002$  & 
  $1.01\pm 0.01$  & $1.00\pm 0.04$  & $0.97\pm 0.09$  \\
9 & $0.555\pm 0.003$  & $0.221\pm 0.002$  & 
  $1.03\pm 0.01$  & $1.06\pm 0.04$  & $1.09\pm 0.11$  \\
10 & $0.603\pm 0.003$  & $0.237\pm 0.002$  & 
  $1.03\pm 0.01$  & $1.02\pm 0.03$  & $1.01\pm 0.07$  \\
\hline
\end{tabular}
\caption[Mean multiplicities and robust observables for bins
of pbar multiplicities each containing 10\% of the events.]
{\label{t:rpb}Mean multiplicities and robust observables for bins
of pbar multiplicities each containing 10\% of the events.  The pbar 
multiplicity increases with increasing bin number 1-10.}
\end{table}

\clearpage
\chapter{Conclusions}

MiniMax ran successfully with many different detector configurations for a 
period of about two years.  During this time, much was 
learned which should prove useful for the operation and design of future 
detectors in the forward region.  More than $1.5\times 10^6$ events with
lead in and almost $10^6$ events with lead out from clean runs when the
detector was running properly have been analyzed for the work presented here.

The GEANT simulation does not accurately represent the amount of 
background seen in the wire chambers from interactions in material 
surrounding the detector.  Even with all of the material included in the 
simulation, the mean number of hits in the MWPC's is low by about a factor of 
two.  However, the PYTHIA/GEANT output is useful for setting upper limits on 
how well the data analysis tools are able to reconstruct events.  Based on the 
efficiencies and numbers of fakes discussed in Sec. \ref{sec:vertexer}, the 
tracker and vertexer appear to work quite well.

The simulations also show that the robust observables of Chapter 5
are insensitive to detector effects and physics complications, as advertised.
The PYTHIA/GEANT simulation, which includes a photon conversion efficiency of 
slightly better than 50\%, an efficiency of 80\% for detecting converted 
photons, resonance production, simulations of detector effects, and many other 
features, matches the predictions of a simple binomial model.
At least for DCC modeled as in Sec. \ref{sec:dccgen}, the robust observables
should be useful in distinguishing DCC from generic production.
 
We have found no evidence for DCC in the total sample of events or in
diffractive events.  Limits on the amount of DCC which could be present
without being detected depend on the models of DCC (e.g. the $p_T$ of the
DCC pions) and of combining DCC with generic production.

Raw measurements of $dN_{ch}/d\eta$ and $dN_\gamma/d\eta$ have been made in a 
previously unexplored region of phase space, from which actual distributions
can be derived, taking into account detection efficiencies and fakes, and 
trigger efficiencies.  Further work on modeling fakes in the data is necessary
before this can be done.

\vfill\newpage
\appendix
\chapter{Uncertainty Calculations}
The normalized factorial moments are defined as
\begin{eqnarray} F_{j,l}={\left< n_c(n_c-1)\dots(n_c-j+1)n_g(n_g-1)
\dots(n_g-l+1)\right>
  \over \left< n_c\right>^j\left< n_g\right>^l} \end{eqnarray}
where $\left< {\cal O}\right>={1\over N}\sum_{n_c,n_g}{\cal O}
{\cal N}(n_c,n_g)$,
$N=\sum_{n_c,n_g}{\cal N}(n_c,n_g)$, and 
${\cal N}(n_c,n_g)$ is the number of events with $n_c$ charged 
tracks and $n_g$ gammas. 

We assume that $\sigma^2_{{\cal N}(n_c,n_g)}={\cal N}(n_c,n_g)$.
The uncertainty in $F_{j,l}$ is then given by
\begin{eqnarray*}\sigma^2_{F_{j,l}}={1\over N}\left[ 
{\left< [n_c\dots(n_c-j+1)n_g\dots(n_g-l+1)]^2\right>
\over \left< n_c\right>^{2j}\left< n_g\right>^{2l}}\right.
-(j+l-1)^2{F_{j,l}}^2 
\end{eqnarray*}\begin{eqnarray*} \qquad
+j^2{\left< n_c^2\right>\over \left< n_c\right>^2}{F_{j,l}}^2 
+l^2{\left< n_g^2\right>\over \left< n_g\right>^2}{F_{j,l}}^2 
+2jl{\left< n_c n_g\right>\over \left< n_c\right>\left< n_g\right>}{F_{j,l}}^2
\end{eqnarray*}\begin{eqnarray*} \qquad
-2j{\left< n_c[n_c\dots(n_c-j+1)n_g\dots(n_g-l+1)]\right>\over 
\left< n_c\right>^{j+1}\left< n_g\right>^l}F_{j,l}
\end{eqnarray*}\begin{eqnarray}\left. \qquad
-2l{\left< n_g[n_c\dots(n_c-j+1)n_g\dots(n_g-l+1)]\right>\over 
\left< n_c\right>^j\left< n_g\right>^{l+1}}F_{j,l}
\right] \end{eqnarray}

In order to find the correlated uncertainty for any two $F$'s we define
\begin{eqnarray}x=F_{j,l}+F_{i,m}\end{eqnarray}
which has uncertainty given by
\begin{eqnarray}\sigma_x^2=\sigma^2_{F_{j,l}}+\sigma^2_{F_{i,m}}+
2\sigma_{F_{j,l}F_{i,m}}\end{eqnarray}
Writing the $F$'s in terms of the ${\cal N}(n_c,n_g)$, we get
\begin{eqnarray*}\sigma_x^2=\sum_{n_c',n_g'}\left[
{n_c'\dots(n_c'-j+1)n_g'\dots(n_g'-l+1)\over
 N\left< n_c\right>^j\left<n_g\right>^l}
+(j+l-1){F_{j,l}\over N}\right.
\end{eqnarray*} \begin{eqnarray*}\qquad\qquad
-j{n_c'F_{j,l}\over N\left< n_c\right>} -l{n_g'F_{j,l}\over N\left< n_g\right>}
\end{eqnarray*} \begin{eqnarray*}\qquad\qquad
+{n_c'\dots(n_c'-i+1)n_g'\dots(n_g'-m+1)\over 
N\left< n_c\right>^i\left< n_g\right>^m} +(i+m-1){F_{i,m}\over N}\end{eqnarray*}
\begin{eqnarray}\qquad\qquad \left. -i{n_c'F_{i,m}\over N\left< n_c\right>} 
-m{n_g'F_{i,m}\over N\left< n_g\right>} \right]^2 {\cal N}(n_c',n_g')
\end{eqnarray}

and we can pick out the correlation terms to find
$$\sigma_{F_{j,l}F_{i,m}}={1\over N}\left[
{\left< [n_c\dots(n_c-j+1)n_g\dots(n_g-l+1)]
[n_c\dots(n_c-i+1)n_g\dots(n_g-m+1)]\right>
\over \left< n_c\right>^{j+i}\left< n_g\right>^{l+m}} 
\right. $$\begin{eqnarray*}\qquad
-i{\left< n_c[n_c\dots(n_c-j+1)n_g\dots(n_g-l+1)]\right>
\over \left< n_c\right>^{j+1}\left< n_g\right>^l}
F_{i,m} \end{eqnarray*}\begin{eqnarray*}\qquad
-m{\left< n_g[n_c\dots(n_c-j+1)n_g\dots(n_g-l+1)]\right>
\over \left< n_c\right>^j\left< n_g\right>^{l+1}}
F_{i,m} \end{eqnarray*}\begin{eqnarray*}\qquad
-j{\left< n_c[n_c\dots(n_c-i+1)n_g\dots(n_g-m+1)]\right>
\over \left< n_c\right>^{i+1}\left< n_g\right>^m}
F_{j,l} \end{eqnarray*}\begin{eqnarray*}\qquad
-l{\left< n_g[n_c\dots(n_c-i+1)n_g\dots(n_g-m+1)]\right>
\over \left< n_c\right>^i\left< n_g\right>^{m+1}}
F_{j,l} \end{eqnarray*}\begin{eqnarray*}\qquad
+ij{\left< n_c^2\right>\over \left< n_c\right>^2}F_{j,l}F_{i,m} 
+lm{\left< n_g^2\right>\over \left< n_g\right>^2}F_{j,l}F_{i,m}
\end{eqnarray*}\begin{eqnarray}\left. \qquad
+(jm+il){\left<n_cn_g\right>\over\left< n_c\right>\left< n_g\right>}
F_{j,l}F_{i,m}
-(j+l-1)(i+m-1)F_{j,l}F_{i,m} \right] \end{eqnarray}
\vskip6ex
The generalized $r$ defined by
\begin{eqnarray}r_{j,l}={F_{j,l}\over F_{j+l,0}}\end{eqnarray}
has uncertainty given by
\begin{eqnarray}\sigma^2_{r_{j,l}}=
\left( {1\over F_{j+l,0}}\right)^2\sigma^2_{F_{j,l}}
+\left( -{r_{jl}\over F_{j+l,0}}\right)^2\sigma^2_{F_{j+l,0}}
+2\left( -{r_{jl}\over F_{j+l,0}^2}\right) \sigma_{F_{j,l}F_{j+l,0}}
\end{eqnarray}
\vfill\newpage
\chapter{Track fitting code}
\section{Track fitter: fit.f}
{\footnotesize
\begin{verbatim}
        SUBROUTINE fit_init

        IMPLICIT NONE

        INTEGER I,NN
        REAL sigsq,wiresp,pi,uang,sum,cu,su
        REAL U0(24),V0(24),ang(24),Z(24)

        common/cham_param/NN,sigsq,wiresp,pi,uang,sum,cu,su,U0,V0,ang,Z

C  position of the center of wire 0 and angle of rotation in (x,y,z)
        real zz(24)/123.69,126.57,129.69,132.57,
     +          135.69,138.69,141.69,144.69,
     +          167.09,169.89,172.84,176.09,
     +          178.84,181.84,185.09,189.65,
     +          194.29,197.29,200.29,203.29,
     +          206.16,209.16,212.41,215.41/
        real xx0(24)/2.806,10.117,0.853,11.792,
     +          9.997,0.818,10.076,1.252,
     +          1.692,10.090,10.970,1.535,
     +          11.199,12.360,11.424,3.366,
     +          11.602,11.327,2.814,12.648,
     +          12.056,3.560,3.140,2.681/
        real yy0(24)/11.653,11.032,10.870,9.255,
     +          10.910,9.193,11.700,1.813,
     +          2.568,12.305,11.702,3.508,
     +          11.781,10.771,11.812,1.695,
     +          12.036,12.524,2.963,11.415,
     +          12.114,1.613,3.048,3.748/
        real aangle(24)/-1.0315,-2.3422,-0.7575,-2.5953,
     +          -2.3318,-0.5044,-2.0857,0.8029,
     +          0.7959,-2.1084,-2.3370,0.5794,
     +          -2.3370,-2.5063,-2.3405,0.9669,
     +          -2.3370,-2.2201,0.8029,-2.4574,
     +          -2.3527,1.0420,0.7837,0.6266/


        sigsq=1./12.  !uncer in location of track (wire spacings) for each hit
        wiresp=0.1    !wire spacing in inches
        NN=4
        pi=3.141592654
        uang=45.43*pi/180.  !taken as mean angle of "u chambers"
        cu=cos(uang)
        su=sin(uang)

        do i=1,24
        U0(i)=xx0(i)*cu+yy0(i)*su    !u,v of center of wire 0
        V0(i)=-xx0(i)*su+yy0(i)*cu
        Z(i)=zz(i)
        if (aangle(i).lt.2*pi) aangle(i)=aangle(i)+2*pi
        ang(i)=aangle(i)-uang
        ang(i)=ang(i)-pi/2.
        enddo

        RETURN
        END
 
c----------------------------------------------------------------

        SUBROUTINE fit_wires

        IMPLICIT NONE

        INTEGER I,J,K,L,num
        INTEGER icham,ichlast,ncham,iwire,nwires(24),iicham(24)
        REAL wire
        REAL awires(24,128)
        REAL*8 X(4,4),XX(100,4),Y(4),YY(100),a(4),XINV(4,4)
        REAL*8 INDEX(4),D

        INTEGER NN
        REAL sigsq,wiresp,pi,uang,sum,cu,su
        REAL U0(24),V0(24),ang(24),Z(24)
        common/cham_param/NN,sigsq,wiresp,pi,uang,sum,cu,su,U0,V0,ang,Z

        INTEGER HLIM,HBSIZE
        PARAMETER(HBSIZE=500000)
        COMMON/PAWC/HLIM(HBSIZE)

        include 'wires.inc'
        include 'fit.inc'
 
        include 'track_dat.inc'
        include 'dst.inc'

        integer ib(24),dmin,newwire
        include 'EVENT.INC'    !to find missing hit in front-8 tracks
 
        DO I=1,ntrack
100     continue

C  will fit track parameters 'a' in matrix eq Xa=Y

        DO J=1,4
          DO K=1,4
          X(J,K)=0
          ENDDO
          Y(J)=0
        ENDDO

C  will need to know which chambers and which wires in each chamber
C  were hit to get chisq
        ncham=0
        ichlast=0
        if (nnwire(i).gt.100) write(6,*)'nnwire=',nnwire(i)
        DO J=1,nnwire(i)                     !go through hit wires from dst
            icham=int((wires(I,J)-1)/128.)+1
            iwire=wires(I,J)-(icham-1)*128
          IF (icham.ne.ichlast) THEN
            ncham=ncham+1
            iicham(ncham)=icham
            nwires(ncham)=1
          ELSE
            nwires(ncham)=nwires(ncham)+1
          ENDIF
          awires(ncham,nwires(ncham))=iwire
          ichlast=icham

C  fit XX(i,j)a(j)=YY(i) for all chambers i to get a(j)
C  min chisq -> Y=Xa, a=(Xinv)Y
C  (take same uncertainty for all hits siqsq=1/12 (wiresp^2))
          
          XX(J,1)=-sin(ang(icham))
          XX(J,2)=-Z(icham)*sin(ang(icham))
          XX(J,3)=cos(ang(icham))
          XX(J,4)=Z(icham)*cos(ang(icham))
          YY(J)=-U0(icham)*sin(ang(icham))+V0(icham)*cos(ang(icham))+
     +          iwire*wiresp
          DO L=1,4
            DO K=1,4
              X(L,K)=X(L,K)+XX(J,L)*XX(J,K)
            ENDDO
            Y(L)=Y(L)+XX(J,L)*YY(J)
          ENDDO
999     continue
        ENDDO
      
C  invert X
        DO J=1,4
          DO K=1,4
            XINV(J,K)=0
          ENDDO
          XINV(J,J)=1
        ENDDO
C  Numerical Recipes subroutines to invert a matrix
        CALL LUDCMP(X,NN,INDEX,D)
        DO K=1,4
          CALL LUBKSB(X,NN,INDEX,XINV(1,K))
        ENDDO
      
        DO J=1,4
          a(J)=0
          DO K=1,4
            a(J)=a(J)+Y(K)*XINV(J,K)    !get track parameters
          ENDDO
        ENDDO

C  fill fit.inc

        a_u(I)=a(1)
        b_u(I)=a(2)
        a_v(I)=a(3)
        b_v(I)=a(4)


C  get chisq (if more than one wire in a chamber is part of the track,
C   use residuals from each wire
C   total number of wires hit = num)
        chisq(I)=0
        num=0
        DO J=1,ncham
          K=iicham(J)
          wire=(-(a_u(I)+b_u(I)*Z(K)-U0(K))*sin(ang(K))
     +    	+(a_v(I)+b_v(I)*Z(K)-V0(K))*cos(ang(K)))/wiresp
          DO L=1,nwires(J)
            num=num+1
            chisq(I)=chisq(I)+(wire-awires(J,L))**2
          ENDDO
        ENDDO
        chisq(I)=chisq(I)/(sigsq*(num-4))
      
C  get covariance matrix
        DO J=1,4
        DO K=1,4
          covar(I,J,K)=(sigsq*0.01)*XINV(J,K)
        ENDDO
        ENDDO
      
 	enddo
 	endif

        ENDDO
      
        RETURN
        END
      
C  -----------------------------------------------------

      SUBROUTINE LUDCMP(A,N1,INDX,D)    !LU decomposition
        IMPLICIT NONE
        integer n,i,j,k,imax,N1
        real*8 nmax,tiny,aamax,sum,dum	!need double precision to get XinvX=1
        PARAMETER (NMAX=46,TINY=1.0E-20) 
        REAL*8 NP,D,A(4,4),INDX(4),VV(4)

        N=N1
        D=1.
        DO 12 I=1,N
        AAMAX=0.
        DO 11 J=1,N
          IF (ABS(A(I,J)).GT.AAMAX) AAMAX=ABS(A(I,J))
11      CONTINUE
        IF (AAMAX.EQ.0.) PAUSE 'Singular matrix.'
        VV(I)=1./AAMAX
12      CONTINUE
        DO 19 J=1,N
        IF (J.GT.1) THEN
          DO 14 I=1,J-1
            SUM=A(I,J)
            IF (I.GT.1)THEN
              DO 13 K=1,I-1
                SUM=SUM-A(I,K)*A(K,J)
13            CONTINUE
              A(I,J)=SUM
            ENDIF
14        CONTINUE
        ENDIF
        AAMAX=0.
        DO 16 I=J,N
          SUM=A(I,J)
          IF (J.GT.1)THEN
            DO 15 K=1,J-1
              SUM=SUM-A(I,K)*A(K,J)
15          CONTINUE
            A(I,J)=SUM
          ENDIF
          DUM=VV(I)*ABS(SUM)
          IF (DUM.GE.AAMAX) THEN
            IMAX=I
            AAMAX=DUM
          ENDIF
16      CONTINUE
        IF (J.NE.IMAX)THEN
          DO 17 K=1,N
            DUM=A(IMAX,K)
            A(IMAX,K)=A(J,K)
            A(J,K)=DUM
17        CONTINUE
          D=-D
          VV(IMAX)=VV(J)
        ENDIF
        INDX(J)=IMAX
          IF(A(J,J).EQ.0.)A(J,J)=TINY
        IF(J.NE.N)THEN
          DUM=1./A(J,J)
          DO 18 I=J+1,N
            A(I,J)=A(I,J)*DUM
18        CONTINUE
        ENDIF
19      CONTINUE
        RETURN
        END



      SUBROUTINE LUBKSB(A,N1,INDX,B)     !back substitution
        IMPLICIT NONE
        integer n,ii,i,ll,j,N1
        REAL*8 sum,NP,A(4,4),INDX(4),B(4)
        II=0
        N=N1 
        DO 12 I=1,N
        LL=INDX(I)
        SUM=B(LL)
        B(LL)=B(I)
        IF (II.NE.0)THEN
          DO 11 J=II,I-1
            SUM=SUM-A(I,J)*B(J)
11        CONTINUE
        ELSE IF (SUM.NE.0.) THEN
          II=I
        ENDIF
        B(I)=SUM
12      CONTINUE
        DO 14 I=N,1,-1
        SUM=B(I)
        IF(I.LT.N)THEN
          DO 13 J=I+1,N
            SUM=SUM-A(I,J)*B(J)
13        CONTINUE
        ENDIF
        B(I)=SUM/A(I,I)
14      CONTINUE
        RETURN
        END


C-----------------------------------------------------
C  wires.inc  (filled by eg vertexer.f)
        integer ntrack,nnwire(200),wires(200,100)
        common/wiresforfit/ntrack,nnwire,wires

C  (must be filled by user unless using the vertexer)
C  ntrack=# tracks
C  nnwire(i)=# wires in track i
C  wires(i,j)=jth wire of ith track in usual format (icham-1)*128+iwire


C-----------------------------------------------------
C  fit.inc (filled by fit.f)
        REAL a_u(200),b_u(200),a_v(200),b_v(200),chisq(200)
        REAL COVAR(200,4,4)
        integer current_evnum
        common/track_param/a_u,b_u,a_v,b_v,chisq,covar,current_evnum

C(determined by fit.f)
C  u=a_u+b_u*z, v=a_v+b_v*z
C
C  for covariance matrix, indices go like
C  1 = a_u, 2 = b_u, 3 = a_v, 4 = b_v
\end{verbatim}}
\vfill\newpage
\section{Code for correcting tracks with double hits in the non-$u$ chambers:
refit.f}
{\footnotesize
\begin{verbatim}
        SUBROUTINE refit

        IMPLICIT NONE

        INTEGER I,J,K,L,M
        INTEGER icham,ichlast,ncham,iwire,nwires(24),iicham(24)

        INTEGER NN
        REAL sigsq,wiresp,pi,uang,sum,cu,su
        REAL U0(24),V0(24),ang(24),Z(24)
        common/cham_param/NN,sigsq,wiresp,pi,uang,sum,cu,su,U0,V0,ang,Z

        integer itrack
        real av,bv,v(24),u(24),vlead,ucham(24)
        real sz,szz,sv,szv,cth(24),sth(24),denom,predwire,wire(18000,24)
        real sigth,thgam,thtail
        integer nwirepl(24),iwirepl(24,10),nwire,icomb

        integer iw(8),i1,i2,i3,i4,i5,i6,i7,i8
        integer ncomb,ichi,ipoint
        real chi(18000),point(18000),chimin,pointmin

        integer ntr,hit2
        integer nch1,nch2
        common/newnch/nch1,nch2

        include 'wires.inc'
        include 'fit.inc'
 
        include 'track_dat.inc'
        include 'dst.inc'

C  get u chambers
        do j=1,24
        ucham(j)=0
        enddo
        ucham(2)=1
        ucham(5)=1
        ucham(8)=1
        ucham(9)=1
        ucham(11)=1
        ucham(13)=1
        ucham(15)=1
        ucham(17)=1
        ucham(19)=1
        ucham(23)=1
        ucham(24)=1

C  will need orientation of non-u chambers to fit in v
        do k=1,24
          if (ucham(k).eq.0) then
            sth(k)=sin(ang(k))
            cth(k)=cos(ang(k))
          endif
        enddo
 
C  check number of wires hit per plane
C  if 2 wires hit are separated by more than 3 wires in any plane, refit
        do i=1,ntrack
          do k=1,24
            nwirepl(k)=0
          enddo
          do j=1,nnwire(i)
            icham=int((wires(i,j)-1.)/128.)+1
            nwirepl(icham)=nwirepl(icham)+1
            iwirepl(icham,nwirepl(icham))=wires(i,j)
          enddo
          hit2=0
          do k=1,24
            if ((nwirepl(k).eq.2).and.
     +          (abs(iwirepl(k,1)-iwirepl(k,2)).gt.3)) then
              hit2=1
            endif
          enddo
          if (hit2.eq.0) goto 999

C  how many chambers have "double hits" and which wires are involved
        ncham=0
        do k=1,24
          nwirepl(k)=0
        enddo
        ichlast=0
        DO J=1,nnwire(i)
          icham=int((wires(I,J)-1)/128.)+1
          if (ucham(icham).eq.0) then
            if (icham.ne.ichlast) ncham=ncham+1
            iicham(ncham)=icham
            nwirepl(ncham)=nwirepl(ncham)+1
            iwirepl(ncham,nwirepl(ncham))=wires(I,J)-(icham-1)*128
            ichlast=icham
          endif
        ENDDO

C  cycle thru all wires!!!
C  and find all possible combinations with 1 wire per plane
        ncomb=0
        if (ncham.eq.3) then
        do i1=1,nwirepl(1)
          iw(1)=i1
        do i2=1,nwirepl(2)
          iw(2)=i2
        do i3=1,nwirepl(3)
          iw(3)=i3
          ncomb=ncomb+1
          do j=1,ncham
          if (ncomb.le.18000) wire(ncomb,j)=iwirepl(j,iw(j))
          enddo
        enddo
        enddo
        enddo
        endif
        if (ncham.eq.4) then
        do i1=1,nwirepl(1)
          iw(1)=i1
        do i2=1,nwirepl(2)
          iw(2)=i2
        do i3=1,nwirepl(3)
          iw(3)=i3
        do i4=1,nwirepl(4)
          iw(4)=i4
          ncomb=ncomb+1
          do j=1,ncham
          if (ncomb.le.18000) wire(ncomb,j)=iwirepl(j,iw(j))
          enddo
        enddo
        enddo
        enddo
        enddo
        endif
        if (ncham.eq.5) then
        do i1=1,nwirepl(1)
          iw(1)=i1
        do i2=1,nwirepl(2)
          iw(2)=i2
        do i3=1,nwirepl(3)
          iw(3)=i3
        do i4=1,nwirepl(4)
          iw(4)=i4
        do i5=1,nwirepl(5)
          iw(5)=i5
          ncomb=ncomb+1
          do j=1,ncham
          if (ncomb.le.18000) wire(ncomb,j)=iwirepl(j,iw(j))
          enddo
        enddo
        enddo
        enddo
        enddo
        enddo
        endif
        if (ncham.eq.6) then
        do i1=1,nwirepl(1)
          iw(1)=i1
        do i2=1,nwirepl(2)
          iw(2)=i2
        do i3=1,nwirepl(3)
          iw(3)=i3
        do i4=1,nwirepl(4)
          iw(4)=i4
        do i5=1,nwirepl(5)
          iw(5)=i5
        do i6=1,nwirepl(6)
          iw(6)=i6
          ncomb=ncomb+1
          do j=1,ncham
          if (ncomb.le.18000) wire(ncomb,j)=iwirepl(j,iw(j))
          enddo
        enddo
        enddo
        enddo
        enddo
        enddo
        enddo
        endif
        if (ncham.eq.7) then
        do i1=1,nwirepl(1)
          iw(1)=i1
        do i2=1,nwirepl(2)
          iw(2)=i2
        do i3=1,nwirepl(3)
          iw(3)=i3
        do i4=1,nwirepl(4)
          iw(4)=i4
        do i5=1,nwirepl(5)
          iw(5)=i5
        do i6=1,nwirepl(6)
          iw(6)=i6
        do i7=1,nwirepl(7)
          iw(7)=i7
          ncomb=ncomb+1
          do j=1,ncham
          if (ncomb.le.18000) wire(ncomb,j)=iwirepl(j,iw(j))
          enddo
        enddo
        enddo
        enddo
        enddo
        enddo
        enddo
        enddo
        endif
        if (ncham.eq.8) then
        do i1=1,nwirepl(1)
          iw(1)=i1
        do i2=1,nwirepl(2)
          iw(2)=i2
        do i3=1,nwirepl(3)
          iw(3)=i3
        do i4=1,nwirepl(4)
          iw(4)=i4
        do i5=1,nwirepl(5)
          iw(5)=i5
        do i6=1,nwirepl(6)
          iw(6)=i6
        do i7=1,nwirepl(7)
          iw(7)=i7
        do i8=1,nwirepl(8)
          iw(8)=i8
          ncomb=ncomb+1
          do j=1,ncham
          if (ncomb.le.18000) wire(ncomb,j)=iwirepl(j,iw(j))
          enddo
        enddo
        enddo
        enddo
        enddo
        enddo
        enddo
        enddo
        enddo
        endif
        if (ncomb.gt.18000)  then     !if too many combinations, skip it
          write(6,*)'ncomb=',ncomb
          return
        endif

C  get chisq and pointing in v for each combination
        chimin=99999.
        pointmin=9999.
        DO M=1,ncomb
        Szz=0
        Sz=0
        Sv=0
        Szv=0
        DO j=1,ncham
          k=iicham(j)
          u(k)=a_u(i)+b_u(i)*Z(k)
          v(k)=(u(k)*sth(k)-U0(k)*sth(k)+V0(k)*cth(k)+
     +		wire(M,j)*wiresp)/cth(k)
          Szz=Szz+Z(k)**2
          Sz=Sz+Z(k)
          Sv=Sv+v(k)
          Szv=Szv+Z(k)*v(k)
        ENDDO
        denom=ncham*Szz-Sz**2
        av=(Szz*Sv-Sz*Szv)/denom
        bv=(ncham*Szv-Sz*Sv)/denom
        chi(M)=0
        DO J=1,ncham
          K=iicham(J)
          predwire=(-(u(k)-U0(K))*sin(ang(K))
     +          +(av+bv*Z(K)-V0(K))*cth(K))/wiresp
            chi(M)=chi(M)+(predwire-wire(M,J))**2
        ENDDO
        chi(M)=chi(M)/(sigsq*(ncham-2))
        if (chi(M).lt.chimin) then
          chimin=chi(M)
          ichi=M
        endif

        thtail=atan(bv)
        vlead=av+bv*149.8
        thgam=atan(vlead/(149.8+a_u(i)/b_u(i)))
        sigth=sqrt((0.01*sigsq)*ncham/(ncham*Szz-Sz**2))*
     +		cos(thtail)**2
        point(M)=abs(thgam-thtail)/sigth
        if (point(M).lt.pointmin) then
          pointmin=point(M)
          ipoint=M
        endif
        ENDDO

C  require vpoint < 8 and chisq < 7, give preference to best pointing
        if (ncomb.gt.0) then
        icomb=0
        if ((pointmin.lt.8.).and.(chi(ipoint).lt.7)) then
          icomb=ipoint
        else
          if ((chimin.lt.7).and.(point(ichi).lt.8)) then
          icomb=ichi
          endif
        endif
        nwire=nnwire(i)
        nnwire(i)=0
        if (icomb.gt.0) then      !no good track --> nnwire(i)=0
        ichlast=0
        DO J=1,nwire                          !refill list of wires corresp
          icham=int((wires(I,J)-1)/128.)+1    ! to track
          if (ucham(icham).eq.1) then
            nnwire(i)=nnwire(i)+1
            wires(i,nnwire(i))=wires(I,J)
          else
            if (icham.ne.ichlast) then
            do k=1,ncham
              if (iicham(k).eq.icham) then
                nnwire(i)=nnwire(i)+1
                wires(i,nnwire(i))=(icham-1)*128+wire(icomb,k)
              endif
            enddo
            endif
          endif
          ichlast=icham
        ENDDO
        endif
        endif


999     continue
        enddo

C  refill wires.inc
        ntr=ntrack
        ntrack=0
        nch1=0
        do i=1,ntr
          if (nnwire(i).gt.0) then
            ntrack=ntrack+1
            if (i.le.nch) nch1=nch1+1
            nnwire(ntrack)=nnwire(i)
            do j=1,nnwire(i)
              wires(ntrack,j)=wires(i,j)
            enddo
          endif
        enddo

C  refit tracks from new wires.inc (for all tracks is cpu intensive but easier)
        call fit_wires

      
        RETURN
        END
\end{verbatim}}
\chapter{Vertexer code}
\section{Vertexer: vertexer.f}
{\footnotesize
\begin{verbatim}
        SUBROUTINE vertexer

        IMPLICIT NONE

        INTEGER HLIM,HBSIZE
        PARAMETER(HBSIZE=500000)
        COMMON/PAWC/HLIM(HBSIZE)

        integer i,j,k,l,m,keep

        real M1(50,2,2),M1INV(50,2,2),MM(2,2),MMINV(2,2)
        real dchisq,umean,vmean,denom

        integer nmatch,imatch(50),ismatched(100)
        integer nnclust(50),iiclust(50,50),iclust,new,newtrack,nmax,nncl
        integer dstnumb(100),nv

        integer max,imax,nsame,nsameu,maxu,icham,ichlast
        integer nback(100),nb(24)
        real thtail,sigth,thgam,zlead,sigu0

        integer ntr,nhead,ntail
        real upoint,vpoint,upoint2,vpoint2

        integer nch1,nch2
        common/newnch/nch1,nch2

        real uold1,vold1
        real dumin,dvmin,dmin,du,dv,uold,vold
        common/taildist/dumin,dvmin,dmin,du,dv,uold,vold

        include 'fit.inc'
        include 'wires.inc'
        include 'vrtx.inc'
        include 'VRTCS.INC'

        include 'track_dat.inc'
        include 'dst.inc'

        zlead=149.82

C  fill wires.inc with all track segments
        ntrack=0
        DO k=1,nch+nnu
C  get number of wires (don't distinguish betwen heads or tails)
        IF (k.le.nch) THEN
          ntrack=ntrack+1
        if (ntrack.gt.100) write(6,*)ntrack
          dstnumb(ntrack)=k
          nnwire(k)=int(charged(k,6))
          DO J=1,nnwire(ntrack)
            wires(k,j)=wires_c(k,J)
          ENDDO
        ELSE
          DO L=9,24
            nb(L)=0
          ENDDO
          DO j=1,neutral(k-nch,6)
            icham=int((wires_n(k-nch,j)-1)/128.)+1
            nb(icham)=1
          ENDDO
          nback(ntrack+1)=0
          DO L=9,24
            nback(ntrack+1)=nback(ntrack+1)+nb(L)
          ENDDO
          if (nback(ntrack+1).ge.8) then  !can change to exclude mid-8 tracks
            ntrack=ntrack+1
            dstnumb(ntrack)=k
            nnwire(ntrack)=int(neutral(k-nch,6))
            DO J=1,nnwire(ntrack)
              wires(ntrack,j)=wires_n(k-nch,J)
 	    ENDDO
          endif
        ENDIF
        ENDDO

C  fit tracks
        call fit_wires
        call refit

C  get position of tracks at lead and uncertainty in pos
        call vrtx(zlead)

C  remove heads which don't point and tracks which don't hit lead
        ntr=ntrack
        ntrack=0
        nch2=0
        do i=1,ntr
          if (i.le.nch1) then
            upoint=-a_u(i)/b_u(i)
          else
            upoint=0
          endif
C  for now, don't demand tails point, loose on heads
          if ((upoint.gt.-50).and.(upoint.lt.60.)) then
            if (i.le.nch1) then
              thtail=atan(b_v(i))
              thgam=atan(v(i)/(zlead+a_u(i)/b_u(i)))
              sigth=sqrt(covar(i,4,4))*cos(thtail)**2
              vpoint=abs(thgam-thtail)/sigth
            else
              vpoint=0
            endif
          if ((vpoint.lt.8.).and.(chisq(i).lt.20.).and.
     +	        (u(i).gt.3.3).and.(u(i).lt.11.3).and.
     +          (v(i).gt.-4.0).and.(v(i).lt.4.0)) then
C  refill wires.inc with kept track segments
            ntrack=ntrack+1
            dstnumb(ntrack)=dstnumb(i)
            if (i.le.nch1) nch2=nch2+1  !number of heads which point
            nnwire(ntrack)=nnwire(i)
            do j=1,nnwire(i)
              wires(ntrack,j)=wires(i,j)
            enddo
            a_u(ntrack)=a_u(i)
            b_u(ntrack)=b_u(i)
            a_v(ntrack)=a_v(i)
            b_v(ntrack)=b_v(i)
            do j=1,4
            do k=1,4
            covar(ntrack,j,k)=covar(i,j,k)
            enddo
            enddo
            chisq(ntrack)=chisq(i)
            u(ntrack)=u(i)
            v(ntrack)=v(i)
            siguu(ntrack)=siguu(i)
            siguv(ntrack)=siguv(i)
            sigvv(ntrack)=sigvv(i)
          endif
          endif
        enddo

	    
C-------------------------------------------------
C  get i,j for tracks to match
        nvrtcs=0

C	increase uncer in pos of tails to account for multiple scattering
        do i=1,ntrack
          ismatched(i)=0
          if (i.gt.nch2) then
            if (siguu(i).lt.0.007) siguu(i)=.007
            if (sigvv(i).lt.0.092) sigvv(i)=.092
          endif
        enddo

C  first match heads to other segments (smaller uncertainty)
        do i=1,ntrack
          if (i.gt.nch2) goto 555
          nmatch=0
        do j=1,ntrack
          if (j.eq.i) goto 557
          CALL match(i,j,dchisq)
          if ((dchisq/2.).lt.5.)  then
            nmatch=nmatch+1  !number of matches to segment i
            imatch(nmatch)=j
            ismatched(j)=1
          endif
557     continue
        enddo	!do j
          imatch(nmatch+1)=i  !(include track i in list of matched tracks)


C  if this head has already been found to match other segments, 
C   combine all matches
        new=1
        do k=1,nvrtcs
        do l=1,nnclust(k)
        do j=1,nmatch+1	
          if (imatch(j).eq.iiclust(k,l)) then
            new=0
            iclust=k
          endif
        enddo
        enddo
        enddo
        if (new.eq.1) then
          nvrtcs=nvrtcs+1
          iclust=nvrtcs
          nnclust(iclust)=0
        endif
        nmax=nnclust(iclust)
        do j=1,nmatch+1
        newtrack=1
        do l=1,nmax
          if (imatch(j).eq.iiclust(iclust,l)) newtrack=0
        enddo
        if (newtrack.eq.1) then
          nnclust(iclust)=nnclust(iclust)+1
          iiclust(iclust,nnclust(iclust))=imatch(j)
        endif
        enddo

555     continue
        enddo	!do i


C  now match tail-tail to get gammas
        uold1=999.
        do i=1,ntrack
          if ((i.le.nch2).or.(ismatched(i).eq.1)) goto 556
          nmatch=0
          uold1=u(i)
          vold1=v(i)
          dmin=999.
          dumin=999.
          dvmin=999.
        do j=1,ntrack
          if ((j.gt.nch2).and.(j.ne.i).and.(ismatched(j).ne.1)) then
          CALL match(i,j,dchisq)
          if ((dchisq/2.).lt.5.) then
            nmatch=nmatch+1
            imatch(nmatch)=j
          endif
          endif	! j ne i
        enddo	!do j
          imatch(nmatch+1)=i

        new=1
        do k=1,nvrtcs
        do l=1,nnclust(k)
        do j=1,nmatch+1	
          if (imatch(j).eq.iiclust(k,l)) then
            new=0
            iclust=k
          endif
        enddo
        enddo
        enddo
        if (new.eq.1) then
          nvrtcs=nvrtcs+1
          iclust=nvrtcs
          nnclust(iclust)=0
        endif
        nmax=nnclust(iclust)
        do j=1,nmatch+1
        newtrack=1
        do l=1,nmax
          if (imatch(j).eq.iiclust(iclust,l)) newtrack=0
        enddo
        if (newtrack.eq.1) then
          nnclust(iclust)=nnclust(iclust)+1
          iiclust(iclust,nnclust(iclust))=imatch(j)
        endif
        enddo

556     continue
        enddo	!do i

        if (uold1.lt.11.25) then
          uold=uold1
          vold=vold1
        endif


C  fill VRTCS.INC with found vertices
        nv=0
        do k=1,nvrtcs
          nverts(k,1)=0
          nverts(k,2)=0
          do j=1,nnclust(k)
            i=iiclust(k,j)
            if (i.gt.nch2) then
              nverts(k,2)=nverts(k,2)+1
            else
              nverts(k,1)=nverts(k,1)+1
            endif
          enddo
          if ((nverts(k,1).eq.1).and.(nverts(k,2).gt.0)) goto 123
                           !if (i,n>0) charged track, pointing is already ok
          nncl=nnclust(k)
          nnclust(k)=0
          nhead=nverts(k,1)
          ntail=nverts(k,2)
          nverts(k,2)=0
          nverts(k,1)=0
          do j=1,nncl
            i=iiclust(k,j)
            upoint=-a_u(i)/b_u(i)
            thtail=atan(b_v(i))
            thgam=atan(v(i)/(zlead+a_u(i)/b_u(i)))
            sigth=sqrt(covar(i,4,4))*cos(thtail)**2
            vpoint=abs(thgam-thtail)/sigth
            if (((upoint.gt.-40).and.(upoint.lt.50).and.
     +          (vpoint.lt.8)).or. 
     +		((nhead.eq.0).and.(ntail.gt.1).and.
     +		(upoint.lt.50))) then
              keep=1
              if ((nhead.eq.0).and.(ntail.eq.1)) then  !tighter on (0,1)'s
        	keep=0
        	do l=9,24
        	  nb(l)=0
        	enddo
        	do l=1,nnwire(i)
        	  icham=int((wires(i,l)-1)/128.)+1
        	  nb(icham)=1
        	enddo
        	icham=0
        	do l=9,24
        	  icham=icham+nb(l)
        	enddo
        	if (icham.ge.14) keep=1
              endif

              if (keep.eq.1) then
              nnclust(k)=nnclust(k)+1
              iiclust(k,nnclust(k))=i
              if (i.gt.nch2) then
                nverts(k,2)=nverts(k,2)+1
              else
                nverts(k,1)=nverts(k,1)+1
              endif
              endif
            endif
          enddo
123       continue
          if (nnclust(k).gt.0) then
            nv=nv+1
            nverts(nv,1)=nverts(k,1)
            nverts(nv,2)=nverts(k,2)
            vrtxpos(nv,1)=0
            vrtxpos(nv,2)=0
        if (nnclust(k).gt.20) write(6,*)'nnclust=',nnclust(k)
C  get mean position of vertex
        MM(1,1)=0
        MM(1,2)=0
        MM(2,2)=0
            do j=1,nnclust(k)
              i=iiclust(k,j)
              nverts(nv,j+2)=dstnumb(i)
              nverts(nv,j+52)=i
        	M1(j,1,1)=siguu(i)
        	M1(j,2,2)=sigvv(i)
        	M1(j,1,2)=siguv(i)
        	M1(j,2,1)=M1(j,1,2)
        	denom=M1(j,1,1)*M1(j,2,2)-M1(j,1,2)*M1(j,2,1)
        	M1INV(j,1,1)=M1(j,2,2)/denom
        	M1INV(j,2,2)=M1(j,1,1)/denom
        	M1INV(j,1,2)=-M1(j,1,2)/denom
        	M1INV(j,2,1)=-M1(j,2,1)/denom
        	MM(1,1)=MM(1,1)+M1INV(j,1,1)
        	MM(1,2)=MM(1,2)+M1INV(j,1,2)
        	MM(2,2)=MM(2,2)+M1INV(j,2,2)
            enddo
            MM(2,1)=MM(1,2)
            denom=MM(1,1)*MM(2,2)-MM(1,2)*MM(2,1)
            MMINV(1,1)=MM(2,2)/denom
            MMINV(2,2)=MM(1,1)/denom
            MMINV(1,2)=-MM(1,2)/denom
            MMINV(2,1)=-MM(2,1)/denom
            umean=0
            vmean=0
            do j=1,nnclust(k)
              i=iiclust(k,j)
              umean=umean+MMINV(1,1)*M1INV(j,1,1)*u(i)+
     +          MMINV(1,2)*M1INV(j,2,1)*u(i)+
     +          MMINV(1,1)*M1INV(j,1,2)*v(i)+
     +          MMINV(1,2)*M1INV(j,2,2)*v(i)
              vmean=vmean+MMINV(2,1)*M1INV(j,1,1)*u(i)+
     +          MMINV(2,2)*M1INV(j,2,1)*u(i)+
     +          MMINV(2,1)*M1INV(j,1,2)*v(i)+
     +          MMINV(2,2)*M1INV(j,2,2)*v(i)
            enddo
            vrtxpos(nv,1)=umean
            vrtxpos(nv,2)=vmean
          endif
        enddo
        nvrtcs=nv

        RETURN
        END

C----------------------------------------------------------
C get u,v,uncer at lead

        SUBROUTINE vrtx(zlead)

        include 'fit.inc'
        include 'wires.inc'
        include 'vrtx.inc'

        include 'track_dat.inc'
        include 'dst.inc'


        thu=45.43*3.141592654/180.
        cu=cos(thu)
        su=sin(thu)

        do i=1,ntrack
        u(i)=a_u(i)+b_u(i)*zlead
        v(i)=a_v(i)+b_v(i)*zlead
        siguu(i)=covar(i,1,1)+2*zlead*covar(i,1,2)+zlead**2*covar(i,2,2)
        sigvv(i)=covar(i,3,3)+2*zlead*covar(i,3,4)+zlead**2*covar(i,4,4)
        siguv(i)=covar(i,1,3)+zlead*(covar(i,1,4)+covar(i,2,3))+
     +	  zlead**2*covar(i,2,4)
        enddo

        RETURN
        END


C----------------------------------------------------------
C  get chisq of match between pair of tracks

        SUBROUTINE match(i,j,dchisq)

        IMPLICIT NONE

        include 'fit.inc'
        include 'vrtx.inc'

        integer i,j
        real C1(2,2),C2(2,2),C1INV(2,2),C2INV(2,2),C(2,2),CINV(2,2)
        real dchisq,umean,vmean,denom

        integer nch1,nch2
        common/newnch/nch1,nch2

        real dist
        real dumin,dvmin,dmin,du,dv,uold,vold
        common/taildist/dumin,dvmin,dmin,du,dv,uold,vold

        real upoint,vpoint,upoint2,vpoint2
        real thtail,sigth,thgam,zlead,sigu0

        INTEGER HLIM,HBSIZE
        PARAMETER(HBSIZE=500000)
        COMMON/PAWC/HLIM(HBSIZE)

        C1(1,1)=siguu(i)
        C1(2,2)=sigvv(i)
        C1(1,2)=siguv(i)
        C1(2,1)=C1(1,2)
        C2(1,1)=siguu(j)
        C2(2,2)=sigvv(j)
        C2(1,2)=siguv(j)
        C2(2,1)=C2(1,2)
C invert
        denom=C1(1,1)*C1(2,2)-C1(1,2)*C1(2,1)
        C1INV(1,1)=C1(2,2)/denom
        C1INV(2,2)=C1(1,1)/denom
        C1INV(1,2)=-C1(1,2)/denom
        C1INV(2,1)=-C1(2,1)/denom
        denom=C2(1,1)*C2(2,2)-C2(1,2)*C2(2,1)
        C2INV(1,1)=C2(2,2)/denom
        C2INV(2,2)=C2(1,1)/denom
        C2INV(1,2)=-C2(1,2)/denom
        C2INV(2,1)=-C2(2,1)/denom
        C(1,1)=C1INV(1,1)+C2INV(1,1)
        C(2,2)=C1INV(2,2)+C2INV(2,2)
        C(1,2)=C1INV(1,2)+C2INV(1,2)
        C(2,1)=C(1,2)
        denom=C(1,1)*C(2,2)-C(1,2)*C(2,1)
        CINV(1,1)=C(2,2)/denom
        CINV(2,2)=C(1,1)/denom
        CINV(1,2)=-C(1,2)/denom
        CINV(2,1)=-C(2,1)/denom
        umean=CINV(1,1)*(C1INV(1,1)*u(i)+C2INV(1,1)*u(j))+
     +        	CINV(1,2)*(C1INV(2,1)*u(i)+C2INV(2,1)*u(j))+
     +        	CINV(1,1)*(C1INV(1,2)*v(i)+C2INV(1,2)*v(j))+
     +        	CINV(1,2)*(C1INV(2,2)*v(i)+C2INV(2,2)*v(j))
        vmean=CINV(2,1)*(C1INV(1,1)*u(i)+C2INV(1,1)*u(j))+
     +        	CINV(2,2)*(C1INV(2,1)*u(i)+C2INV(2,1)*u(j))+
     +        	CINV(2,1)*(C1INV(1,2)*v(i)+C2INV(1,2)*v(j))+
     +        	CINV(2,2)*(C1INV(2,2)*v(i)+C2INV(2,2)*v(j))
        dchisq=(u(i)-umean)*C1INV(1,1)*(u(i)-umean)
     +	      +2*(u(i)-umean)*C1INV(1,2)*(v(i)-vmean)
     +	      +(v(i)-vmean)*C1INV(2,2)*(v(i)-vmean)
     +	      +(u(j)-umean)*C2INV(1,1)*(u(j)-umean)
     +	      +2*(u(j)-umean)*C2INV(1,2)*(v(j)-vmean)
     +	      +(v(j)-vmean)*C2INV(2,2)*(v(j)-vmean)

        RETURN
        END



C ----------------------------------------------------------
C  vrtx.inc
        real u(200),v(200),siguu(200),sigvv(200),siguv(200)
        common/lead/u,v,siguu,sigvv,siguv

C  (determined by subroutine vrtx in vertexer.f)
C  position and uncertainty of track at z=zlead

C ----------------------------------------------------------
C       VRTCS.INC
        integer nvrtcs
        integer nverts(100,102)
        real vrtxpos(100,2)
        integer ischarged(100),isgamma(100)

        common/fndvrts/nvrtcs,nverts,vrtxpos,ischarged,isgamma

C  (determined by vertexer.f)
C  nvrtcs=number of vertices at zlead
C  nverts(i,1)=number of heads in vertex i
C  nverts(i,2)=number of tails in vertex i
C  nverts(i,2+j)=dst-track-number of track segment j in vertex i
C  nverts(i,52+j)=fitter-track-number of track segment j in vertex i

C  (determined by chgam.f)
C  ischarged(i)/isgamma(i) =1 if vertex is a charged track/ gamma
C                          =0 otherwise
\end{verbatim}}
\vfill\newpage
\section{Code for classifying vertices as charged tracks or photon conversions:
chgam.f}
{\footnotesize
\begin{verbatim}
        SUBROUTINE CHGAM
 
        IMPLICIT NONE
 
        include 'EVENT.INC'
        include 'fit.inc'
        include 'wires.inc'
        include 'VRTCS.INC'

        integer nch1,nch2
        common/newnch/nch1,nch2
        INTEGER NN
        REAL sigsq,wiresp,pi,uang,sum,cu,su
        REAL U0(24),V0(24),ang(24),Z(24)
        common/cham_param/NN,sigsq,wiresp,pi,uang,sum,cu,su,U0,V0,ang,Z

        INTEGER I, J, IOS, k, ihit, icham, nb(24), keep,ii,jj,kk,nu,nv
        INTEGER nkeep,ikeep(20),imin
        REAL dist,dmin,r
        real mean,sigmean,upoint,sigupoint,wire
        integer itr,itr0,keepcham,keepwire,keeppoint,npoint

        do j=1,nvrtcs
          ischarged(j)=0
          isgamma(j)=0

C  Cyrus's acceptance
        if ((vrtxpos(j,1).gt.4.25).and.(vrtxpos(j,1).lt.10.25).and.
     +          (vrtxpos(j,2).gt.-3.).and.(vrtxpos(j,2).lt.3.)) then
        r=sqrt((vrtxpos(j,1)-7.25)**2+vrtxpos(j,2)**2)
        if (r.lt.4.) then

C  (1,n>0) charged track
          if ((nverts(j,1).ge.1).and.(nverts(j,2).gt.0)) ischarged(j)=1

C  (1,0)'s have to pass tighter cuts
          if ((nverts(j,1).ge.1).and.(nverts(j,2).eq.0)) then
            keep=0

C  require hit all 8 chambers
            do i=1,nverts(j,1)
            itr0=nverts(j,52+i)
              keepcham=0
              do k=1,8
                nb(k)=0
              enddo
              do k=1,nnwire(nverts(j,52+i))
        	nb(int((wires(nverts(j,52+i),k)-1)/128.)+1)=1
              enddo
              icham=0
              do k=1,8
                if (nb(k).eq.0) then
                wire=nint( (-(a_u(itr0)+b_u(itr0)*Z(K)-U0(K))*sin(ang(K))
     +          +(a_v(itr0)+b_v(itr0)*Z(K)-V0(K))*cos(ang(K)))/wiresp )
                dmin=128
                do jj=1,nhits
                  if ((mod(chamb_num(jj),100).eq.k).and.
     +	              (abs(wire_num(jj)-wire).lt.dmin)) 
     +	            dmin=abs(wire_num(jj)-wire)
                enddo
                if (dmin.le.1) then
                  nb(k)=1
                  nnwire(itr0)=nnwire(itr0)+1
                  wires(itr0,nnwire(itr0))=wire+(k-1)*128
                endif
                endif
 	        icham=icham+nb(k)
              enddo
              if (icham.eq.8)keepcham=1

C  don't share wires in all 3 u or all 3 v chambers with other head
              keepwire=1
              do jj=1,nvrtcs
              if (jj.ne.j) then
              do ii=1,nverts(jj,1)+nverts(jj,2)
              if (nverts(jj,52+ii).le.nch2) then
              itr=nverts(jj,52+ii)
              nv=0
              nu=0
              do kk=1,nnwire(itr)
              do k=1,nnwire(itr0)
                if (wires(itr,kk).eq.wires(itr0,k)) then
   	         icham=int((wires(itr0,k)-1)/128.)+1
 	         if ((icham.eq.1).or.(icham.eq.3).or.(icham.eq.6)) 
     +	           nv=nv+1
 	         if ((icham.eq.2).or.(icham.eq.5).or.(icham.eq.8)) 
     + 	           nu=nu+1
                endif
              enddo
              enddo
              if ((nv.eq.3).or.(nu.eq.3)) keepwire=0
              endif
              enddo	!ii
              endif
              enddo	!jj

C  point to mean z of other charged tracks
              if (i.eq.1) then
                mean=0
                sigmean=0
                npoint=0
                do jj=1,nvrtcs
                if ((nverts(jj,1).ge.1).and.(nverts(jj,2).ge.1)) then
                do ii=1,nverts(jj,1)+nverts(jj,2)
                if (nverts(jj,52+ii).le.nch2) then
                  npoint=npoint+1
                  itr=nverts(jj,52+ii)
                  upoint=-a_u(itr)/b_u(itr)
                  sigupoint=( covar(itr,1,1)+(a_u(itr)/b_u(itr))**2*
     +		  covar(itr,2,2)-2*(a_u(itr)/b_u(itr))*covar(itr,1,2) )
     +		  /b_u(itr)**2
                mean=mean+upoint/sigupoint
                sigmean=sigmean+1./sigupoint
                endif
                enddo	!ii
              endif
              enddo	!jj
              if (npoint.gt.0) then
                sigmean=1./sigmean
                mean=mean*sigmean
              endif
              endif	!i=1
              keeppoint=1
              if (npoint.gt.0) then
                dist=((-a_u(itr0)/b_u(itr0))-mean)
     + 	          /sqrt(sigmean+( covar(itr0,1,1)+(a_u(itr0)/
     +         	  b_u(itr0))**2*covar(itr0,2,2)-2*(a_u(itr0)/
     +            b_u(itr0))*covar(itr0,1,2) )/b_u(itr0)**2 )
                if (abs(dist).gt.2.) keeppoint=0
            endif	!npoint>0
            if ((keepcham.eq.1).and.(keepwire.eq.1).and.
     +	      (keeppoint.eq.1)) keep=1
          enddo	!i

          if (keep.eq.1) ischarged(j)=1

          endif	! (n,0)

C  (0,n) gamma
          if ((nverts(j,1).eq.0).and.(nverts(j,2).gt.0)) isgamma(j)=1

        endif	!vrtxpos
        endif	!r<4
        enddo	!j


      RETURN
      END
\end{verbatim}}



\vfill\newpage
\begin{thebibliography}{99}
\bibitem{QCD}R. D. Field, {\em Applications of Perturbative QCD}, Redwood City,
USA: Addison-Wesley (1989).
\bibitem{goldst}J. Goldstone, Nuovo Cimiento {\bf 19}, 154 (1961).
\bibitem{sigma} M. Gell-Mann and M. Levy, Nuovo Cimiento, {\bf 16}, 705 (1960);
B. W. Lee, ``Chiral Dynamics'', ed Gordon and Breach, 1972.
\bibitem{kapusta}J. I. Kapusta and A. M. Srivastava, Phys. Rev. D {\bf 50},
5379 (1994). 

\bibitem{dcc1}A. A. Anselm, Phy. Lett. B {\bf 217}, 169 (1989).
\bibitem{dcc2}A. A. Anselm and M. G. Ryskin, Phy. Lett. B {\bf 266}, 482 (1991).
\bibitem{dcc3}J. D. Bjorken, Int. J. Mod. Phys. A {\bf 7}, 4189 (1992).
\bibitem{dcc4}J. D. Bjorken, Acta Phys. Pol. B {\bf 23}, 561 (1992).
\bibitem{dcc5}J.-P. Blaizot and A. Krzywicki, Phys. Rev. D {\bf 46}, 246 (1992).
\bibitem{whitepaper}K. L. Kowalski and C. C. Taylor, Report No. CWRUTH-92-6, 
hep-ph/9211282 (unpublished).
\bibitem{dcc6} K. Rajagopal and F. Wilczek, Nucl. Phys. {\bf B399}, 395 (1993).
\bibitem{bj}J. D. Bjorken, K. L. Kowalski, and C. C. Taylor, in
{\em Proceedings of Les Recontres de la Vall\'{e}e D' Aoste}, La Thuile, Italy,
1993, edited by M. Greco (Editions Frontieres, Gif-sur-Yvette, France, 1993),
p. 507.
\bibitem{dcc7} M. Martinis, V. Mikata-Martinis, A. \u{S}varc, and 
J. \u{C}rnugelj, Phys. Rev. D {\bf 51}, 2482 (1995); Fizika {\bf B3}, 197 
(1994); M. Martinis, V. Mikata-Martinis, and J. \u{C}rnugelj, Phys. Rev. C 
{\bf 52} 1073 (1995).
\bibitem{dcc8} R. D. Amado {\em et al.}, Phys. Rev. Lett. {\bf 72}, 970 (1994).
\bibitem{dcc9} B. M\"{u}ller, Rep. Prog. Phys. {\bf 58}, 611 (1995).
\bibitem{raj} A good review can be found in K. Rajagopal, in {\em Quark-Gluon 
Plasma 2}, edited by R. Hwa, World Scientific, Singapore, (1995), p. 454.
\bibitem{sqrtf1} I. V. Andreev, JETP Lett. {\bf 33},  67 (1981).
\bibitem{sqrtf2} V. Karmanov and A. Kudrjavtsev, Report No. ITEP-88, 1983, 
(unpublished).
\bibitem{sqrtf3}D. Horn and R. Silver, Ann. Phys. (N.Y.) {\bf 66}, 509 (1971).

\bibitem{centauro}L. T. Beradzei {\em et al.} (Chacaltaya-Pamir Collaboration),
Nucl. Phys. {\bf B370}, 365 (1992); C. M. G. Lattes, Y. Fujimoto and S. 
Hasegawa, Phys. Rep. {\bf 65}, 151 (1980).
Chacaltaya-Pamir Collaboration, ICR-Report-258-91-27 (unpublished) and 
references cited therein.
\bibitem{hasegawa} S. Hasegawa, (Chacaltaya-Pamir Collaboration), 
ICR-Report-151-87-5 (unpublished).
\bibitem{Goulianos}See, for example,
%
K. Goulianos, ``A Comment on Centauro Production'', Comm. on Nucl. and
Par. Phys. {\bf 17}, 195 (1987);
%
K. Goulianos, ``Exotic Diffraction Dissociation in Cosmic Rays and Hadron 
Colliders'', {\em Proceedings of VIIth Int. Symp. on Very High
Energy Cosmic Ray Interactions}, edited by L. Jones, Ann Arbor, June 1992.
%
\bibitem{diffr}K. Goulianos, Phys. Rep. {\bf 101}, 169 (1983). 
\bibitem{jacee}J. J. Lord and J. Iwai, {\em Intl. Conference on High Energy 
Physics}, Dallas, TX, August 1992; J. Iwai, Report No. UWSEA 92-06; Y. Takahashi
(JACEE Collaboration), {\em Proceedings of VIIth Int. Symp. on Very High
Energy Cosmic Ray Interactions}, edited by L. Jones, Ann Arbor, June 1992;
H. Wilczynski (JACEE Collaboration), talk given at the IXth International 
Symposium on Very High Energy Cosmic Ray Interactions, Karlsruhe, Germany, 
August 1996.
\bibitem{CDFcent}P. L. Melese (CDF Collaboration), FERMILAB-CONF-96-205-E,
July 1966. 
\bibitem{UA5cent}K. Alpg{\aa}rd {\em et al.}, Phys. Lett. B {\bf 115}, 71 
(1982); G. J. Alner {\em et al.}, Phys. Lett. B {\bf 180}, 415 (1986).
\bibitem{UA5rep}G. J. Alner {\em et al.}, Phys. Rep. {\bf 154}, 247 (1987).
\bibitem{UA1cent}G. Arnison {\em et al.}, Phys. Lett. B {\bf 122}, 189 (1983).
\bibitem{CDFmult}F. Abe {\em et al.}, Phys. Rev. D {\bf 41},
2330 (1990).
\bibitem{UA1mult}G. Arnison {\em et al.}, Phys. Lett. B {\bf 123}, 108 (1983).
\bibitem{p238}C. J. Liapis, PhD thesis, Yale University, New Haven, CT, 1995;
R. Harr, {\em et al.}, hep-ex/9703002, submitted to Phys. Lett. B.
\bibitem{UA5mult}R. E. Ansorge {\em et al.}, Z. Phys. C {\bf 43}, 75 (1989).
\bibitem{UA5sim}G. J. Alner {\em et al.}, Nucl. Phys. {\bf B291}, 445 (1987).
\bibitem{Bev}P. R. Bevington, B. D. Anderson, R. J. Barrett, and F. H. Cverna, 
Nucl. Inst. Meth. {\bf 129}, 373 (1975).
\bibitem{HERA}Based on private communication from HERA-B to Bjorken.
\bibitem{PDG}R. M. Barnett {\em et al.}, Phys. Rev. D {\bf 54}, 1 (1996).
\bibitem{CDF} F. Abe {\em et. al}, Phys. Rev. D, {\bf 50}, 5550 (1994);
F. Abe {\em et. al}, Phys. Rev. D, {\bf 50}, 5518 (1994);
F. Abe {\em et. al}, Phys. Rev. D, {\bf 50}, 5535 (1994).
\bibitem{Michcal}The calorimeter cells made by the University of Michigan are
based on a design described in H. Fessler {\em et al.}, 
Nucl. Instr. Meth. A {\bf 240}, 284 (1985).
\bibitem{wsucal}
K. Barish {\em et al.}, to be submitted to NIM; 
C. Pruneau (E864 Collaboration),
{\em Proceedings of the VI International Conference on Calorimetry in High 
Energy Physics}, Fracati, Italy, July 1996 (in print);
C. Pruneau (E864 Collaboration),
{\em Proceedings of the 12th Winter Workshop on Nuclear Dynamics}, 
Snowbird, Utah, Feb 3-10, 1996, published in {\em Advance in Nuclear Dynamics 
2}, edited by W.Bauer and G.D.Westfall, Plenum, p. 199. 
\bibitem{FAD}J. D. Bjorken, ``A Full Acceptance Detector for SSC Physics at Low
and Intermediate Mass Scales'', SSC EoI-19, SLAC-PUB-5545 (May 1991).
\bibitem{MAX}Fermilab Proposal P-864 (J. Bjorken and M. Longo, spokespersons),
``MAX'', unpublished.
\bibitem{pyth} T. Sjostrand and M. Bengtsson, Computer Physics Commun. 
{\bf 43}, 367 (1987); H.-U. Bengtsson and T. Sjostrand, Computer Physics 
Commun. {\bf 46}, 43 (1987).
\bibitem{pythman} T. Sjostrand, CERN-TH.7112/93 (1993).
\bibitem{parton}Parton-model phenomenology is described in
M. E. Peskin and D. V. Schroeder, {\em An Introduction to 
Quantum Field Theory}, Reading, USA: Addison-Wesley (1995).
\bibitem{lund}B. Andersson, G. Gustafson, G. Ingelman and T. Sjostrand,
Phys. Rep. {\bf 97}, 31 (1983).
\bibitem{geant}GEANT -- Detector Description and Simulation Tool, CERN, 
PM0062 (1993).
\bibitem{Pumplin}J. Pumplin, Phys. Rev. D {\bf 50}, 6811 (1994).
\bibitem{alice}The Forward Muon Spectrometer of ALICE -- Addendum to the 
Technical Proposal for A Large Ion Collider at the CERN LHC,  CERN/LHCC 96-32,
LHCC/P3-Addendum 1, 1996.
\bibitem{trackers}
The combinatorial tracker was written by K. DelSignore.
Another tracker used a hougher algorithm and is documented in the senior 
thesis of E. Kangas, CWRU, 1995.
A third tracker written by T. Jenkins was based on a track-following algorithm.
\bibitem{Bevstat}P. R. Bevington, {\em Data Reduction and Error Analysis
for the Physical Sciences}, New York, McGraw-Hill (1969).
\bibitem{gf1}P. Carruthers and C. C. Shih, Int. J. Mod. Phys. A {\bf 2}, 1447 
(1987).
\bibitem{gf2}C. Geich-Gimbel, Int. J. Mod. Phys. A {\bf 4}, 1527 (1989).
\bibitem{gf3}I. M. Dremin, Mod. Phys. Lett. A {\bf 8}, 2747 (1993); Pis'ma Zh. 
Eksp. Teor. Fiz. {\bf 59}, 561 (1994) [JETP Lett. {\bf 59}, 585 (1994)].
\bibitem{gf4}I. M. Dremin and R. C. Hwa, Phys. Rev. D {\bf 49}, 5805 (1994).
\bibitem{gf5}For a review, see E. A. DeWolf, I. M. Dremin and W. Kittel, 
Phys. Rep. {\bf 270}, 1 (1996).
\bibitem{gf6}G. H. Thomas and B. R. Webber, Phys. Rev. D {\bf 9}, 3113 (1974).
\bibitem{gf7}L. Di\'{o}si, Nucl. Instrum. Methods {\bf 138}, 241 (1976).
\bibitem{gf8}L. Di\'{o}si, Nucl. Instrum. Methods {\bf 140}, 533 (1977).
\bibitem{gf9}L. Di\'{o}si and B. Luk\'{a}cs, Phys. Lett. B {\bf 206}, 707 
(1988).
\bibitem{robust} T. C. Brooks {\em et al.}, accepted for publication in 
Phys. Rev. D, tenatively to appear in May 1997. 
\bibitem{GR}I. S. Gradshteyn and I. M. Ryzhik, {\em Table of Integrals, Series, 
and Products}, Academic Press, Orlando, 1980, p. 20.
\bibitem{spyr} M. Spyropoulou-Stassinaki {\em et al.}, Nucl. Phys. {\bf A525},
487c, (1991).

\end{thebibliography}
\end{document}